UNIVERSITE DE LORRAINE - INSTITUT NATIONAL POLYTECHNIQUE DE LORRAINE
ECOLE DOCTORALE EMMA

# Mémoire d'Habilitation à Diriger des Recherches

présenté par

# Sébastien ALLAIN

# Comportement mécanique des aciers : des mécanismes fondamentaux à la déformation macroscopique

Soutenu publiquement le 6 Décembre 2012 à l'Institut Jean Lamour, devant le jury composé de :

| | | |
|---|---|---|
| Mr Marc FIVEL | Directeur de Recherche CNRS, SIMAP Grenoble | Rapporteur |
| Mr Xavier FEAUGAS | Professeur, LEMMA La Rochelle | Rapporteur |
| Mr Armand COUJOU | Professeur, CEMES Toulouse | Rapporteur |
| Mr Javier GIL SEVILLANO | Professeur, CEIT-TECNUM San Sebastian | Examinateur |
| Mr Michel VERGNAT | Professeur, IJL Nancy | Examinateur |
| Mr Mikhaïl LEBEDKIN | Directeur de Recherche CNRS, LEM3 Metz | Examinateur |
| Mr Olivier BOUAZIZ | Docteur HDR, Arcelormittal Maizières les Metz | Examinateur |
| Mr Alain JACQUES | Directeur de Recherche CNRS, IJL Nancy | Parrain Scientifique |
| Mr Thierry IUNG | Docteur, Arcelormittal Maizières les Metz | Invité |

Disciplines: Milieux Denses et Matériaux, Chimie des Matériaux, Mécanique

Arcelormittal Maizières Research SA, Voie Romaine, BP 30320 F-57283 Maizières les Metz

*A ma Femme, Nathalie,
mes Filles, Fanny, Enora
et Chloé, mon petit ange*

# Remerciements

Je suis tout d'abord très reconnaissant envers les rapporteurs de ce mémoire, Armand Coujou, Xavier Feaugas et Marc Fivel d'avoir accepté très spontanément cette charge et m'avoir apporté par leurs grandes expériences de nouveaux questionnements sur mes travaux. J'aimerais aussi remercier chaleureusement Javier Gil Sevillano pour l'intérêt qu'il a pu manifester pour mes travaux, son accueil à de nombreuses reprises au CEIT et notre fructeuse collaboration. Merci aussi à Mikhail Lebedkin pour sa participitation à mon jury, sa confiance au cours de ces dernières années et sa patience de pédagogue avec moi sur les problématiques de viellissement dynamique. Je voudrais également témoigner de toute ma gratitude à Michel Vergnat pour avoir accepté de présider le jury et dirigé les discusssions lors de la soutenance. Merci aussi à Alain Jacques pour sa confiance dans mon potentiel de chercheur, ses conseils et pour son engagement dans ma démarche d'habiliation.

Je souhaite exprimer toute ma gratitude et mon amitié à Jean-Philippe Chateau-Cornu et Olivier Bouaziz, que je considère mes mentors en sciences des matériaux, qui m'ont dirigé en DEA et en thèse, puis accompagné et enfin encouragé dans ma démarche de recherche depuis maintenant 12 ans. Merci aussi à Colin Scott et Mohamed Gouné leurs amitiés, leurs soutiens dans les moments les plus durs et nos grandes discussions en métallurgie au cours de ces dernières années.

Je voudrais aussi remercier Thierry Iung pour m'avoir offert un cadre dès mon arrivée chez Arcelor pour poursuivre mes travaux scientifiques dans un contexte industriel et soutenu dans ma démarche d'habilitation d'un point de vue académique. Sa relecture attentive et la pertinence des questions soulevées m'ont permis d'améliorer significativement le mémoire. Je tenais aussi à remercier Jean-Hubert Schmitt et David Embury, qui m'ont apporté une certaine motivation exterieure et auprès de qui j'ai toujours pu toujours trouver une écoute attentive et de nouvelles idées ces dernières années.

Comme l'a souligné Olivier lors de la soutenance, nous avons eu la chance de vivre avec le développement des aciers TWIP à Arcelor une véritable « aventure », en toute insouciance, sans anticiper et comprendre les implications de nos travaux au début des années 2000. Ces travaux m'ont apporté depuis une certaine reconnaissance de la communauté et m'ont permis de rencontrer et collaborer avec des chercheurs formidables, que je souhaiterais remercier ici. Je pense à Nicolas Guelton, Philippe Cugy et Michel Faral et à l'équipe FeMn. J'ai aussi une pensée particulière pour Maurita Roscini et son combat.

# Table des Matières

# Sigles et localisation des entités de recherche citées

**AM** : Arcelormittal, Arcelormittal Maizières Research SA, Maizières les Metz, FRANCE
   Anciennement IRSID, Institut de Recherche SIDérurgique

**NSC** : Nippon Steel Corporation, Steel Research Laboratories, Futtsu, JAPON

**IJL** : Institut Jean Lamour, Université de Lorraine, Nancy, FRANCE

**LPM** : Laboratoire de Physique des Matériaux, Ecole des Mines de Nancy, Nancy, FRANCE
   Maintenant intégré dans l'IJL

**LSGS** : Laboratoire de Science et Génie des Surfaces, Ecole des Mines de Nancy, Nancy, FRANCE
   Maintenant intégré dans l'IJL

**LEM3** : Laboratoire d'Etude des Microstructures et Mécanique des Matériaux, Université Paul Verlaine, Metz, FRANCE

**LETAM** : Laboratoire d'Etude des Textures et Application aux Matériaux, Université Paul Verlaine, Metz, FRANCE
   Maintenant intégré dans le LEM3

**LPMM** : Laboratoire de Physique et Mécanique des Matériaux, Université Paul Verlaine, Metz, FRANCE
   Maintenant intégré dans le LEM3

**CEIT-TECNUN** : Centro de Estudios e Investigaciones Técnicas de Guipúzcoa, Technological Campus of the University of Navarra, San Sebastian, ESPAGNE

**GPM** : Groupe de Physique des Matériaux, Université de Rouen, Saint Etienne du Rouvray, FRANCE

**HK University** : The University of Hong Kong, Department of Mechanical Engineering, Hong Kong, CHINE

**ICMCB** : Institut de la Matière Condensée de Bordeaux, Université de Bordeaux, Pessac, FRANCE

**KUL** : Katholieke Universiteit Leuven, Department of Metallurgy and Materials Engineering, Heverlee, BELGIQUE

**LPMTM** : Laboratoire des Propriétés Mécaniques et Thermodynamiques des Matériaux, Université Paris XIII, Villetaneuse, FRANCE

**McMaster University** : McMaster University, Department of Materials Science and Engineering, Hamilton, CANADA

**Monash University** : Monash University, Department of Materials Engineering, Clayton, AUSTRALIE

**MPI** : Max Planck Institute, Düsseldorf, ALLEMAGNE

**SIMAP** : Laboratoire de Science et Ingénierie des Matériaux et Procédés, Université Joseph Fourier, St Martin d'Hères, FRANCE

**UBC** : The University of British Columbia, Department of Materials Engineering, Vancouver, CANADA

**UCL** : Université Catholique de Louvain, Département des Sciences des Matériaux et des Procédés, Louvain la Neuve, BELGIQUE

# Résumé

Mon activité de recherche scientifique concerne principalement la compréhension et la modélisation du comportement mécaniques des aciers ; des mécanismes fondamentaux à la déformation macroscopique. Ce mémoire est consacré en particulier à l'effet TWIP (TWinning Induced Plasticity) des aciers austénitiques FeMnC à haute teneur en manganèse et l'effet Dual-Phase des aciers Ferrite-Martensite.

L'effet TWIP est un mécanisme d'écrouissage spécifique des aciers austénitiques FeMnC lié à un processus de maclage mécanique, mécanisme de déformation compétitif au glissement des dislocations. L'accumulation de ces macles, défauts plans d'épaisseur nanométrique, crée au cours de la déformation une microstructure enchevêtrée et difficilement franchissable par les dislocations mobiles à l'intérieur des grains austénitiques. Au cours de nos travaux, ces microstructures ont été expliquées et quantifiées à différentes échelles. Nous avons ainsi pu modéliser la double contribution du maclage à l'écrouissage grâce à une augmentation de densité de dislocations statistiquement stockées et à une contribution de nature cinématique, associée à l'incompatibilité de déformation entre macles et matrice. L'influence de ce mécanisme a en conséquence été mieux comprise lors de trajets de mise en forme complexes. Afin de pouvoir optimiser le comportement mécanique de ces aciers TWIP, notre second axe de recherche a porté sur l'effet de leurs compositions chimiques sur ces mécanismes d'écrouissage, en particulier au travers de la relation entre maclage mécanique et énergie de défaut d'empilement (EDE). Ces travaux ont débouchés sur l'identification d'un « paradoxe carbone » que nous sommes en passe de résoudre.

Mes travaux ultérieurs de modélisation du comportement des aciers Dual-Phase (DP) Ferrite-Martensite se sont aussi attachés à décrire systématiquement les effets de fraction et de tailles des microstructures. Ils ont eu différentes finalités :

- l'extension en plasticité polycristalline d'un modèle monophasé analytique pour des applications en rhéologie appliquée (prévision des surfaces de charges sous sollicitations complexes).
- le développement d'un modèle biphasé générique pour des utilisations en « alloy-design » métallurgique. Le modèle intègre en outre nos travaux les plus récents sur les aciers martensitiques (Approche Composite Continu) et a été ajusté sur une large base de données issues de la littérature.
- l'approfondissement de nos connaissances sur les effets de morphologie et de topologie de la microstructure DP sur le comportement et la rupture de ces aciers composites. Il passe par le développement d'une chaîne de simulation à champs locaux par Eléments Finis (EF), allant de la numérisation aux calculs sur Volume Elémentaire Représentatif (VER) de la microstructure, sensibles aux gradients de déformation, et intégrant les mécanismes d'endommagements pertinents. L'approche est encore incomplète mais permet de traiter des questions au premier ordre comme l'aspect néfaste d'une structure en bandes sur l'endommagement.

# 1. Introduction – démarche scientifique

> *« Les hommes construisent trop de murs et pas assez de ponts »*
> *Isaac Newton*

Mon activité de recherche scientifique concerne principalement la compréhension et la modélisation du comportement mécaniques des aciers ; des mécanismes fondamentaux à la déformation macroscopique. Cette connaissance du lien entre microstructure et propriété est stratégique pour un sidérurgiste. En effet, celui-ci produit dans ses usines des microstructures et des revêtements particuliers, mais il vend et garantit des propriétés d'emploi et d'usage durables dans le temps.

Ma formation initiale en métallurgie physique, de « plasticien », m'a donc mis à l'interface entre « métallurgistes » et « mécaniciens » et le sens de mon action a finalement été de construire des ponts entre ces différents métiers. Mes travaux, à la fois expérimentaux et de modélisation, ont toujours visé à mettre en évidence quels sont le ou les paramètres microstructuraux pertinents pouvant expliquer les comportements mécaniques macroscopiques, et ce, avec le souci constant de valider la démarche à plusieurs échelles (structuration de la plasticité) ou sous plusieurs angles (contraintes internes, effets de la vitesse de déformation ou de la température).

Mes principales productions « pratiques » ont ainsi été :
- soit des outils pratiques à destination des « métallurgistes », et utilisables par des non initiés en mécanique, pour des travaux d'optimisation (par exemple, un modèle d'énergie de défaut d'empilement en fonction de la composition chimique pour les aciers austénitiques FeMnC ou un modèle générique de comportement des aciers DP à des fins « d'alloy design »),
- soit des modèles de comportement à base microstructurale pour les « mécaniciens » et les rhéologues (par exemple, des modèles de comportement à composantes isotrope et cinématique pour les aciers ferrito-perlitiques ou un modèle de sensibilité à la vitesse pour les aciers ferritiques).

Ces travaux s'appuient sur une bonne connaissance des mécanismes fondamentaux de la plasticité. En fonction des travaux antérieurs de la littérature, j'oriente et dirige des activités de recherche de nature plus expérimentale pour élucider la nature des mécanismes d'écrouissage (par exemple, mes travaux sur la microstructure de maclage à différentes échelles de la microscopie électronique à transmission (MET) à la microscopie optique (MO)).

Mes centres d'intérêt actuels concernent la compréhension fine de phénomènes hautement couplés comme le paradoxe carbone dans les aciers TWIP ou dans lesquels les hétérogénéités spatio-temporelles de contraintes et de déformation jouent un rôle particulier (les mécanismes de localisation et d'endommagement des aciers DP, les mécanismes d'écrouissage des aciers martensitiques ou les mécanismes de vieillissement dynamique).

Cette position particulière à l'interface de nombreuses spécialités explique la diversité de mes activités de recherche. Le premier support de ces études a été pour moi les aciers austénitiques à hautes teneurs en manganèse présentant un effet TWIP (pour Twinning Induced Plasticity) au cours de ma thèse. Ces aciers «redécouverts» depuis les années 1990 par les grands sidérurgistes présentent de formidables potentiels pour la construction automobile en particulier grâce à des mécanismes de déformation et d'écrouissage multiples et en interactions complexes (glissement des dislocations, maclage mécanique, transformation martensitique induite, vieillissement dynamique). Cette grande richesse en fait un système d'étude intéressant à différentes échelles, des interactions entre dislocations / solutés aux problématiques de mise en forme et de rupture en passant par la nature de la microstructure de maclage mécanique.

Depuis mon arrivée en 2004 à l'IRSID, le plus grand centre de Recherche et Développement d'Usinor (maintenant AM), j'ai aussi eu l'opportunité de diversifier mon champ d'applications et de connaissances en travaillant sur le comportement mécanique des nombreuses phases ferritiques « basses températures » qu'offre le système FeMnC (comme les aciers bainitiques, martensitiques, perlitiques ou tout simplement ferritiques).

A titre d'illustration, la Figure 1 montre le positionnement des sujets de mes publications dans des revues à comité de lecture par rapport aux différentes familles d'aciers utilisés dans la construction automobile. Elles sont représentées traditionnellement dans un plan résistance mécanique / allongement à rupture.

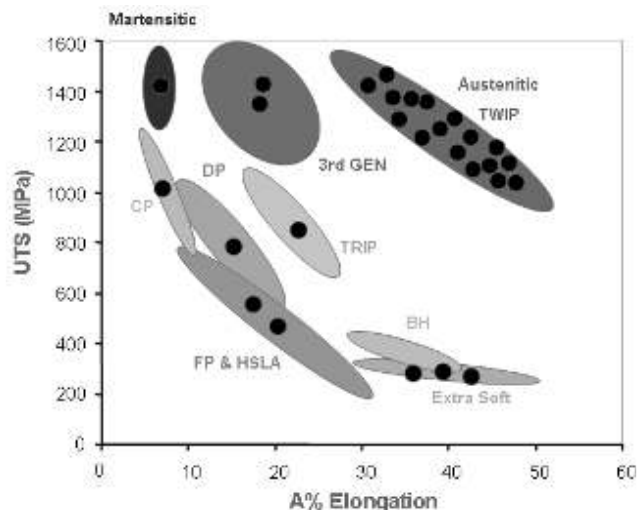

Figure 1 : Positionnement relatif des sujets de mes publications internationales depuis 2002 par rapport aux grandes familles d'aciers utilisés dans le domaine automobile.

Ma position de chercheur-ingénieur de recherche dans le secteur privé m'a amené aussi à diriger des projets de recherche de développement produit court et moyen terme. En tant que chef de projet, ma mission est d'affecter les ressources, assurer le suivi technique et le management de la qualité des projets. Cette position particulière m'a aussi permis de conduire des discussions avec des experts de différents domaines de la sidérurgie (de l'acier liquide à la mise en forme chez les clients constructeurs automobiles). A ce titre, je suis actuellement très impliqué dans l'industrialisation de produits relevant des aciers THR (Très Haute Résistance) dits « de troisième génération » (3rd Gen sur Figure 1) et dans la coordination technique et scientifique des activités associées à ces métallurgies spécifiques (Quenching and Partionning steels, Carbide-Free bainitic steels, TRIP steels with annealed martensitic matrix par exemple). Cette activité s'est concrétisée par la rédaction de deux brevets produits et procédés sur ces aciers. Cependant, mes travaux « publics » relevant de cette thématique sont encore peu nombreux et concernent principalement des études métallurgiques sur les aciers bainitiques sans carbure. J'ai fait le choix de ne pas détailler ces travaux dans ce mémoire, mais le lecteur pourra se reporter à quelques références bibliographiques ([ALLAIN 2008_3][HELL 2011_1][HELL 2011_2]).

Pour ce mémoire, j'ai choisi une grille de lecture de mes travaux associée non pas à des mécanismes ou des échelles métallurgiques, mais à des microstructures particulières.
Ce mémoire est donc divisé en deux grands chapitres :
- le premier est dédié à la mise en perspectives de mes travaux sur les aciers austénitiques FeMnC TWIP dans un contexte international très actif.
- le second est consacré aux aciers ferritiques Dual-Phase. C'est aussi une occasion pour présenter mes travaux récents et en cours sur le comportement des aciers martensitiques.

Ce choix permet de maintenir une unité dans les concepts micromécaniques utilisés, même si de nombreuses similitudes existent entre ces deux systèmes, comme je le démontrerais. Cet ordre de présentation résulte d'une mise en perspective chronologique de mes travaux et non de la complexité des systèmes étudiés, bien au contraire. Il permettra par contre de montrer comment certains concepts développés pour les aciers TWIP ont ainsi été adaptés aux aciers DP.

Ces travaux qui ont pour la plupart fait l'objet de publications scientifiques sont bien entendu des œuvres collectives, réalisées en partie en association avec le monde académique et universitaire et en partie avec certains chercheurs du secteur privé, collègues d'Arcelormittal ou concurrents, que je tiens à remercier ici encore. Les références aux travaux que j'ai dirigés ou auxquels j'ai été associé seront indiquées en bleu dans la suite.

La rédaction de ce mémoire a aussi été pour moi l'occasion de finaliser et discuter certaines études non publiées à ce jour. Je pense en particulier aux comparaisons entre modèles cinématiques et composites de l'effet TWIP (cf. chapitre 2.3.5 page 52), au modèle biphasé de l'effet DP (cf. chapitre 3.3 page 107) ou d'autres disponibles uniquement dans des travaux de thèse comme l'étude des changements de trajets sur les aciers TWIP et la modélisation des DP par EF. Ces résultats importants d'un point de vue scientifique et révélateurs de ma démarche scientifique ont donc été détaillés dans ce manuscrit. Je suis conscient que ce choix peut malheureusement nuire dans certaines parties à l'esprit de synthèse qui doit animer un tel mémoire.

# 2. Microstructure et comportement des aciers austénitiques FeMnC à effet TWIP

> *« C'est l'inconnu qui m'attire. Quand je vois un écheveau bien enchevêtré je me dis qu'il serait bien de trouver un fil conducteur »*
> *Pierre Gilles de Gennes*

## 2.1. Introduction

### 2.1.1. Contexte technique et industriel

Les aciers austénitiques TWIP (TWinning Induced Plasticity) à haute teneur en manganèse font actuellement l'objet d'un engouement important dans la communauté des matériaux et de la métallurgie. Cet intérêt est non seulement dû à des perspectives d'application structurale importantes dans le domaine de construction automobile mais aussi à la richesse et la complexité de questions scientifiques posées par cette métallurgie.

Concrètement, il se traduit depuis ces 5 dernières années par une augmentation importante des publications scientifiques et de brevets dédiés à ces aciers comme le montre la Figure 2. Une première conférence dédiée spécifiquement à ces aciers a même récemment été organisée en Corée, soutenue par un sidérurgiste d'envergure internationale (1st International Conference on High Manganese Steels - HMnS 2011 - Séoul) et des numéros spéciaux de grandes revues scientifiques ont été consacrés à cette métallurgie ([VIEWPOINT 2012]).

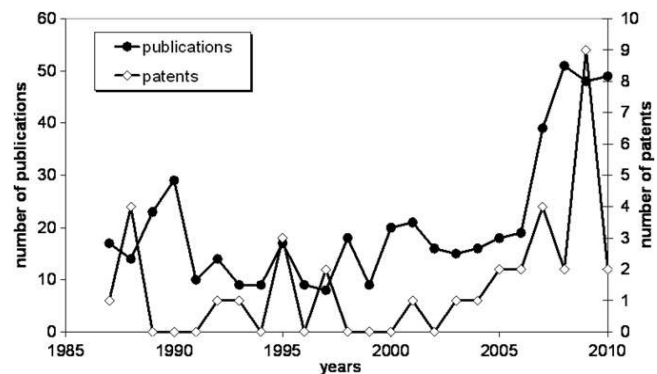

**Figure 2 : Evolution du nombre de publications et de brevets concernant les aciers FeMnC TWIP de 1985 à 2010**
[BOUAZIZ 2011_1]

Les performances de ces aciers « austénitiques » sont uniques et bien supérieures à la fois en termes de résistance et de ductilité aux aciers à matrice « ferritiques » généralement utilisés dans le domaine de la construction automobile. Depuis peu, on parle d'ailleurs de ces aciers comme étant la seconde génération des aciers THR (Très Haute Résistance) pour l'automobile (cf. Figure 3).

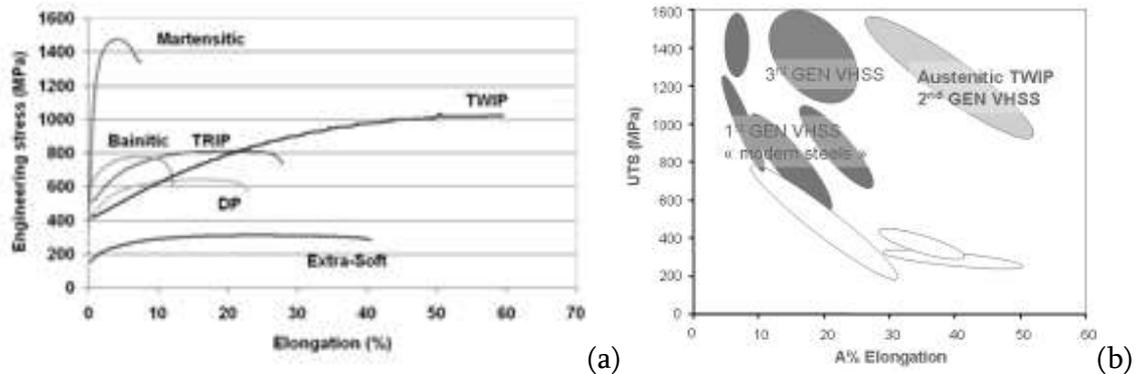

Figure 3 : (a) Courbes de traction conventionnelles typiques d'aciers ferritiques (aciers actuellement utilisés dans la construction automobile) (b) Positionnement dans un plan résistance mécanique / allongement à rupture des 3 grandes générations d'aciers dits à THR (Très Haute Résistance).

Ces performances en traction sont comparables à celles des aciers austénitiques inoxydables comme le montre la Figure 4. Leur intérêt techno-économique provient donc principalement de leur coût sur des marchés de grande diffusion, comme l'automobile, et non de performances spécifiques. En effet, le manganèse comme matière première présente des prix bien inférieurs à ceux du nickel (de l'ordre de grandeur de ceux du chrome) tout en ayant un pouvoir de stabilisation des phases austénitiques (effet dit gammagène) supérieur.

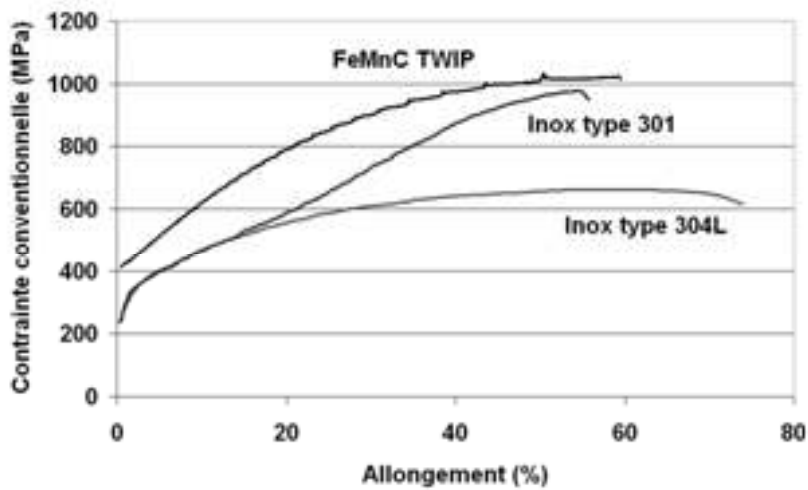

Figure 4 : Courbes de traction conventionnelles typiques d'aciers FeMnC TWIP, de deux aciers austénitiques FeNiCr inoxydables de type 301 (à effet TRIP – TRansformation Induced Plasticity) ou 304L (stable) d'après [TAL YAN 1998].

Grâce au développement de cette nouvelle famille d'aciers, une mutation rapide des pratiques dans le design et la construction automobile était attendue avant 2020 [CORNETTE 2005][SCOTT 2006], mais ce changement n'a pas eu lieu. L'introduction des aciers TWIP reste à l'heure actuelle confidentielle. La Figure 5 montre par exemple la première implémentation d'un acier TWIP1000 produit par POSCO sur véhicule FIAT [MAGGI 2012]. De façon surprenante, il s'agit d'une pièce anti-intrusion (poutre de pare-choc) nécessitant plutôt des produits à fortes limites d'élasticité (comme les aciers martensitiques).

Les principales raisons qui ont conduit au retard dans le développement et l'implémentation industrielle de ces produits sont de plusieurs ordres :
- problématique process : élaboration nécessitant une filière compliquée et des capabilités proches de la production des aciers inoxydables (élaboration d'acier liquide, coulée continue, laminage par exemple)
- problématique d'usage : soudage hétérogène difficile entre les aciers ferritiques et ces aciers TWIP [BEAL 2011] sur véhicules automobiles lié aux compatibilités de propriétés thermiques (liquidus, conductivité),
- problématique de durée de vie : casses différées et de fragilisation par l'hydrogène, problématique bien connue des aciers austénitiques inoxydables,
- problématique de valorisation : faible potentiel d'allègement de ces nuances pour des applications en substitution pure, car les limites d'élasticité des aciers austénitiques sont faibles. Des grandes déformations sont nécessaires sur pièces pour pouvoir bénéficier du potentiel de durcissement de ces alliages. Les pièces et fonctions doivent donc être entièrement repensées. Dans l'état actuel du marché automobile, peu de constructeurs sont prêts à investir sur ces nouvelles possibilités. Il existe toutefois des possibilités métallurgiques mais onéreuses pour améliorer ce paramètre (précipitation, pré-déformation) [BOUAZIZ 2011_1].

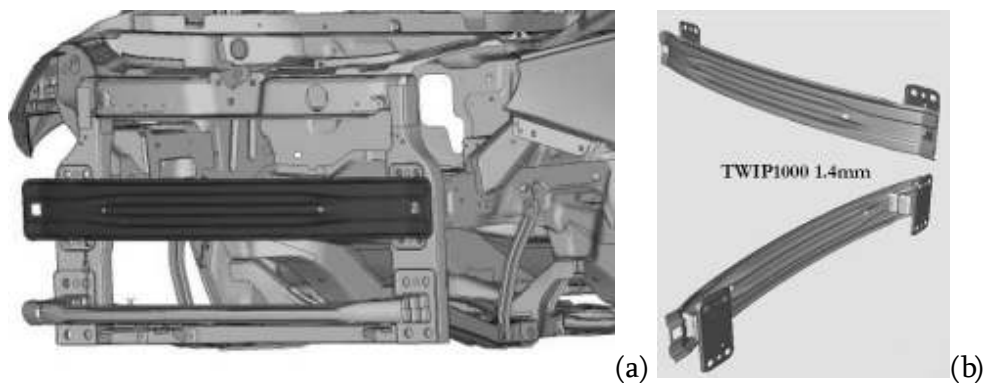

(a) (b)

**Figure 5 : Première implémentation des aciers TWIP dans le domaine automobile - Poutre pare-choc mono-coquille et « crash box » en TWIP 1000 fourni par POSCO sur véhicule de série FIAT (véhicule actuellement commercialisé) [MAGGI 2012]**

### 2.1.2. Mise en perspective des travaux personnels et collectifs

Ces aciers ont été découverts par Sir Robert Hadfield en 1888. Beaucoup de travaux depuis cette époque ont été consacrés à l'identification puis à la compréhension des mécanismes de durcissement de ces alliages. L'identification du maclage mécanique comme mécanisme de plasticité dans ces structures date des années 1960 et la première interprétation de l'effet TWIP est proposée par Rémy dans les années 1970 [REMY 1975]. Suite à ces travaux, ces métallurgies ont principalement été considérées pour applications cryogéniques ou de résistance à la corrosion grâce à de forts ajouts d'aluminium. Les premiers alliages destinés spécifiquement au marché automobile sont dus au sidérurgiste POSCO à la fin des années 1990. Le lecteur pourra se reporter à l'historique détaillé dans l'article de revue récent de notre équipe [BOUAZIZ 2011_1].

J'ai eu la chance de participer au début de l'aventure du développement des aciers TWIP pour AM dès 1999, au cours de mon DEA puis de ma thèse au LPM, intitulée :
- « Caractérisation et modélisation thermomécaniques multi-échelles des mécanismes de déformation et d'écrouissage d'aciers austénitiques à haute teneur en manganèse – application à l'effet TWIP » [ALLAIN 2004_1]

Ces travaux ont ensuite donné lieu à deux thèses financées par AM que j'ai directement co-encadrées :
- Thèse de A. Dumay, au LPM, intitulée : « Amélioration des propriétés physiques et mécaniques d'aciers TWIP FeMnXC : influence de la solution solide, durcissement par précipitation et effet composite » [DUMAY 2008_1].
- Thèse de D. Barbier, au LETAM/LPMM, intitulée : « Etude du comportement mécanique et évolutions microstructurales de l'acier austénitique Fe22Mn0.6C à effet TWIP sous sollicitations complexes. Approche expérimentale et modélisation » [BARBIER 2009_1].

J'ai aussi eu l'occasion de collaborer à la thèse de N. Shiekhelsouk au LPMM (directeurs de thèse : M. Cherkaoui, V. Favier) et à la thèse de K. Renard à l'UCL (directeur de thèse : P. Jacques). En parallèle de ces thèses, cette thématique de recherche m'a aussi permis de coopérer directement ou indirectement avec de nombreux chercheurs, je pense en particulier à M. Lebedkin, T. Lebedkina, A. Roth (LEM3), A. Deschamps (SIMAP), D. Embury, A. Zurob (Mc Master university), C. Sinclair (UBC), Y. Estrin (Monash university), et M. Huang (Hong-Kong university).

### 2.1.3. Problématique

La composition de référence à laquelle la plupart de mes travaux ont été consacrés est une nuance austénitique à haute teneur en manganèse et carbone Fe22Mn0.6C. Elle permet d'atteindre des résistances mécaniques élevées supérieures à 1 GPa associées à une ductilité supérieure à 50%. Les meilleures performances ductilité / résistance mécaniques sont obtenues à température ambiante, grâce à la combinaison d'un glissement faiblement thermiquement activé et à l'activation d'un mécanisme de déformation compétitif au glissement, le maclage mécanique. A plus basse température, la transformation martensitique ε se substitue au maclage, permettant d'atteindre des résistances mécaniques plus élevées au détriment de la ductilité alors qu'à haute température, le maclage mécanique est inhibé à cause d'une énergie de défaut d'empilement (EDE) plus élevée. Les résistances mécaniques sont alors réduites. D'autres hypothèses sur l'origine du fort taux d'écrouissage de ces nuances, comme le vieillissement dynamique, sont proposées dans la littérature et seront discutées dans ce mémoire.

La Figure 6 présente ainsi les évolutions des paramètres du comportement en traction en fonction de la température d'essai pour cette nuance avec une taille de grain d'environ 3 µm. L'allongement total (en rouge) passe par un optimum à température ambiante quand le maclage mécanique est activé et la limite d'élasticité conventionnelle (en vert) atteint un palier athermique. A haute température, les résistances mécaniques (en bleu) sont faibles et à basse température, les capacités d'écrouissage sont réduites (faible différence entre la limite d'élasticité et la résistance mécanique). Les procédures suivies pour la réalisation de ces essais sont détaillés dans [ALLAIN 2004_1]. La Figure 7 montre des micrographies en microscopie électronique à transmission (MET) après 10% de déformation de cet alliage à différentes températures caractéristiques, pour illustrer les possibles mécanismes de déformation. De gauche à droite, on reconnaîtra une microstructure après transformation martensitique ε induite par la déformation (obtenue à 77K), une microstructure de maclage mécanique (à 298 K) et enfin une distribution homogène de dislocations (à 673K). Nous reviendrons dans le chapitre suivant sur l'origine et la caractérisation de cette forte dépendance des mécanismes de plasticité à la température.

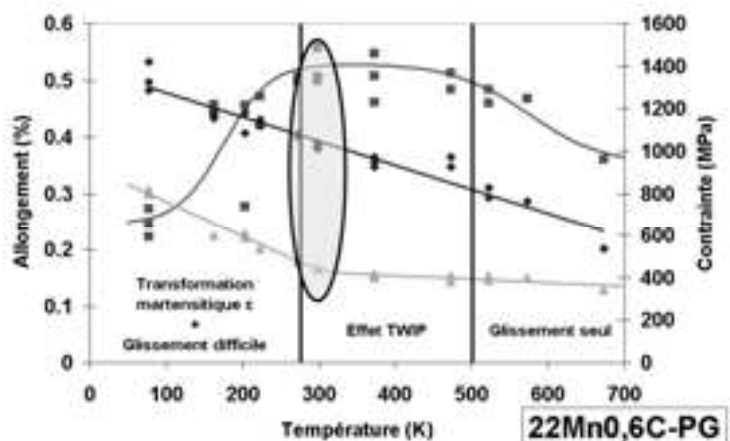

Figure 6 : Evolution des paramètres de traction d'un acier Fe22Mn0.6C (taille de grain = 3 µm), résistance mécanique (bleu), limite d'élasticité (vert) et allongement à rupture (rouge), en fonction de la température d'essai. Sont indiqués en outre les mécanismes d'écrouissage actifs [ALLAIN 2004_1].

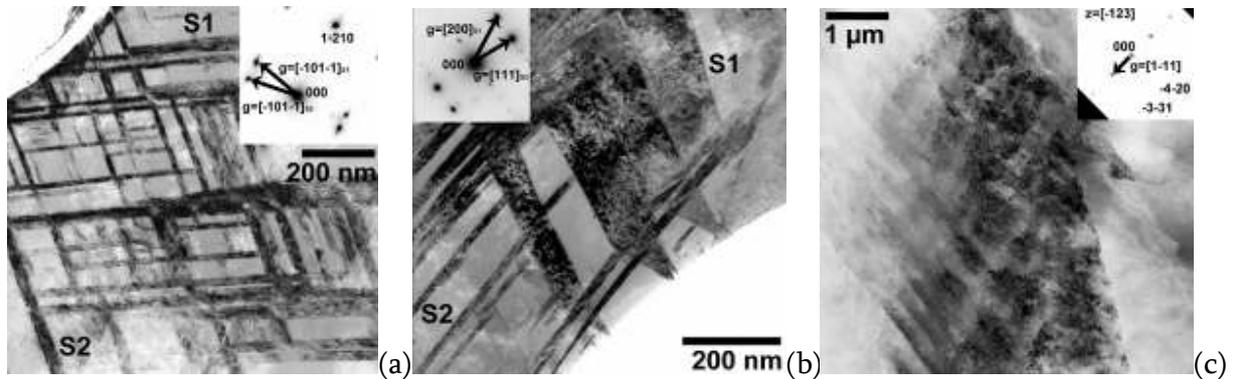

Figure 7 : Micrographies en MET en champ clair de la nuance de référence après 10% de déformation en traction à différentes températures, mettant en lumière les mécanismes de déformation actifs (a) martensite ε à 77 K (b) maclage mécanique à 298 K (c) glissement à 673 K [ALLAIN 2004_1].

L'effet TWIP est un mécanisme d'écrouissage spécifique de ces aciers lié à l'apparition de nanomacles au cours de leur déformation. Ces défauts plans d'épaisseur nanométrique créent une microstructure enchevêtrée et difficilement franchissable par les dislocations mobiles à l'intérieur des grains austénitiques. Il en résulte une double contribution au durcissement, la première de nature isotrope (augmentation de densité de dislocations statistiquement stockées) et la seconde de nature cinématique (mécanisme de type composite associé à l'incompatibilité de déformation entre macle et matrice). Dans ce mémoire, nous aborderons aussi largement le cas de la transformation martensitique ε, transformation présentant de très fortes analogies avec le maclage mécanique, et dont les contributions à l'écrouissage sont similaires.

Le processus de maclage mécanique est contrôlé par l'énergie de défaut d'empilement (EDE) de l'alliage considéré (composition). Contrairement aux éléments d'alliage substitutionnels, la teneur en carbone augmente cette EDE et donc serait défavorable pour le maclage mécanique, mais on observe au contraire qu'elle contribue à accroître le taux d'écrouissage de ces aciers. Dans la littérature, un grand rôle sur le comportement de ces aciers était attribué au carbone par certains auteurs à travers un mécanisme de vieillissement dynamique. Cette contribution indépendante du maclage mécanique s'est avérée être négligeable. Deux hypothèses sont actuellement investiguées pour résoudre ce paradoxe en considérant la relation entre carbone et maclage mécanique :
- Le carbone inhiberait les processus de relaxation plastique dans les macles conduisant à une augmentation de l'effet composite.
- Le carbone modifie très sensiblement la mobilité des dislocations parfaites via un processus thermiquement activé et la planéité du glissement (glissement dévié). Ces conditions favoriseraient le maclage mécanique en abaissant la contrainte critique de maclage, en compétition avec le glissement.

Dans ce mémoire, on se propose donc de discuter du comportement de ces alliages selon trois axes :
- Le premier concerne la morphogénèse de la microstructure de maclage mécanique, c'est-à-dire de la structuration spatiale et la distribution de la microstructure de maclage ainsi que ses conditions d'apparition (travaux principalement de nature expérimentale).
- Le second axe concerne la modélisation proprement dire de l'effet TWIP, c'est-à-dire de l'impact de la microstructure de maclage sur les propriétés mécaniques. On s'intéressera en particulier à la nature de l'écrouissage dû à l'effet TWIP et son rôle lors des chargements thermomécaniques présentant des complexités croissantes, de plus en plus représentatives de chargement réels (emboutissage).
- Le troisième concerne les effets de la composition chimique sur le comportement et le rôle paradoxal du carbone dans ces aciers.

Le cas des aciers austénitiques sans carbone FeMnSiAl TWIP ou les alliages à mémoire de forme FeMnSi sera aussi bien entendu évoqué pour mieux comprendre les effets complexes du carbone dans les systèmes ternaires FeMnC.

## 2.2. Morphogénèse de la microstructure de maclage

Les travaux détaillés dans ce chapitre concernent l'étude de la microstructure de maclage et sont tirés de mon travail de thèse et de celles de D. Barbier (EBSD, MET, RX) et d'A. Dumay (MET). Après un court rappel bibliographique, nous discuterons des caractéristiques clefs de cette microstructure ayant une conséquence pour l'effet TWIP : la fraction maclée, les épaisseurs de macles et la topologie.

### 2.2.1. Introduction : maclage mécanique, transformation martensitique ε et EDE

Le maclage mécanique apparaît dans la plupart des matériaux cristallins (structures cubiques centrées CC, cubiques à faces centrées CFC, hexagonaux compactes HC ou de structures plus complexes) sous certaines conditions de déformations. Ce mécanisme ne joue souvent qu'un rôle mineur dans les matériaux ductiles, mais se révèle être indispensable dans les matériaux dans lesquels le nombre de système de glissement est limité [FISCHER 2003]. Il compense le manque de systèmes de glissement indépendants pour accommoder la déformation dans les agrégats polycristallins de structures hexagonales [CAHN 1953] (cf. Figure 8(a)) ou les alliages ordonnés par exemple [JIN 1995][FARENC 1993]. Le maclage mécanique joue cependant un rôle important dans certains matériaux ayant de plus nombreuses symétries comme les CC Cu-Ag [SUZUKI 1958], Fe-Si [YANEZ 2003] (appelées aussi bandes de Neuman, sur la Figure 8(b)), structures martensitiques lenticulaires FeNiC (dites martensite vierges [MAGEE 1971]) ou les CFC comme les alliages Cu-Si [COUJOU 1983], Cu-Al [MORI 1980], Co-Ni [REMY 1978] et mêmes les aciers inoxydables austénitiques [LECROISEY 1972] ou les aciers FeMnC Hadfield [KARAMAN 2001]. Une revue bibliographique très complète a été proposée par Christian et Mahajan [CHRISTIAN 1995] dans laquelle le lecteur pourra trouver de nombreux détails et commentaires sur le maclage mécanique dans différentes structures cristallographiques et en particulier dans les alliages CFC.

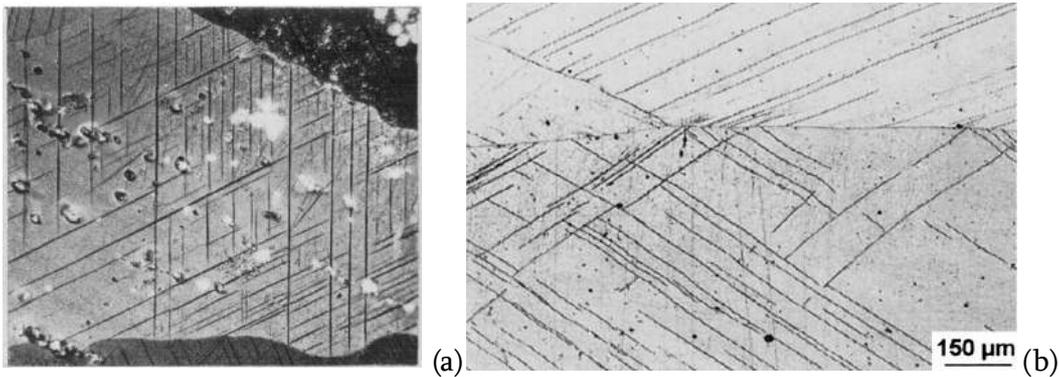

Figure 8 : Exemples de microstructures de maclage dans des alliages métalliques. Micrographies optiques issues de la littérature (a) dans l'uranium α d'après [CAHN 1953] (b) dans un acier dit électrique, Fe3Si, d'après [YANEZ 2003] (microstructures connues sous le nom de « Bandes de Neuman »).

Bien que ce mémoire soit principalement dédié aux aciers austénitiques, nous reviendrons à la fin de ce chapitre sur l'analogie qu'il peut exister avec les structures à faibles nombres de systèmes de glissement. Nous donnerons aussi quelques pistes de recherche pour comprendre pourquoi ces aciers présentent ce mécanisme compétitif au glissement et comment la composition chimique semble contrôler l'activation de ce mécanisme.

### 2.2.1.1. Maclage mécanique

Le maclage mécanique et la transformation martensitique ε sont des mécanismes de déformation compétitifs au glissement des dislocations. Ils sont très similaires du point de vue de leur mécanisme de germination et de croissance (en termes de réactions entre dislocations), mais aussi de celui de la morphologie résultante. Cette similitude est illustrée par exemple sur la Figure 7, qui montre les micrographies en MET des deux types de microstructures de déformation.

Dans les alliages CFC à faible EDE, il est maintenant bien établi que le maclage mécanique est le résultat d'un glissement collaboratif de dislocations partielles de Shockley a/6<112> tous les plans parallèles {111} successifs, définissant la direction de maclage et le plan de maclage respectivement. La nature du défaut traîné par cette dislocation partielle de tête est alors supposée de nature intrinsèque. Des modèles de germination supposent que la nature peut être extrinsèque nécessitant alors de faire passer les dislocations partielles de tête tous les deux plans atomiques. Les macles mécaniques apparaissent donc au cours de la déformation de l'alliage. Elles ne doivent pas être confondues avec les macles de recuit qui apparaissent lors de la recristallisation des alliages CFC.

La structure reconstruite entre les plans fautés, les joints de macle, est toujours de structure cristallographique CFC mais en orientation de macle Σ3 par rapport à la matrice comme le montre la Figure 9.

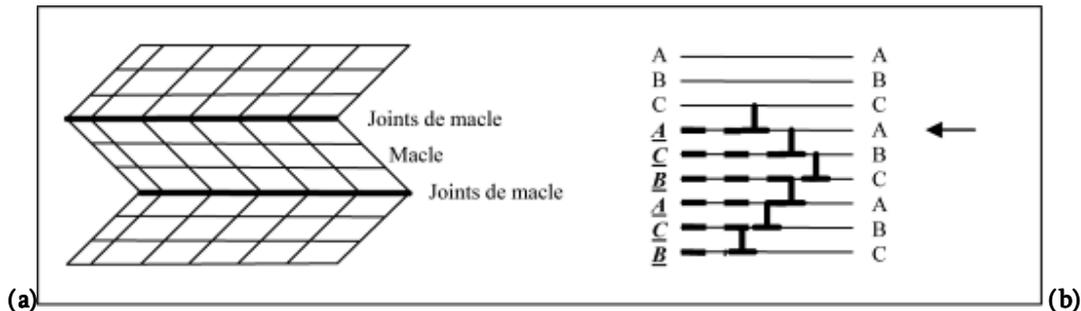

Figure 9 : (a) représentation schématique d'une macle dans un arrangement périodique. Les joints de macles sont des plans de symétrie (b) Représentation schématique typique du réarrangement des plans compacts autour d'une macle mécanique créée par le déplacement de dislocations partielles intrinsèques tous les plans atomiques. La flèche indique la position du joint de macle plan miroir pour le réseau CFC.

Les éléments cristallographiques du maclage dans un réseau CFC sont, selon les conventions habituelles :
- K1 le plan de maclage de type {111} correspondant au plan de glissement des dislocations partielles de Shockley (correspondant aussi aux joints de macle),
- η1 la direction de cisaillement de type <112> correspondant à la direction des vecteurs de Burgers des dislocations partielles,
- K2 le plan non distordu au cours du cisaillement, de type {111},
- η2 l'intersection des plans K2 et K1 de type <112>.

Le glissement produit par une macle parfaite d'origine intrinsèque dans la direction η1 est égale à $\gamma_T = b_{112} / d_{111}$ avec $d_{111}$ la distance réticulaire entre deux plans denses {111} et $b_{112}$ le vecteur de Burgers des dislocations de type $a/6<112>$. Si l'on considère par contre un mécanisme mettant en jeu des défauts d'origine extrinsèques, le glissement résultant est divisé par 2 et est égal à $\gamma_T / 2$ [COUPEAU 1999].

Certains auteurs ont souligné que ce processus ne permet pas de reproduire tout à fait à l'identique la maille de la matrice dans le cas des alliages ternaires. En effet, Adler *et al.* ont montré que les atomes de carbone se situaient préférentiellement dans les sites interstitiels octaédriques de la structure CFC et que ces sites étaient en fait transformés en sites tétraédriques lors du processus de maclage [ADLER 1986]. D'après les auteurs, ce processus de « pseudo-maclage » induit un durcissement important, responsable de l'excellent taux d'écrouissage dans ces alliages. Toutefois, cette approche ne permet en aucun cas d'expliquer les taux d'écrouissage des alliages sans carbone comme les aciers FeMnSiAl étudiés par Grässel *et al* [GRASSEL 2000] par exemple.

Concernant le processus de maclage mécanique dans les structures CFC, il ressort aussi de la littérature que :
- la propagation des macles est extrêmement rapide, de l'ordre de grandeur de la vitesse du son dans le matériau [LUBENETS 1985][REMY 1975]. L'étape de germination est donc le processus limitant,
- la germination des macles est hétérogène à partir d'une configuration particulière de dislocations [REMY 75][VENABLES 1964][LUBENETS 1985][CHRISTIAN 1995] [CHRISTIAN 1969][1],
- Il nécessite en outre l'activation préalable de plusieurs systèmes de glissements [CHRISTIAN 1969].

### 2.2.1.2. *Transformation martensitique ε et α'*

La transformation martensitique ε est un mécanisme très similaire au maclage et se produit quand des dislocations partielles de Shockley de nature intrinsèque se propagent tous les deux plans atomiques {111} ou tous les plans atomiques dans le cas de défaut de nature extrinsèque [IDRISSI 2009]. Contrairement au maclage, ce mécanisme collectif de dislocations ne reconstruit pas une structure CFC en volume mais une structure hexagonale compacte (HC). Les plans {111} d'habitat de la martensite dans l'austénite correspondent aux plans denses de la structure hexagonale. Maclage et transformation martensitique sont si proches qu'ils peuvent co-exister au sein d'une même bande de déformation comme l'ont observé certains auteurs [BRACKE 2007_2].

Maclage mécanique et transformation martensitique ε sont fortement reliés à l'Energie de Défaut d'Empilement (EDE) de l'alliage qui contrôle le coût énergétique de la création des défauts d'empilement et donc la distance de dissociation entre les dislocations partielles bordant ce défaut [FRIEDEL 1964][FERREIRA 1998][BUYN 2003][KARAMAN 2000_1]. Une relation existe d'ailleurs entre cette énergie et l'enthalpie libre de transformation γ → ε $\Delta G^{\gamma \to \varepsilon}$. Cette formule due à Hirth [HIRTH 1970] a été reprise et rendue populaire par les travaux de Olson et Cohen. Cette approche consiste à établir une équivalence entre un défaut d'empilement de nature intrinsèque et une plaquette de martensite ε ayant une épaisseur de deux plans atomiques compacts et créant deux nouvelles interfaces γ/ε. Il en résulte :

---

[1] un processus de maclage homogène tel que discuté par Lubenets *et al* [LUBENETS 1985] nécessiterait des contraintes de l'ordre de 20% du module d'Young.

$$\text{EDE}_{\text{int}} = 2\rho_{111} \Delta G^{\gamma \to \varepsilon} + 2\sigma^{\gamma/\varepsilon} \qquad (1)$$

Avec $\rho_{111}$ la densité surfacique molaire sur les plans d'habitat {111} et $\sigma^{\gamma/\varepsilon}$ l'énergie par unité de surface des interfaces γ/ε.

Un second type de transformation martensitique est aussi observé dans cet alliage correspondant à une transformation γ → α' de structure tétragonale centrée. A l'instar de la transformation martensitique ε, elle peut avoir lieu au cours d'une trempe (transformation thermique) ou de la déformation de ces alliages (transformation induite mécaniquement). Cette transformation martensitique particulière touche les alliages les moins riches en carbone et/ou Mn ou lors de déformation à basse température [SCHUMANN 1972].

Ces transformations sont très largement documentées dans la littérature car elles correspondent à celles rencontrées traditionnellement dans les aciers ferritiques (on pourra se reporter par exemple à la revue de Krauss [KRAUSS 1999]). Nous reviendrons sur les propriétés de cette phase à la section 3.3.2 (page 111). Dans le cas des aciers austénitiques, cette transformation au cours de la déformation induit un fort effet TRIP (Transformation Induced Plasticity) dans ces alliages qui peut conduire à des résistances mécaniques élevées (comme par exemple l'acier inoxydable de type 301 sur la Figure 4). Par contre, ces aciers montrent alors des ductilités généralement moins élevées que les aciers TWIP et sont sujets à une forte sensibilité à la casse différée [DAGBERT 1996]. C'est pourquoi, ces alliages austénitiques instables sont moins regardés actuellement dans la littérature et seront peu souvent abordés dans le cadre de ce mémoire.

Une confusion est souvent faite dans la littérature entre EDE et occurrence de la transformation martensitique α'. Il n'est en effet pas rare de retrouver des germes de martensite α' aux intersections de plaquettes de martensite ε de différents variants ou de bandes de cisaillement [BRACKE 2007_1][OLSON 1975][STRINGFELLOW 1992][TOMITA 1995]. Cependant, la nature de cette transformation ne fait pas intervenir nécessairement de réaction γ → ε. La coïncidence entre occurrence de la martensite α' et les très faibles EDE est donc purement « fortuite » et ne peut être systématisée.

### 2.2.1.3. *EDE et mécanismes de déformation*

Pour une composition donnée, Rémy a démontré que l'EDE est nécessairement une fonction croissante de la température au travers de la relation suivante :

$$\frac{d\,\text{EDE}_{\text{int}}}{dT} = -\frac{8}{\sqrt{3}} \frac{1}{a_{\text{CFC}}^{2} \aleph} \Delta S^{\gamma \to \varepsilon} \qquad (2)$$

avec $a_{CFC}$ le paramètre de maille de l'austénite, $\aleph$ le nombre d'Avagadro et $\Delta S^{\gamma \rightarrow \varepsilon}$ la variation d'entropie au cours de la transformation γ → ε. Si la température est supérieure à Es, la température de transformation martensitique ε spontanée au refroidissement, $\Delta S^{\gamma \rightarrow \varepsilon}$ est nécessairement négative et conduit donc à une sensibilité positive de l'EDE à la température.

Cette relation est intéressante car elle permet d'expliquer en grande partie l'évolution des microstructures de déformation en fonction de la température de déformation présentée sur la Figure 7. Cette séquence dans l'apparition des mécanismes de déformation compétitifs au glissement est valable pour une composition donnée.

A haute température par rapport à Es, l'EDE est grande. Le coût énergétique de la dissociation de dislocations parfaites est grand et donc défavorable. Le seul mécanisme possible est donc le glissement de dislocations (comme observé par de nombreux auteurs sur la nuance de référence à 400°C). Les hautes températures et les EDE élevées sont favorables au glissement dévié et à l'activation de nombreux systèmes de glissement [RAUCH 2004].

Quand la température diminue, le glissement des dislocations devient de plus en plus planaire, car le glissement dévié devient défavorable (ce processus sera discuté en détail au chapitre 2.4.3.4 page 89). En dessous d'une certaine valeur d'EDE, de larges dissociations de dislocations parfaites sont possibles à un coût énergétique faible [BYUN 2003] et l'apparition du maclage mécanique est favorisée. Actuellement, le mécanisme réactionnel entre dislocations conduisant à former un germe de macle ne fait pas l'objet d'un consensus dans la communauté. Les mécanismes de croissance et d'épaississement ne semblent pas non plus évidents. Des expériences récentes *in situ* au MET [IDRISSI 2010_1] suggèrent que le mécanisme de germination puisse être une réaction polaire avec déviation comme proposé par Cohen and Weertman [COHEN 1963_1] [COHEN 1963_2] ou par Miura, Takamura et Narita [MIURA 1968] alors que la croissance de la macle (épaississement) soit due à un mécanisme polaire simple comme proposé initialement par Venables [VENABLES 1974].

A plus basse température, la transformation martensitique ε remplace le maclage mécanique. La transition entre ces deux mécanismes de plasticité est expliquée par une évolution du caractère des défauts d'empilement responsables de leur germination (intrinsèque et extrinsèque pour une macle et la martensite respectivement) comme l'a montré récemment Idrissi *et al* [IDRISSI 2009]. Le calcul proposé par Lecroisey et Pineau [LECROISEY 1972] soutient cette hypothèse, dans la mesure où l'EDE extrinsèque est environ 1,5 fois plus élevée que l'EDE intrinsèque. Des températures très basses sont donc nécessaires pour permettre la transformation martensitique ε par rapport au maclage. Comme dans le cas du maclage mécanique, la transformation martensitique ε est un mécanisme compétitif au glissement. Ce dernier assure toujours la plus grande contribution de la déformation macroscopique, mais peut être significativement inhibé par l'ajout d'éléments d'alliage[2].

Ainsi, pour un acier de composition donné, différents mécanismes de déformation peuvent être activés en fonction de la température de sollicitation mécanique. De même, il est possible en modifiant la composition chimique de l'alliage d'activer certains mécanismes de déformation à une température donnée. Ce constat permet d'envisager des stratégies « d'alloy design » tout à fait excitantes.

La première étude majeure du système ternaire FeMnC date de 1972 permettant de relier la composition chimique et la stabilité de l'austénite à température ambiante vis-à-vis des transformations martensitiques ε et α' et a été réalisée par Schumann [SCHUMANN 1972]. Il a établi la première carte du plan Mn/C définissant les compositions stables au cours de la déformation (incluant le maclage mécanique, difficile à identifier à l'époque) et instables (subissant une transformation martensitique ε). La limite entre ces deux domaines peut s'écrire :

$$\%Mn = -20\%C + 32 \tag{3}$$

Avec %Mn et %C les teneurs massiques en éléments d'alliage en pourcents.

Cette formule montre par exemple que les alliages Fe35Mn ou Fe22Mn0.6C (acier de référence de cette étude) ou Fe12Mn1.1C (acier Hadfield) sont 100% austénitiques à température ambiante et ne présenteront pas de transformation martensitique ε au cours de leurs déformations.

Depuis Schumann, nous avons contribué à l'instar de nombreuses équipes à élaborer des relations moins empiriques entre compositions, EDE et mécanismes de déformation, sans réel consensus. Saeed-Akbari *et al.* [SAEED-AKBARI 2009] ont récemment réalisé une revue bibliographique de ce sujet et montré que la plupart des auteurs concluait que cette limite définie par Schumann correspondait à une iso-valeur d'EDE d'environ 20 mJ.m$^{-2}$. Nous reviendrons dans le chapitre 2.4.1 page 68 sur les stratégies de mesure et de modélisation de cette EDE, et sur le rôle particulier joué par le carbone sur cette valeur.

---

[2] Le glissement est largement inhibé par l'ajout massif d'éléments d'alliage comme le silicium. C'est le cas par exemple des Alliages à Mémoire de Forme (AMF) Fe30Mn6Si, austénitiques à température ambiante, qui se déforment principalement par transformation martensitique ε [MYAZAKI 1989]. Dans ce cas, cette dernière transformation martensitique est massive et contribue de façon majoritaire à la déformation macroscopique de l'alliage (cf. Figure 58 page 93).

### 2.2.2. Microstructure de maclage : nanomacles et faisceaux

Le maclage mécanique dans ces structures austénitiques adopte une structuration multi-échelle particulière dont les caractéristiques de taille, de morphologie et de distribution spatiale expliquent en grande partie l'effet TWIP. Une partie importante de nos travaux a donc été consacrée à la description et la compréhension de cette structure, dans la nuance de référence ou des compositions proches.

Nous avons caractérisé ces microstructures de déformation à plusieurs échelles, de celle de la microscopie électronique à transmission (MET) pour la caractérisation des macles individuelles à la microscopie optique (MO) pour la structuration des systèmes de maclage en passant par la microscopie électronique à balayage (MEB) couplée à des analyses en mode Electron Back Scattering Diffraction (EBSD) pour l'analyse systématique des relations entre maclage mécanique et cristallographie. Le couplage de ces différentes techniques permet de donner un aperçu statistiquement admissible de la microstructure de maclage et de son évolution au cours de la déformation.

Les macles mécaniques observées dans ces alliages sont en réalité des nanomacles, c'est-à-dire des objets très minces dont l'épaisseur est de l'ordre de quelques dizaines de nanomètres comme le montre la Figure 10. Le nombre de dislocations partielles impliquées dans le processus reste donc limité. A titre d'ordre de grandeur, environ 27 dislocations sont nécessaires pour former une macle de 100 nm d'épaisseur.

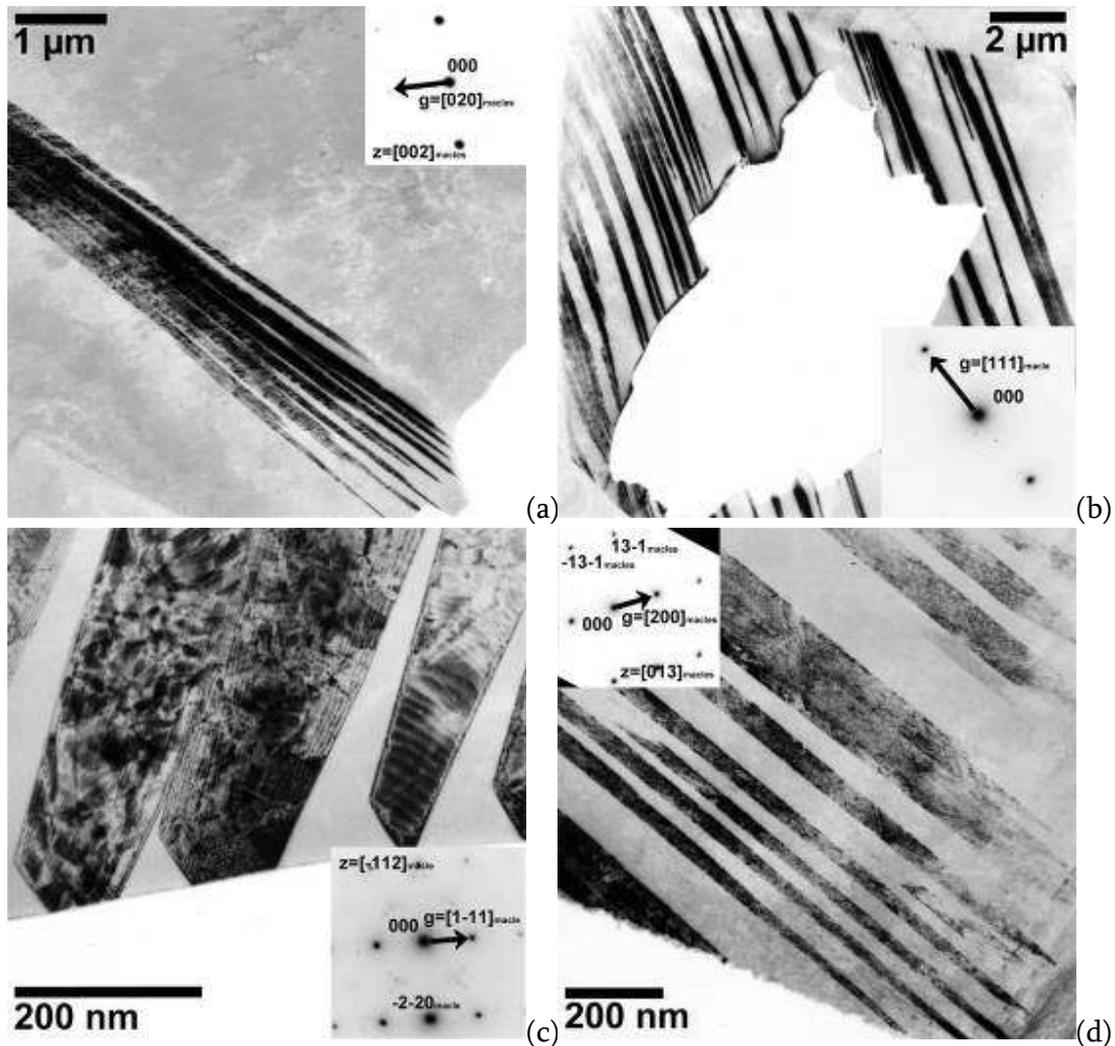

Figure 10 : (a), (b), (c) et (d) Nanomacles organisées en faisceaux (paquets) observée par MET en champ sombre dans la nuance de référence après déformation [ALLAIN 2004_1].

Peu d'études ont été consacrées à la mesure de l'épaisseur de macles de façon systématique et statistique. Cette lacune s'explique par la nécessité de réaliser ce travail en MET après indexation systématique des zones (normale au plan K1 perpendiculaire au faisceau incident) sans aucune garantie d'obtenir une netteté suffisante à cause des densités importantes de dislocations sur les interfaces [IDRISSI 2010_2]. Cette carence dans la littérature n'en reste pas moins paradoxale car l'épaisseur des macles est un paramètre clef dans la compréhension des mécanismes d'écrouissage (cf. chapitre 2.3.5.1 page 52). Ces fines épaisseurs expliquent que les zones maclées supportent des contraintes élevées de l'ordre de plusieurs GPa sans possibilité de relaxation. [SEVILLANO 2012].

Afin de pallier ce manque, nous avons développé une approche pour prédire l'épaisseur des macles en fonction de paramètres cristallographiques et thermochimiques [ALLAIN 2004_1] [ALLAIN 2004_4]. Ce modèle est basé sur une extension des travaux de Friedel [FRIEDEL 1964]. Celui-ci assimile une macle à un assemblage de boucles de dislocations partielles circulaires et concentriques et calcule son facteur de forme S en fonction de la cission appliquée $\tau_{app}$.

$$S = \frac{e}{D} = \frac{2d_{111}}{\mu b_{112}} \tau_{app} \tag{4}$$

Avec e l'épaisseur de la macle considérée, $d_{111}$ l'espacement réticulaire entre les plans {111}, $b_{112}$ le vecteur de Burgers des dislocations partielles de Shockley, $\mu$ le module d'élasticité en cisaillement. D est le rayon de la macle. Nous avons étendu ce concept en introduisant une contrainte critique d'émission qui dépend de l'EDE intrinsèque. Cet ajout permet de définir l'épaisseur minimale des macles $e_{min}$ issue d'un processus de germination (avant épaississement) :

$$e_{min} = \frac{2d_{111}}{\mu b_{112}} \left( \frac{EDE_{int}}{b_{112}} \right) D = J_{isolé}^{analytique} \left( \frac{EDE_{int}}{b_{112}} \right) D \tag{5}$$

Cette formule montre que l'épaisseur minimum des macles doit augmenter avec l'EDE et son diamètre (son libre parcours moyen). Numériquement, avec $\mu$ = 62 GPa, EDE = 20 mJ.m$^{-2}$, $b_{112}$ = 0,147 nm, le facteur de forme minimal d'une macle isolée est de $3,1 \times 10^{-3}$ selon ce modèle. Pour une taille de grain de 15 µm, ceci correspond à une épaisseur de 47 nm, en accord avec les ordres de grandeurs observés [ALLAIN 2004_1] et ceux rapportés dans la littérature [RENARD 2012].

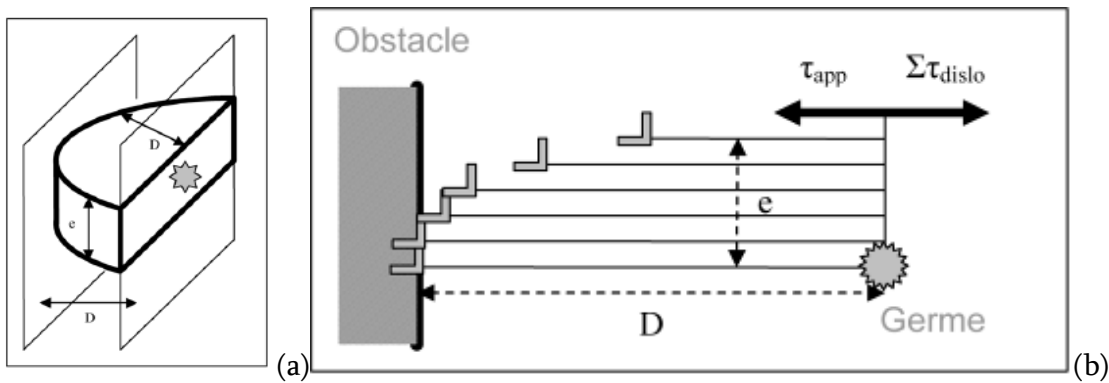

Figure 11 : Schéma de principe du modèle 2D de dynamique discrète des dislocations (DDD) simplifié pour simuler le fonctionnement d'un germe de micromacle et la formation d'un « pseudo-empilement » en front de macle sur un obstacle (a) configuration 3D (b) grandeurs caractéristiques du modèle [ALLAIN 2004_1].

Les résultats de ce modèle analytique ont été confirmés par la suite à l'aide d'un modèle simplifié de dynamique discrète des dislocations (DDD), schématisé sur la Figure 11. Il permet de simuler le fonctionnement d'un germe de macle et l'établissement d'une configuration d'équilibre sous contrainte de ce « pseudo-empilement ». Ce modèle est détaillé dans [ALLAIN 2004_1][ALLAIN 2004_4] et permet de retrouver la même sensibilité à l'EDE et à la taille des macles, mais la constante de proportionnalité J est sensiblement plus élevée (environ 7,57 $10^{-5}$ MPa$^{-1}$ au lieu de 2,28 $10^{-5}$ MPa$^{-1}$). L'épaisseur prévue pour les micromacles par cette simulation DDD est de 158 nm avec les valeurs usitées ci-dessus.

La forme du front de macle calculée par ce modèle de DDD simplifié est très proche d'un empilement de dislocations. En effet, seule une faible proportion des dislocations partielles formant la macle arrive au contact de l'obstacle. La plupart reste distribuée sur l'interface à l'instar d'un empilement. Cette constatation est importante à double titre :
- l'approximation proposée par Mullner et al [MULLNER 2002] assimilant un front de macle à un dipôle de disclinations est acceptable à grande échelle mais ne reproduit pas fidèlement le champ de contraintes autour d'un front de macle. Elle ne peut rendre compte de la disposition particulière des dislocations le long du front observée par de nombreux auteurs [FARENC 1993][CHRISTIAN 1995].
- les dislocations d'interface sont donc nombreuses et contribuent à rendre ce défaut infranchissable en pratique par des dislocations incidentes, même si de nombreuses combinaisons d'interactions macle/glissement sont théoriquement possibles [REMY 1975][IDRISSI 2010_2].

Une autre spécificité est que ces nanomacles sont groupées généralement en paquets. Cette structuration que nous avons qualifiée de « faisceaux » est particulièrement visible sur les premiers pourcents de déformation et devient difficile à distinguer à plus grande déformation. La Figure 10(a) montre un bel exemple de cette configuration. Ces structures déjà mises en évidence par Rémy [REMY 1975], sont assimilées à des zones entièrement maclées lors d'études à plus faibles grossissements (MEB ou MO). Cette assimilation sans précaution conduit alors à une surestimation de la fraction maclée. La Figure 12 montre les grandeurs caractéristiques des nanomacles et des faisceaux observées dans la nuance de référence après déformation.

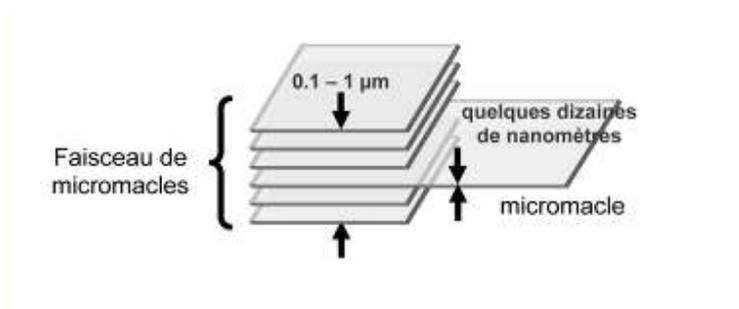

**Figure 12 : Grandeurs caractéristiques des nanomacles et des faisceaux observés dans la nuance de référence après déformation.**

### 2.2.3. Structuration multi-échelle du maclage au cours de la déformation (MET, EBSD, MO) en relation avec la texture

Au delà de la structuration en faisceaux, la microstructure de maclage présente un second niveau d'organisation spatiale qui apparaît au cours de la déformation à l'échelle du grain austénitique. Les macles se développent en effet sur plusieurs plans d'habitats parallèles ou sécants, correspondant à une ou deux familles de plans {111}. La première famille de plans sur lesquels apparaissent les macles est souvent qualifiée de primaire, et les autres de secondaires (par abus de langage, on parlera de systèmes de maclage primaires et secondaires). La Figure 10 (a) montre par exemple un grain d'austénite contenant un seul et unique système de maclage. Quand un système de maclage secondaire est activé, les macles de ce nouveau système ne peuvent s'étendre sur l'ensemble du grain mais restent généralement bloquées par les macles du système primaire. Cela conduit à la formation de structures caractéristiques dites en échelle comme le montre la Figure 13.

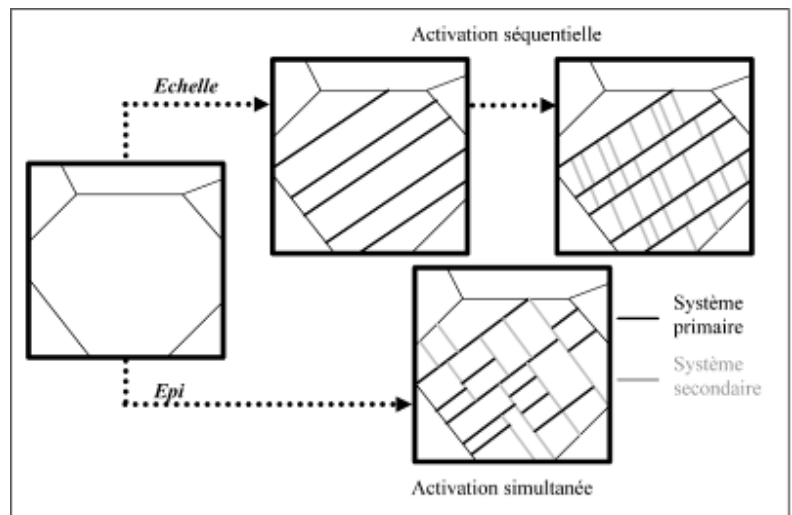

Figure 13 : Schéma de principe de la mise en place de la microstructure de maclage. Configurations après activation séquentielle ou simultanée des systèmes de maclage [ALLAIN 2004_1].

Si les deux systèmes de maclages apparaissent non pas de manière séquentielle, mais de manière concomitante, alors les deux systèmes de maclage se retrouvent enchevêtrés de manière complexe, comme l'illustre aussi la Figure 13[3].

La mise en place de ces structures particulières à l'échelle des grains dans la nuance de référence a été observée en microscopique optique pour les grains de relativement grande taille (20 µm), en EBSD ou en microscopie MET. Les observations en MET de la Figure 7(b) et de la Figure 14, confirment toutes que les micromacles sont des obstacles forts pour l'extension de micromacles incidentes sécantes[4]. Ce constat semble être aussi valable dans le cas de la transformation martensitique ε comme le montre la Figure 7(a).

---

[3] Dans la littérature, la plupart des auteurs rapporte l'activation de deux systèmes de maclage pour des déformations en traction. De rares cas de trois systèmes ont toutefois été rapportés [RENARD 2012].

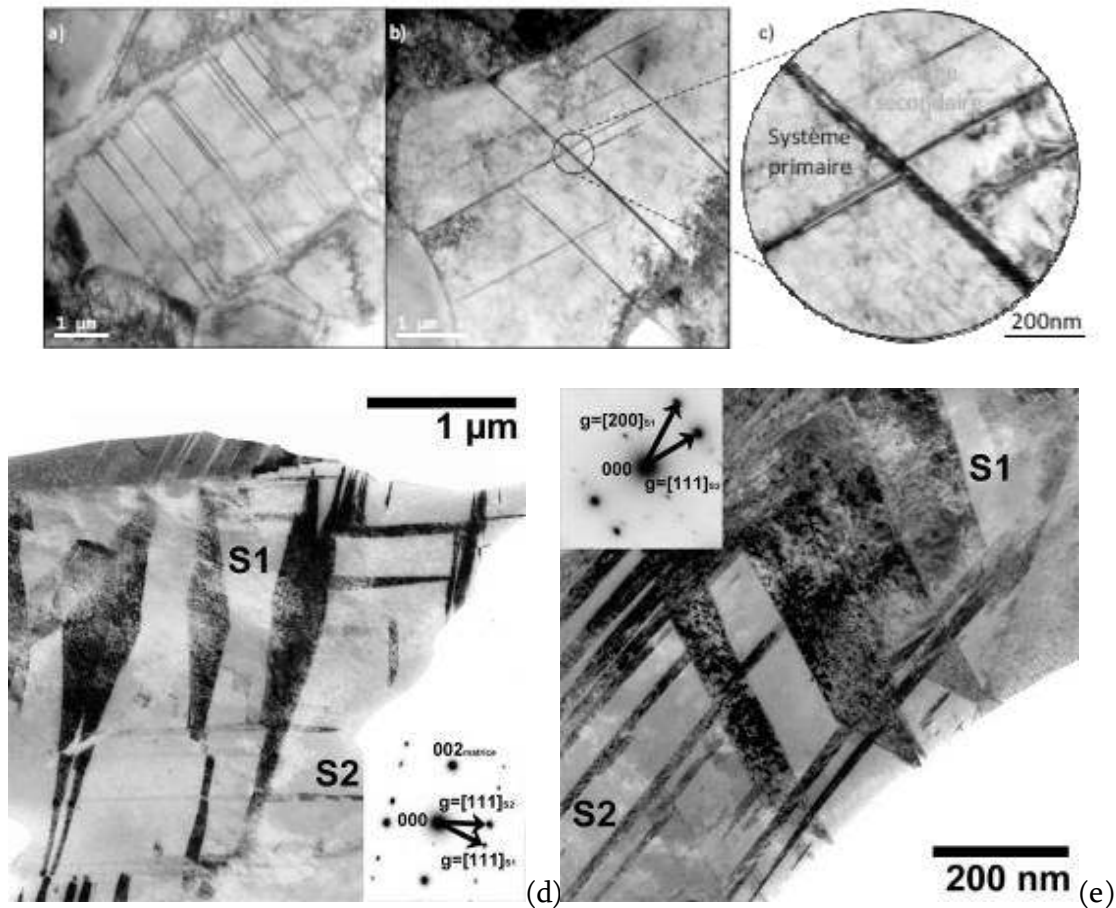

Figure 14 : Micrographies MET de la microstructure maclage de la nuance de référence (a), (b) et (c) en champ clair après 5% de déformation en traction – grains maclés par un ou deux systèmes – agrandissement de l'interaction entre deux systèmes de maclage [BARBIER 2009_1] (d) et (e) en champs sombres après 33% de déformation en traction– mise en évidence de l'enchevêtrement des deux systèmes de maclage par la sélection simultanée des conditions de diffraction des deux systèmes [ALLAIN 2004_1].

L'observation de l'évolution des microstructures de maclage à de moindres grossissements en MO permet de quantifier l'activation séquentielle des systèmes de maclage, comme le montre par exemple les Figure 15 et Figure 16. Elles mettent en lumière que tous les grains ne présentent pas une microstructure de maclage identique pour un niveau de déformation donnée. Certains grains ne sont pas ou peu maclés après 30% de déformation alors que d'autres présentent depuis 10% de déformation plusieurs systèmes de maclage. Cette différence de comportement entre les grains s'explique par une forte sensibilité du maclage à l'orientation cristallographique du grain matrice par rapport à la direction de sollicitation.

---

[4] Rémy a prouvé que certaines configurations de transmissions sont toujours théoriquement possibles à des coûts énergétiques élevés dans des structures maclées modèles [COUJOU 1992]. Dans le cas de ces aciers, les interfaces des micromacles saturées de dislocations incidentes ou de réactions comme l'a montré récemment [IDRISSI 2010_2], ce qui rend encore plus difficile la possibilité d'interactions entre deux micromacles sécantes. Par contre, il est tout à fait possible de considérer qu'une intersection de micromacles agisse comme un concentrateur de contraintes suffisant pour participer à la germination d'une nouvelle micromacle, donnant l'illusion à une échelle trop grande d'un franchissement direct. La nouvelle micromacle sera donc considérée comme un mécanisme potentiel de relaxation. C'est probablement le cas de la Figure 14(c).

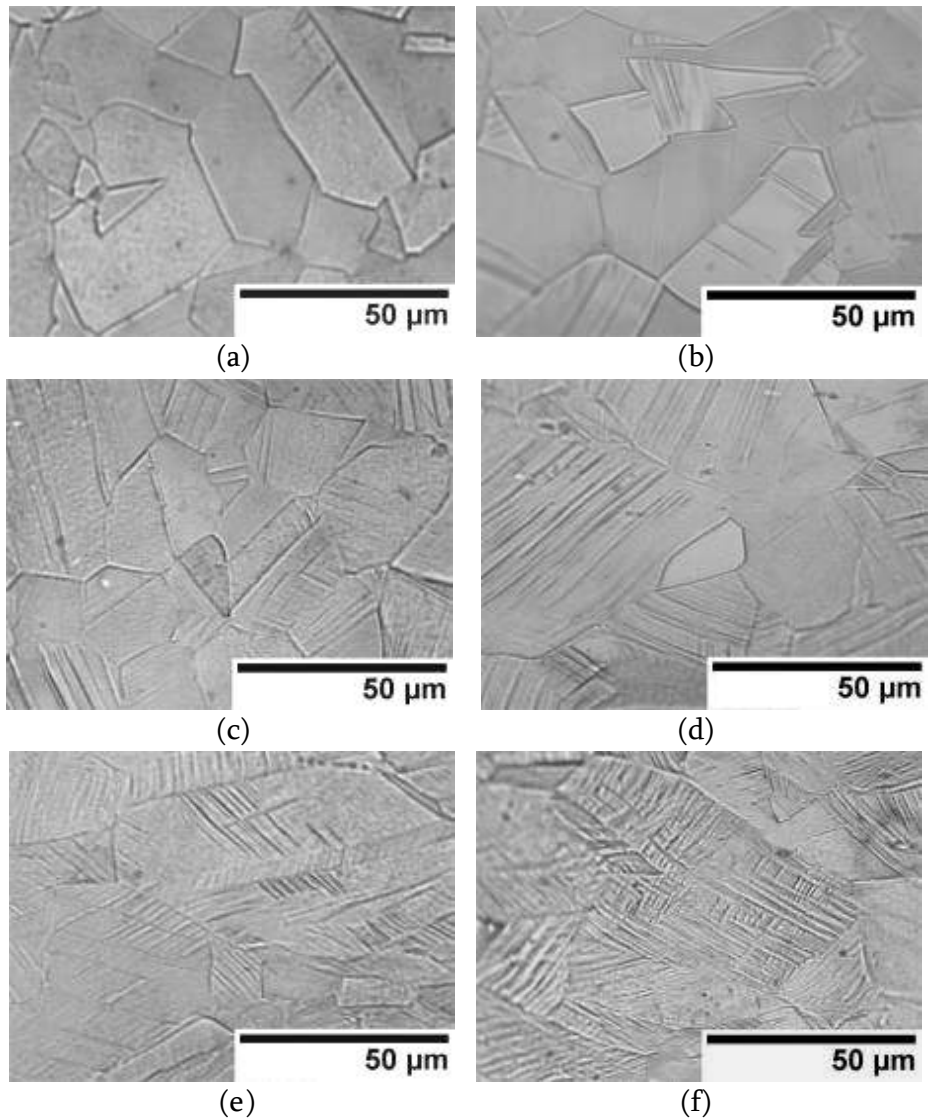

Figure 15 : Micrographies de la nuance de référence (taille de grain environ 20 µm) en microscopie optique après attaque électrolytique. Les échantillons ont été déformés de (a) 4.8 %, (b) 9.5 %, (c) 13.9 %, (d) 18.2 %, (e) 26.3 % et (f) 33.6 % respectivement. La direction de traction est horizontale [ALLAIN 2004_1].

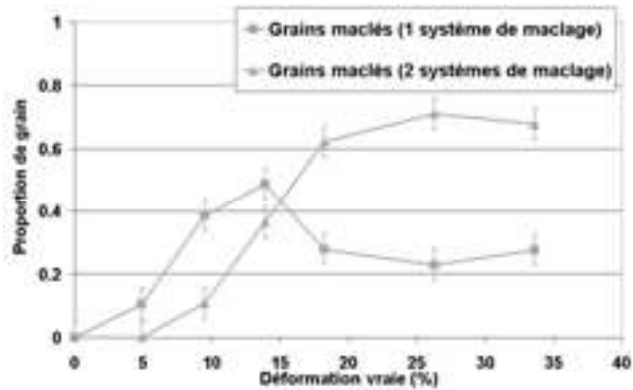

Figure 16: Evolution de la proportion de grains maclés par un ou deux systèmes de maclage mesurée en microscopie optique sur la nuance de référence (taille de grain environ 20 µm) [ALLAIN 2004_1].

Dès 2004, grâce à une étude en EBSD [ALLAIN 2004_1], nous avons vérifié statistiquement pour un grain donné que les macles du système primaire apparaissaient sur le plan (111) ayant le système de dislocations partielles de Shockley avec le meilleur facteur de Schmid par rapport à la direction de sollicitation (par abus de langage, on parlera dans la suite du facteur de Schmid du système de maclage). Ce résultat a été approfondi dans le cadre de la thèse de David Barbier [BARBIER 2009_1][BARBIER 2009_3]. Cette étude confirme que la sélection des systèmes de maclage dans un grain dépend bien d'une loi de Schmid et que par conséquent la texture de déformation joue un rôle important sur l'évolution de la fraction de macle en fonction de la déformation. Par abus de langage, nous parlerons de cinétique de maclage dans la suite de ce mémoire.

La traction suivant un axe DT (Direction Transverse) de la nuance de référence produit une texture très prononcée caractérisée par quatre composantes principales de déformation ; laiton {110}<112>, cuivre tournée {112}<110> appartenant à la fibre <111>//DT, Goss tournée {110}<110> et cube {001}<100> appartenant à la fibre <100>//DT. Ces deux dernière fibres sont particulièrement visibles (grains en bleu et en rouge) sur la Figure 17 après 30 % de déformation. Ces différentes composantes se mettent en place dès les premiers pourcents de déformation et se renforcent au cours de la déformation progressivement, aux dépens des orientations Goss et Cuivre présentes initialement (absence de texture sur la nuance de référence après traitement de recristallisation). Cette texture de déformation est typique des aciers austénitiques à faible EDE.

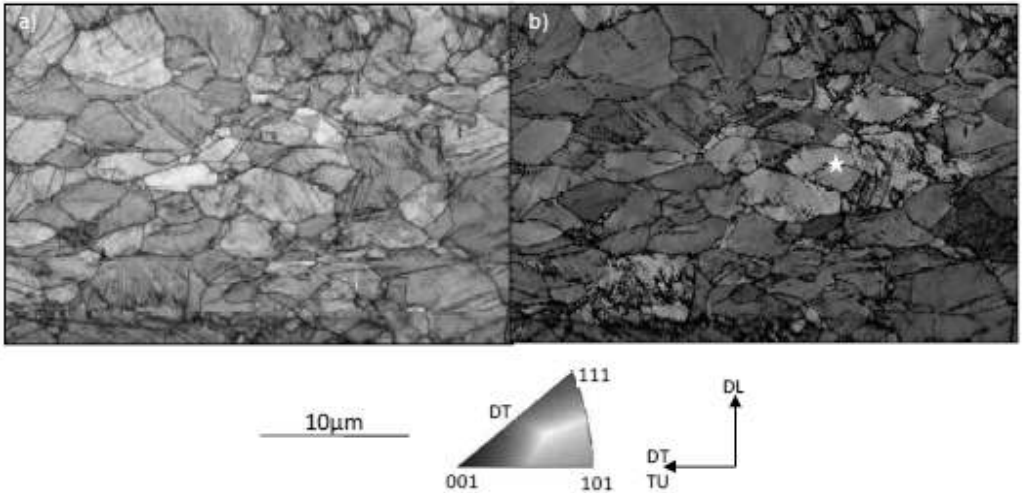

Figure 17 : Cartographie EBSD de la nuance de référence après 30% de déformation vraie (a) en contraste de bandes (b) en orientation selon les couleurs du triangle standard. Mise en évidence des fibres <111>//DT (en bleu) et <100>//DT (en rouge) [BARBIER 2009_1].

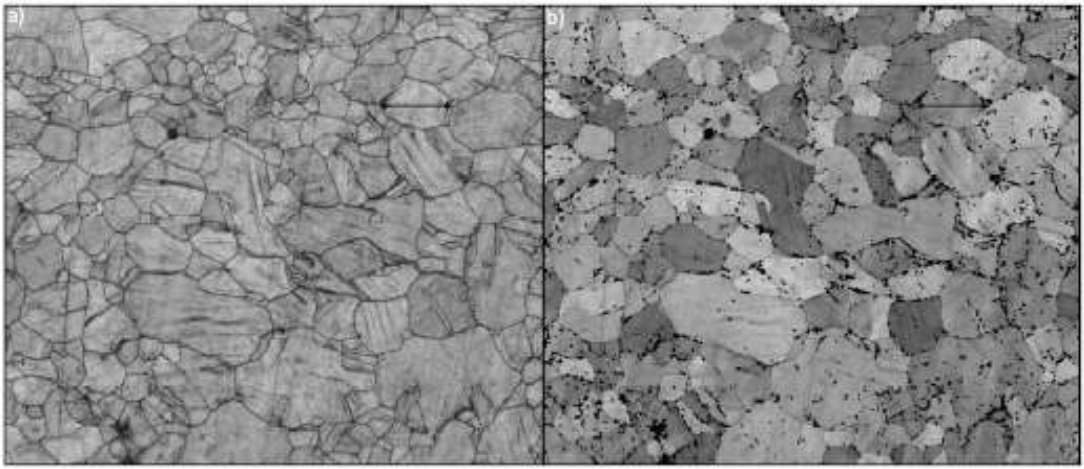

Figure 18 : Cartographie EBSD de la nuance de référence après 30% de déformation vraie (a) en contraste de bandes (b) en orientation selon les couleurs du triangle standard. La texture est alors peu prononcée et 50% des grains sont déjà maclés comme le montre (a) en particulier ceux de la fibre <101>//DT [BARBIER 2009_1].

Dans la nuance de référence, les premières macles apparaissent très tôt au cours de la déformation. L'activation des macles dépend alors de deux facteurs, l'orientation cristallographique et la taille de grain. Les grains ayant une orientation proche des composantes Goss et Cuivre (axe <101>//DT, en vert sur la Figure 18) et une taille suffisante (>5µm) présentent des macles mécaniques. Cette fibre présente des facteurs de Schmid favorable au maclage mécanique comme le montre la Figure 19 [BARBIER 2009_1][SATO 2011].

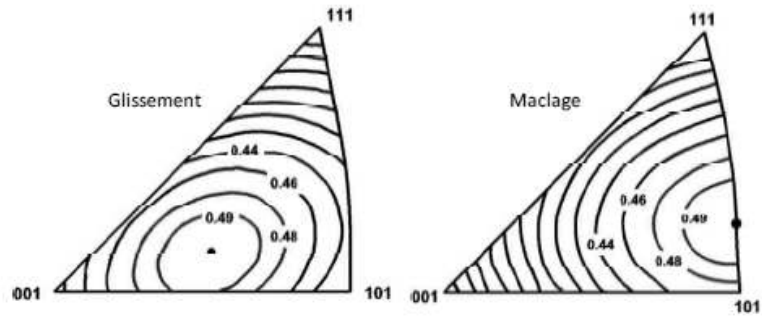

Figure 19 : Figures de pôles inverses des facteurs de Schmid en traction pour le glissement et le maclage. Les lignes d'iso-intensité <0.44 varient au pas de 0.02 (d'après [BARBIER 2009_1]).

Ces deux dernières composantes disparaissent au cours de la déformation au delà de 30 % de déformation. Par contre, les grains appartenant aux orientations principales de la fibre <111>//DT, possèdent également des facteurs d'orientations plus favorables au maclage qu'au glissement cristallographique. Ainsi, l'intensification de la fibre <111>//DT sera plutôt favorable au maclage contrairement à l'intensification de la fibre <100>//DT.

Les orientations cristallographiques locales influencent donc la formation de la microstructure de maclage. Réciproquement, le maclage génère de nouvelles orientations cristallographiques susceptibles de modifier la texture de déformation. Toutefois, les macles formées dans les grains de la fibre <111>//DT ou <110>//DT contribuent à renforcer la fibre <100>//DT et inversement, celles formées dans les grains de la fibre <100>//DT renforcent la fibre <111>//DT. Ce transfert entre fibres préexistantes n'ajoute pas de nouvelles fibres au cours de la déformation. Les aciers TWIP en traction ne développent donc pas de texture particulière par rapport aux aciers austénitiques stables !

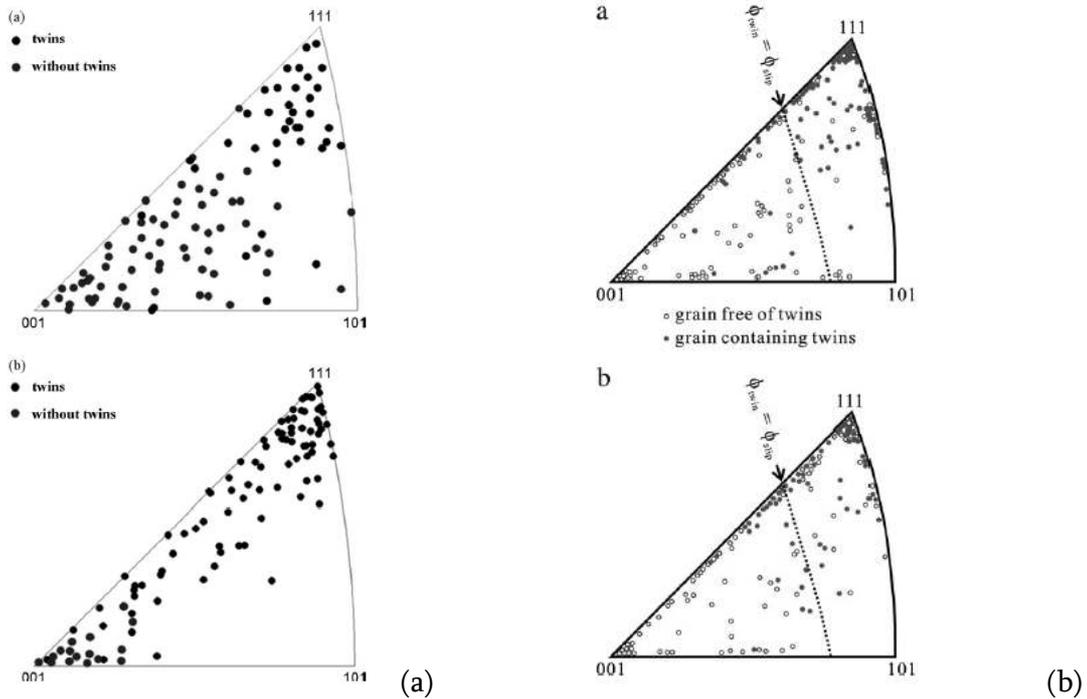

Figure 20: (a) Figures de pôle inverse pour l'axe de traction de la nuance de référence après 5 et 30% de déformation vraie montrant les grains maclés (points noirs) et non maclés (en rouge) [GUTIERREZ 2010]. (b) Figures équivalentes pour 2 aciers TWIP après 20% déformation [SATO 2010].

Ces différentes constatations permettent d'expliquer parfaitement les observations rapportées systématiquement dans la littérature montrant qu'après déformation, les grains non maclés généralement présentent une orientation <100>//DT (en rouge) et les grains maclés une orientation <111>//DT (en bleu) sur les cartographies EBSD précédentes. Cette répartition particulière intervient très tôt au cours de la déformation (dès 5% de déformation comme le montre la Figure 20(a)) [GUTIERREZ 2010][SATO 2010]. Ce résultat suggère que le maclage est bien un mécanisme supplétif au glissement car dans l'orientation <111>//DT le facteur de Schmid du glissement est très faible (0,3) du même ordre de grandeur que celui du maclage. C'est dans cette fibre en particulier que les deux mécanismes sont susceptibles de rentrer en compétition. Si les grains sont bien orientés vis-à-vis du glissement, l'acier ne semble pas avoir besoin de macler pour assurer la vitesse de déformation imposée. Nous reviendrons sur cette compétition probable entre glissement et maclage mécanique au chapitre 2.4.3.3 page 88.

Cette discussion détaillée dans le cas de la traction uniaxiale s'applique aussi à d'autres modes de chargement [BARBIER 2009_1]. La structuration du maclage, les fractions et textures résultantes vont donc beaucoup dépendre du mode de chargement (cisaillement, traction, expansion…). Nous discuterons de ces implications dans la suite de ce mémoire.

### 2.2.4. Discussion

#### 2.2.4.1. *Notion de contrainte critique de maclage*

Le fait que l'apparition des systèmes de maclage primaires puis secondaires suit une loi de Schmid conforte nécessairement l'existence d'une cission résolue critique pour le maclage mécanique, i.e. une cission seuil pour la germination des macles, aussi appelée contrainte critique de maclage [GUTIERREZ 2010] [BRACKE 2009] [CHRISTIAN 1995] [KARAMAN 2000_1] [MEYERS 2001].

Cette contrainte critique pour l'apparition du maclage a été observée et mesurée aussi bien dans des mono que des polycristaux de différents éléments purs ou alliages métalliques. Cette valeur dépend au premier ordre de l'EDE pour les austénitiques mais peut dépendre aussi du type de charge (traction/compression), de la température, de la pré-déformation, de la taille de grain ou de la vitesse de déformation. Une récente revue de ce sujet a été publiée par Meyers *et al.* [MEYERS 2001].

Si l'on considère un mécanisme générique de formation d'une micromacle reposant sur la cission nécessaire pour émettre une première boucle de dislocation partielle de Shockley traînant un défaut d'empilement, la cission critique de maclage s'exprime sous la forme suivante :

$$\tau^{C-int} = \frac{EDE_{int}}{b_{112}} + \frac{\mu b_{112}}{R_C} \qquad (6)$$

avec $R_c$ un rayon critique d'émission de boucle de dislocation [LUBENETS 1985][CHRISTIAN 1969]. Les dislocations suivantes constitutives de la micromacle ne ressentent pas la création du défaut et sont donc émises rapidement à la suite. C'est ce processus qui a été utilisé dans le modèle de DDD décrit ci-dessus. En considérant une EDE de 20 mJ.m$^{-2}$, un rayon Rc de 50 nm, et un facteur de Taylor de 3, la contrainte critique de maclage calculée est de l'ordre de 1200 MPa, bien supérieure à la contrainte d'écoulement de ces alliages. Ces valeurs de contraintes ne sont atteintes qu'après plus de 30 % de déformation alors qu'un maclage déjà intense est observé dans la nuance de référence à gros grain.

Les valeurs déterminées par cette équation sont donc bien supérieures aux valeurs de contraintes critiques de maclage mesurées à l'échelle macroscopique. Il existe un accord dans la littérature pour suggérer que la germination des macles est bien un processus assisté par des concentrations de contraintes comme des empilements de dislocations permettant localement de dépasser la contrainte critique locale définie ci-dessus. Le premier modèle incluant ces différents mécanismes a été proposé en 1964 par Venables sur la base d'un mécanisme polaire [VENABLES 1964]

$$\left[\frac{1}{3} + \frac{(1-\upsilon)L_{pile}}{1,84\,\mu b_{110}}\tau^{C-twin}\right]\tau^{C-twin} = \frac{EDE_{int}}{b_{112}} \qquad (7)$$

avec $L_{pile}$ la longueur caractéristique des empilement de dislocations parfaites, $\upsilon$ le coefficient de Poisson, $b_{110}$ le vecteur de Burgers des dislocations parfaites. Depuis ce premier modèle, des versions plus sophistiquées ont été proposées se distinguant principalement par le modèle de réactions entre dislocations pour la germination des macles, comme Karaman *et al.* reprenant un modèle par glissement dévié ou Buyn basé sur la simple extension de fautes d'empilement [KARAMAN 2000_1] [MEYERS 2001][GUTIERREZ 2010][BUYN 2003].

Au-delà de son aspect fonctionnel, cette équation suggère la relation profonde qu'il existe entre maclage mécanique et la nature du glissement dans ces structures. Nous reviendrons au chapitre 2.4.3 page 85 de manière détaillée sur cette relation complexe. La question de la germination possible des macles sur d'autres sites privilégiés comme des précipités est discutée en Annexe 6 page 163.

### 2.2.4.2. *Mesure de la fraction maclée*

Il ressort aussi de cette présentation une difficulté centrale et inhérente à l'étude de l'effet TWIP, celle de la mesure de la fraction de phase maclée. Les deux raisons principales en sont :
- la zone reconstruite dans la macle est de structure CFC donc non différentiable de la matrice d'un point de vue cristallographique. La fraction de macle ne peut dont être mesurée par une analyse volumique classique par diffraction des rayons X, comme on pourrait l'envisager dans le cas d'une transformation martensitique $\varepsilon$ ou $\alpha'^{5}$.
- les nanomacles ne peuvent être résolues qu'à l'échelle de la microscopie électronique à transmission (MET). Compte tenu du volume des zones analysées par cette technique, on peut mettre en doute rapidement sa représentativité statistique pour estimer de façon fiable des fractions. Aucunes des techniques à plus grandes échelles comme le MEB, l'EBSD ou pire encore la MO qui pourraient permettre une meilleure statistique ne peuvent garantir une résolution suffisante permettant de distinguer les nanomacles de leur faisceaux ou de révéler toutes les micromacles isolées (plus minces que la résolution théorique de la microscopie optique).

---

[5] Cette problématique a été abordée dans le cadre de la thèse de J.L. Collet à l'ESRF de Grenoble financée par AM [COLLET 2009]. Ces travaux ont confirmé l'intérêt des rayonnements X à haute énergie pour l'étude du maclage dans ces aciers. Toutefois, les résultats sont soumis à interprétation par le biais d'un modèle d'interaction entre rayonnement et défauts d'empilement. En conséquence, les cinétiques de maclage « mesurées » sont donc très dépendantes de l'ajustement de ces multiples paramètres et ne peuvent donc être considérées comme absolues.

Pour pallier cette difficulté d'ordre expérimentale, plusieurs méthodes alternatives indirectes ont été développées au cours de nos travaux. La première consiste à utiliser la relation stéréologique de Fullman [FULLMAN 1953], relation essentielle pour la compréhension de ces aciers, qui relie la distance moyenne entre macles t, leur épaisseur moyenne e et la fraction maclée F.

$$\frac{1}{t} = \frac{1}{2e}\frac{F}{(1-F)} \tag{8}$$

Cette distance moyenne entre macles t peut être déduite de la mesure expérimentale L du libre parcours moyen dans la microstructure, mesure intégrant joints de macles et de grains (D la taille de grain efficace).

$$\frac{1}{L} = \frac{1}{D} + \frac{1}{t} \tag{9}$$

Ces deux relations permettent d'estimer la fraction maclée F en fonction de la mesure expérimentale de L (par la technique des interceptes par exemple) et de e (en MET).

$$F = \frac{1}{1 + \frac{1}{2e}\left(\frac{1}{D} - \frac{1}{L}\right)} \tag{10}$$

Cette relation peut être affinée en tenant compte de la structuration en faisceaux du maclage [ALLAIN 2004_1].

$$F = \left(\frac{e}{e + \lambda_{faisceau}}\right)\frac{1}{\left(\frac{LD}{2N_{faisceau}(e + \lambda_{faisceau})(L-D)}\right) + 1} \tag{11}$$

Avec $N_{faisceau}$ le nombre moyen de micromacles par faisceaux et $\lambda_{faisceau}$ l'espacement inter-macles moyen dans les faisceaux. Appliquée au cas de la nuance de référence, cette relation permet de confirmer que la fraction de phase maclée reste faible même après 50 % de déformation (environ 12%) (cf. Figure 21).

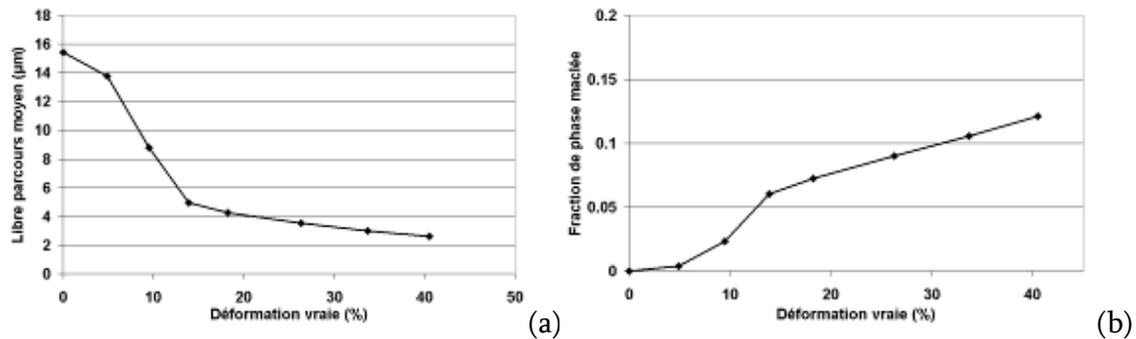

Figure 21 : (a) Evolution du libre parcours moyen L par la méthode des interceptes au cours de la déformation en traction de la nuance de référence (mesures en MO après attaque) (b) Evolution de la fraction de macles déduite de l'équation (11) [ALLAIN 2004_1].

Depuis maintenant une dizaine d'année, la technique EBSD en MEB-FEG a permis de faire des progrès dans la mesure des fractions maclées grâce à des temps d'acquisition raisonnables. Cette technique permet d'analyser de grandes plages en conservant des distances entre pas d'analyse inférieures à 100 nm, progrès auquel nous avons pu contribuer [BARBIER 2009_1][BARBIER 2009_2]. Toutes les macles ne peuvent être indexées compte tenu de la résolution ou tout simplement révélées, rendant inadéquate l'utilisation directe des informations concernant les orientations cristallographiques. Pour contourner cette difficulté, Renard a utilisé les informations supplémentaires contenue dans les cartographies de contrastes de bandes. Les difficultés locales d'indexation sont utilisées alors comme marqueurs de la présence de macles [RENARD 2012]. Cette méthode reste bien sûr plus précise que des mesures en microscopie optique et élimine tous les biais liés à l'attaque métallographique, mais est toujours imparfaite.

En jouant sur cette difficulté d'indexation des nanomacles par EBSD, Barbier a proposé une méthode de mesure tout à fait originale et automatisable. Cette technique repose sur les constats suivants :
- l'EBSD permet de déterminer l'orientation des grains mais ne permet pas de résoudre les macles. Pour une orientation donnée, seules les matrices non maclées participeront à la mesure.
- L'analyse des densités de distribution d'orientations obtenues par DRX intègre les contributions des matrices et des macles de façon indifférenciées.
- Pour une orientation donnée, la fraction de phase maclée est donc la différence entre la mesure par DRX (matrice+macle) $F_{DRX}$ et par EBSD (matrice uniquement) $F_{EBSD}$.
- L'augmentation de la fibre <100>//DT est principalement due au maclage au cours de la déformation

On peut donc alors écrire

$$F = F_{DRX}^{<100>//DT} - F_{EBSD}^{<100>//DT} \qquad (12)$$

La validité de l'approche pourrait être améliorée en intégrant ces constats sur de nombreuses orientations. En l'état, les mesures pour l'acier de référence et la seule orientation <100> //DT sont représentées pour différents niveaux de déformation sur la Figure 22. Ces valeurs restent raisonnablement proche des résultats de l'autre mesure indirecte présentés sur la Figure 21(b).

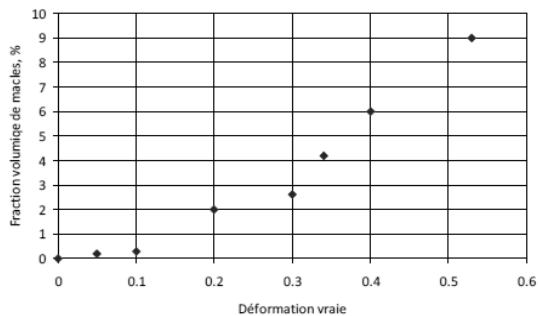

Figure 22 : Evolution de la fraction volumique de macle en fonction de la déformation vraie estimée grâce à l'équation (12) sur la nuance de référence déformée en traction [BARBIER 2009_1].

Malgré un constat d'échec partagé dans la littérature, voire une certaine résignation, peu de travaux ont été dédiés spécifiquement à cette question et l'on peut s'en étonner compte tenu des enjeux. Il serait probablement intéressant de redécouvrir des techniques comme la microscopie à force atomique pour y répondre. Cette technique utilisée pour l'étude du maclage par Coupeau *et al.* [COUPEAU 1999] possède une résolution spatiale suffisante pour résoudre la présence de micromacles incidente à une surface polie. Toutefois, il est indispensable de connaître l'orientation cristallographique du grain considéré pour définir de manière univoque l'épaisseur de la macle et son degré de perfection (fautes résiduelles). Il serait donc intéressant de coupler cette technique avec une mesure préalable par EBSD des orientations des grains analysés.

### 2.2.5. Conclusions intermédiaires

La morphogénèse de la microstructure de maclage dans les aciers TWIP fait maintenant l'objet d'un large consensus dans la littérature scientifique, auquel nous avons contribué, en particulier sur les aspects suivants :
- l'identification des nanomacles, leur taille et organisation en faisceaux.
- L'activation séquentielle de différents systèmes de maclage suivant une loi de Schmid et ses conséquences sur la texture de déformation.

On notera paradoxalement que les progrès accomplis dans le domaine de la mesure de la fraction de phase maclée restent marginaux. En conséquence les modèles de cinétique de maclage depuis la thèse de Rémy ont peu progressé en l'absence de données expérimentales fiables. Cette évolution de fraction de phase maclée reste souvent le dernier paramètre d'ajustement des modèles structure-propriétés dont nous allons maintenant discuter.

### 2.3. L'effet TWIP : La relation entre maclage mécanique et comportement

Compte tenu de son fort potentiel pour le marché automobile, il existe une forte attente des « mécaniciens » pour comprendre les spécificités du comportement mécanique de ces alliages, en particulier leurs très forts taux d'écrouissage, de nature principalement cinématique ou la forme particulière de leurs surfaces de charge. Pour répondre à ce besoin, nous avons développé des lois de comportement à base physique pour la mise en forme et l'usage de ces aciers (comportement à haute vitesse de déformation par exemple pour des applications en conditions de crash), en prenant en compte les principaux mécanismes élémentaires de la plasticité, glissement et maclage.

Ces travaux nous ont naturellement amenés à considérer des chargements de plus en plus complexes pour se rapprocher des conditions d'utilisation réalistes (des essais de traction aux essais d'emboutissage en passant par les changements de trajets thermomécaniques). Nous mettrons en évidence dans ce mémoire l'origine et les conséquences des contraintes internes dans ces résultats.

Mes travaux de thèse et au-delà ont principalement permis de mieux appréhender et quantifier l'impact du maclage mécanique et son interaction avec le glissement sur le comportement mécanique macroscopique, appelé maintenant communément effet TWIP. Je tiens d'ailleurs ici à rendre hommage aux travaux précurseurs du Professeur L. Rémy dans les années 1970 dans ce domaine, qui m'ont fourni une base précieuse et dont les interprétations restent plus que jamais d'actualité.

Au cours de mes travaux de thèse, j'ai mis en évidence que l'effet TWIP pouvait être interprété comme un mécanisme d'effet « Hall et Petch » dynamique. Nous avons en effet pu relier les propriétés d'écrouissage de ces aciers à la mise en place de la microstructure de maclage, en particulier leur sous-structures nanométriques et l'activation de multiples systèmes non coplanaires au cours de la déformation, créant des défaut planaires infranchissables pour les dislocations et induisant une rapide diminution de leurs libres parcours moyens.

Plus récemment, avec mes collègues, O. Bouaziz et C.Scott, nous avons développé un modèle micromécanique basé sur les travaux de Sinclair *et al.* (extension du modèle de Mecking-Kocks-Estrin) [SINCLAIR 2006] intégrant une composante cinématique sensible aux effets de taille et une description phénoménologique de la microstructure de maclage. Ce nouveau modèle a permis de décrire de façon complète non seulement les effets de tailles de grain sur le comportement en traction mais aussi la composante cinématique de l'écrouissage de ces aciers, associée à la microstructure de maclage. Ces travaux ont ensuite été intégrés comme un prolongement naturel dans le cadre de la thèse de D. Barbier.

Dans cette partie, nous considérerons implicitement que seul le maclage mécanique contribue au durcissement des aciers TWIP FeMnC. Les autres mécanismes potentiels de durcissement comme le vieillissement dynamique seront discutés au chapitre suivant. Nous utiliserons les notions classiques d'essai Bauschinger, d'écrouissage cinématique ou isotrope et de surface de charge. Toutefois, nous ne reviendrons pas sur ces concepts. Le lecteur pourra se référer par exemple au mémoire d'Habilitation à Diriger des Recherches d'O. Bouaziz pour plus de détails [BOUAZIZ 2005].

### 2.3.1. Premier modèle de l'effet TWIP : Rôle précurseur de L. Remy

La première modélisation du rôle du maclage mécanique sur l'écrouissage de ces aciers a été proposée par Rémy en 1978 [REMY 1978]. Afin de rendre compte de l'effet des larges empilements de dislocations sur les joints de macles, il propose par analogie avec une approche de type « Hall et Petch » :

$$\Sigma(E) = \sigma_{matrice}(\varepsilon_{matrice}) + K_T \left(\frac{1}{t}\right)^{r_T} \tag{13}$$

avec $\Sigma$ et $E$ les contraintes et déformations macroscopiques, $\sigma_{matrice}$ et $\varepsilon_{matrice}$ les contraintes et déformations de la matrice sans macle mécanique, $t$ la distance moyenne entre macles sur lesquelles se stockent les dislocations et $K_T$ et $r_T$ deux constantes. $r_T$ pouvant prendre classiquement des valeurs entre 1 et ½, en fonction du mécanisme d'écrouissage considéré. La distance moyenne entre macle $t$ est alors évaluée grâce à la relation de Fullman (cf. équation (10)).

Afin d'estimer le terme $K_T$, Rémy propose une autre modélisation de la contrainte basée sur l'équilibre à grandes distances entre contraintes en retour d'un empilement de $n_T$ dislocations et la contrainte appliqué $\tau_{app}$ de façon analogique à l'approche de Friedel [FRIEDEL 1964] présentée au chapitre 2.2.2 page 28 :

$$\tau_{app} = n_T \frac{\mu b_{110}}{2} \frac{1}{t} \Rightarrow \frac{(\Sigma - \sigma_{matrice})}{M} = n_T \frac{\mu b_{110}}{2} \frac{1}{t} \tag{14}$$

Avec M un facteur de Taylor. Cette relation conduit alors à la relation suivante :

$$\Sigma(E) = \sigma_{matrice}(\varepsilon_{matrice}) + n_T \frac{M\mu b_{110}}{2}\left(\frac{1}{t}\right)^{r_T} \tag{15}$$

Dans les travaux de Rémy sur les alliages Ni-Co, $n_T$ prend des valeurs relativement grandes de l'ordre de 30-50 dislocations. Stricto sensu, cette équation est un modèle de durcissement intracristallin et ne permet pas de prendre en compte de façon paradoxale l'effet de taille de grain initial. Par contre, en considérant le champ de contraintes à longue distance des empilements de dislocations, ce modèle d'écrouissage est le premier d'essence purement cinématique.

### 2.3.2. Premières extensions du modèle Mecking-Kocks-Estrin

A la suite de ces travaux précurseurs, Karaman *et al.* [KARAMAN 2000_2] et Bouaziz et Guelton [BOUAZIZ 2001_1] ont proposé au début des années 2000, des approches de l'effet TWIP considérant une extension du modèle de Mecking-Kocks-Estrin [MECKING 1981][ESTRIN 1984]. La nature de ces approches est purement isotrope. Le maclage mécanique est supposé réduire le libre parcours moyen des dislocations et donc favoriser l'augmentation de la densité de dislocations statistiquement stockées (DSS).

$$\frac{d\rho_{DSS}}{d\varepsilon_M} = M.\left(\frac{1}{b_{110}L} + \frac{k}{b_{110}}\sqrt{\rho_{DSS}} - f.\rho_{DSS}\right) \tag{16}$$

Avec L donné par

$$\frac{1}{L} = \frac{1}{D} + \frac{1}{2e}\frac{F}{(1-F)} \tag{17}$$

Avec $\rho_{DSS}$ la densité de DSS, $\varepsilon_M$ la déformation plastique dans la matrice non maclée, k et f sont deux paramètres caractérisant respectivement un mécanisme d'écrouissage latent et un mécanisme de restauration dynamique.

La contrainte d'écoulement est alors déduite d'un modèle de Taylor (durcissement de la forêt de DSS) et s'écrit :

$$\Sigma = \sigma_0 + \alpha M\mu b_{110}\sqrt{\rho_{DSS}} \tag{18}$$

avec $\alpha$ une constante (durcissement de la forêt), M un factor de Taylor moyen et $\sigma_0$ une contrainte de friction dépendant uniquement de la solution solide. On peut prendre en compte dans ces approches la contribution du maclage à la déformation macroscopique :

$$d\Gamma = (1-F).d\gamma_M + \gamma_T.dF \qquad (19)$$

avec $\Gamma$ le cisaillement macroscopique, $\gamma_M$ le cisaillement dû au glissement des dislocations, $\gamma_T$ le cisaillement produit par une macle et F la fraction de macles. $\gamma_T$ est une constante qui dépend uniquement de la nature des macles (cf. chapitre 2.2.1.1 page 22).

Cette dernière équation permet de déterminer la déformation totale en introduisant le facteur de Taylor. La contribution totale des macles à la déformation en traction $\varepsilon_T$ peut donc s'écrire :

$$\varepsilon_T = \frac{F\gamma_T}{M} \qquad (20)$$

Si l'on considère une fraction de phase maclée de l'ordre de 10% (soit environ la fraction de phase maclée de la nuance de référence vers 50% de déformation – cf. Figure 21), cette contribution vaut alors à peine 2%. Ce calcul d'ordre de grandeur montre pourquoi ce terme est souvent négligé dans les modèles micromécaniques les moins élaborés.

### 2.3.3. Extension polycristalline

A l'instar de Karaman *et al.* [KARAMAN 2001_2], nous avons étendu ce modèle isotrope dans le cadre de la plasticité polycristalline en considérant que les équations précédentes étaient applicables pour tous les systèmes de glissement d'un grain [ALLAIN 2002][ALLAIN 2004_1][ALLAIN 2004_2]. Nous avons introduits deux originalités :
- la matrice décrivant les interactions latentes entre les systèmes de glissement (matrice de type Franciosi) s'est révélée insuffisante pour décrire l'écrouissage dû à l'effet TWIP. Nous avons donc rajouté un terme supplémentaire correspondant à un terme d'auto-écrouissage des systèmes de glissement. Nous avons qualifié ce terme de contribution « d'Hall et Petch » (HP) car il était aussi nécessaire pour décrire les effets de taille de grain et correspondait dans l'esprit à une contrainte en retour due à des empilements de dislocations (back-stress). Nous n'avons pas fait à l'époque explicitement le lien avec une contribution de nature cinématique.
- une matrice d'intersections entre les systèmes de maclage et de glissement, qui permet de calculer la réduction du libre parcours moyen d'un système de glissement donné seulement par les micromacles sécantes à ce système.

Ces différentes lois ont été intégrées dans un schéma polycristallin viscoplastique uniaxiale selon trois modèles de transition d'échelle simples, Taylor relaxé (iso-déformation entre grains), Sachs (iso-contrainte) ou isoW (iso-travail dissipé entre les grains) [BOUAZIZ 2002].

Un modèle de cinétique de maclage empirique mais basé sur la notion de contrainte critique de maclage a été introduit pour chaque système de maclage.

La Figure 23 montre les courbes de traction prédites par le modèle avec 100 grains (orientations aléatoires) selon les trois modèles de transition d'échelle retenus. Les deux hypothèses polycristallines les plus réalistes donnent des résultats satisfaisants et permettent de décrire assez finement l'évolution du taux d'écrouissage.

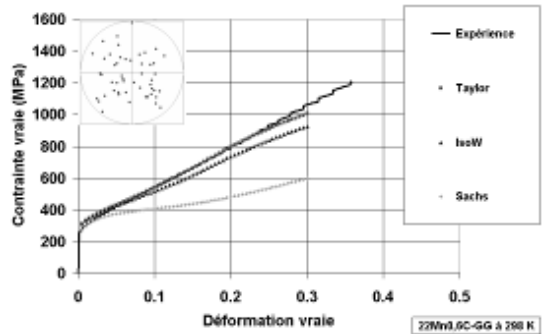

Figure 23 : Courbes de traction expérimentales de la nuance de référence à 298 K et simulées selon les trois modèles de transition d'échelles [ALLAIN 2004_1].

Cette approche est d'autant plus intéressante qu'elle permet de décrire avec un bon accord les observations expérimentale de l'évolution des microstructures, comme le montre la Figure 24. La Figure 24(a) montre l'évolution du libre parcours moyen dans la microstructure pour des dislocations entre obstacles forts (joints de grains et de macles) et la Figure 24(b) la proportion de grains maclés par 1 ou 2 systèmes.

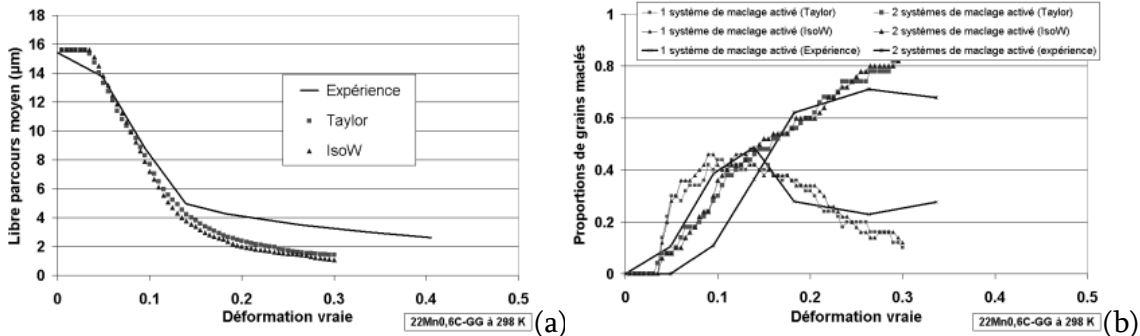

Figure 24 : (a) Evolution du libre parcours moyen dans la microstructure pour des dislocations entre obstacles forts (joints de grains et de macles) et (b) proportion de grains maclés par 1 ou 2 systèmes de maclage en fonction de la déformation vraie prévues par le modèle selon deux modèles de transition d'échelle [ALLAIN 2004_1].

Cette description de la microstructure de maclage et de la réduction du libre parcours moyen des dislocations par les macles en plasticité cristalline a ensuite été utilisée dans les modélisations micromécaniques viscoplastiques des thèses de N. Shiehkelshouk [SHIEKHELSOUK 2006] et de D. Barbier [BARBIER 2009_1].

### 2.3.4. Distinction et couplage des composantes cinématiques et isotropes

La principale limitation des modèles décrits précédemment et développés avant 2008 par notre équipe est de prévoir un écrouissage de nature principalement isotrope (excepté la dispersion de contraintes d'écoulement entre les grains dans le modèle de plasticité polycristalline). Même si le terme de type HP correspond à un mécanisme de durcissement cinématique, il n'a jamais été exploité comme tel. Ces premiers modèles, y compris les extensions en plasticité cristalline [BARBIER 2009_1], ont donc été mis en défaut par les premières mesures d'effet Bauschinger [BOUAZIZ 2008_1]. Depuis, ces mesures ont été confirmées plusieurs fois en particulier par l'équipe du CEIT [SEVILLANO 2009][SEVILLANO 2012] montrant que l'écrouissage des aciers TWIP est principalement d'origine cinématique, comme envisagé initialement par Rémy.

Le premier modèle incluant les deux composantes d'écrouissage a été proposé par notre équipe en 2008 afin d'expliquer le fort effet Bauschinger de ces aciers, mais aussi de bien reproduire simultanément le simple effet « Hall et Petch » (évolution de la limite d'élasticité avec la taille de grain austénitique) de ces nuances, généralement mal capté par tous les modèles précédents (y compris celui de Rémy).

Le principal apport de ce modèle par rapport au modèle de Bouaziz et Guelton est d'introduire explicitement la notion de dislocations stockées sur les obstacles forts de la structure (dislocations géométriquement nécessaires DGN) et de définir la contrainte en retour résultante.

Cette contrainte en retour $\sigma_b$ s'écrit de façon très analogue à la proposition de Rémy (cf. équation (14)) :

$$\sigma_b = M \frac{\mu . b_{110}}{L} n \qquad (21)$$

avec n le nombre moyen de dislocations mobiles stoppées sur des joints de grains ou de macles (L tient compte des deux types d'obstacles forts)[6]. Cette approche est justifiée par les observations par EBSD des très forts gradients locaux de déformation observés par D. Barbier sur les structures maclées. La Figure 25 montre par exemple les gradients cumulés et locaux observés dans un grain maclé de la nuance de référence après seulement 20% de déformation.

---

6 Ce modèle repose aussi sur l'hypothèse que les joints de macles sont instantanément saturés par des DGN. Si ce n'était pas le cas, les nouvelles macles créées ne contribueraient pas autant au durcissement cinématique.

D'un bout à l'autre du grain, la désorientation varie de 12° et localement près des macles de 3° sur des distances très faibles.

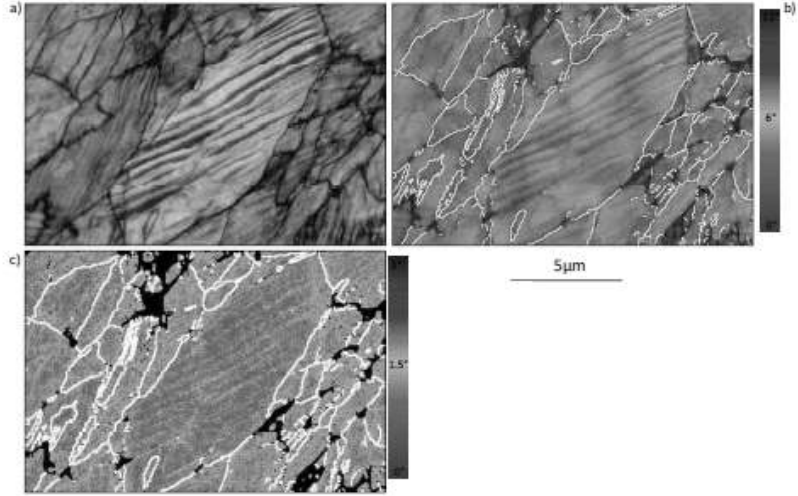

Figure 25 : Cartographie EBSD des désorientations intragranulaires de la nuance de référence déformée de 20% en traction (a) en contraste de bandes (b) en gradient d'orientations à longue distance par rapport à la moyenne du grain (c) en gradient d'orientations à courte distance (désorientation angulaire entre pixels voisins) [BARBIER 2009_1].

Le flux de dislocations arrivant sur les joints par bandes de glissement s'écrit :

$$\frac{dn}{d\varepsilon_{matrice}} = \frac{\lambda}{b_{110}}\left(1 - \frac{n}{n_0}\right) \tag{22}$$

avec $\lambda$ l'espace moyen entre bandes de glissement, $n_0$ le nombre maximum de boucles de dislocations arrêtées sur les joints de grain et de macle. Le rapport $\lambda/b_{110}$ représente le nombre de dislocations par bandes de glissement nécessaire pour fournir la déformation. Au-delà de $n_0$ les joints de grains ne stockent plus de DGN car des mécanismes de relaxation sont activés, comme l'émission de dislocations derrière l'obstacle ou l'activation de glissement sur des plans différents. L'équation de MKE utilisée par Bouaziz et Guelton [BOUAZIZ 2001_1] est alors modifiée pour tenir compte de la saturation et de la relaxation possible des joints. La densité peut augmenter localement mais le vecteur de Burgers total résultant reste constant. Cette nouvelle équation est directement inspirée des travaux de Sinclair *et al.* [SINCLAIR 2006][BOUAZIZ 2008_2] :

$$\frac{d\rho_{DSS}}{d\varepsilon_{matrice}} = M.\left(\frac{\left(1 - n/n_0\right)}{b_{110}.L} + \frac{k}{b_{110}}\sqrt{\rho_{DSS}} - f.\rho_{DSS}\right) \tag{23}$$

Par extension de l'équation (18), la contrainte d'écoulement macroscopique s'écrit :

$$\Sigma = \sigma_0 + \sigma_F + \sigma_b = \sigma_0 + \alpha M \mu b_{110} \sqrt{\rho_{DSS}} + M \frac{\mu \cdot b_{110}}{L} n \quad (24)$$

La cinétique de maclage dans ce modèle est encore choisie de manière empirique, en l'absence de modèle à base physique dans la littérature :

$$\text{pour } \varepsilon > \varepsilon_{init}, \quad F = F_0 \left(1 - e^{-\beta_T (\varepsilon - \varepsilon_{init})}\right)^{m_T} \quad (25)$$

Ou $\varepsilon_{init}$, $F_0$, $\beta_T$ et $m_T$ sont 4 paramètres à identifier. $F_0$ correspond à la fraction totale de phase maclées et $\varepsilon_{init}$ une déformation minimum avant l'apparition des premières macles.

Cette approche après ajustement des cinétiques de maclages permet de décrire simultanément :
- les courbes de comportement en traction pour différentes tailles de grains [SCOTT 2006], avec une bonne description en particulier des limites d'élasticité,
- le comportement estimé de l'alliage sans maclage et pour une grande taille de grain [ALLAIN 2004_1] (cf. chapitre 2.3.6.3 page 65),
- l'écrouissage cinématique en fonction de la déformation qui a la particularité d'être non saturant.

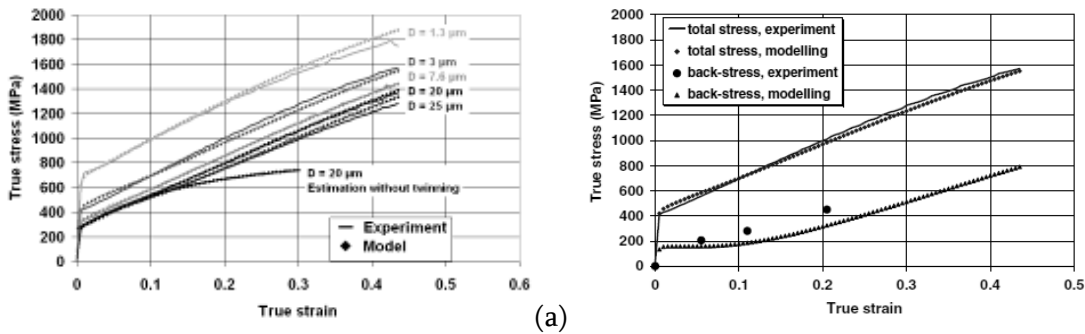

Figure 26 : (a) Courbes de traction expérimentales et simulées pour différentes tailles de grains [SCOTT 2006] et sans effet TWIP [ALLAIN 2004_1] et (b) prédiction de l'écrouissage cinématique en fonction de la déformation – « back-stress » mesurée par des essais Bauschinger en fonction de la pré-déformation [BOUAZIZ 2008_1].

Seul le paramètre $\varepsilon_{init}$ a été ajusté pour prédire individuellement les courbes de la Figure 26. Les 3 autres paramètres du maclage sont pris constants pour tous les aciers. Ce décalage de déformation pour l'apparition d'un maclage effectif est reporté sur la Figure 27 ainsi que la contrainte d'écoulement macroscopique correspondante pour chaque acier étudié. Afin de reproduire le comportement expérimental, le décalage en déformation augmente avec la taille de grain mais la contrainte d'apparition semble constante, de l'ordre de 550 MPa. Ceci est cohérent avec la notion de contrainte critique de maclage.

Toutefois, ce résultat suggère aussi que les macles apparaissent à de plus faibles niveaux de déformation dans les aciers à petits grains. Ce résultat est contradictoire avec les observations de nombreux auteurs, y compris D. Barbier, qui montrent que le maclage est plus susceptible d'apparaitre dans les grains de grande taille, c'est-à-dire ceux dont la contrainte est la moins élevée. Pour cette question, l'élément clef reste la mesure expérimentale de la cinétique de maclage. Ces différents modèles n'amélioreront leurs capacités de prédiction que si des progrès sont faits pour la modélisation à base physique des cinétiques de maclage.

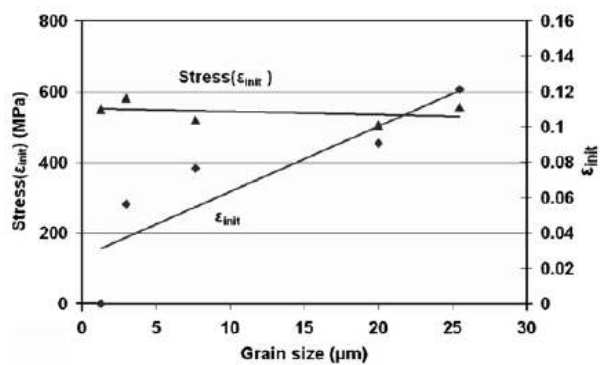

**Figure 27 :** Paramètres $\varepsilon_{init}$ ajustés en fonction de la taille de grains austénitiques et contraintes d'écoulement respectives [BOUAZIZ 2008_1].

Nous reviendrons principalement au chapitre 2.4.3 page 85 sur l'utilisation et le calage de ce modèle sur une base de données d'aciers FeMnC.

### 2.3.5. Approches composites et contraintes dans les macles

#### 2.3.5.1. *Analogie entre les approches*

Sevillano *et al.* [SEVILLANO 2009][SEVILLANO 2012] ont proposé récemment une nouvelle interprétation de l'effet TWIP basée sur une approche « composite »[7]. Cette dernière relève d'une vision « mécanicienne » mais nous allons montrer que cette nouvelle approche et celle développée par notre équipe, découlant d'une vision « plasticienne », sont en fait très proches et duales.

---

[7] Une autre approche introduisant un aspect « composite » a été proposée par Kim *et al.* [KIM 2009]. Cette approche ne vise pas à considérer les macles comme des objets durs dans une matrice molle, mais à considérer deux populations de grains, les grains non maclés et les autres sur-durcis par le maclage mécanique (cf. Figure 20).

Sevillano *et al.* distinguent deux contributions à l'écrouissage de la microstructure de maclage :
- la diminution de la taille de grain effective par les joints de macles conduisant à un sur-durcissement isotrope de la matrice du composite,
- la présence de macles considérées comme les secondes phases « dures » du composite, qui apparaissent dynamiquement au cours de la déformation. Ces secondes phases contribuent à l'écrouissage en supportant des contraintes internes élevées avant relaxation.

Dans le cadre de cette approche, la contrainte d'écoulement macroscopique du composite s'écrit :

$$\Sigma = (1-F)\sigma_M + F\sigma_T \qquad (26)$$

avec $\sigma_M$ et $\sigma_T$ les contraintes dans la matrice et les macles respectivement. Ces deux valeurs permettent de définir les contraintes internes de la manière suivante :

$$\begin{aligned}(\sigma_{int})_T &= \Sigma - \sigma_T \\ (\sigma_{int})_M &= \Sigma - \sigma_M\end{aligned} \qquad (27)$$

La somme des contraintes internes pondérée par les fractions respectives de phase maclée F et de matrice (1-F) est bien entendu nulle.

$$F(\sigma_{int})_T + (1-F)(\sigma_{int})_M = 0 \qquad (28)$$

Cette configuration classique est schématisée sur la Figure 28 en reprenant les notations de Sevillano *et al.*.

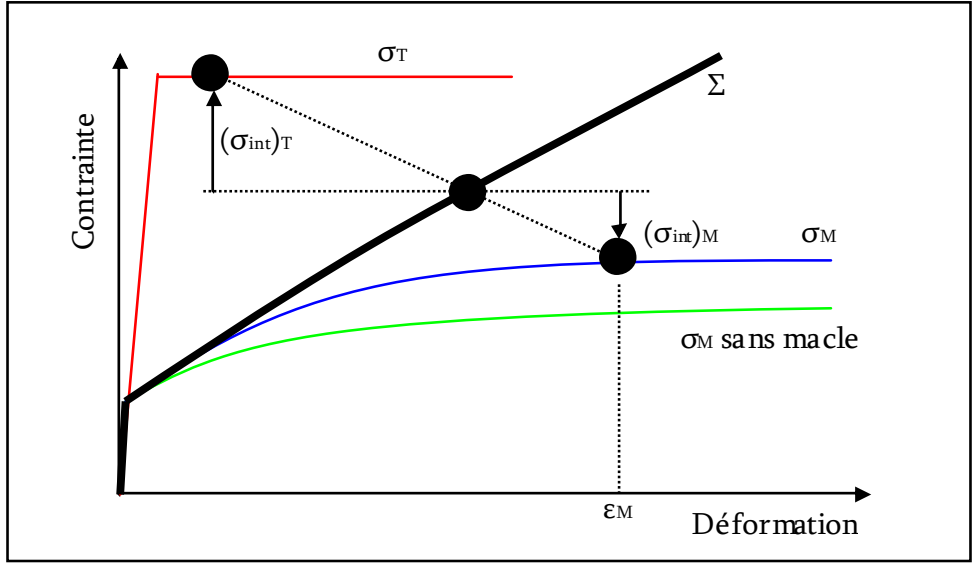

**Figure 28 : Représentation schématique du mécanisme d'écrouissage par les macles selon une approche composite (comportement sans macle, comportement de la matrice durcie par les macles σ$_M$, contraintes dans les macles σ$_T$, contrainte macroscopique Σ).**

Par rapport à l'équation (24), le terme σ$_M$ est tout à fait analogue au terme σ$_0$ + σ$_F$. La seule approximation réalisée à ce stade est (1-F) ≈ 1. Nous allons maintenant montrer que le terme σ$_b$ est en fait équivalent au terme Fσ$_T$.

D'après notre modèle, la contrainte de back-stress σ$_b$ est définie comme :

$$\sigma_b = M \frac{\mu b_{110}}{L} n = M\mu b_{110} \left( \frac{1}{D} + \frac{1}{2e} \frac{F}{(1-F)} \right) n_0 \left( 1 - \exp\left( -\frac{\varepsilon_M \lambda}{n_0 b_{110}} \right) \right) \quad (29)$$

Comme discuté précédemment, les macles sont très rapidement saturées d'empilements de dislocations, c'est-à-dire que le terme exponentiel devient vite négligeable en quelques pourcents de déformation. De même, dès la formation des premières macles, le libre parcours moyen L devient indépendant de D. Ces deux approximations successives donnent alors :

$$\sigma_b = M \frac{\mu b_{110}}{L} n = M\mu b_{110} \left( \frac{1}{2e} \frac{F}{(1-F)} \right) n_0 = F \frac{M\mu b_{110}}{2e} \frac{n_0}{1-F} \approx F \frac{M\mu b_{110}}{2e} n_0 = F\sigma_T^{LB} n_0 \quad (30)$$

Ce terme est très similaire à celui proposé par Sevillano *et al.* du point de vue de la forme et de la signification. Ce terme est proportionnel à F et dépend d'un mécanisme de relaxation des empilements au niveau de la macle, c'est-à-dire faisant intervenir soit le comportement plastique des macles soit des mécanismes de relaxation. Le seul terme additionnel est n$_0$ qui caractérise la longueur maximum des empilements de dislocations sur les macles et représente donc un facteur de concentration de contraintes dans les macles elles-mêmes. Ce terme permet d'expliquer pourquoi Sevillano *et al.* ont estimé des contraintes dans les macles bien supérieures à la borne inférieure $\sigma_T^{LB}$ [SEVILLANO 2012].

En retenant la valeur des paramètres pour la nuance de référence, cette borne inférieure serait de 800 MPa seulement (e = 30 nm, µ = 65 GPa, b = 2.5x10$^{-10}$ m, M = 3), très rapidement inférieure à la contrainte d'écoulement moyenne de l'acier. Cela signifie nécessairement que les mécanismes de relaxation sont plus difficiles à activer que ne le laisse penser cette borne inférieure et que des mécanismes concentrateurs de contraintes sont requis pour activer ce processus. On pourra penser aux très nombreuses dislocations sessiles observées par Idrissi *et al.* [IDRISSI 2009] pour expliquer cette différence. On retrouve en tout cas dans ces deux modèles l'importance de la finesse de macles comme un probable frein pour le mécanisme de relaxation plastique au niveau des macles.

### 2.3.5.2. *Contraintes internes et essais Bauschinger*

Dans notre approche développée au chapitre 2.3.4, nous avons comparé directement les valeurs du back-stress $\sigma_b$ à une composante cinématique d'écrouissage X mesurée par essai Bauschinger (cf. Figure 26(b)). Cette comparaison était clairement motivée par la volonté de se placer dans un cadre mécanique classique de type Lemaitre et Chaboche [LEMAITRE 2004], définissant l'écrouissage macroscopique comme la somme de termes isotrope et cinématique. Par contre, d'un point de vue micromécanique, cette comparaison est abusive.

Il est souvent délicat de relier certaines grandeurs micromécaniques, comme la contrainte dans les macles, à une valeur de contrainte interne unique mesurées par des essais de type Bauschinger (écrouissage cinématique macroscopique X) [FEAUGAS 1999]. En effet, ce type de changement trajet est un processus complexe, qui fait intervenir même dans les cas les plus simples plusieurs mécanismes successivement (relaxation des boucles de dislocations [ALLAIN 2010_3], émission de nouvelles dislocations, …) et surtout un processus transitoire vers un nouvel état stationnaire [ALLAIN 2012] (mis à part peut-être l'effet Bauschinger permanent).

La mesure de limite d'élasticité lors de ce changement de trajet, donc la mesure de X, dépend aussi très fortement du choix fait de la valeur de décalage en déformation plastique (le « plastic strain onset »). La valeur déterminée d'effet Bauschinger peut varier du simple au double comme le montre expérimentalement Sevillano *et al.* [SEVILLANO 2012]. La Figure 29 illustre cette difficulté [ALLAIN 2010_3].

Mesurer rigoureusement une valeur d'écrouissage X au travers d'un essai Bauschinger présente donc une vraie difficulté expérimentale et théorique. Par contre, cela reste un des seuls essais mécaniques possibles pour investiguer les contraintes internes à l'échelle des microstructures.

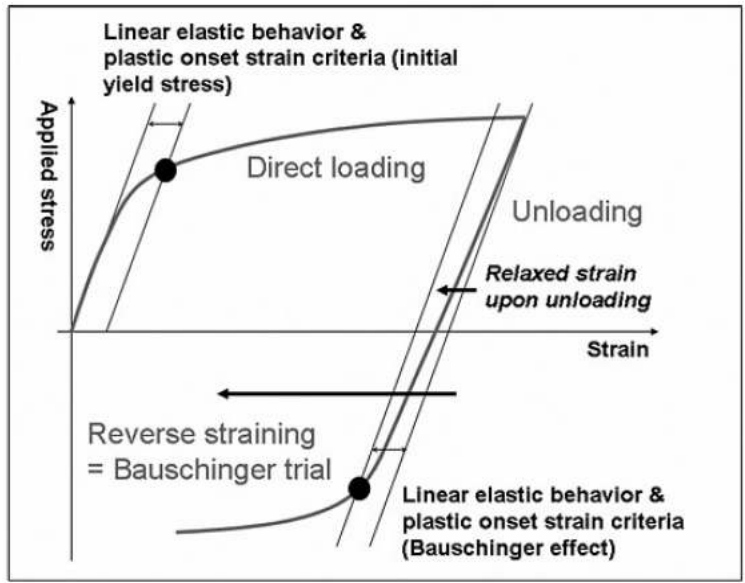

Figure 29 : Détermination de l'effet Bauschinger, illustrant l'ambigüité de cette mesure sur la base d'un critère de « plastic strain onset » [ALLAIN 2010_3].

Dans leur première publication, Sevillano *et al.* relient cette composante X à la valeur de $(\sigma_{int})_M$, hypothèse rigoureusement dérivée d'un modèle composite de type Masing. Ce choix signifie que lors du changement de trajet on choisit de définir la nouvelle limite d'élasticité du composite comme étant la limite d'élasticité de la phase la plus « molle » (celle en compression lors du chargement initiale). Cette hypothèse nécessite alors de définir X pour de faibles valeurs du décalage de déformation plastique. Dans leurs travaux postérieurs, ils reviennent sur cette hypothèse stricte en spécifiant que $(\sigma_{int})_M$ représente en fait une large fraction de X.

Nous avons déjà été confrontés à cette question au cours d'une étude sur des aciers ferrite-perlite que l'on pourrait qualifier de microstructure composite « modèle ». Afin de pouvoir utiliser les valeurs conventionnellement mesurées par des essais Bauschinger (avec un « onset » conventionnel de 0,2%), nous avons proposé une relation reliant cette composante cinématique X mesurée à des valeurs micromécaniques dans chacune des phases 1 et 2 [ALLAIN 2008_1] :

$$X = F_1 X_1 + F_2 X_2 + F_1 F_2 |\sigma_1 - \sigma_2| \qquad (31)$$

Avec $F_i$, $\sigma_i$ et $X_i$, les fractions, contraintes et composantes cinématiques propres de la phase i (i valant 1 ou 2). Cette relation est phénoménologique mais permet de reproduire toutes les tendances connues d'évolution de la composante cinématique d'écrouissage des composites et a été validée expérimentalement dans le cadre des aciers ferrito-perlitiques (pour des valeurs « d'onset » conventionnelles).

Dans le cas de cette étude, cette expression de X peut se réduire au terme non linéaire d'interactions entre les phases, c'est-à-dire entre la matrice et les macles. En effet, la composante cinématique de durcissement dans les macles est probablement nulle ($X_2 = 0$) car elles restent principalement élastiques au cours de la déformation. La composante cinématique de la matrice austénitique dépend dans le cas général de l'effet taille de grain [FEAUGAS 1999]. Dans le cas de valeurs « d'onset » conventionnelles et de grandes tailles de grain austénitique, ce terme $X_1$ peut être négligé par rapport au terme composite.

Il vient alors :

$$X = F(1-F)(\sigma_T - \sigma_M) \qquad (32)$$

avec

$$\Sigma = (1-F)\sigma_M + F\sigma_T \qquad (33)$$

On peut donc en déduire les composantes de contraintes dans chacune des phases.

$$\sigma_T = \Sigma + \frac{X}{F} \text{ et } \sigma_M = \Sigma - \frac{X}{1-F} \qquad (34)$$

On retrouve dans cette expression selon les notations de Sevillano *et al.* que la contrainte interne $(\sigma_{int})_M$ est bien une valeur négative, proche de X dans la mesure où (1-F) reste limitée :

$$(\sigma_{int})_M = \sigma_M - \Sigma = \frac{-X}{1-F} < 0 \qquad (35)$$

Par contre, cette valeur $(\sigma_{int})_M$ est nécessairement supérieure en valeur absolue à la composante d'écrouissage cinématique X mesurée expérimentalement. On notera aussi que la composante cinématique s'écrit alors en fonction de la contrainte dans les macles :

$$X = F(\sigma_T - \Sigma) = F\sigma_T + F\Sigma \qquad (36)$$

Si $\sigma_T \gg \Sigma$, on retrouve alors l'approximation que nous avons utilisée précédemment, c'est-à-dire que le back-stress peut être comparé à $X \approx \sigma_b = F\sigma_T$.

Un cas d'application concret de ces relations s'est présenté dans les travaux de thèse de K. Renard [RENARD 2012][BOUAZIZ 2013][8]. Au cours de ce travail, des essais Bauschinger ont été réalisés par cisaillement réversible et les fractions de macles correspondantes mesurées pour deux niveaux de prédéformation. Le Tableau 1 répertorie pour ces deux niveaux de pré-déformation les valeurs de contraintes, de fraction maclées et d'effet Bauschinger, ainsi que les contraintes locales déduites des équations précédentes.

---

[8] Malheureusement, ce type d'analyse n'a pu être mené après les travaux de D. Barbier pour deux raisons principales :
- les fractions maclées n'ont pas été mesurées lors des essais Bauschinger et il semble que les taux de maclage soient fonction du mode de chargement (principalement pour des raisons de texture).
- la taille de grain initiale est faible (3 µm) ce qui pourrait contribuer significativement à l'effet Bauschinger en début de déformation. Le terme $X_1$ serait alors non négligeable. Cette contribution expliquerait les mesures de X de 310 MPa pour des contraintes d'écoulement de 720 MPa après 10% de déformation alors que seulement 30% des grains sont faiblement maclés.

Dans ce cas, il conviendra de conserver dans l'équation précédente la contribution au durcissement cinématique de la matrice $X_M$, sachant que ce terme va saturer très rapidement après quelques pourcents de déformation :

$$X = F(1-F)(\sigma_T - \sigma_M) + (1-F)X_M$$

| Paramètres | Signification | Valeurs | | |
|---|---|---|---|---|
| E | Déformation macroscopique équivalente | 0 | 0.1 | 0.2 |
| F | Fraction de macle estimée | 0 | 0.09 | 0.11 |
| Σ | Contrainte macroscopique équivalente | 332 | 600 | 800 |
| $X_{0.2\%}$ | Effet Bauschinger (onset de 0.2%) | 0 | 170 | 275 |
| $X_{0.5\%}$ | Effet Bauschinger (onset 0.5%) | 0 | 115 | 200 |
| $\sigma_M^{0.2\%} = \Sigma - \dfrac{X_{0.2\%}}{1-F}$ | Contrainte dans la matrice (onset de 0.2%) | 332 | 413 | 489 |
| $\sigma_M^{0.5\%} = \Sigma - \dfrac{X_{0.5\%}}{1-F}$ | Contrainte dans la matrice (onset de 0.5%) | 332 | 474 | 574 |
| $\varepsilon_M$ | Déformation plastique dans la matrice | 0 | 0.06 | 0.16 |
| $\sigma_T^{0.2\%} = \Sigma + \dfrac{X_{0.2\%}}{F}$ | Contrainte dans les macles (onset de 0.2%) | - | 2489 | 1878 |
| $\sigma_T^{0.5\%} = \Sigma + \dfrac{X_{0.5\%}}{F}$ | Contrainte dans les macles (onset 0.5%) | - | 3191 | 2539 |

Tableau 1 : Contraintes dans les macles et la matrice après déformation de la nuance de référence, déduites des mesures d'effet Bauschinger (deux niveaux de plastiques strain onset – 0.2 et 0.5%) – données initiales issues de [RENARD 12].

Le Tableau 1 montre que les contraintes dans les macles prennent des valeurs comprises entre 2 et 3 GPa, en cohérence avec les valeurs proposées par Sevillano et donc avec les épaisseurs de macles rapportées dans la littérature.

La Figure 30 montre la comparaison de ces valeurs de contraintes dans la matrice avec les courbes de comportement d'un acier austénitique binaire Fe30Mn [BOUAZIZ 2011_2][HUANG 2011] et Fe36Mn0.6C stables sans effet TWIP. A la limite d'élasticité près (effet de la taille de grains), le comportement de l'austénite sans maclage déduite de cette analyse est comparable aux autres aciers de référence. Ce nouveau résultat (Fe22Mn0.6C(II)) est aussi comparé avec la courbe de comportement de l'acier de référence Fe22Mn0.6C sans maclage déduite par une autre méthode indirecte (détaillé au chapitre 2.3.6.3 page 65) (Fe22Mn0.6(I)).

De façon plus surprenante, il apparaît sur la Figure 30 que le carbone n'influe pas ou peu sur l'écrouissage de ces structures austénitiques. Ce résultat confirme notre choix dans la modélisation micromécanique des aciers TWIP de ne pas faire varier le paramètre de restauration dynamique avec la teneur en carbone [BOUAZIZ 2011_2].

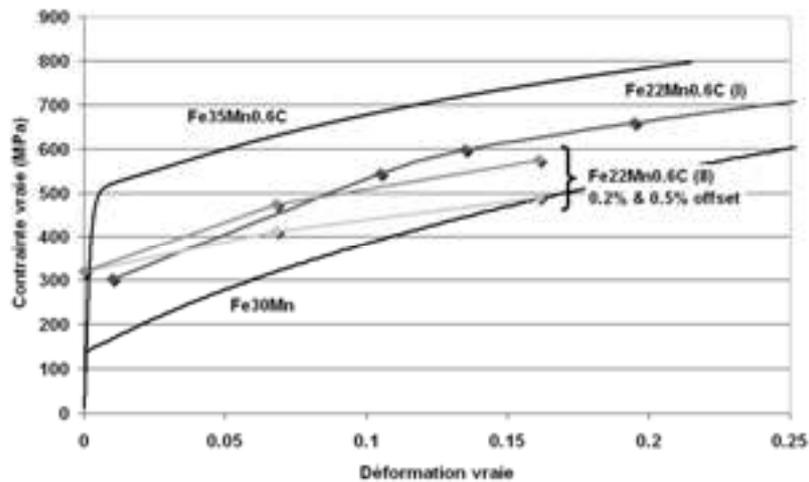

Figure 30 : Courbes de comportement en traction d'aciers Fe30Mn et Fe36Mn0.6C stables sans effet TWIP. Comparaison avec le comportement estimé d'un acier Fe22Mn0.6C sans maclage par des changements de trajets thermomécanique (voir ci-dessous) (I) ou l'analyse des essais Bauschinger de K. Renard (II) en fonction de la déformation vraie dans la matrice.

### 2.3.6. Influence du mode de chargement

#### 2.3.6.1. *Influence du mode de chargement (trajets monotones)*

Une des difficultés majeures dans l'analyse des contraintes internes au cours d'essais Bauschinger dans ces alliages, est que la microstructure de maclage (fraction et morphogénèse) semble très sensible aux conditions de chargement, y compris le long de trajet direct. Par conséquent, on peut s'attendre à des différences notables de comportement mécanique équivalent entre traction uni-axiale et cisaillement dans ces alliages. Il conviendra par exemple de pratiquer avec prudence des comparaisons entre des données d'essais Bauschinger en cisaillement réversible et des données en traction.[9]

---

[9] Cette précaution n'a par exemple pas été prise en compte dans le cadre de notre modèle [BOUAZIZ 2008_1] présenté ci-dessus. Toutefois, les cinétiques de maclage ont été adaptées pour la modélisation les différents essais de traction et la mesure des contraintes internes X en cisaillement uniquement approchée en termes de forme par le modèle.

Contrairement aux aciers TRIP montrant une transformation γ→α' [JACQUES 2007], cette différence de comportement ne réside pas dans une cinétique de transformation (de maclage) contrôlée par la valeur de triaxialité du chargement car le processus de maclage ne s'accompagne pas de variation de volume mesurable. Par contre, au cours de sa thèse, D. Barbier a montré que les microstructures de maclage en traction ou en cisaillement simple n'évoluaient pas de manière similaire [BARBIER 2009_2]. Au cours d'un essai de traction, deux systèmes de maclages apparaissent séquentiellement au cours de la déformation, alors qu'un seul système est généralement activé par grain en cisaillement. Cette différence s'explique notamment par des différences de texture de déformation, qui contrôle les systèmes de glissement et de maclage activés. De manière quantitative, les fractions de phase maclées semblent être équivalentes après 30 % de déformation équivalente, voire supérieure en traction. Ce résultat est contradictoire avec les résultats récents de K. Renard [RENARD 2012] dans un alliage Fe20Mn1.2C qui au contraire montrent une cinétique de maclage très élevée dans le cas du cisaillement et une saturation précoce au cours de la déformation. Ils confirment la prévalence d'un seul système de maclage par grain. On pourra peut-être expliquer cette différence par des tailles de grains austénitiques sensiblement différentes entre les deux études.

Ces microstructures de maclage expliquent probablement les différences observées de comportement équivalent en cisaillement et en traction dans le cas de l'utilisation de surfaces de charge simple (Von Mises ou Hill 48, figures tirées de [ALLAIN 2004_1] et [BARBIER 2009_1] reprises sur la Figure 31). On notera que dans ces deux cas (petits et gros grains de la nuance de référence) les courbes de comportement divergent seulement après 20%-30% de déformation. Cette observation viendrait plutôt confirmer les conclusions de D. Barbier sur la sensibilité de la microstructure de maclage au mode de sollicitation, discutées ci-dessus.

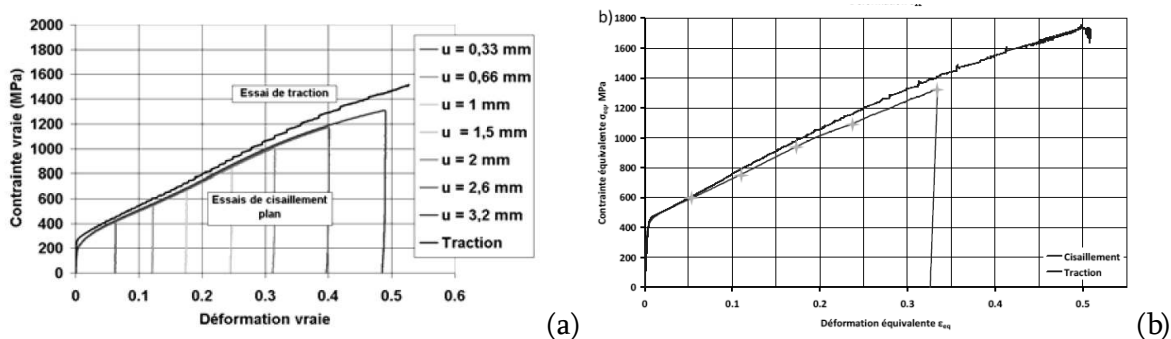

Figure 31 : Comparaison entre le comportement équivalent en traction et en cisaillement (a) dans le cas d'une surface de charge de type Mises (nuance de référence à gros grains) [ALLAIN 2004_1] (b) dans le cas d'une surface de charge de type Hill48 (nuance de référence à petits grains) [BARBIER 2009_1].

Selon le trajet de chargement direct, les microstructures de maclage produites sont différentes avec un fort impact sur les taux d'écrouissage. Les modèles de plasticité polycristalline les plus récents de l'effet TWIP basés sur des considérations de texture de déformation sont susceptibles de reproduire ce processus (avec des cinétiques de maclage ajustées). Toutefois, aucun des modèles de la littérature ainsi ajusté n'est capable de prédire le comportement lors de trajets alternés type Bauschinger, comme le montre la synthèse récente de Favier *et al.* [FAVIER 2012]. La forme de la surface de charge des aciers TWIP est donc aussi curieuse et inattendue comme le montre l'étude récente de [CHUNG 2011], reproduite sur la Figure 32.

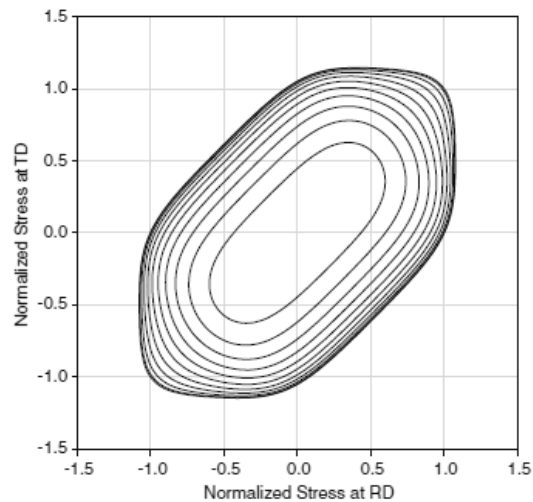

**Figure 32 : Surface de charge (écrouissage purement isotropes) d'un acier TWIP [CHUNG 2011].**

La formulation empirique retenue pour l'écrouissage est purement isotrope et nécessiterait l'introduction d'une composante cinématique. Ce résultat confirme aussi de manière indirecte le rôle prépondérant du maclage mécanique dans l'interprétation du comportement de ces aciers par rapport à un mécanisme supposé de vieillissement dynamique.

### 2.3.6.2. *Changement de trajets de chargement mécanique*

L'apparition de la microstructure de maclage conduit à une évolution profonde de contraintes internes révélées macroscopiquement lors de changements de trajets dits alternés (essais Bauschinger). La morphogénèse de cette structure est orientée aussi suivant une loi de Schmid. Cela suggère que les aciers TWIP présentent une sensibilité très particulière aux autres changements de trajets mécaniques (par exemple, essai de traction d'une éprouvette pré-déformée en cisaillement). Différents cas de figures ont donc été étudiés dans le cadre de la thèse de D. Barbier.

La Figure 33 montre les courbes de contraintes et déformations équivalentes obtenues selon les différents scenarios suivants :
- Traction monotone

- Cisaillement monotone
- traction plane DT puis cisaillement DL ($\theta = 0$ – changement de trajet dur)
- traction large DT (4% ou 8% de déformation équivalente) puis traction DL ($\theta = -0.5$)
- expansion equi-biaxiée (9% et 17 % de déformation équivalente) puis traction DL ($\theta = +0.5$)

L'angle ($\theta$) correspond à l'indicateur de la dureté d'un changement de trajet au sens de Schmitt :

$$\theta = \frac{\varepsilon_1 : \varepsilon_2}{\sqrt{\varepsilon_1 : \varepsilon_1}\sqrt{\varepsilon_2 : \varepsilon_2}} \qquad (37)$$

avec $\varepsilon_1$ et $\varepsilon_2$ définissant classiquement les tenseurs de déformation du trajets de pré-déformation (1) et de caractérisation (2) respectivement [SCHMITT 1994].

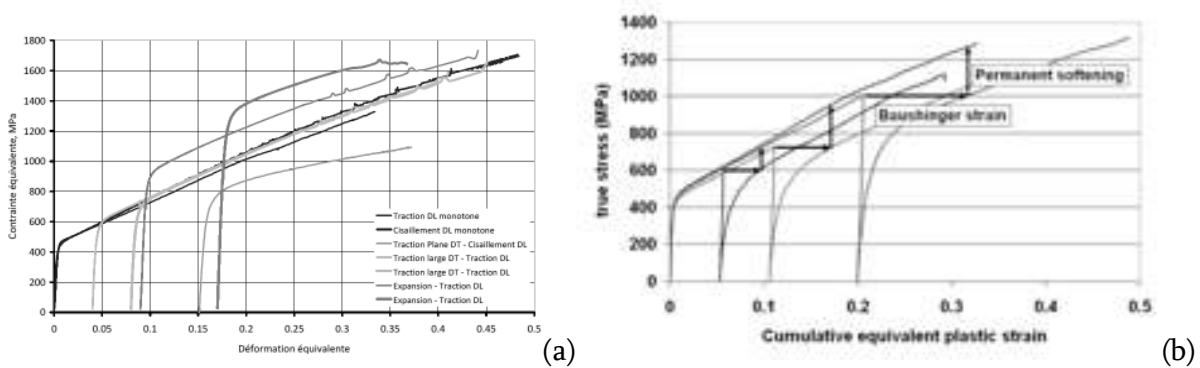

Figure 33 : (a) Comportement mécanique de la nuance de référence après les différents types de pré-déformation – représentation en contrainte et déformation équivalente décalée du niveau de pré-déformation [BARBIER 2009_1] (b) Essais Bauschinger en cisaillement réversible de la nuance de référence représentés en déformation cumulée [BOUAZIZ 2008_1].

Les courbes obtenues après pré-déformation sont représentées en tenant compte de ce décalage en déformation équivalente générée lors de la sollicitation initiale. Les comportements équivalents ont été déduits d'un critère de Hill 48, en l'absence de critère plus adapté. Ce choix n'affecte toutefois pas les conclusions ci-dessous. Dans cette analyse, ont aussi été intégrés bien entendu les essais de cisaillement alternés discutés ci-dessus et correspondant à un angle $\theta = -1$ caractéristique des changements de trajet Bauschinger. Ces essais ont été repris sur la Figure 33(b) permettant de mettre en valeur non seulement les contraintes internes (écrouissage cinématique) mais aussi le fort effet Bauschinger permanent.

La Figure 33 met en évidence que selon la nature du changement de trajets,
- les niveaux de contraintes à la recharge peuvent être très supérieurs ou inférieurs à la courbe de comportement de référence pour un niveau de pré-déformation donnée.

- par contre, après une faible déformation, les courbes de comportement retrouvent le taux d'écrouissage de la nuance de référence. Les essais Bauschinger en cisaillement de la Figure 33(b) présentent en conséquence un fort adoucissement permanent. Ce phénomène est aussi clairement visible lors des essais d'expansion-traction.

Les rapports $R_\theta$ entre contrainte équivalente à la recharge et contrainte de référence ont été définis pour chaque essai selon la procédure représentée sur la Figure 34(a). Ces rapports sont ensuite reportés sur la Figure 34(b) en fonction de l'angle θ des différents essais. Ces valeurs relatives ont été comparées à celles d'un acier doux [SCHMITT 1994] et au comportement attendu d'un acier Dual-Phase [DILLIEN 2010_2].

L'acier doux « classique » présente un faible effet Bauschinger (uniquement liés à la taille de grain) mais un durcissement important lors des changements de trajets durs (cf. Figure 35(a)). Ce phénomène est lié à la structure orientée et polarisée de cellules de dislocations générées lors de la pré-déformation. Les cellules ainsi créées induisent un sur-durcissement important lors d'un chargement orthogonal ultérieur. Ce sur-durcissement n'est que transitoire car les cellules de dislocations subissent rapidement un remodelage.

Les aciers DP a contrario présentent un fort effet Bauschinger lié à la présence d'une phase dure (martensite) dans une matrice molle (ferrite) comme nous le verrons au chapitre 3. Les fortes contraintes internes conduisent à un retard dans l'apparition des structures de cellules polarisées [GARDEY 2005]. En conséquence, le durcissement lors d'un changement de trajet dur est inexistant dans les aciers DP (cf. Figure 35(b))

Le cas des aciers TWIP est donc particulièrement intéressant. Les essais de cisaillement réversible montrent bien entendu un fort effet Bauschinger, par contre, les mesures avec des angles θ proches de 0 montrent que ces aciers présentent certainement un durcissement important lors des changements de trajet durs[10]. Ce sur-durcissement est probablement dû à la structure de maclage qui se trouve être orientée par rapport à la sollicitation initiale et donc non propice pour le second chargement[11]. Contrairement aux aciers DP, contraintes internes et genèse d'une structure fortement polarisée en relation avec la cristallographie ne sont donc pas incompatibles dans les aciers TWIP.

Le deuxième résultat fondamental de cette étude est que ces effets dus au pré-chargement sont permanents dans le cas des aciers TWIP (adoucissement permanent) contrairement aux aciers DP ou IF. Cette permanence s'explique simplement par l'impossibilité de remodeler la microstructure de maclage au cours du second chargement, contrairement aux structures de cellules de dislocations.

---

[10] Un essai de changement de trajet dur a aussi été réalisé dans le cadre de la thèse de D. Barbier (traction plane suivie d'un cisaillement). Toutefois, les conclusions de cet essai sont assez incohérentes. Ce résultat unique et isolé n'a donc pas été présenté dans ce mémoire.

[11] Ce comportement spécifique ne peut être expliqué uniquement par les changements de facteur de Taylor comme l'a démontré D. Barbier dans sa thèse, mais bien à la microstructure de maclage.

Ce résultat est tout à fait original par rapport à la littérature et fera bientôt l'objet d'une publication spécifique.

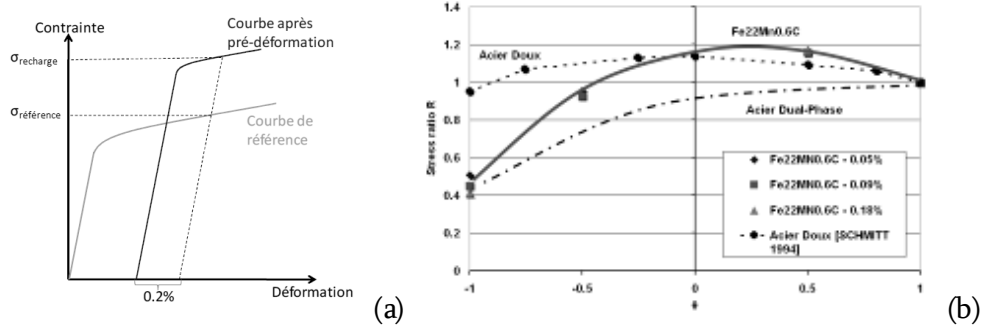

Figure 34 : (a) Méthode de mesure de la contrainte équivalente après pré-déformation pour le calcul du rapport de contrainte Rθ pour chaque essai (b) Représentation du rapport Rθ en fonction de l'angle θ caractérisant la dureté du changement de trajet pour l'acier de référence et les comportements attendus typiques d'aciers doux et Dual-Phase [BARBIER 2009_1].

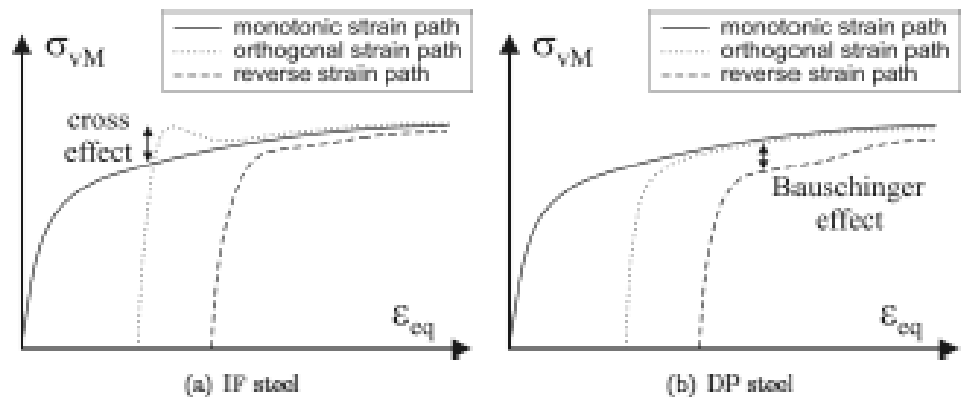

Figure 35 : Comportements typiques attendues lors de changements de trajets de déformations (a) d'acier doux (IF) (b) d'acier DP [DILLIEN 2010_2].

### 2.3.6.3. *Changement de trajets thermomécaniques*

Dès mes travaux de thèse, nous avons envisagé un autre type de changement de trajets, que nous avons qualifié de thermomécanique, consistant à changer la température entre pré-déformation et essai de traction proprement dit. La pré-déformation permet d'introduire une microstructure initiale différente d'un trajet direct. Des essais de pré-déformation ont été par exemple réalisés à 400°C puis les éprouvettes ont été déformées à température ambiante afin d'évaluer l'impact d'une densité importante de dislocations parfaites (seul mécanisme de déformation actif à 400°C) sur le maclage mécanique (à température ambiante). Ces différents essais sont représentés sur la Figure 36.

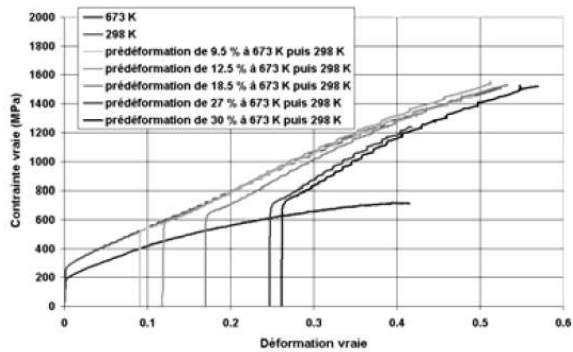

Figure 36 : Essais de traction à température ambiante après différents taux de pré-déformation en traction à 400°C. Les courbes de comportement attendues à ces deux températures sont aussi représentées (en bleu et rouge respectivement) [ALLAIN 2004_1].

Cette expérience originale a permis de mettre en lumière deux résultats fondamentaux au sujet de l'effet TWIP :
- Les lieux des limites d'élasticité à la recharge à température ambiante correspondent aux contraintes d'écoulement qu'aurait l'austénite après déformation mais en l'absence de macles mécaniques (cette contrainte tenant compte aussi de la contrainte effective). La courbe d'évolution de ces contraintes en fonction des pré-déformations représente donc la courbe de comportement de l'alliage sans maclage. Cette analogie repose sur l'hypothèse que les densités de dislocations générées à 400°C pour une déformation donnée sont identiques à celles pouvant être générées à température ambiante sans maclage. La comparaison entre ce comportement estimé sans maclage et la courbe de comportement de l'alliage de référence est représentée sur la Figure 37(a). Ces résultats présentent une parfaite concordance avec les courbes de comportement de l'alliage binaire Fe30Mn ou le ternaire Fe35Mn0.6C stables sans effet TWIP présentées sur la Figure 30 au chapitre 2.3.5.2. Ce résultat vient naturellement confirmer le rôle important du maclage mécanique sur l'écrouissage de ces aciers.
- Le second résultat majeur est révélé en comparant les taux d'écrouissage de la nuance selon un trajet direct ou après déformation pour une contrainte donnée. Pour les faibles déformations (inférieures à 15%), les courbes de comportement sont confondues. Par contre, au-delà, comme le montre le Figure 37(b), les courbes divergent. L'écrouissage des aciers TWIP est donc contrôlé par deux variables d'état[12], très probablement la densité de dislocations et la microstructure de maclage, comme nous l'avons implicitement supposé dans notre modèle cinématique.

Les résultats d'un essai mécanique sur ces aciers vont donc être très sensibles à l'histoire thermomécanique de l'échantillon, contrairement aux aciers doux par exemple [RAUCH 1997].

---

[12] Ces conclusions sont sensiblement différentes de celle présentées dans une de nos publications [BOUAZIZ 2010].

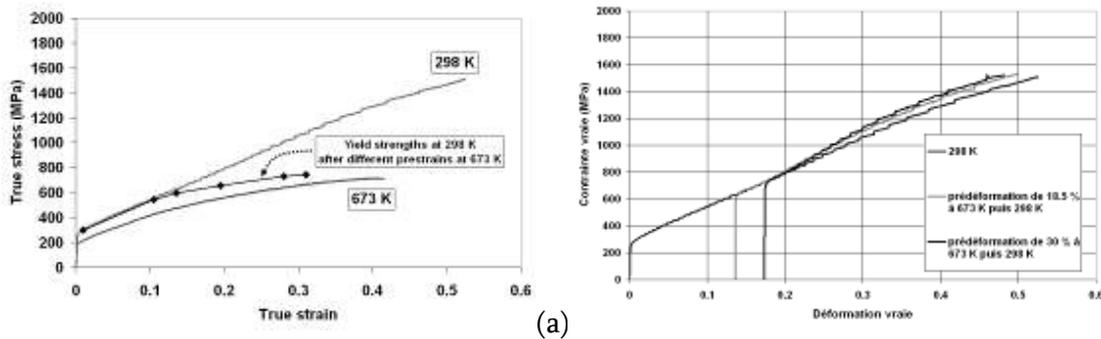

Figure 37 : (a) Lieux des limites d'élasticité à température ambiante en fonction des niveaux de pré-déformation à 400°C permettant d'estimer le comportement de la nuance de référence sans effet TWIP. Les courbes de comportement attendues à 25°C et à 400°C sont aussi représentées [BOUAZIZ 2010]. (b) Courbes de comportement après pré-déformation à 400°C recalées en déformation pour faire correspondre leur limite d'élasticité avec la courbe de comportement à température ambiante [ALLAIN 2004_1].

### 2.3.7. Conclusion intermédiaire

Depuis les travaux précurseurs de L. Rémy, notre principale contribution scientifique a été finalement de découpler et quantifier les contributions d'origine cinématique et isotrope de l'effet TWIP. La première contribution s'apparente à un durcissement dû aux joints de grains (empilements de dislocations). Ce processus est dynamique car la densité de ces joints augmente au cours de la déformation. Ce terme peut avoir une interprétation duale si l'on considère les macles et la matrice comme un composite à inclusions dures. Le second terme isotrope est lié à la morphogénèse de la microstructure de maclage (texture).

Ces contributions relatives sont révélées en particulier lors des changements de trajets de sollicitations, comme nous l'avons montré expérimentalement. Ces résultats originaux s'expliquent par la morphogénèse de microstructure de maclage et sa persistance lors de changements de trajets. Ils ont des conséquences importantes pour l'utilisation de ces aciers (sollicitations mécaniques de pièces après mise en forme par exemple).

Les meilleurs modèles micromécaniques polycristallins de l'effet TWIP à l'heure actuelle sont susceptibles de décrire assez correctement des effets de textures mais s'avèrent encore insuffisants pour décrire correctement les changements de trajets de ces aciers et reposent toujours sur des cinétiques de maclage empiriques.

## 2.4. Effet de la composition chimique : le rôle particulier du carbone

Comprendre et mettre en évidence les mécanismes d'influence de la composition chimique, et en particulier du carbone en solution solide, sur le comportement mécanique de ces aciers a toujours été un objectif important pour moi. Cet axe de travail permet de fournir aux « métallurgistes » des outils et des règles métier pour optimiser les performances mécaniques de ces alliages [SCOTT 2006].

Je me suis intéressé dans un premier temps à l'influence de la composition chimique de l'alliage sur son EDE, et j'ai élaboré dans le cadre de ma thèse un modèle thermodynamique complet pour la prédiction de cette énergie intégrant pour la première fois les effets du carbone dans le système FeMnC. Ce modèle a ensuite été étendu dans le cadre de la thèse d'A. Dumay à d'autres éléments d'alliages comme l'aluminium ou le cuivre. Ce paramètre thermochimique permet de prédire les mécanismes de déformation activables dans ces alliages mais il est insuffisant pour comprendre et prédire l'écrouissage de ces alliages. Nous montrerons en particulier le rôle paradoxal du carbone.

Dans un second temps, mon intérêt s'est donc porté sur la compréhension du rôle spécifique du carbone sur le comportement mécanique des aciers TWIP ternaires. En effet, il ne se réduit pas à augmenter l'EDE, mais contribue aussi à un mécanisme de vieillissement dynamique, modifie la mobilité des dislocations et leur structuration indépendamment des macles. On reviendra aussi sur son rôle présumé dans le processus de maclage.

Ces différents travaux ont principalement été menés en collaboration avec l'IJL de Nancy et le LEM3 de Metz.

### 2.4.1. Composition, EDE et écrouissage

#### 2.4.1.1. *Modélisation thermochimique de l'EDE*

La mesure directe de l'EDE d'un acier est un exercice de MET particulièrement délicat, reposant sur l'analyse de la configuration de jonctions triples de dislocations parfaites [REMY 1975], la méthode de référence, ou des distances de dissociations entre dislocations partielles [BRACKE 2007_2]. Ces différentes méthodes sont sujettes à de nombreux biais (l'épinglage par exemple des dislocations partielles par des atomes de carbone au cours de traitements thermiques [REMY 1975]). Elles sont donc rares dans la littérature et souvent sources de controverses, comme le jeu de mesures de Volosevitch *et al.* [VOLOSEVITCH 1976]. Pour une composition d'acier donnée comme les aciers Hadfield (typiquement Fe12Mn1.2C), les EDE mesurées à température ambiante peuvent varier du simple au double selon les auteurs (23 mJ.m$^{-2}$ pour Karaman *et al.* [KARAMAN 2000_1] ou 50 mJ.m$^{-2}$ pour Dastur et Leslie par exemple [BAYRAKTAR 2004][DASTUR 1981])

C'est donc un exercice auquel nous n'avons pas directement participé ! Notre équipe s'est plutôt consacrée à la modélisation de cette EDE sur des bases thermodynamiques par une méthode indirecte, les résultats et paramètres étant ensuite ajustés en fonction des températures de transformation martensitique de ces alliages [COTES 1998].

Ces calculs reposent sur l'existence d'une relation entre EDE et forces motrice de transformation martensitique ε, reprise ci-dessous :

$$EDE_{int} = 2\rho_{111} \Delta G^{\gamma \to \varepsilon} + 2\sigma^{\gamma/\varepsilon} \tag{38}$$

Avec $\rho_{111}$ la densité molaire surfacique des plans denses {111} et $\sigma^{\gamma/\varepsilon}$ l'énergie de surface γ/ε.

Dans ce type de calcul, la détermination de l'énergie de Gibbs pour la transformation volumique γ→ε reste le point clef. De nombreux auteurs ont participé à la description du système binaire FeMn [MIODOWNIK 1998][HUANG 1989][REMY 1975][LIN 1997] ou du système ternaire FeMnSi [COTES 1998], mais nous avons été parmi les premiers à proposer une approche complète intégrant de manière semi-empirique les effets du carbone [ALLAIN 2004_3][ALLAIN 2004_1]. Ces travaux ont ensuite été repris et étendu à des systèmes plus complexes comme aux FeMnAlC par Saeed-Akbari *et al.* [SAEED-AKBARI 2009], FeMnCuSiAlC par Dumay *et al.* [DUMAY 2008_2] ou aux FeMnCN par Curtze *et al.* [CURTZE 2010]. La révision la plus récente et la plus complète a été proposée par Nakano *et al.* [NAKANO 2010] avec, pour la première fois, une approche globale, compatible avec le diagramme des phases stables du système FeMn binaire complet et cohérent avec l'approche du SGTE (Scientific Group Thermodata Europe).

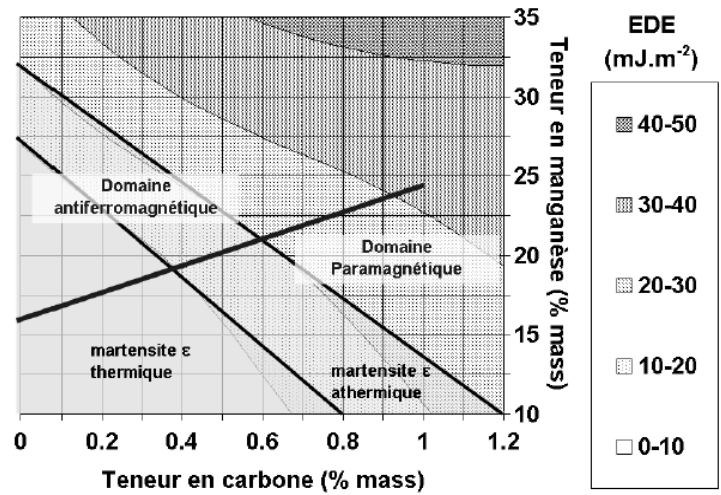

Figure 38: Cartographie d'EDE prédite par notre modèle en fonction de la teneur massique en carbone et manganèse. Sont identifiés les domaines d'apparition de la martensite ε au refroidissement (thermique) et au cours de la déformation (athermique) définis par Schumann [SCHUMANN 1972] et les domaines d'états magnétiques calculés (en rouge) [ALLAIN 2004_1][ALLAIN 2004_3].

Notre modèle reste très souvent utilisé en pratique par d'autres équipes car il ne nécessite pas l'utilisation de logiciels spécifiques (type THERMOCALC) et fonctionne sur un simple tableur. La Figure 38 montre le calcul des EDE en fonction des teneurs en carbone et manganèse des alliages FeMnC à température ambiante. Le modèle a été ajusté pour que les lignes d'iso-EDE suivent les limites des domaines de transformation martensitique ε. Sont reportés en outre les domaines d'états magnétiques (séparés par la ligne rouge). Les deux éléments Mn et C contribuent donc à augmenter l'EDE, mais dans des proportions différentes à température ambiante. La Figure 39 montre l'influence de l'ajout d'autres éléments d'alliage substitutionnels comme le Cr, Al, Si ou le Cu d'après le modèle de A. Dumay *et al.* [DUMAY 2008_2].

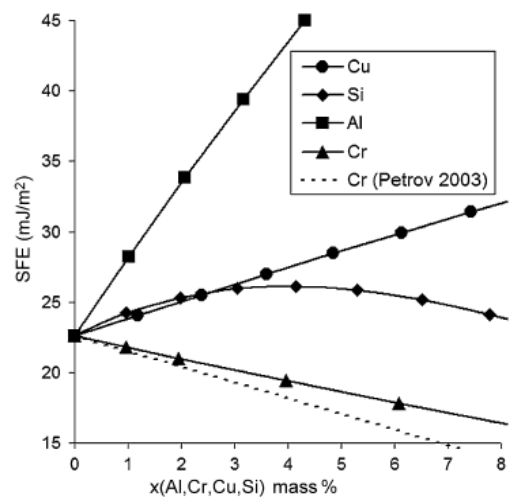

Figure 39 : Influence des éléments d'alliage substitutionnels sur l'EDE relativement à la nuance de référence Fe22Mn0.6C [DUMAY 2008_2]

Le dernier terme de l'équation précédente, correspondant à l'énergie d'interface γ/ε est souvent considéré comme une variable d'ajustement de ces modèles. Saeed-Akbari *et al.* [SAEED-AKBARI 2009] a proposé une revue pertinente des différentes valeurs trouvées dans la littérature pour ce paramètre (parfois variable en fonction de la chimie). On peut attendre beaucoup dans l'avenir des calculs *ab initio* pour définir cette énergie sur des bases plus physiques ([KIBEY 2006] par exemple).

### 2.4.1.2. *Etat magnétique et module d'élasticité*

Ces études récentes dédiées à l'estimation thermodynamique de l'EDE reposent en partie sur la connaissance de l'état magnétique (para- ou antiferro-magnétique) de l'austénite, information clef sur la stabilité des phases. Dans certains cas, la contribution magnétique en excès peut contrebalancer le terme chimique, et conduit à une stabilisation à elle seule de l'austénite. Contrairement à la plupart des aciers austénitiques inoxydables, la température critique de transition entre ces deux états magnétiques (appelé température de Néel) est supérieure à 250K dans la plupart des alliages FeMnC. Ce changement d'état magnétique a donc des conséquences directes et significatives sur leur comportement à température ambiante.

Cette transition entre état paramagnétique de haute température et antiferromagnétique de basse température est connue de longue date dans ces alliages par les « physiciens » (comme le montre la revue de L. Rémy) [REMY 1975]. Elle a pourtant souvent été négligée par les « métallurgistes » malgré de très forts effets attendus sur les transformations de phases et les propriétés mécaniques. Le module d'élasticité en cisaillement µ de la nuance de référence Fe22Mn0.6C mesuré par Dynamical Mechanical Thermal Analysis (DTMA) en fonction de la température est représenté sur la Figure 40(a). Cette évolution présente une anomalie évidente autour de la température de Néel. A basse température, les modules d'élasticité sont faibles et quasiment insensibles à la température comme dans les alliages Invar. Au dessus de la température de Néel, dans l'état paramagnétique, il diminue quasi linéairement avec la température comme attendu par le modèle empirique de Ghosh et Olson [GHOSH 2002], et ce après une très forte augmentation vers 290 K (+15 GPa en 50°C environ). On notera que la température de Néel définie d'un point de vue magnétique correspond toujours au milieu de la transition comme l'a remarqué expérimentalement Rémy. Cette anomalie de module d'élasticité est d'importance car toutes les composantes de la contrainte d'écoulement sont des fonctions de ce module, y compris les termes de contrainte effective ou de frottement de réseau.

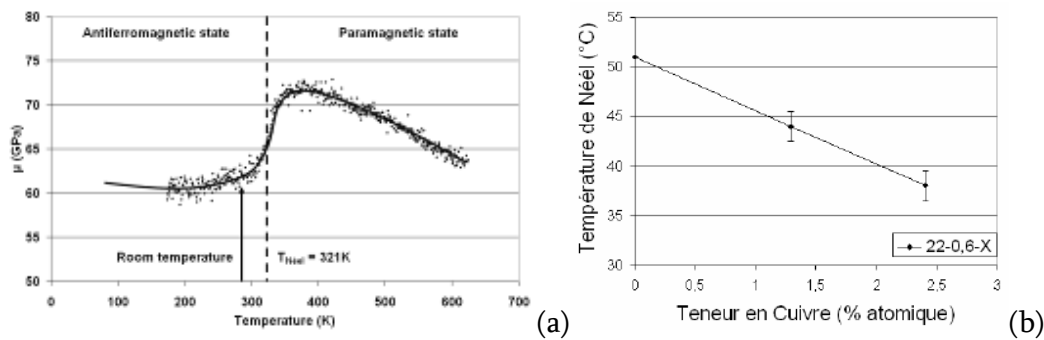

Figure 40 : (a) Evolution du module d'élasticité en cisaillement µ de la nuance de référence en fonction de la température dans les domaines antiferro- et paramagnétiques [ALLAIN 2010_2] (b) Effet de l'ajout de cuivre dans la nuance de référence sur la température de Néel de l'alliage [DUMAY 2008_1]

La contribution magnétique en excès à l'EDE est généralement calculée en utilisant la méthode de Hillert et Jarl [INDEN 1981][HILLERT 1978]. Cette équation requiert les températures de Néel et les moments magnétiques de chacune des phases (austénite γ et martensite ε) en fonction de la composition des alliages. Dans le cadre de la thèse de A. Dumay, les températures de Néel de nombreux alliages ont été mesurées pour déterminer les effets relatifs du cuivre, manganèse et carbone en solution solide. La Figure 40(b) montre par exemple l'évolution de la température de Néel de la nuance de référence en fonction de sa teneur en cuivre. Ces différents résultats nous ont permis de définir une équation statistique de cette température $T^{\gamma}_{Néel}$ critique pour l'austénite en fonction de sa composition chimique [HUANG 1989][MIODOWNIK 1998][ZHANG 2002][DUMAY 2008_1] :

$$T^{\gamma}_{Néel}[K] = 250\ln(x_{Mn}) - 4750x_C x_{Mn} - 222x_{Cu} - 2.6x_{Cr} - 6.2x_{Al} - 13x_{Si} + 720 \quad (39)$$

avec $x_i$ la fraction atomique de l'élément chimique i considéré et $T^{\gamma}_{Néel}$ en K.

Par contre, peu de progrès ont été faits dans la mesure de moments magnétiques de ces alliages. Ces termes restent la plupart du temps des paramètres d'ajustement du modèle d'EDE [ALLAIN 2004_2][COTES 1998]. Là encore, des progrès significatifs sont attendus des calculs *ab initio* dans le futur.

### 2.4.1.3. *EDE et écrouissage : le paradoxe carbone*

Comme nous l'avons discuté dans la section précédente, la susceptibilité d'une nuance FeMnC à macler est fortement corrélée à son EDE. Cette énergie contrôle en effet les possibilités de dissociation des dislocations parfaites et la contrainte critique de maclage, comme le montre l'équation (7).

La conséquence directe de cette observation est que l'ajout d'éléments d'alliage augmentant l'EDE devrait avoir pour effet de ralentir la cinétique de maclage et donc diminuer globalement le taux d'écrouissage d'un acier donné. Ce résultat a été confirmé expérimentalement par A. Dumay en étudiant l'ajout de Cu dans la nuance de référence. Le cuivre augmente l'EDE de 3.5 mJ.m$^{-2}$.%Cu$^{-1}$ selon la Figure 39. Les courbes de traction de 3 aciers Fe22Mn0.6CxCu sont représentées sur la Figure 41(a), montrant clairement la chute du taux d'écrouissage pour les nuances les plus riches en cuivre à partir de 15% de déformation [DUMAY 2008_1].

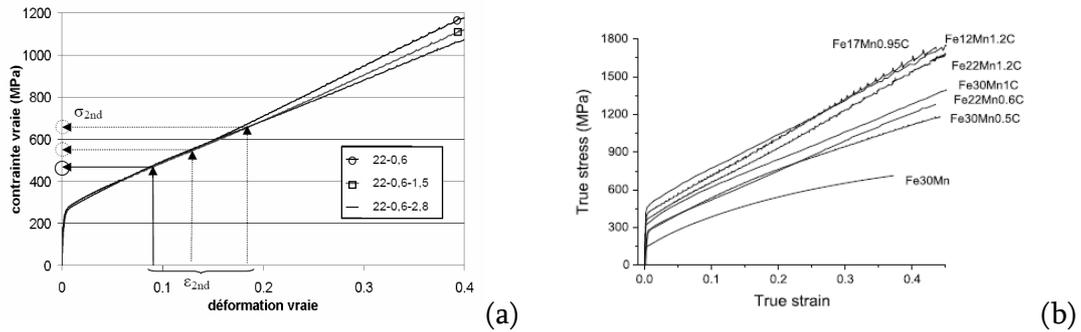

Figure 41 : (a) Effet de l'ajout de cuivre sur le comportement en traction de la nuance de référence [DUMAY 2008_1] (b) Courbes de traction rationnelles de différents alliages ternaires FeMnC TWIP [HUANG 2011][BOUAZIZ 2011].

De façon très surprenante, c'est l'effet inverse qui est observé avec l'ajout de carbone, nous parlerons donc dans la suite de « paradoxe carbone ». Nous avons compilé dans [HUANG 2011][BOUAZIZ 2011] les courbes de traction de différents aciers FeMnC reprises sur la Figure 41. Toutes les courbes de traction présentent globalement le même taux d'écrouissage initial, très similaire à celui d'une austénite à faible EDE sans maclage, comme l'acier binaire Fe30Mn de la Figure 30. Les différences apparaissent après 5% à 10% de déformation. Les nuances les plus riches en carbone voient leur taux d'écrouissage augmenter, avec un changement de courbure dans certains cas (comme les aciers Fe17Mn0.95C ou Fe30Mn1.0C).

Comme nous l'avons montré précédemment, la contrainte d'écoulement des aciers TWIP peut s'écrire :

$$\Sigma(E) \approx \sigma_0 + \sigma_M(\varepsilon_M) + F(E)\sigma_T(\varepsilon_T) \tag{40}$$

Avec $\sigma_0$ la contrainte de friction dû aux éléments en solution solide, $\sigma_M$ la contrainte dans la matrice austénitique et $F\sigma_T$ la contrainte supportée par la fraction maclée (la contrainte de back-stress $\sigma_b$) et $\varepsilon_T$ la déformation plastique dans les macles (très faible). Dans la mesure où $\sigma_M$ ne semble pas dépendre de la composition en carbone, seule la contribution à l'écrouissage de nature cinématique due au maclage peut expliquer la différence entre ces alliages.

Ce second terme apparaît empiriquement comme une fonction polynomiale simple de la déformation plastique et de la composition en manganèse et carbone [BOUAZIZ 2011] :

$$F(E)\sigma_T(\varepsilon_T) \approx (60661z - 261874z^2)E \quad \text{avec} \quad z = \frac{C}{Mn - 5} \tag{41}$$

De façon très surprenante et paradoxale, il apparaît alors que les iso-valeurs de z ainsi déterminées suivent dans le plan Mn/C des lignes quasiment orthogonales aux iso-valeurs EDE (cf. Figure 42). Autrement dit, l'écrouissage des nuances TWIP ne dépend pas de l'EDE !

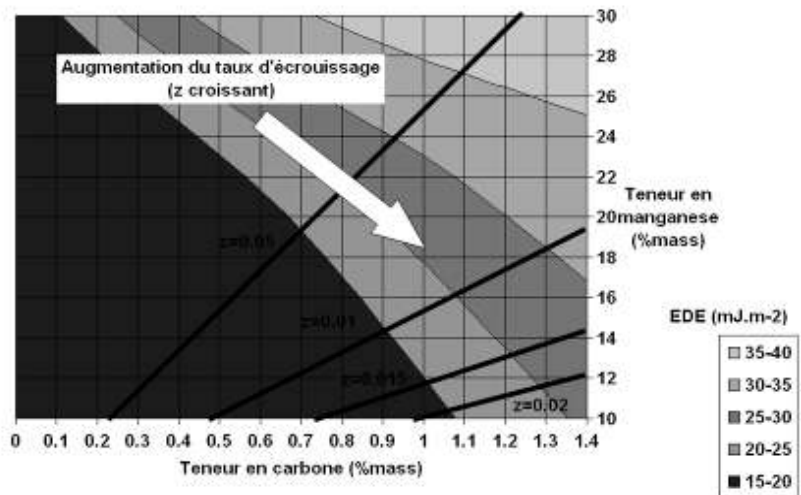

Figure 42 : Cartographie d'EDE en fonction de la teneur en carbone et en manganèse. Surimposition des lignes iso-z, valeur représentative du taux d'écrouissage des alliages à grande déformation. Qualitativement, les lignes iso-z sont normales aux lignes iso-EDE.

Dans la littérature, on attribue aussi au carbone un grand rôle sur le comportement de ces alliages au travers d'un mécanisme de vieillissement dynamique. Comme nous le verrons, cette contribution directe est probablement négligeable et ne permet pas d'expliquer le paradoxe ci-dessus. Par contre, la teneur en carbone va modifier sensiblement la mobilité des dislocations et leur structuration au cours de la déformation, donc avoir un effet indirect en retour sur la microstructure de maclage et sur les propriétés d'écrouissage. Dans la section suivante, nous allons donc présenter nos travaux sur l'activation thermique du glissement et sur le mécanisme de vieillissement dynamique dans ces aciers. Nous présenterons et discuterons alors un certain nombre d'hypothèses permettant d'expliquer cette relation complexe entre carbone, glissement et maclage mécanique et donc effet TWIP.

### 2.4.2. Activation thermique du glissement, vieillissement dynamique et leurs conséquences

Dans un premier temps, on s'intéressera en particulier à l'influence des éléments d'alliage et du carbone en particulier sur la dynamique du glissement sur une large gamme de température et vitesse de déformation. Ces travaux sont basés sur une analyse thermique de type interaction « dislocation / obstacle ponctuel fixe ». Cette étude fondamentale de la mobilité des dislocations a souvent été négligée dans la communauté au profit des microstructures de déformation observables comme le maclage ou les transformations martensitiques qui fournissent une explication directe de l'écrouissage de ces nuances.

Compte tenu de la vitesse de diffusion du carbone dans la structure CFC de ces aciers, il existe des domaines de températures et de vitesses de déformation plus restreints où des interactions dynamiques entre ces atomes mobiles et dislocations sont possibles, c'est-à-dire, un mécanisme de vieillissement dynamique. C'est en particulier le cas de la nuance de référence Fe22Mn0.6C à température ambiante lors d'essais quasi-statiques. De nombreuses études ont par contre été dédiées à ce processus. Nous reviendrons dans un second temps sur ce mécanisme, ses conséquences supposées et avérées.

### 2.4.2.1. *Eléments d'alliage et mobilité des dislocations*

Le point de départ de cette recherche a été une large étude bibliographique sur l'évolution de la limite d'élasticité en traction de ces nuances en fonction de la température d'essai et de la vitesse de déformation [ALLAIN 2010_1]. Ce paramètre est supposé être représentatif de la mobilité des dislocations (contrainte d'écoulement en l'absence de macles mécaniques dans des structures recristallisées à grandes tailles de grains). Le Tableau 2 reprend les compositions des alliages considérés, les tailles de grains, températures de Néel estimées et les conditions d'essais tirées des publications respectives [ALLAIN 2004_1][REMY 1975][TOMOTA 1986][CHOI 99][KIM 1986][KUNTZ 2007][ADLER 1986]. La Figure 43(a) montre l'évolution des limites d'élasticité en traction (contraintes d'écoulement pour des décalages de déformation plastique faibles de 0.2% à 1%) en fonction de la température d'essai. La Figure 43(b) présente les mêmes données mais normalisées par la contrainte d'écoulement à température ambiante de chacun des alliages. Cette normalisation permet de se départir d'une contribution structurale liée à la taille de grain.

| Studied steels | | | | | Tensile procedure | |
|---|---|---|---|---|---|---|
| wt.% Mn | wt.% C | Others | grain size (µm) | TNéel (K) | $\dot{E}$ | $\varepsilon_p$ |
| 31 | 0.001 | | ≥ 20 | 432 | 0.00330 | 0.2% |
| 36 | 0.004 | | ≥ 20 | 470 | 0.00330 | 0.2% |
| 24 | 0.120 | | ≥ 50 | 361 | 0.00100 | 1.0% |
| 26 | 0.200 | | ≥ 50 | 375 | N.S. | 0.2% |
| 30 | 0.300 | 0.1 Nb | 80 | 403 | N.S. | N.S. |
| 30 | 0.300 | 5 Al 0.1 Nb | 80 | 340 | N.S. | N.S. |
| 22 | 0.600 | | 2.3 | 313 | 0.00070 | 1.0% |
| 22 | 0.600 | | ≥ 10 | 292 | 0.00380 | 0.2% |
| 22 | 1.000 | | 2.3 | 292 | 0.00070 | 1.0% |
| 13 | 1.200 | | 220 | 173 | 0.00083 | 0.2% |

**Tableau 2 : Aciers ternaires FeMnC étudiées d'après [Allain 2004_1][Rémy 1975][Tomota 1986][Choi 99][Kim 1986][Kuntz 2007][Adler 1986].**

Les aciers austénitiques à haute teneur en manganèse présentent le comportement typique des structures CFC durcies par des interstitiels [KOCKS 1995] lors de sollicitations quasi-statiques, c'est-à-dire :

- Une rapide diminution de la contrainte d'écoulement avec la température sous la température ambiante, définissant le domaine du glissement thermiquement activé. Autrement dit, le glissement des dislocations nécessite le franchissement d'obstacles locaux dont la barrière enthalpique est trop grande pour la température considérée. Leur franchissement nécessite l'application d'une contrainte supplémentaire, appelée contrainte effective.
- Un plateau au-dessus d'une température critique, proche de la température ambiante (en réalité, la limite d'élasticité continue à diminuer faiblement avec la température et cette diminution s'explique comme nous le verrons par une baisse linéaire du module de cisaillement). Ce plateau correspond à un régime athermique et est souvent associé à un processus de vieillissement dynamique. Dans ce domaine de température, la contrainte effective est quasiment nulle.

La Figure 44 montre par exemple le comportement d'autres alliages austénitiques avec des teneurs en carbone variables, des alliages NiC et FeNiC, qui présentent un comportement tout à fait similaire, mais des températures critiques de transition entre régimes différentes (environ 150K pour les alliages de nickel).

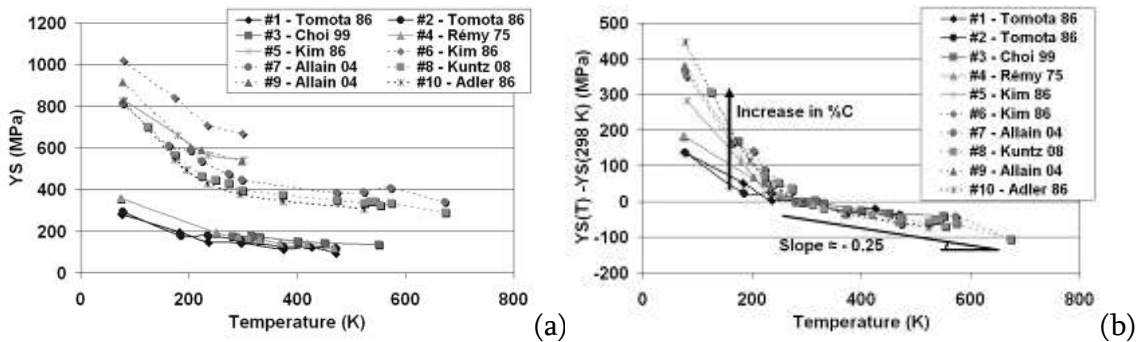

Figure 43 : (a) Evolution des limites d'élasticité en traction (contraintes d'écoulement pour des décalages de déformation plastique faibles de 0.2% à 1%) en fonction de la température d'essais pour les aciers détaillés dans le Tableau 2 (b) mêmes données que (a) mais normalisées par la contrainte d'écoulement à température ambiante [ALLAIN 2010_1].

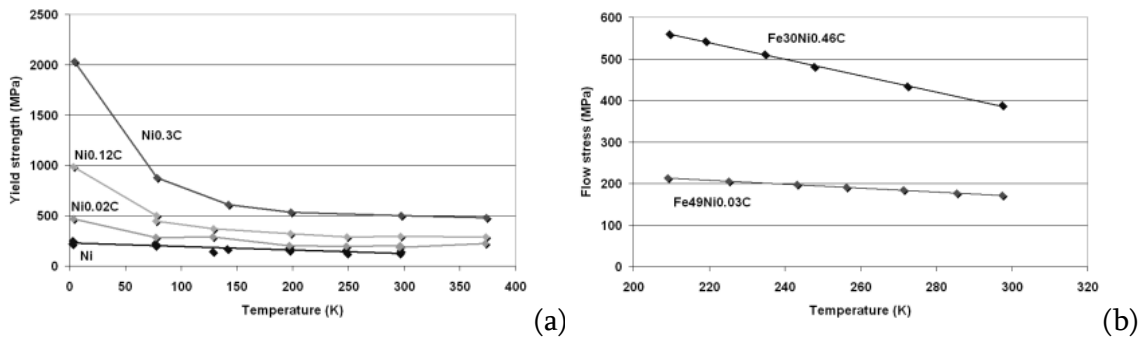

Figure 44 : Evolution des limites d'élasticité en traction en fonction de la température d'essai pour différents alliages (a) NiC [NAKADA 1971] et (b) FeNiC [KOCKS 1995].

La contrainte effective $\sigma_{effective}$ peut se modéliser grâce à un potentiel viscoplastique tiré de la théorie des interactions entre dislocations et défauts discrets fixes [ALLAIN 2009][ALLAIN 2010_1] :

$$\sigma_{effective}(T,\dot{E}) = \frac{1}{M}\sinh^{-1}\left(\frac{\dot{E}}{2M\rho_m b_{110}^2 \nu_{Debye}}\exp\left(\frac{\Delta G_0}{k_B T}\right)\right)\frac{k_B T}{V^*} \qquad (42)$$

Avec M un facteur de Taylor moyen, E la déformation, $\rho_m$ la densité de dislocations mobiles, $\nu_{Debye}$ la fréquence de Debye, $k_B$ la constante de Boltzmann, $V^*$ et $\Delta G_0$ les volumes et énergies d'activation apparents du processus.

Ce modèle a été appliqué pour décrire les données de la Figure 43. La procédure d'ajustement est décrite dans [ALLAIN 2010_1]. Tous les alliages présentent les mêmes températures de transition entre régime athermique et thermiquement activé et donc la même énergie d'activation apparente. Par contre, le volume d'activation varie très fortement avec la teneur en carbone de l'alliage comme le montre la Figure 45.

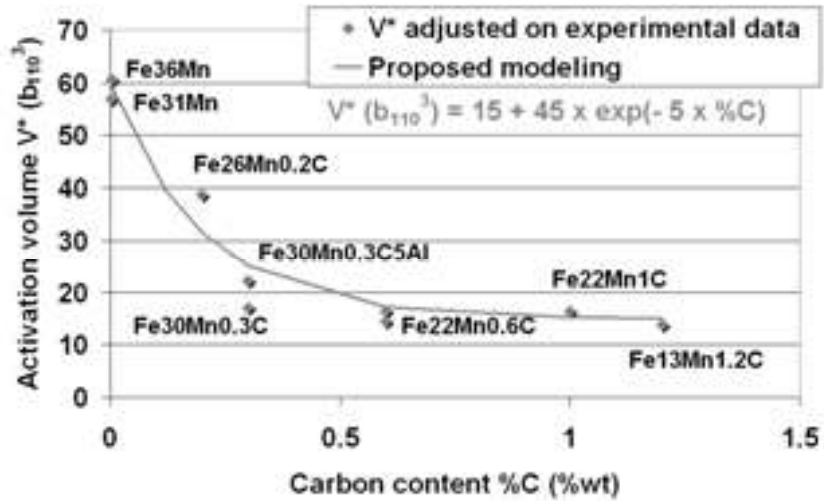

Figure 45 : Evolution du volume d'activation apparent de la contrainte effective en fonction de la teneur en carbone des alliages considérés [ALLAIN 2010_1].

Cette figure démontre l'effet négligeable de l'état magnétique (antiferromagnétique ou paramagnétique), de la teneur en Mn, en Al ou microalliage sur ce processus d'activation thermique du glissement. Dans le cas de l'acier de référence il est possible de déterminer la transition entre les deux régimes car le volume de données disponibles est important (thèse personnelle et thèse M. Kuntz [KUNTZ 2008]) mais aussi car les limites d'élasticité peuvent être décorrélées de l'évolution des modules d'élasticité connus par ailleurs en fonction de la température (cf. Figure 40)

La Figure 46 montre l'évolution du rapport entre la limite d'élasticité et le module de cisaillement de la nuance de référence en fonction de la température. Cette figure montre que dans le domaine défini comme athermique précédemment, la contrainte d'écoulement est constante jusqu'à des températures élevées (700K). Par contre, la température de transition est significativement supérieure à la température ambiante (environ 350K). Le glissement dans l'acier de référence est donc « thermiquement activé » à température ambiante. La contrainte effective à température ambiante est de l'ordre de 70 MPa pour des essais à $7 \times 10^{-4}$ et $4 \times 10^{-3}$ s$^{-1}$. Cette valeur est très cohérente par rapport aux essais de relaxation, que nous avons réalisés par ailleurs, qui donnent une valeur d'environ 60 MPa [ALLAIN 2011] (cf. Figure 46(b)).

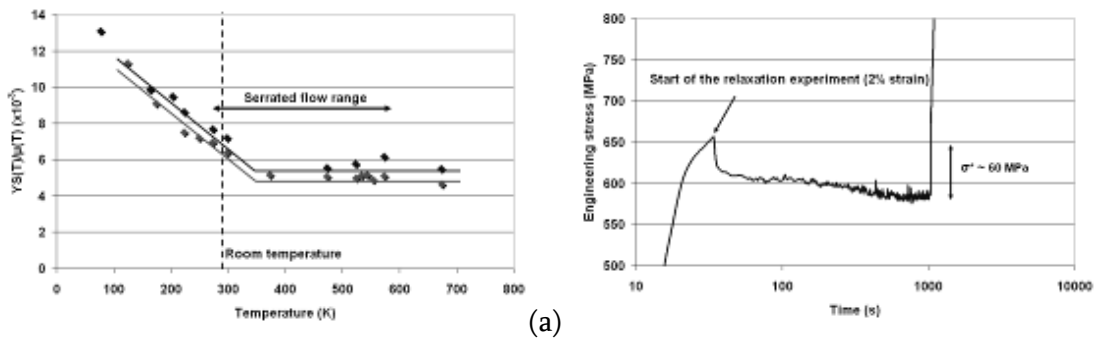

Figure 46 : (a) évolution du rapport entre la limite d'élasticité et le module de cisaillement de la nuance de référence en fonction de la température. Données de [ALLAIN 2010_2] en bleu, données de [KUNTZ 2008] en rouge (b) Mesure de la contrainte effective à température ambiante par un essai de relaxation sur la nuance de référence [ALLAIN 2011].

Le même comportement en température a été observé dans des nuances stables sans carbone comme les aciers Fe25Mn3Si3Al. Ceux-ci présentent une température critique bien supérieure à celles des aciers ternaires FeMnC (400 K environ, révélateur d'une énergie d'activation apparente plus élevée) mais un volume d'activation plus faible que la nuance de référence Fe22Mn0.6C [GRASSEL 2000]. Compte tenu des résultats précédents, on peut penser que ce sont les atomes de silicium qui contrôlent le processus et jouent un rôle similaire au carbone (d'où les énergies et volume d'activation différents). Ce résultat explique en grande partie le comportement des Alliages à Mémoire de Forme (AMF) FeMnSi. Le glissement des dislocations est presque complètement inhibé dans les nuances Fe30Mn6Si, au profit de la seule transformation martensitique ε [MYAZAKI 1989].

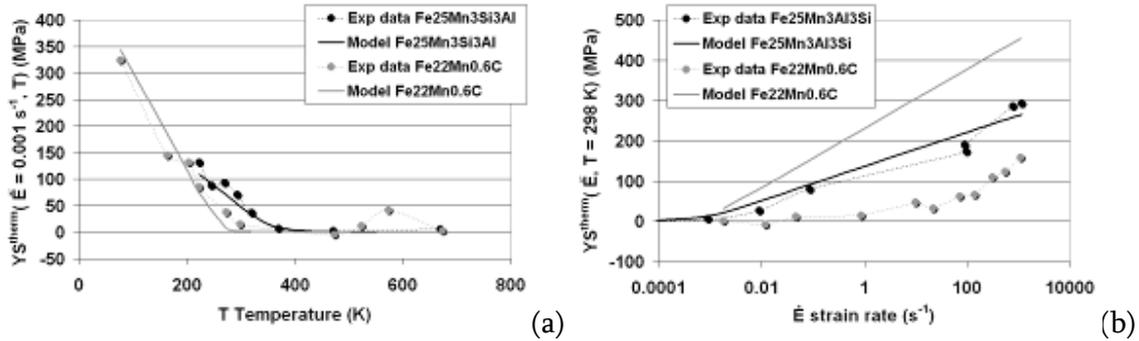

Figure 47 : (a) Evolution expérimentale et modélisée de la contrainte effective en fonction de la température pour la nuance de référence et une nuance sans carbone Fe25Mn3SiSAl [GRASSEL 2000] (b) Evolution expérimentale et modélisée de la contrainte effective en fonction de la vitesse de déformation pour les mêmes nuances que (a) [ALLAIN 2010_1].

Dans le cas des aciers Fe25Mn3Si3Al sans carbone, l'équation(42) permet de d'écrire simultanément les effets de la vitesse de déformation et de température sur la contrainte effective comme le montre la Figure 47. Par contre, elle se révèle malheureusement insuffisante dans le cas des aciers FeMnC. En effet, ces aciers à température ambiante sont le siège d'un mécanisme de vieillissement dynamique qui modifie profondément la mobilité des dislocations (vitesse de glissement interdites) et rend négative la sensibilité des alliages à la vitesse de déformation.

Aux faibles vitesses de déformations, le paramètre classique $m_{vitesse}$ de sensibilité à la vitesse de déformation vaut pour la nuance de référence :

$$m_{vitesse} = \frac{\partial \ln(\dot{E})}{\partial \ln(\Sigma)}\bigg|_{T=298K} \approx \frac{\Delta \ln(\dot{E})}{\Delta \ln(\Sigma)}\bigg|_{T=298K} = -170 \qquad (43)$$

Dans le cas de la nuance de référence, ce mécanisme spécifique d'interactions n'agit qu'aux faibles vitesses de déformation (< 1 s$^{-1}$). La Figure 47(b) montre d'ailleurs qu'au-delà, la variation de la contrainte avec la vitesse de déformation est bien décrite par le modèle (pente). Les données à hautes vitesses peuvent donc être exploitées pour estimer le profil en contrainte des obstacles au glissement des dislocations, dont le franchissement est le processus thermiquement activé. Suivant la théorie de Kocks [KOCKS 1995], l'aire apparente d'activation des obstacles s'exprime en fonction de la contrainte :

$$\Delta a(\Sigma) = \frac{m_{vitesse}(\Sigma) k_B T}{b_{110} \Sigma} \qquad (44)$$

La Figure 48 représente les contraintes d'écoulement en fonction des aires apparentes d'activation pour les essais réalisés à des vitesses de déformation supérieures à 1 s$^{-1}$ de la

Figure 47. Quelle que soit la valeur de contrainte, le glissement est thermiquement activé avec des aires d'activation apparentes très faibles (échelle du processus de franchissement de l'obstacle) de l'ordre de 30 b² cohérente avec un durcissement par atomes interstitiels. La cohérence de cette analyse (mesure et interprétation) est renforcée par l'extrapolation aux faibles aires. Cette extrapolation montre une valeur de « mechanical threshold » de 1000 MPa en parfait accord avec les valeurs estimées de limite d'élasticité à 0K de 950 MPa (cf. Figure 48 qui reprend les données de la Figure 46(a) prolongées linéairement à 0K).

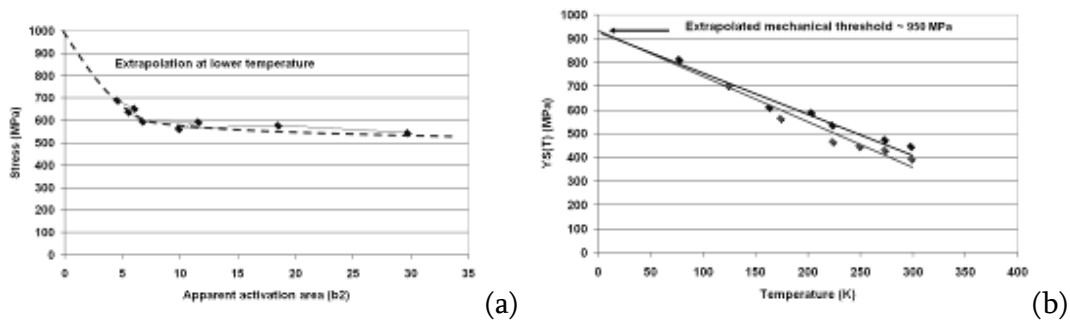

Figure 48 : (a) Profil apparent en contrainte des obstacles au glissement des dislocations en fonction de l'aire d'activation déterminé grâce aux essais à haute vitesse de déformation à température ambiante (hors vieillissement dynamique) établi dans le cas d'une interaction « dislocations / obstacles ponctuels fixes ». Définition d'un seuil mécanique (« mechanical threshold ») par extrapolation à 0K. (b) Extrapolation à 0K de l'évolution des limites d'élasticité de la nuance de référence avec la température [ALLAIN 2010_2].

On regrettera toutefois l'absence de données expérimentales qui permettrait un ajustement du modèle sans avoir à considérer le domaine de vieillissement dynamique, c'est-à-dire des résultats d'essais en vitesses à basse température principalement. Ces données permettraient de dresser un profil d'obstacle complet en l'absence d'interactions dynamiques avec les atomes de carbone et ce avec une approche conventionnelle, comme dans le cas des alliages FeMnSiAl. Dans un second temps, ce modèle pourrait être complété par la prise en compte de la contribution du vieillissement dynamique, comme le propose Hong par exemple [HONG 1986] ou par une approche de type Kocks pour les interactions dynamiques avec des solutés mobiles [KOCKS 1995].

2.4.2.2.    *Caractéristiques du processus de vieillissement dynamique*

Comme le montre la Figure 49, les courbes de traction de la nuance de référence Fe22Mn0.6C présentent des instabilités en contrainte typiques à des températures proches de l'ambiante. La forme de ces perturbations de contraintes est similaire à celle reportée dans le cas de l'effet Portevin-Le Chatelier (PLC) dans des alliages d'aluminium (morphologie de type A, B ou C par exemple). Ces instabilités, tout comme la sensibilité négative de la contrainte à la vitesse de déformation, sont les symptômes d'un mécanisme de vieillissement dynamique lié au à l'interaction dynamique entre atomes de carbone mobile et glissement des dislocations.

En effet, les aciers austénitiques sans carbone ne présentent jamais de courbes de traction perturbées, y compris les nuances à très faible EDE [REMY 1975]. Le phénomène est d'ailleurs réduit dans les alliages contenant de l'aluminium, qui est supposé réduire l'activité du carbone [ZUIDEMA 1987][SHUN 1992] ou dans les alliages avec substitution partielle du carbone par de l'azote [BRACKE 2007_1]. Ce mécanisme et ses conséquences macroscopiques ont fait l'objet de nombreuses études ces dernières années, sans toutefois faire émerger un consensus sur la nature du mécanisme d'interaction dislocation/carbone.

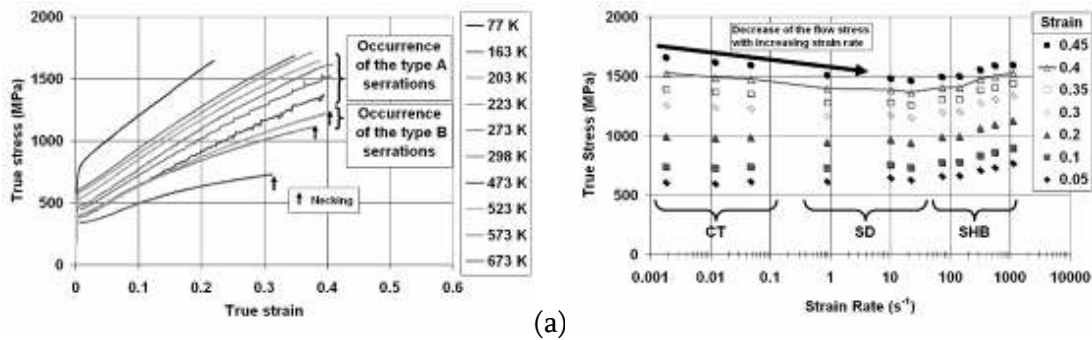

Figure 49: (a) Courbes de traction rationnelles de la nuance de référence à différentes températures – les domaines d'apparition d'instabilités sur les courbes sont indiqués (b) Evolution des contraintes d'écoulement à différents niveaux de déformation en fonction de la vitesse de traction – sensibilité négative à la vitesse de déformation dans le domaine des basses vitesses (CT : Machine de traction conventionnelle, SD : Machine de traction à grande vitesse, SHB : Barres d'Hopkinson) [ALLAIN 2008_2].

La Figure 49 présente l'évolution des contraintes d'écoulement de la nuance de référence à température ambiante après des taux de déformation données en traction sous différentes vitesses de déformations (CT = machine de traction conventionnelle, SD = machine de traction grande vitesse, SHB = Barre d'Hopkinson). Elle confirme la sensibilité négative à la vitesse de déformation de cette nuance à température ambiante et aux faibles vitesses de déformation. C'est dans ce domaine que l'on observe la présence d'instabilités sur les courbes de traction. Ces domaines de températures et vitesses de déformation (domaine de sensibilité négative à la vitesse de déformation, incluant le domaine d'apparition des instabilités) sont relativement bien identifiés dans la littérature, non seulement pour la nuance de référence mais aussi pour d'autres nuances comme les aciers Hadfield [DASTUR 1981][ALLAIN 2004_1][BRACKE 2007_1][SHUN 1992][RENARD 2010].

La Figure 49(a) montre que selon ces conditions de température et de vitesse de déformation, les instabilités ne semblent apparaître qu'après une certaine déformation critique, caractéristique aussi commune avec l'effet PLC, et donc essentielle, car il s'agit d'un processus thermiquement activé [SHUN 1992]. Toutefois, la détermination précise de cette déformation critique est difficile, comme le montre nos travaux sur l'analyse multi-fractal des courbes de comportement [CHATEAU 2009] ou les mesures récentes de champs de déformation locaux de Roth *et al.* [ROTH 2012][LEBYODKIN 2012]. Ces deux travaux montrent que malgré l'absence d'instabilités formées sur la courbe de traction, les éprouvettes de traction sont le siège de fluctuations locales de vitesses de déformation corrélées de façon spatio-temporelle. Cette difficulté de mesure explique en grande partie la dispersion dans la littérature des valeurs d'énergie d'activation apparente de ce processus. Kuntz, Allain and Bracke rapportent des valeurs proches de 15 kJ.mol$^{-1}$ pour des nuances Fe-22Mn-CN alors que Dastur *et al* et Shun *et al.* proposent des valeurs de 146 et 60 kJ.mol$^{-1}$ respectivement pour des aciers Hadfield. La découverte de ces fluctuations de déformations préalables aux instabilités en contrainte ouvre un champ d'investigation nouveau et pertinent pour mieux quantifier ce processus de vieillissement dynamique dans les aciers TWIP et tenter de rétablir un consensus sur la nature du mécanisme.

Les instabilités sur les courbes de traction sont dues comme dans le cas du PLC à des hétérogénéités très locales de vitesse de déformation dans les éprouvettes, sous la forme de bandes de déformation. Nous avons été parmi les premiers à identifier et caractériser la dynamique de ces bandes dans les aciers TWIP ([CHEN 2007][ALLAIN 2008_2][LEBEDKINA 2009]) et ce en utilisant différentes techniques. La Figure 50 montre par exemple des observations de bandes de déformation et leurs déplacements sur le fût d'une éprouvette au cours d'un essai de traction par des techniques de thermographie infrarouge (IR) ou des mesures de champs locaux par corrélation d'images digitales [ALLAIN 2008_2]. Elles illustrent la propagation des bandes de déformation dans le cas d'instabilités de type A.

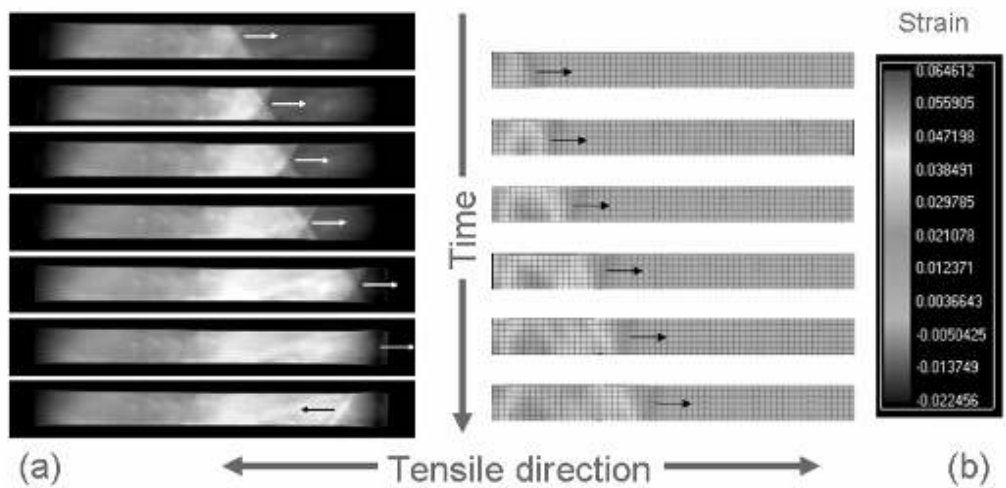

Figure 50 : Observation de la propagation de bandes de déformation corrélées aux instabilités sur les courbes de traction (a) en caméra infrarouge (mesure de l'échauffement) et (b) en corrélation d'images digitales (mesure de champs locaux) [ALLAIN 2008_2].

Les caractéristiques spatio-temporelles de ces bandes ont été documentées dans de nombreuses nuances par ces mêmes techniques [LEBEDKINA 2009],[ZAVATIERRI 2010] [DECOOMAN 2009][ALLAIN 2008_2][RENARD 2010] [CANADINC 2008][CHEN2007]. Les conclusions de ces différentes études sont de façon surprenante très cohérentes entre elles :

- La vitesse des bandes de déformation augmente avec la vitesse de déformation mais diminue avec la déformation
- La déformation dans les bandes de déformation augmente avec la déformation macroscopique
- La vitesse de déformation dans les bandes est très supérieure de 10 à 20 fois à la vitesse de déformation macroscopique appliquée.

La Figure 51 tirée de [LEBEDKINA 2009] montre par exemple ces évolutions dans le cas de la nuance de référence.

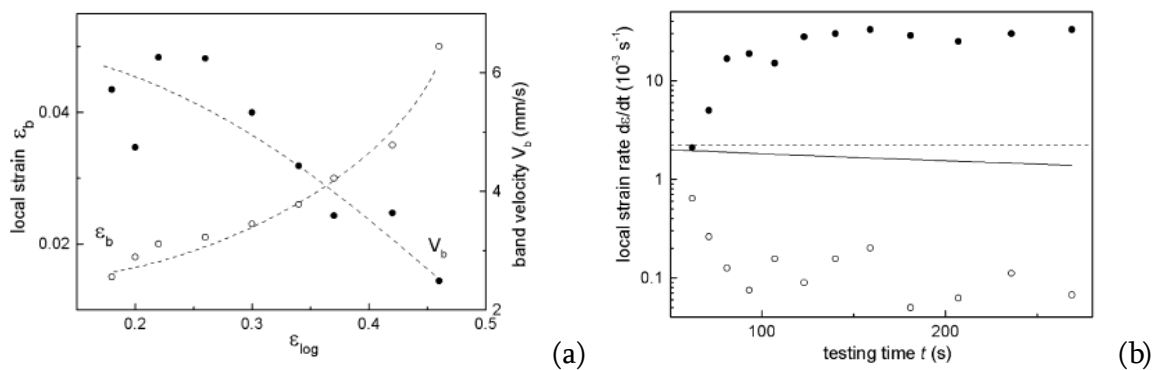

Figure 51 : (a) Evolution de la déformation locale dans les bandes et vitesse de propagation de ces bandes sur l'éprouvette de traction en fonction de la déformation (b) Vitesse de déformation en-dehors et dans les bandes de déformation en fonction du temps d'essai [LEBEDKINA 2009] lors d'essais de traction à $2.2 \times 10^{-3}$ s$^{-1}$.

La plupart des études sur les aciers TWIP FeMnC rapporte des instabilités de type A selon la classification conventionnelle du PLC. Elles se caractérisent par la formation de bandes de déformation relativement minces mais mobiles. Ces bandes se forment lors des sauts de contraintes et se déplacement lors des « plateaux », à contrainte quasiment constante. Des instabilités de type B ou C sont moins souvent décrites, sauf à plus hautes températures ou aux très faibles vitesses de déformation [RENARD 2012]. Contrairement à un effet PLC conventionnel, le domaine de vitesse de déformation dans lequel les instabilités sont de type A est beaucoup plus étendu que dans le cas des alliages d'aluminium [LEBEDKINA 2009]. On pourrait ainsi soupçonner que vieillissement dynamique et fort écrouissage dû au maclage mécanique interagissent pour stabiliser ce processus. Le processus de vieillissement dynamique dans ces aciers reste encore peu compris et présente donc les caractéristiques d'un PLC atypique.

Les conditions d'apparition des instabilités restent ainsi un phénomène surprenant comme le montre l'analyse du comportement après des essais de relaxation sur la Figure 52. De façon inattendue, les essais de relaxation réalisés après différents taux de pré-déformation ont déclenché après recharge des instabilités alors que la déformation critique était loin d'être atteinte pour ces conditions de température et vitesse de déformation. Ce résultat reste pour l'instant inexpliqué, mais pourrait être lié à une intensification « catastrophique » des fluctuations observés par Roth *et al.*.

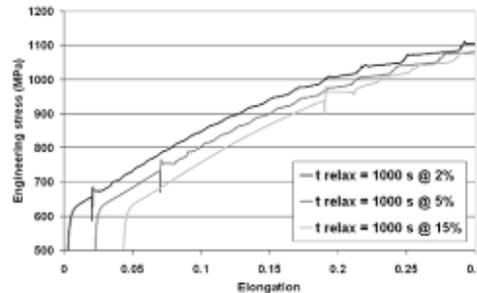

**Figure 52 : Courbes de traction conventionnelles suite à des essais de relaxation (1000 s) après différents niveaux de pré-déformation. Les courbes ont été décalées en déformation pour des raisons de lisibilité [ALLAIN 2011_2].**

La contribution relative du vieillissement dynamique à l'écrouissage et au comportement mécanique de cette famille d'aciers est certainement ce qui fait le plus débat actuellement dans la littérature. Cette position proposée initialement par Dastur et Leslie et suivie par de nombreux autres auteurs [DASTUR 1981][ZUIDEMA 1987][LAI 1989_1][LAI 1989_2] suppose que le seul mécanisme de vieillissement dynamique peut expliquer le fort taux d'écrouissage de ces aciers, de manière indépendante du maclage mécanique et de la transformation martensitique ε. Sur la base de calculs *ab initio*, Owen et Grujicic ont suggéré de plus le rôle particulier de dipôles MnC dans ce processus [OWEN 1999][13].

Cette explication a le mérite en effet d'expliquer simplement le résultat de la Figure 41, montrant le rôle particulier du carbone sur le comportement. Elle explique en outre pourquoi certains alliages très chargés en carbone (avec une forte EDE) présentent un fort taux d'écrouissage en l'absence de maclage.

A contrario, notre équipe s'est plutôt toujours tournée vers des explications relevant de la microstructure de déformation (maclage mécanique ou transformation martensitique) pour expliquer ce comportement, scénario largement détaillé dans le chapitre précédent. Cette position rend par contre plus difficile l'explication des effets du carbone. Les quatre principales raisons motivant ce choix sont les suivantes :
- Dès mes travaux de thèse, nous avions estimé la différence de comportement de l'acier de référence avec et sans maclage dans des conditions de vieillissement dynamique identiques qui prouve l'existence d'un durcissement d'origine structural important (cf. Figure 37).
- Les mécanismes associés à des interactions locales avec le carbone sont nécessairement des contributions isotropes à l'écrouissage. Il serait impossible d'expliquer les forts effets Bauschinger dans ces structures (cf. Figure 26).

---

[13] Cette explication très spécifique rend difficile l'interprétation du mécanisme de vieillissement dynamique aussi observé dans les alliages FeNiC, comme le Fe22Ni0.6C [ALLAIN 2008_2].

- La forme particulière de la surface de charge lors de trajets monotones directs tels que représentés sur la figure ne peut s'expliquer que par des effets de la texture au travers du maclage mécanique (cf. Figure 32).
- L'évaluation de la contribution du vieillissement dynamique par l'équipe de Y. Estrin, montre que la contribution au durcissement isotrope est additive de l'ordre de 20 MPa au mieux dans une nuance Fe18Mn0.6C1.5Al [KIM 2009].

Ce scénario n'exclut cependant pas que le processus de vieillissement dynamique ait un effet sur le maclage mécanique et donc un effet sur le comportement macroscopique. Cette discussion sur les effets indirects du carbone sur l'effet TWIP au travers du glissement est l'objet de la section suivante.

### 2.4.3. Les effets indirects du carbone, de la structuration des dislocations au maclage mécanique.

Contrairement aux éléments substitutionnels qui augmentent l'EDE et contribue donc à un adoucissement de ces aciers, le carbone augmente le taux d'écrouissage de ces nuances, en jouant très probablement sur l'intensité de sa contribution cinématique [BOUAZIZ 2011]. La plupart des nuances ternaires FeMnC est sujette à température ambiante à un processus de vieillissement dynamique. Ce processus ne contribue que très peu à l'écrouissage de l'austénite, par contre cette interaction entre carbone et dislocations contribue à la contrainte d'écoulement sous la forme d'une contrainte effective. Cette dernière contribution est significative à température ambiante et dépend de la teneur en carbone.

A ce jour, il n'existe dans la littérature aucun consensus sur le lien entre carbone, mécanismes de déformation et écrouissage. Les scénarios d'explications présentés ci-dessous ne sont donc que des hypothèses de travail documentées qu'il serait intéressant de détailler et d'approfondir dans de futures études.

Si l'on considère l'équation (40), reprise ci-dessous

$$\Sigma(E) \approx \sigma_0 + \sigma_M(\varepsilon_M) + F(E)\sigma_T(\varepsilon_T) \qquad (45)$$

Les différents résultats montrent que $\sigma_M(\varepsilon_M)$ ne dépend que peu de la teneur carbone (cf. Figure 30). Selon cette équation, il ne reste alors que deux façons possibles d'expliquer une augmentation du taux d'écrouissage des aciers TWIP avec leur teneur en carbone :
- l'augmentation de la contrainte supportée par les macles $\sigma_T$
- l'augmentation de la cinétique de maclage (propension au maclage) F

### 2.4.3.1. *Carbone et contraintes dans les macles*

Le premier scénario a été suivi par Huang *et al.* [HUANG 2011] pour expliquer les courbes de comportements présentées sur la Figure 41. A l'aide de notre modèle de durcissement cinématique [BOUAZIZ 2008_1], les courbes de traction ont été simulées en ajustant uniquement le terme $n_0$ en fonction de la teneur en carbone, sans changer la cinétique de maclage d'un acier à l'autre. Les valeurs de $n_0$ ainsi ajustées évoluent linéairement avec la teneur en carbone, comme le montre la Figure 53(a). La figure (b) reprend les courbes de traction simulées et expérimentales de deux aciers Fe22Mn0.6C et Fe22Mn1.2C à titre d'exemple.

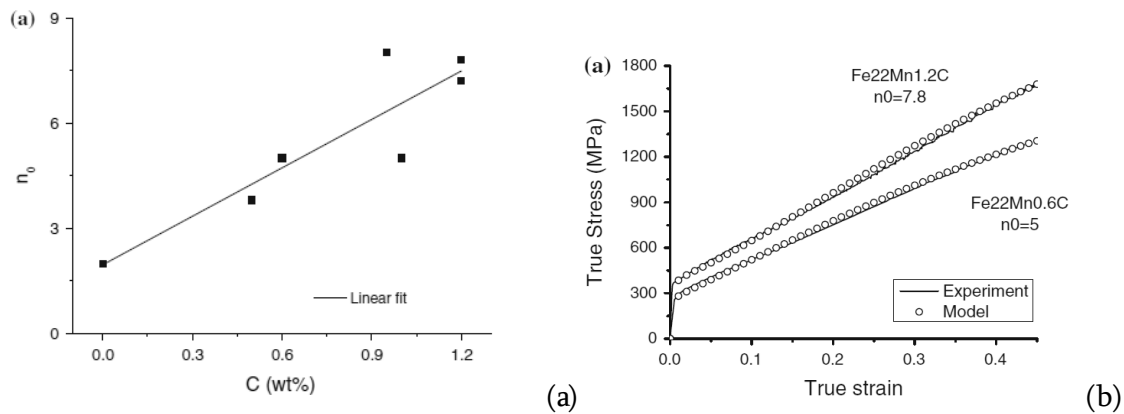

**Figure 53 : (a) Evolution du paramètre $n_0$ de notre modèle de durcissement cinématique pour reproduire l'effet carbone sur le comportement des aciers ternaires FeMnC (b) exemple d'ajustement du modèle pour reproduire la courbe de traction de la nuance de référence et d'un acier Fe22Mn1.2C [HUANG 2011].**

Considérer que le terme $n_0$ augmente avec la teneur en carbone revient à penser que les processus de relaxation des empilements sur les macles s'activent plus difficilement (ou que la contrainte supportée par une macle est plus grande, selon une vision composite). L'explication mise en avant par ce mécanisme est basée sur les observations expérimentales récentes en MET de Idrissi *et al.* [IDRISSI 2009] de l'interface des macles sur des aciers TWIP. Les joints de macle des aciers chargés en carbone contiendraient de nombreuses dislocations sessiles a contrario des aciers sans carbone. En conséquence, les empilements de dislocations sur ces interfaces seraient facilités.

### 2.4.3.2. *Carbone et cinétique de maclage*

Le second scenario repose sur une plus grande facilité des nuances au carbone à macler, autrement dit, la teneur en carbone serait susceptible de jouer sur la contrainte critique et sur la cinétique de maclage des aciers.

Il n'existe malheureusement pas dans la littérature d'étude comparée complète des cinétiques de maclages (y compris avec toutes les réserves discutées dans la section précédente) pour des teneurs en carbone variables. La Figure 54 montre par contre les propriétés de traction et les mécanismes de déformation d'aciers ternaires issus de différentes études mais présentant une EDE faible mais identique, c'est-à-dire une même propension théorique à macler. La figure (a) représente les limites d'élasticité conventionnelles et contraintes d'écoulement à rupture de ces aciers FeMnC à gros grains (> 20 µm) et la figure (b) leurs positionnements relatifs dans une carte Mn/C de Schumann par rapport à la droite délimitant les domaines d'austénite stables et instables. Ce positionnement leur confère des EDE estimées très proches.

La figure (a) confirme le « paradoxe carbone ». La limite d'élasticité de ces aciers augmente peu avec la teneur en carbone, avec une sensibilité de 250 MPa/% environ. Cette valeur est relativement comparable à celle retenue par Bouaziz *et al.* de 187 MPa.% [BOUAZIZ 2011]. Cette évolution s'explique quasiment pour moitié par l'augmentation de la contrainte effective à température ambiante (cf. Figure 43) qui dépend fortement de la teneur en carbone. L'autre contribution est probablement d'ordre structural. Par contre, la contrainte d'écoulement augmente elle très sensiblement avec la teneur en carbone (taux d'écrouissage).

Le résultat le plus surprenant est que ces 5 aciers ne présentent pas les mêmes mécanismes de déformation. L'acier binaire Fe30Mn ne présente ni maclage ni transformation martensitique ε induite, alors que les 4 autres aciers ternaires FeMnC se déforment en partie par maclage mécanique et leurs courbes de traction présentent des instabilités en contraintes caractéristiques, manifestation d'un mécanisme de vieillissement dynamique.

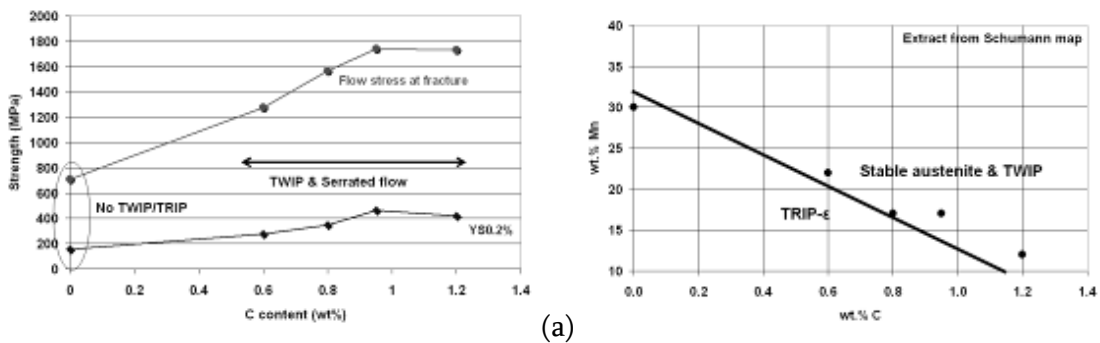

Figure 54 : (a) Limites d'élasticité et contrainte vraie à rupture en fonction de la teneur en carbone d'aciers FeMnC présentant la même EDE estimée (b) Position des aciers de (a) dans une carte de Schumann Mn/C [ALLAIN 2010_2].

Nous avons retrouvé ce type de comportement dans des alliages « cousins » des aciers austénitiques FeMnC TWIP, les aciers austénitiques FeNiC TRIP α'. La Figure 55 tirée de Dagbert *et al.* [DAGBERT 1996] montre l'évolution des limites d'élasticité et résistances mécaniques d'aciers FeNiC dont les ratios de composition en Ni et C ont été choisis judicieusement afin que tous les alliages présentent la même température Ms de transformation martensitique α' (environ – 50°C). Cette dernière température est comme l'EDE un indicateur de la stabilité d'un point de vue thermochimique des alliages. De manière analogue aux aciers FeMnC, les aciers FeNiC peu chargés en carbone (C< 0.3%) présentent des écrouissages faibles (différence entre Rm et Re) sans effet TRIP alors que les alliages fortement chargés (C>0.3%) présentent un effet TRIP et donc des écrouissages importants, et ce, en présence concomitante d'un mécanisme de vieillissement dynamique.

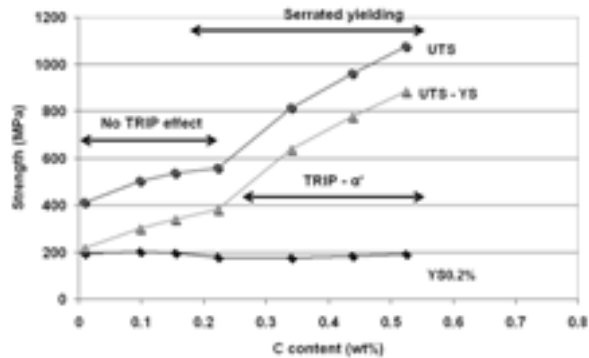

Figure 55 : Limites d'élasticité (YS) et résistances mécaniques (UTS) en fonction de la teneur en carbone d'aciers FeNiC présentant la même Ms (température de transformation martensitique α'). Domaine d'apparition des mécanismes d'écrouissage (TRIP et vieillissement dynamique) [DABGERT 1996]

Pour des EDE faibles, la susceptibilité à macler semble donc augmenter avec le carbone. On notera en revanche qu'un alliage Fe35Mn0.6C présentant une EDE forte (40 mJ/m$^2$) ne macle pas malgré une teneur en carbone importante. Pour expliquer ce résultat, plusieurs hypothèses et mécanismes peuvent être invoqués.

### 2.4.3.3. *Activation thermique du glissement et maclage mécanique*

Certains auteurs comme le groupe de Koyama *et al.* [KOYAMA 2012] suggèrent que le vieillissement dynamique explique la propension à macler ou à la transformation martensitique ε. Leur explication est basée sur un différentiel de mobilité supposé entre dislocations partielles de tête et de queue bordant une faute d'empilement en présence d'un processus de vieillissement dynamique. Cette hypothèse pourrait paraître convaincante mais elle ne peut permettre d'expliquer que des alliages FeMnSiAl, bien étudiés par ailleurs, présentent de fortes cinétiques de maclage [GRASSEL 2010]. La seule caractéristique commune entre alliages FeMnC et FeMnSiAl susceptible d'expliquer cette propension au maclage semble être finalement une contrainte effective élevée due à un processus de glissement thermiquement activé, comme noté aussi par [CHUMLYAKOV 2002].

En rendant à l'échelle locale le glissement difficile, le maclage peut être vu comme un mécanisme supplétif au glissement localement car de nombreuses dislocations mobiles sont bloquées par des atomes de carbone et ne peuvent assurer la vitesse de glissement [ALLAIN 2010_2]. Les macles assureraient donc localement des « bouffées » de vitesses de déformation suffisantes (rôle adoucissant des macles mis en évidence par N. Shiekhelshouk [SHIEKHELSHOUK 2006] par exemple ou les bandes de cisaillement de Gutierrez [GUTIERREZ 2010]). Ce scénario est assez cohérent aussi avec les observations de macles principalement dans les grains de la fibre <111>//DT lors d'une déformation en traction. On pourrait aussi envisager un modèle reposant sur un différentiel de mobilité des dislocations partielles, thermiquement activé, de manière analogue à la proposition de Farenc *et al.* dans un alliage TiAl [FARENC 1993].

On retrouve aussi cette idée finalement dans les calculs de Meyers *et al.* [MEYERS 2001] d'une contrainte critique de maclage macroscopique, basée sur une compétition entre contrainte effective pour le glissement et le maclage. Pour expliquer et quantifier cette constatation expérimentale décisive, un modèle d'interaction locale à l'échelle des dislocations reste à construire.

### 2.4.3.4. *Glissement planaire et maclage mécanique*

Le second élément d'explication possible est que le carbone favorise un glissement planaire. Cette planéité du glissement a pour conséquence des empilements plus grands et donc conduit à abaisser la contrainte critique de maclage effective de l'acier. Cela entraîne naturellement une propension plus importante au maclage mécanique, comme le montre l'équation (7) de Venables.

Nous avons observé bien entendu les caractéristiques de ce glissement planaire lors de nos études en MET de la nuance de référence. La Figure 56 montre par exemple une comparaison de la structuration des dislocations dans l'acier de référence Fe22Mn0.6C après 5% de déformation [BARBIER 2009_1] et un acier Fe30Mn binaire après 20 % de déformation [HUANG 2011]. Dans le premier cas, les segments de dislocations s'organisent selon des structures très anisotropes (de type Taylor) caractéristiques d'un glissement planaire [GUTIERREZ 2012], alors que dans le second, la distribution en cellules est très isotrope.

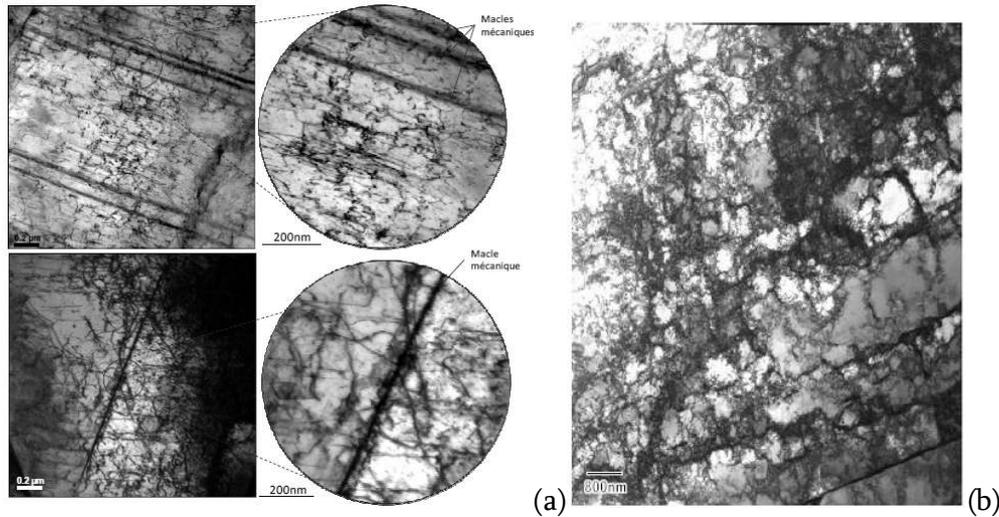

Figure 56 : Micrographies MET en champ clair des structures de dislocations (a) dans la nuance de référence après 5% de déformation [BARBIER 2009] (b) dans une nuance binaire Fe30Mn après 20% de déformation [HUANG 2011].

Le glissement planaire dans les structures austénitiques s'explique par une inhibition des mécanismes de glissement dévié. Cette inhibition peut avoir quatre causes distinctes :
- Une EDE faible qui rend énergétiquement défavorable la constriction des fautes d'empilement et la recombinaison des dislocations partielles en segment vis avant glissement dévié (processus de constriction [PUSCHL 2002]). C'est bien entendu le cas des aciers austénitiques TWIP. Par contre, comme l'EDE augmente avec la teneur en carbone, le glissement dévié devrait être facilité par cet élément !
- un mécanisme de mise en ordre à courte distance (« Short Range Ordering : SRO »). C'est une hypothèse envisagée depuis peu par DeCooman *et al* [DECOOMAN 2009] et suggérée par les travaux de Gerold *et al* [GEROLD 1989] et les calculs d'interactions sur les couples MnC de Owen et Grujicic [OWEN 1999]. Aucune évidence expérimentale de ce mécanisme n'a toutefois été produite à ce jour dans les aciers ternaires FeMnC (appariement de dislocations). Gutierrez et Raabe ont même montré l'absence de ce mécanisme dans des alliages encore plus complexes comme les FeMnAlC [GUTIERREZ 2012]. C'est donc une cause que nous n'envisageons pas.
- Une friction de réseau élevée. La probabilité d'un évènement de glissement dévié s'exprime souvent en fonction de la distance d'annihilation entre segments vis notée ys [GEROLD 1989] lors d'un processus de restauration dynamique :

$$ys = \mu b \frac{\sin(\theta_{cross})}{2\pi}\left(\tau_0 - S_{cross}\tau_{app} \pm \tau_{interne}\right) \qquad (46)$$

Avec $\theta_{cross}$ et $S_{cross}$ deux facteurs constants dépendant de l'orientation cristalline. $\tau_0$, $\tau_{app}$ et $\tau_{interne}$ sont respectivement la friction de réseau, la contrainte appliquée et un terme de contrainte interne. Cette relation montre que plus la friction de réseau $\tau_0$ est élevée, plus la distance ys est grande, plus les évènements de glissement dévié seront rares. La contrainte effective due à l'interaction avec les atomes de carbone aura un effet similaire. Dans certaines orientations cristallines, ce phénomène sera amplifié par les très fortes contraintes internes développées dans ces aciers.

- Processus de constriction perturbé par les atomes en solution solide [GUTIERREZ 2012]. Andrews *et al.* [ANDREWS 2000] a montré que l'interaction entre atomes de carbone et dislocations vis conduisait à une augmentation de l'énergie de constriction. Les atomes de carbone contribueraient donc à inhiber le processus de glissement dévié.

Parmi ces quatre explications possibles, les deux dernières suggèrent que l'ajout de carbone peut rendre le glissement planaire, en inhibant localement le processus de glissement dévié ; et ce de façon antagoniste à l'augmentation d'EDE.

### 2.4.3.5. *PLC, structuration du glissement et contraintes locales*

On ne peut pas non plus totalement exclure une contribution possible du vieillissement dynamique à l'activation du maclage, comme le suggère les figures précédentes. En effet, Hong a montré qu'en présence de vieillissement dynamique, les dislocations mobiles, donc géométriquement nécessaires, s'accumulaient dans des zones de fortes contraintes internes au cours de la déformation [HONG 2000]. Ce mécanisme pourrait expliquer l'observation de très nombreuses régions, mal indexées en EBSD, sans pour autant être des macles, dans les grains après déformation et correspondant à des zones de forts gradients de déformation. Ces régions ressemblent donc à des sous-joints mais apparaissent très tôt au cours de la déformation (20% de déformation dans le cas présenté sur la Figure 57). Ces régions pourraient à elles seules constituer des zones de concentration de contraintes nécessaires pour l'activation du maclage, et qui ne pourraient pas être relaxées par l'émission continue de dislocations parfaites, mais nécessiteraient une émission coordonnée (germe de macle).

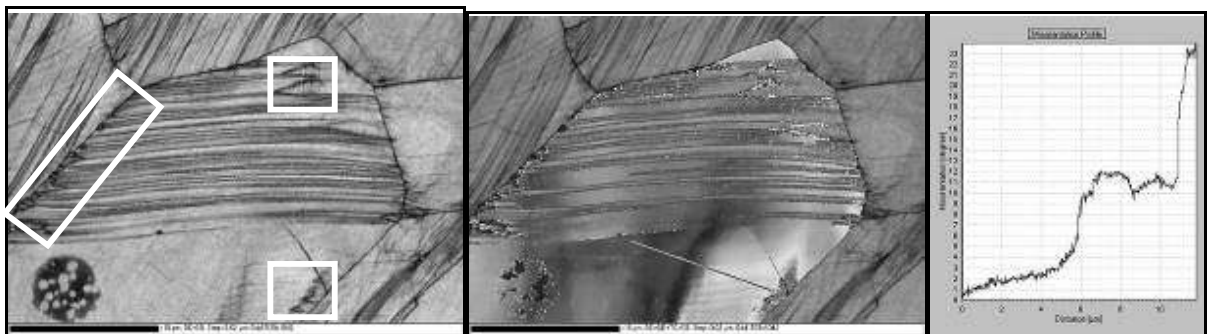

Figure 57 : Cartographie en contraste de bandes. Le grain principal comporte un système de maclage. En rouge : désorientation de 60° autour de <111>. Les carrés blancs indiquent la présence de zones montrant une forte désorientation et mal indexées. Représentation du gradient de désorientation au sein du grain et tracé du profil de désorientation suivant la ligne noire. La couleur bleu représente l'orientation majoritaire du grain [BARBIER 2008_1].

### *2.4.3.6. Discussion*

Il est possible que ces trois mécanismes stimulent simultanément le maclage, comme processus localement substitutif au glissement. Le premier contribue à rendre le glissement plus difficile alors que les deux seconds contribuent à diminuer la contrainte critique de maclage. On ne peut donc réduire ce processus à une simple transition entre glissement et maclage comme le propose Meyers *et al.* [MEYERS 2001], car le maclage n'est jamais observé seul, sans glissement préalable.

Tous les mécanismes décrits ci-dessus restent en l'état des hypothèses, qu'il serait pertinent d'approfondir à la fois en terme de mécanismes locaux (à l'échelle des interactions entre dislocations) et d'effet quantifié sur la cinétique de maclage.

## 2.5. Conclusions et Perspectives

Les aciers TWIP sont des matériaux passionnants qui sont loin d'avoir livré tous leurs secrets malgré une intense recherche académique et industrielle. Ils couplent les difficultés des aciers inoxydables austénitiques, des aciers au carbone et des alliages d'aluminium (vieillissement dynamique). En partie grâce à nos travaux, l'effet TWIP est compris dans les grandes lignes mais les challenges scientifiques et techniques restent nombreux :

- Axe de recherche 1 : Mesure expérimentale de la fraction de phase maclée (cinétique de maclage) par une technique statistique et reproductible. C'est la source principale d'ambigüités dans l'interprétation du comportement de ces aciers. Le couplage entre la microscopie à force atomique et la technique d'indexation en EBSD devrait permettre cette quantification sans biais.
- Axe de recherche 2 : Identification et quantification des processus d'interactions entre carbone et dislocations (processus de vieillissement dynamique atypique). Des essais de traction à basse températures et à différentes vitesses devraient permettre de fournir des informations originales pour modéliser ce processus.
- Axe de recherche 3 : Calcul de l'EDE. Les méthodes thermodynamiques actuelles reposent sur de nombreux paramètres d'ajustement. Des progrès sont attendus dans le futur grâce aux techniques *ab initio*.
- Axe de recherche 4 : Définition d'un modèle pour la cinétique de maclage, basée sur une contrainte critique de maclage intégrant les notions d'orientations mais aussi les caractéristiques du glissement (planaire et thermiquement activé)
- Axe de recherche 5 : Modélisation micromécanique des changements de trajet (trajet Bauschinger et changement de trajets dur), capable de prédire la surface de charge et son évolution (écrouissage cinématique et isotrope). A terme, le formalisme devrait pouvoir expliquer le comportement en fatigue.

Ces différents travaux ne pourront qu'être enrichis par des comparaisons avec des alliages austénitiques « cousins » comme les FeNiC, NiC ou simplement FeMnSi, qui présentent des comportements mécaniques et des mécanismes de déformation similaires, et sont très largement étudiés dans la littérature.

## 2.6. Pour le plaisir des yeux

Pour conclure cette partie, je ne résiste au plaisir de proposer sur la Figure 58 quelques micrographies particulièrement « esthétiques » obtenues grâce à ces aciers FeMnC TWIP

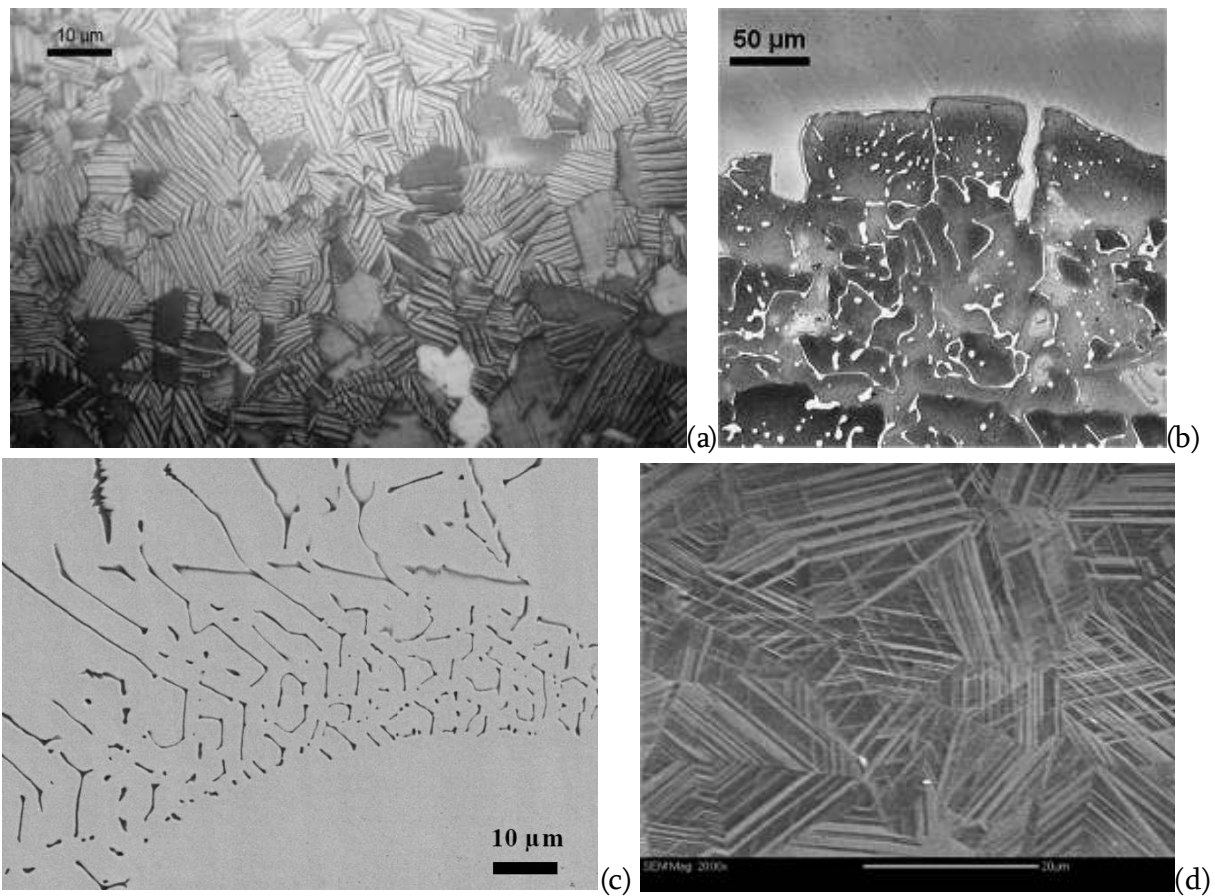

Figure 58 : (a) Artefact d'attaque Nital sur la nuance de référence – les figures linéaires sont prises parfois à tort pour des macles par certains auteurs ! [DUMAY 2008] (b) Phénomène d'infiltration de cuivre liquide dans la nuance de référence (fragilisation par métal liquide, du cuivre en l'occurence) [DUMAY 2008][BEAL 2011] (c) Solidification eutectique de TiC (sur-réseau hexagonal) dans la nuance de référence [DUMAY 2008] (d) Transformation martensitique ε intense après déformation d'un alliage à mémoire de forme Fe30Mn6Si, obtenue dans le cadre du stage de master de Roney Lino.

# 3. Comportement des aciers Dual-Phase ; des composites modèle ?

> *« En science, la phrase la plus excitante que l'on peut entendre, celle qui annonce des nouvelles découvertes, ce n'est pas Eureka mais c'est « drôle ». »*
> *Isaac Asimov*

## 3.1. Introduction

### 3.1.1. Contexte historique et technique

Contrairement aux aciers TWIP FeMnC, les aciers Dual-Phase (DP) sont maintenant industrialisés depuis les années 1990-1995 et largement utilisés dans le domaine de la construction automobile. Ils constituent la majorité des aciers THR dit modernes ou de première génération. Ces aciers présentent généralement des niveaux de résistances mécaniques compris entre 600 et 1200 MPa (cf. Figure 1 page 12), et présentent d'excellentes qualités de mise en forme (emboutissabilité ou pliabilité).

Le terme Dual-Phase signifie que ces aciers sont constitués principalement de deux phases, de la ferrite de structure cubique centrée (CC) de morphologie généralement polygonale et de la martensite α' CC ou de structure tetragonale centrée (TC), riche en carbone, en fraction très variable. De façon caricaturale, la ferrite est considérée alors comme une phase molle, et la martensite comme une phase dure. C'est cet aspect composite qui leur confère leurs excellentes propriétés mécaniques, compromis entre résistance et formabilité.

La Figure 59 montre des exemples de pièces automobiles réalisées grâce à des aciers DP de différents grades et épaisseurs : pare-choc en acier DP 1180 et pied milieu en acier DP 980 fine épaisseur pour des applications anti-intrusion et une roue de style en DP600 forte épaisseur.

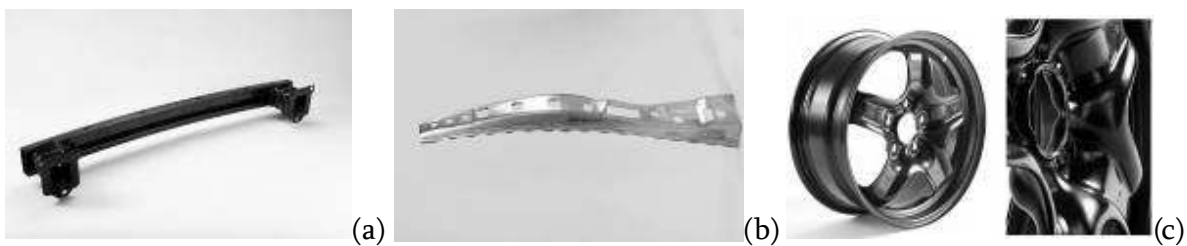

Figure 59 : Exemple d'application des aciers Dual-Phase dans le domaine de la construction automobile (a) Poutre de pare-choc en DP1180, (a) Pied-milieu en DP980 fine épaisseur et (c) roue de style en DP600 forte épaisseur. Les valeurs indiquées se rapportent traditionnellement au niveau de résistance mécanique (www.arcelormittal.com).

Les aciers DP industriels contiennent généralement d'autres phases comme la perlite ou la bainite. La quantité de ces secondes phases peut excéder la fraction de martensite mais ces aciers restent par abus de langage commercialisés sous le nom de DP. Dans la suite de ce mémoire, on ne s'intéressa bien entendu qu'aux microstructures réellement Ferrite-Martensite.

Ces microstructures particulières peuvent être obtenues dans les plupart des aciers FeC hypoeutectoïdes, à partir soit d'une décomposition de l'austénite (germination de la ferrite au refroidissement et transformation martensitique de l'austénite résiduelle sous Ms) comme dans le cas d'un procédé de laminage à chaud, soit dans le cas d'un recuit intercritique comme dans le cas d'un procédé de production de bandes de fine épaisseur laminées à froid.

La Figure 60 montre par exemple dans des diagrammes des transformations température-temps des chemins de production d'aciers DP en produits laminés à chaud ou à froid (recuit de recristallisation).

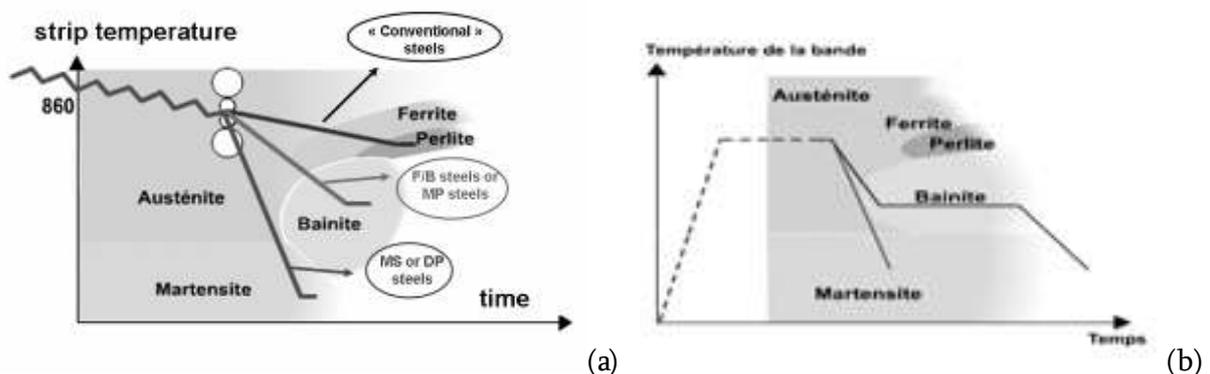

(a) (b)
Figure 60 : Traitements thermomécaniques typiques pour la production industrielle (en rouge) d'aciers Dual-Phase (a) sous forme laminée à chaud et (b) sous forme laminée à froid recuits représentés dans des diagrammes temps – température. Sont indiqués les domaines typiques d'apparition des différents produits de décomposition de l'austénite au refroidissement.

Dans la suite de cet exposé, nous ne reviendrons que peu sur les procédés d'obtention des microstructures et les processus de morphogénèse qui font encore l'objet de recherches dans le domaine des transformations de phase. Le lecteur pourra se reporter par exemple aux travaux récents de thèse dédiée à ces questions [KREBS 2009][VIARDIN 2008]. On se limitera aussi à l'étude du comportement mécanique des aciers Ferrite-Martensite « fraîche », bien que les aciers DP dit « revenu » aient aussi une grande importance pratique et industrielle [PUSHKAREVA 2009][BOUAZIZ 2011_3].

### 3.1.2. Bibliographie succincte et problématique

#### 3.1.2.1. *Lien microstructure-propriétés*

Comme la montre la Figure 3 page 12, les aciers DP présentent un comportement mécanique en traction caractérisé par une faible limite d'élasticité sans palier de Lüders Re et un taux d'écrouissage initial important. Cette combinaison de propriétés leur permet d'atteindre des allongements répartis élevés, et donc de grandes résistances mécaniques Rm. Leur rapport Re/Rm est généralement proche de 0.5.

Cette limite d'élasticité basse est due à la faible contrainte d'écoulement de la ferrite[14] et se visualise très bien en l'absence de palier de Lüders. L'absence de palier est surprenante en comparaison des aciers HSLA (High Strength Low Alloyed) et en raison de leur teneur en carbone élevé. On l'attribue à la présence d'une densité importante de dislocations mobiles à l'interface martensite/ferrite. Celle-ci s'explique par la transformation martensitique à basse température, par nature displacive accompagné d'une importante dilatation volumique. La Figure 61 montre par exemple une micrographie MET près de l'interface F/M d'un acier DP à l'état non vieilli.

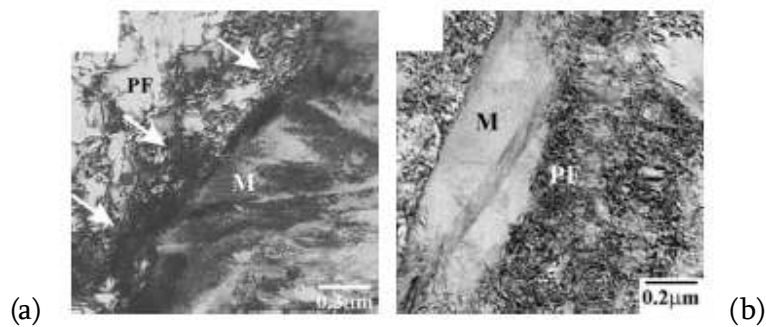

**Figure 61 : Micrographie MET des structures de dislocations dans la ferrite autour des ilots de martensite d'un acier Dual-Phase (a) dans l'état initial brut de fabrication et (b) après 10% de déformation, d'après [TIMOKHINA 2007] – M = Martensite et PF = Ferrite.**

Par analogie avec les composites à matrice métallique, les taux d'écrouissage des aciers DP s'expliquent par la présence de phases dures dans une matrice molle (l'effet DP). On considère souvent, à tort, que la martensite est suffisante « dure » pour rester élastique au cours du chargement. Nous montrerons que cette idée est généralement fausse et que le comportement plastique de la martensite joue un grand rôle sur le comportement macroscopique des aciers

---

[14] Certains auteurs attribuent la faible limite d'élasticité des aciers DP aux contraintes internes dues à cette transformation. Toutefois, ces contraintes internes n'ont jamais pu être mises en évidence par des anisotropies de comportement traction/compression. Il est par contre, plus raisonnable de penser comme nous le montrerons que la contrainte d'écoulement de la ferrite augmente avec ces densités importantes de dislocations.

DP. De cette vision composite, découle naturellement que l'écrouissage des aciers DP est contrôlé au premier ordre par la fraction de martensite. La Figure 62 montre par exemple les courbes de traction de 3 aciers DP avec des fractions variables de martensite [LIEDEL 2002].

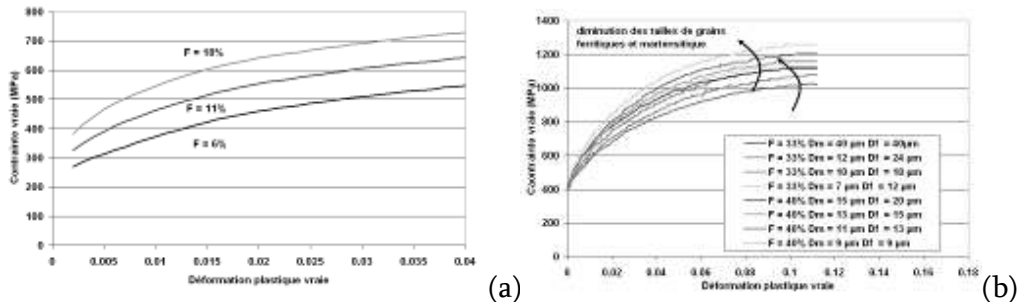

Figure 62 : (a) Courbes de traction rationnelles d'une nuance Fe0.09C1.4Mn0.1Si0.7Cr avec différentes fractions de martensite, d'après [LIEDL 2002] (b) Courbes de traction rationnelles d'une nuance Fe0.2C0.47Mn0.3Si0.1Cr avec différentes fractions de martensite et tailles de structure – grains et ilots de martensite, d'après [JIANG 1992].

Comme toutes les structures composites, les aciers DP vont présenter un composante d'écrouissage cinématique forte due aux différences de contraintes d'écoulement entre les phases [ASARO 1975][ALLAIN 2012]. Ce simple effet composite est en fait très exacerbé par le contraste très important de comportement entre les phases, qui vont générer des incompatibilités importantes de déformation entre les phases [BOUAZIZ 2005][KADKHODAPOUR 2011][KIM 2012]. Les gradients de déformation résultant sont principalement localisés dans la ferrite, la phase molle et sont révélés par l'observation en MET de fortes densités de Dislocations Géométriquement Nécessaires (DGN) autour des ilots de martensite, comme le montre la Figure 61 [TIMOKHINA 2007][KORZEKWA 1984][GARDEY 2005].

L'existence de ces gradients localisés rend le comportement des aciers DP très sensible aux effets de taille, en particulier à la taille de grain de la ferrite recristallisé et à la taille des îlots martensitiques, bien qu'il soit difficile de découpler ces effets expérimentalement des effets de fraction. La Figure 61 montre les courbes de traction d'aciers DP avec des fractions de phases constantes (33% ou 40%) mais des tailles variables [JIANG 1992]. Compte tenu des tailles de grains ferritiques, les limites d'élasticité sont globalement constantes mais les taux d'écrouissage évoluent significativement. Nous reviendrons dans la suite de cette exposé sur les multiples sources de contraintes internes à l'échelle de ces microstructures (joints de grains ferritiques, interfaces ferrite/martensite, gradients de déformation dans la martensite).

Paradoxalement, le comportement en traction des aciers DP dépend peu par contre de la morphologie (forme des structures martensitiques) et de la topologie (distribution spatiale, connectivité) des phases. La Figure 63 montre les courbes de deux aciers élaborés à partir de la même composition mais ayant subis des traitements thermiques de recuit différents.

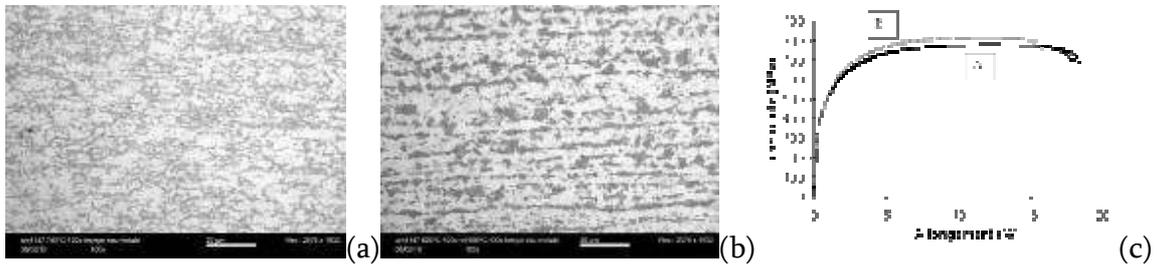

Figure 63 : Micrographie optique après attaque Nital de deux aciers Dual-Phase Fe0.09C2Mn présentant la même fraction de martensite (28% et 27% respectivement) mais des morphologies et topologies différentes (a) distribution isotrope d'îlots et (b) structure en « bandes ». (c) Courbes de traction conventionnelles correspondantes (données de L. Durrenberger / M. Gouné d'AM).

Les morphologies et topologies résultantes de la martensite sont très différentes mais les fractions de phase sensiblement équivalentes. La différence de comportement est faible malgré une morphologie en bande ou très dispersé de la martensite. Par contre, cette morphologie joue un rôle majeur dans les processus d'endommagement comme nous le verrons au chapitre 3.4 page 137.

### 3.1.2.2. *Modélisation du comportement*

Dans la littérature scientifique, on retrouve trois grandes familles de modélisation micromécanique du comportement des aciers Dual-phase.

- Les approches monophasées à champs moyens : le comportement du composite est réduit au comportement de la ferrite durcie par la présence de martensite [SUGIMOTO 1997][BOUAZIZ 2001_2][JIANG 1992][MA 1989]. Ces modèles sont basés sur l'hypothèse implicite que la martensite reste élastique lors du chargement et permettent de rendre compte des effets de taille et de fraction mais sont valables uniquement aux faibles fractions de martensite (pas de percolation du réseau de martensite, teneur en carbone local dans la martensite élevée). Ces modèles sont généralement développés pour décrire et prédire le comportement des matériaux lors de sollicitations simples (traction monotone ou essai alterné). Elles peuvent toutefois être étendues en plasticité polycristalline pour décrire des comportements sous sollicitations complexes (chargement biaxiés, changements de trajets) comme nous l'avons réalisé dans le cadre de la thèse de S. Dillien.
- les approches biphasées à champs moyens : le comportement de l'acier DP résulte de la moyenne pondérée du comportement des deux phases Ferrite et Martensite. Ils reposent souvent dans la littérature sur des modèles d'homogénéisation spécifiques : simples à un paramètre [LIAN 1991], type auto-cohérent avec des comportements locaux elasto-viscoplastiques [BERBENNI 2004], type Mori-Tanaka avec des comportements elasto-plastiques [LANI 2007][ JACQUES 2007], type Tomota avec des comportements elasto-plastiques empiriques [TOMOTA 1992]. Ces approches souffrent généralement de deux défauts majeurs : les lois de comportements locales de chacune des phases sont empiriques, voire réductrices pour la martensite, et les effets induits par les gradients de déformations aux interfaces ferrite/martensite ne sont pas pris en compte. Ces approches ne sont sensibles ni aux effets de taille, ni aux effets de teneur en carbone dans la martensite. Delincé *et al.* [DELINCE 2010] ont bien proposé une approche sensible aux effets de tailles de la ferrite, mais le comportement de la martensite est décrit très sommairement (elasto-parfaitement plastique).

- les approches locales par Eléments Finis (EF) sur Volume Elémentaire Représentatif (VER) de la microstructure : Ces modélisations réalisées sur des microstructures modèles [HUPPER 1999] [AL-ABBASI 2007] [PRAHL 2007] [LIEDL 2002] [UTHAISANGSUK 2011] ou numérisées [LI 1990_2][LI 1990_1] [PAUL 2012] [CHOI 2009] [KUMAR 2007] [SODJIT 2012][PAUL 2013] permettent de décrire les effets de fraction, de dispersion et de morphologie de la martensite. Leurs principales limitations résident dans l'absence complète d'effets de taille (pas d'influence des gradients de déformation). Kadkhodapour *et al* ont cependant proposé une approche par EF originale à coques concentriques autour d'ilots de martensite pour reproduire un gradient de comportement dans la ferrite et ainsi un effet de taille d'ilots [KADKHODAPOUR 2011]. Cette démarche est très similaire à celle envisagée par Pipard et al [PIPARD 2009] pour décrire les effets de tailles dans la ferrite. Ces modèles par EF nécessitent en outre des mises en données généralement longues, en particulier les modèles à microstructure numérisée.

De façon surprenante on retrouve peu dans la littérature de modèle de plasticité polycristalline dédié à l'étude des DP [DILLIEN 2010_2].

### 3.1.3. Mise en perspective des travaux

Connues de longue date, ces microstructures ont fait l'objet de très nombreuses études et de revues dans le domaine académique dans les années 1970-1980 soutenues par des intérêts sidérurgiques et industriels importants dans le cadre de la construction automobile, le transport ou le domaine des tôles fortes. L'engouement pour ces structures s'est ensuite atténué au profit dans les années 1990 de l'étude des aciers ferritiques TRIP. Cela étant, ce dernier effet micromécanique dans les aciers ferritiques au carbone peut être vu comme un effet DP dynamique [PRAHL 2007][PERLADE 2003][LANI 2007].

Mes travaux sur le comportement des aciers Dual-Phase ont débuté en 2005 pour le groupe Arcelormittal et ont principalement été de nature théorique (modélisation micromécanique). Ces travaux internes, et à ce jour partiellement confidentiels [ALLAIN 2005], ont été mené principalement en collaboration avec NSC et ont fait l'objet de peu de publications.

D'un point de vue collaboration scientifique, j'ai co-encadré la thèse de S. Dillien au KUL (directeurs de thèse : P. VanHoutte, M. Seefeld) sur cette thématique, intitulée : « Bridging the Physics-Engineering Gap in Dual Phase Steel Formability »,
J'ai aussi participé directement à la thèse de B. Krebs au LETAM (directeurs de thèse : A. Hazotte, L. Germain) en réalisant des calculs EF avec l'étudiant, et plus indirectement à celle d'I. Pushkareva au LSGS (directeur de thèse : A. Redjaimia) sur leur endommagement et à la thèse de A. Aouafi au LPMTM (directrices de thèse :M. Gaspérini, S. Bouvier) sur la modélisation des effets de taille dans les aciers ferritiques.

Mes travaux ont eu plusieurs finalités dans ce domaine, que nous détaillerons dans la suite de ce mémoire :

- Axe 1 : Extension d'un modèle monophasé en plasticité polycristalline pour des applications en rhéologie appliquée (prévision des surfaces de charges sous sollicitations complexes). Ces travaux ont permis de fournir aux « mécaniciens » une extension du modèle de Teodosiu-Hu pour les aciers DP. Ce chapitre sera aussi l'occasion d'introduire les différents mécanismes d'écrouissages de ces aciers composite (de nature isotrope et cinématique) constituant la base du modèle biphasé.

- Axe 2 : Développement d'un modèle biphasé quasi-analytique pour des utilisations en « alloy-design » métallurgique. Ce modèle à vocation d'être « générique » en décrivant non seulement le comportement des aciers DP, mais aussi les aciers IF (Interstitial Free) 100% ferritiques et les aciers 100% martensitiques, en tenant compte des effets de fraction (fraction et teneur en carbone dans la martensite) et de tailles (taille de grains ferritiques, des ilots martensitiques). Cet outil a été développé à destination des « métallurgistes » et est basé sur une compréhension fine des mécanismes de plasticité de ces deux phases et de leurs interactions. Après une présentation détaillée de ces mécanismes, des exemples concrets d'utilisation de l'approche seront discutés.

- Axe 3 : Approfondissement de nos connaissances dans le domaine de l'endommagement des structures Dual-Phase. J'ai piloté depuis 2008 un projet de développement d'une chaine de simulation numérique par EF du comportement mécanique et d'endommagement des aciers DP (projet de recherche Arcelormittal DPNDI 187). Le synoptique de cette démarche est représentée sur la Figure 64. Elle s'étend de la numérisation de microstructures aux calculs sur VER sensibles aux gradients de déformation et intégrant des mécanismes simulant l'endommagement (éléments cohésifs). Les objectifs scientifiques de cette démarche sont de quantifier et de mieux comprendre les mécanismes d'endommagement de ces structures DP et à terme leurs actionneurs métallurgiques. Ceci nécessite non seulement une connaissance fine et juste des états de contraintes et déformations des différentes phases au cours de la déformation macroscopique mais aussi de pouvoir gérer la compétition entre les mécanismes d'endommagement (relaxation). Cette troisième partie sera consacrée à l'étude grâce à ces outils numériques d'un cas particulier, celui de l'impact néfaste des structures morphologiques dites « en bandes » sur l'endommagement des aciers DP. Les perspectives d'amélioration de la démarche seront enfin exposées.

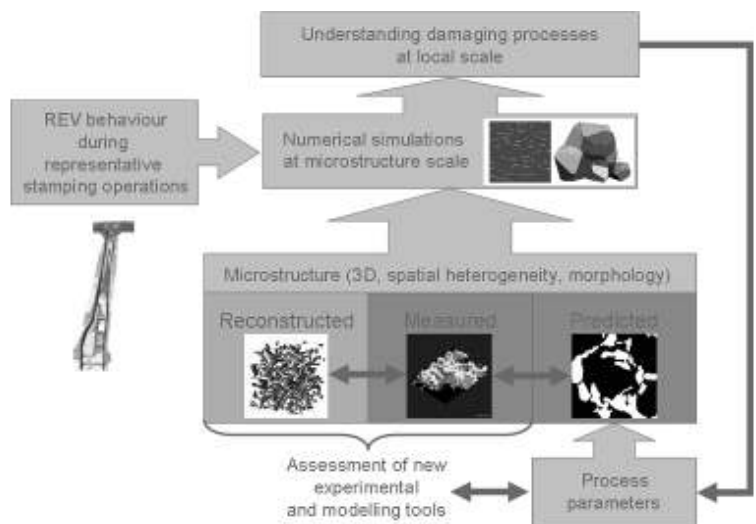

Figure 64 : Synoptique de la démarche de modélisation par EF sur VER du comportement et de l'endommagement des aciers Dual-Phase (Projet de Knowledge Building AM).

## 3.2. Extension d'un modèle monophasé en plasticité polycristalline pour des applications en rhéologie appliquée

Dans cette première partie, nous reviendrons sur les modèles monophasés analytiques du comportement des aciers DP. Ces travaux issus de la métallurgie physique mettent généralement en valeur les effets de fraction et de tailles. Ils sont donc une base indispensable pour notre plateforme générique de modélisation détaillée au chapitre 3.3. Malgré leurs domaines d'utilisation limités en termes de fraction de martensite, il s'agit d'outils de modélisation d'un grand intérêt pratique. Ils sont simples mais permettent quand même la description de composantes d'écrouissage cinématiques pour des calculs de mise en forme par exemple. Nous montrerons en outre qu'ils peuvent être étendus dans le cadre de la plasticité polycristalline et donc fournir un cadre théorique pour des applications en rhéologie appliquée.

### 3.2.1. Composante isotrope de l'effet DP

Cette courte présentation bibliographique ne vise pas l'exhaustivité mais permet d'évoquer les différents mécanismes à la base de l'écrouissage de ces aciers et la nature de cet écrouissage.

En 2001, Bouaziz *et al.* [BOUAZIZ 2001_2][PERLADE 2003] proposent de décrire que la contrainte d'écoulement $\Sigma^{DP}$ d'un acier DP correspond à celle de la ferrite pour de faibles fraction de martensite $F_m$. En suivant un modèle de type Taylor (écrouissage par la forêt des dislocations), il vient alors :

$$\Sigma^{DP}\left(E_p^{DP}\right) = \sigma^\alpha\left(\varepsilon_p^\alpha\right) = \sigma_0^\alpha + \alpha M \mu b_{111} \sqrt{\rho_{DSS} + \rho_{DP}} \qquad (47)$$

avec $\rho_{DSS}$ la densité de dislocations statistiquement stockées dans la ferrite et dont l'évolution va suivre une loi de type Mecking-Kocks-Estrin et $\rho_{DP}$ une densité de dislocations additionnelles dues à la présence de martensite. $E_p^{DP}$ est la déformation plastique macroscopique et $\sigma^\alpha$ et $\varepsilon_p^\alpha$ la contrainte et déformation plastique dans la ferrite. $\sigma_0$ est une friction de réseau qui ne dépend que de la composition chimique. $\alpha$ est facteur de durcissement de la forêt, M le facteur de Taylor, $\mu$ le module d'élasticité en cisaillement et $b_{111}$ le vecteur de Burgers des dislocations parfaites dans la ferrite. Le terme $\rho_{DP}$ représente les densités de Dislocations Géométriquement Nécessaires (DGN) réparties uniformément dans les grains ferritiques. Paradoxalement, il correspond à un mécanisme de durcissement isotrope et est estimé comme les DGN au sens d'Ashby en présence d'une particule dure de taille $d_m$ [JIANG 1992][BOUAZIZ 2001][BOUAZIZ 2005] :

$$\rho_{DP} = \frac{8MF}{b_{111}d_m}\varepsilon_\alpha^p \qquad (48)$$

Il vient alors :

$$\Sigma^{DP}\left(E_p^{DP}\right) = \sigma_0^\alpha + \alpha M \mu \sqrt{b_{111}} \cdot \sqrt{\frac{1-\exp\left(-f.M.\left(E_p^{DP}+\varepsilon_0\right)\right)}{f.d} + 8.\frac{F_m}{d_m}.\frac{1-\exp\left(-r.M.\left(E_p^{DP}+\varepsilon_0\right)\right)}{r}} \qquad (49)$$

Le premier terme $\rho_{SSD}$ est bien entendu sensible à la taille de grain ferritique d et dépend d'un processus de restauration dynamique paramétré par f (le terme d'écrouissage latent est considéré négligeable k = 0 en référence à l'équation (16)). Une pré-déformation $\varepsilon_0$ est introduite pour reproduire les effets de la transformation martensitique qui génère de nombreuses dislocations mobiles (cf. Figure 61) dans la ferrite. Le terme relevant des DGN au sens d'Ashby est considéré comme saturant, avec un paramètre r.

Nous avons étudié par EBSD ces gradients de déformation dans la ferrite dans le cadre de la thèse de S. Dillien [DILLIEN 2010_1][DILLIEN 2010_2]. L'acier étudié est un DP 600 (laminé à chaud) avec une structure très isotrope. La Figure 65 montre des observations en MEB EBSD avant et après 15% de déformation en traction. La martensite est mal indexée et apparait en noir. Le code couleur représente les angles de désorientations par rapport à la moyenne du grain et les grains sont définis grâce à la méthode MMC (Maximized Misorientation Contrast). Les désorientations sont mesurées le long des lignes et les résultats sont représentés sur les figures jointes. La figure (a) montre que même à l'état brut de trempe, la microstructure contient déjà de nombreuses dislocations géométriquement nécessaires distordant la maille. Grâce aux gradients de désorientations, nous avons évalué cette densité à $6 \cdot 10^{13}$ m$^{-2}$, correspondant à une contribution au durcissement de 20 MPa environ. Ces densités augmentent considérablement après déformation ($1.5 \cdot 10^{14}$ m$^{-2}$ mesuré sur la figure (d)) et apportent une contribution significative de 50-100 MPa environ. Cette contribution n'est pas suffisante pour expliquer totalement l'effet DP, c'est pourquoi l'ajout d'une contribution d'origine cinématique est indispensable pour une description cohérente.

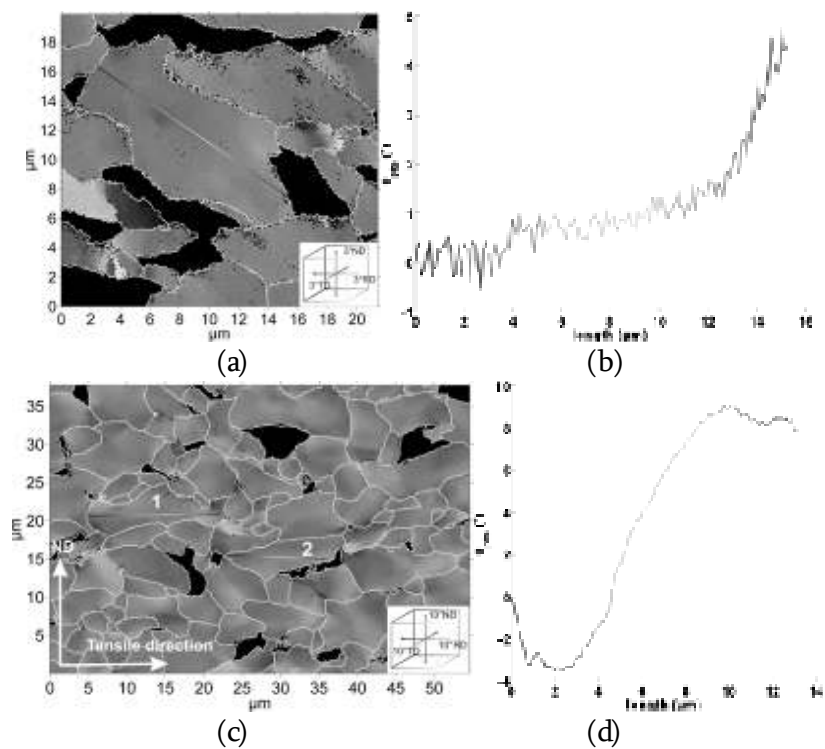

Figure 65 : Cartographie EBSD d'une structure DP600 (laminé à chaud) (a) à l'état initial et (c) après 15% de déformation. Le code couleur représente la désorientation par rapport à la moyenne du grain considéré selon la méthode du MMC (Maximized Misorientation Contrast). Les zones non indexées (en noir) correspondent à des zones martensitiques. Les gradients de désorientations le long de certaines lignes sont représentés en (b) et (d) respectivement [DILLIEN 2010_2].

### 3.2.2. Introduction d'une composante cinématique

Le modèle de nature purement isotrope présenté ci-dessus a été complété par Bouaziz en 2005 [BOUAZIZ 2005] grâce à un terme de nature cinématique. Cette contribution correspond à la contrainte en retour exercée par les particules dures sur la matrice en présence d'une incompatibilité de déformation. Brown, Stobbs et Atkinson [BROWN 1971][ATKINSON 1974] expriment cette contrainte supplémentaire comme :

$$\sigma_{DPX} = 2\,\xi\mu\,F_m\varepsilon_\alpha^{p^*} \tag{50}$$

avec $\xi$ un facteur d'accommodation proche de ½ et $\varepsilon_\alpha^{p^*}$ la déformation plastique non relaxée dans la matrice. Ce terme est similaire à celui associé au durcissement par le maclage mécanique dans une approche type composite (cf. 2.3.5 page 52). Bouaziz propose alors d'écrire :

$$\Sigma(E_{DP}^p) = \sigma_0^\alpha + \alpha M\mu\sqrt{b_{111}}\cdot\sqrt{\frac{1-\exp(-f.M.(E_{DP}^p+\varepsilon_0))}{f.d} + 8.\frac{F_m}{d_m}.\frac{1-\exp(-r.M.(E_{DP}^p+\varepsilon_0))}{r}} + \sigma_X \tag{51}$$

$$\sigma_X = M\mu\sqrt{b_{111}}\cdot\frac{F_m}{\sqrt{d_m}}\sqrt{\frac{1-\exp(-r'.M.E_{DP}^p)}{r'}} \tag{52}$$

$\sigma_X$ correspond à un écrouissage cinématique dérivé de l'équation (50) et saturant avec un seul paramètre r'.

Cette équation en fait très similaire dans sa structure et ses dépendances au modèle proposé par Sugimoto *et al.* en 1997 [SUGIMOTO 1997].

$$\sigma = \sigma_0 + \sigma_{matrice\,ferrite} + \mu\sqrt{\frac{b_{111}F_m\varepsilon_\alpha^p}{2d_m}} + \zeta\frac{7-5\nu}{5(1-\nu)}\mu F\varepsilon_\alpha^{p^*} \tag{53}$$

La seule différence réside dans le découplage entre la contrainte d'écoulement de la ferrite et l'écrouissage dû aux dislocations d'Ashby (troisième terme de l'égalité).

Dans cette famille d'approche monophasée, la modélisation du comportement DP revient à ajouter au comportement de la ferrite deux contributions :
- la première est de nature cinématique et a pour origine les DGN restant au contact des ilots de martensite. Ces dislocations induisent un champ de contrainte en retour à longue distance, qui est directement proportionnel à la fraction de martensite et peut dépendre de la taille des ilots.

- La seconde est de nature isotrope, et correspond à des DGN émises à grandes distances des particules dans les grains de ferrite (de l'ordre du micron toutefois). Les champs qui en résultent sont non polarisés à grande distance et ces dislocations contribuent à l'écrouissage selon un mécanisme de type forêt. Ce terme est proportionnel à la racine carrée de la fraction de martensite et non nul en début de déformation (effet de la transformation displacive avec changement de volume à basse température).

Ces deux contributions vont naturellement saturer au cours de la déformation grâce à des mécanismes de relaxation, comme l'émission de boucles secondaires par exemple [DILLIEN 2010_2].

### 3.2.3. Extension polycristalline à destination des rhéologues

L'approche décrite ci-dessus au chapitre 3.2.2 a été étendue dans le cadre de la thèse de S. Dillien à une modélisation d'agrégat polycristallin en 3D. La finalité de ce travail était la prédiction des surfaces de charges sous sollicitations complexes (trajets non monotones) des aciers DP, sur des bases de métallurgie physique.

En pratique, les travaux ont visé l'extension aux aciers DP du modèle micro-macro pour le comportement de la ferrite développé par Peeters *et al.* [PETEERS 2001][15]. Ce modèle très élaboré gère de multiples effets comme les changements de trajets durs grâce à une description des cellules de dislocations, mais malheureusement pas les effets de taille de grains. Les cissions locales $\tau_s^c$ pour chaque système de glissement s ont été modifié de la façon suivante :

$$\tau_s^c = \tau^0 + \tau^{init} + (1 - F_m)(\tau_s^{BP} + F_m \tau^D) + F_m \tau_s^{BS} \quad (54)$$

avec $\tau^{init}$ une cission due au dislocations initialement présentes dans la ferrite autour des îlots de martensite (zone 3 sur la Figure 66) $\tau^{BP}$ le terme de Bart Peeters (zone 1), $\tau^D$ la cission due au DGN d'Ashby (zone 3) et $\tau^{BS}$ une contrainte de « back-stress » modélisée de manière empirique par une loi de type Voce (zone 2). Cette équation correspond à une extension des équations (52) ou (53) pour chaque système de glissement.

---

[15] Ce modèle, rappelons-le, est la version issue de la métallurgie physique du modèle de rhéologie célèbre de Teodosiu-Hu.

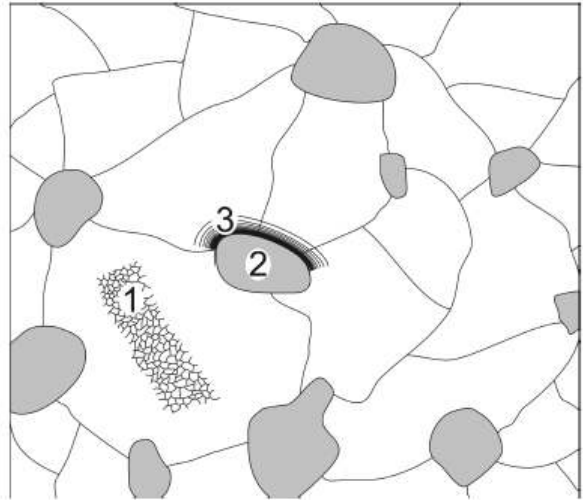

**Figure 66 : Représentation schématique des mécanismes de durcissement considérés et de leur localisation dans une microstructure DP modèle. 1 = Structure de déformation en cellule dans la ferrite (gouvernée par les lois de Peteers *et al*), 2 = Contrainte de back-stress due aux ilots de martensite, 3 = Nuage de dislocations autour des grains de martensite [DILLIEN 2010_2].**

La Figure 67 montre les prédictions du modèle, une fois ajusté, du comportement lors de changements de trajet durs (cross) ou alternés (Bauschinger) d'un acier DP 600. Comme attendu on notera que le changement de trajet dur n'induit pas de durcissement alors que l'acier présente un fort effet Bauschinger. Ce résultat est à comparer à celui de la Figure 35 page 65.

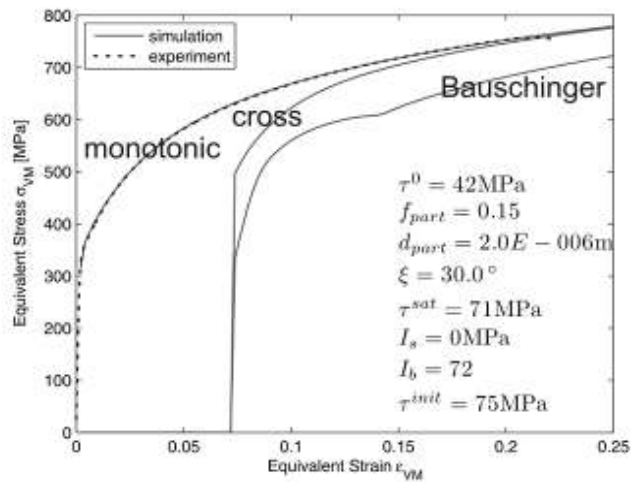

**Figure 67 : Prédiction du comportement par le modèle selon des changements de trajets typiques [DILLIEN 2010_2].**

Ce modèle 3D a donc une grande importance pratique pour définir la forme spécifique des surfaces de charges de ces aciers. Toutefois, il souffre encore de lacunes importantes pour décrire tenir compte des effets de fraction ou de taille, inhérente au modèle de Peeters *et al.*, y compris pour décrire le comportement de la ferrite.

### 3.3. Plateforme de modélisation « générique » des aciers Ferrite-Martensite

Depuis 2005, avec l'aide de mon collègue Olivier Bouaziz, nous développons une modélisation à champs moyens du comportement des aciers DP « générique », c'est-à-dire sensibles aux effets de tailles, de fraction et de composition, pour tous les aciers Ferrite-Martensite, des aciers IF aux aciers martensitiques. Elle fait suite aux développements analytiques monophasés de mon collègue en 2001 et 2005 rappelés au chapitre précédent.

Ce nouveau développement répond non seulement à l'attente des « mécaniciens », c'est-à-dire qu'il produit des lois de comportement faisant la distinction entre les écrouissages isotropes et cinématiques sur des bases microstructurales. Il a aussi pour vocation d'aider les métallurgistes dans des phases de conception produit générique, en étant un outil « d'alloy design » ou d'ingénierie des microstructures. En effet, il outrepasse le domaine de validité des modèles monophasés (restreint aux faibles fractions de martensite) et est applicable à tous les aciers Ferrito-Martensitique.

Cette modélisation repose sur une approche bi-phasée à champs moyens, c'est-à-dire que l'on considère séparément le comportement de la ferrite et de la martensite. Par contre, contrairement aux travaux antérieurs de la littérature, nous avons fait l'effort d'introduire des lois à base microstructurale pour ces deux constituants et pris le parti d'hypothèses d'homogénéisation plus simples. Cette nouvelle philosophie permet de considérer des fractions de martensite élevées en tenant compte de la plasticité de cette phase, mais pas des gradients de déformation aux interfaces ferrite/martensite. Nous avons donc introduit a posteriori des termes de couplage originaux entre les phases pour décrire les dislocations d'Ashby.

Le modèle est suffisamment simple pour fonctionner sur un simple tableur mais permet de traiter de cas d'applications complexes comme nous le montrerons dans une dernière partie.

Dans la suite de cet exposé, les indices ou exposants $\alpha$ et $m$ renverront respectivement aux termes de la ferrite et de la martensite, les lettres capitales ($\Sigma$ ou $E$ par exemple) aux comportements macroscopiques alors que $R$ et $X$ indiqueront des composantes isotropes ou cinématiques (à toutes les échelles).

### 3.3.1. Comportement de la ferrite: Effet de la taille de grain

Le modèle de comportement de la ferrite a été développé dans le cadre de la thèse d'A. Aoufi [BOUAZIZ 2008_2] et est très similaire au modèle de description du comportement de l'austénite dans le cas d'un effet TWIP (cf. chapitre 2.3.4 page 49). Ce cas est cependant plus simple car le libre parcours moyen des dislocations reste constant (taille de grain).

On suppose que la contrainte d'écoulement d'un acier ferritique se décompose de la manière suivante :

$$\sigma^\alpha\left(\varepsilon_p^\alpha\right) = \sigma_0^\alpha + \sigma_R^\alpha + \sigma_X^\alpha \tag{55}$$

avec
- $\sigma_0$ une contrainte de friction du au durcissement en solution solide et qui ne dépend que de la composition chimique :

$$\sigma_0^\alpha = 52 + 33\%\text{Mn} + 80\%\text{Si} + 60\%\text{Cr} + 80\%\text{Mo} \tag{56}$$

- $\sigma_R$ un terme de durcissement isotropes de type Taylor (durcissement de la forêt)
- $\sigma_X$ un terme de durcissement cinématique.
- $\varepsilon_p$ est la déformation plastique (fonction de la déformation totale et de la déformation élastique $\varepsilon_{el}$)

$$\varepsilon_p^\alpha = \varepsilon^\alpha - \varepsilon_{el}^\alpha \tag{57}$$

Le terme de durcissement isotrope s'écrit classiquement avec la densité de dislocations statistiquement stockées (DSS) $\rho_\alpha$

$$\sigma_R^\alpha\left(\varepsilon_p^\alpha\right) = \alpha M \mu b_{111} \sqrt{\rho_\alpha} \tag{58}$$

Par contre, l'évolution de cette densité ne suit plus une loi de type Mecking-Kocks-Estrin (MKE) simplifiée mais une loi d'évolution inspirée de Sinclair *et al* [SINCLAIR 2006], déjà utilisée pour la modélisation des aciers TWIP. Cette approche a l'avantage de pouvoir décrire les effets de taille de grains sur un domaine beaucoup plus étendu que l'approche MKE conventionnelle, en particulier les tailles de grains de l'ordre du micron souvent observés dans les aciers DP.

L'évolution de la densité de DSS avec la déformation plastique résulte de la compétition entre un processus d'accumulation (sur les joints de grains et par écrouissage latent) et de restauration dynamique.

$$\frac{d\rho_\alpha}{d\varepsilon_p^\alpha} = M. \left( \frac{\left(1 - \frac{n_\alpha}{n_0^\alpha}\right)}{b_{111} d_\alpha} + k_\alpha \sqrt{\rho_\alpha} - f_\alpha \rho_\alpha \right) \quad (59)$$

avec $d_\alpha$ la taille de grain ferritique, $k_\alpha$ et $f_\alpha$ deux constantes. $n_\alpha$ et $n_0^\alpha$ représentent le nombre moyen et maximum de dislocations stockés sur les joints de grains respectivement (DGN). Ce terme correspond donc à un effet d'écrantage des joints de grains par la présence de DGN. Ce flux de DGN aux joints arrivant par bande de glissement s'écrit empiriquement :

$$\frac{dn_\alpha}{d\varepsilon_p^\alpha} = \frac{\lambda_\alpha}{b_{111}} \left(1 - \frac{n_\alpha}{n_0^\alpha}\right) \quad (60)$$

avec $\lambda_\alpha$ l'espacement moyen entre les bandes de glissement arrivant sur les joints de grains.

Les DGN stockées sur les obstacles forts (seulement les joints de grains dans ce cas, contrairement aux aciers TWIP) génèrent un champ de contraintes à longue distance dans la microstructure, contribution additive à la contrainte d'écoulement isotrope :

$$\sigma_X^\alpha = \frac{M\mu \, b_{111}}{d_\alpha} n_\alpha \quad (61)$$

Ce terme est très sensible à la taille de grains (en fait, à la densité d'interface). Dans le cadre de la thèse, nous avons ajusté ce modèle sur les courbes de traction d'une très large gamme d'aciers IF (sans interstitiels) avec des tailles de grains très variables de 3.5 µm à 20 µm, représentant bien les microstructures ferritiques attendus des aciers DP [BOUAZIZ 2008_2]. Ce modèle permet de prédire en outre avec précision les niveaux de contraintes internes dans certains de ces aciers (mesurés par essais Bauschinger), comme le montre la Figure 68 et aux réserves près discutés précédemment.

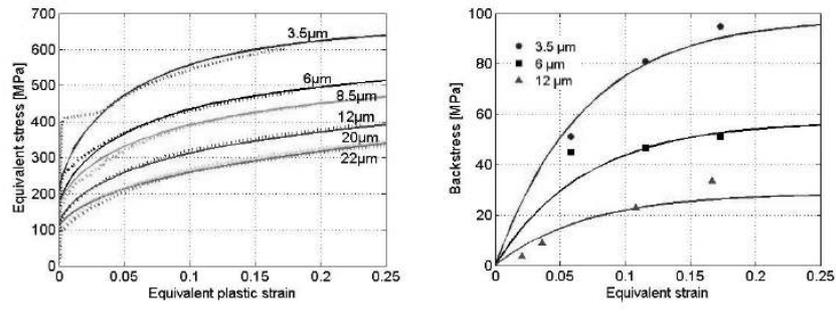

Figure 68 : (a) courbes de traction rationnelles expérimentales et simulées d'aciers ferritiques avec différentes tailles de grains (b) Composantes d'écrouissage cinématique expérimentales et simulées pour 3 aciers ferritiques avec différentes taille de grains en fonction de la déformation équivalente (mesures par essais Bauschinger en cisaillement) [BOUAZIZ 2008_2].

Les valeurs de constantes physiques et paramètres ajustables du modèle sont repris dans le tableau suivant :

|   | *Valeur* | *signification physique* |
|---|---|---|
| α | 0.38 | durcissement de la forêt |
| M | 2.77 | module de cisaillement |
| μ | 80 GPa | facteur de Taylor |
| $b_{111}$ | $2.5\,10^{-10}$ m | vecteur de Burgers |
| $\lambda_\alpha/b_{111}$ | 90 | espacement moyen entre bandes de glissement |
| $n^\alpha_0$ | 6.2 | nombre maximum de dislocations stockées aux joints de grains |
| $k_\alpha$ | 0.007 | écrouissage latent |
| $f_\alpha$ | 1.3 | restauration dynamique |

Tableau 3 : Valeurs des paramètres du modèle de comportement de la ferrite [BOUAZIZ 2008_2]

Pour traiter de la transition élasto-plastique dans cette phase, nous avons retenu un schéma explicite et simplifié pour éviter le recours à des boucles de convergence (résolution du système d'équation en déformation totale) :

$$\text{Si } \varepsilon^\alpha \geq \frac{\sigma_0^\alpha}{Y}, \text{ alors } \varepsilon_p^\alpha = \varepsilon^\alpha - \frac{\sigma_0^\alpha}{Y} \text{ sinon } d\varepsilon_p^\alpha = 0 \text{ and } \varepsilon^\alpha = \varepsilon_{el}^\alpha \qquad (62)$$

et

$$\sigma^\alpha(\varepsilon^\alpha) = \min\left(\sigma^\alpha(\varepsilon_p^\alpha), Y\varepsilon^\alpha\right)$$

Avec Y le module d'Young.
Ce choix ne se posera pas dans le cas du modèle de martensite car la transition entre élasticité et plasticité est décrite naturellement et explicitement.

### 3.3.2. Comportement de la martensite : Effet de la teneur en carbone

*3.3.2.1.    Bilan de la littérature*

La principale difficulté concernant ce constituant est qu'il n'existait pas dans la littérature de loi de comportement à base microstructurale convaincante!

C'est pourtant une phase connue de longue date et beaucoup de travaux lui ont été consacrés pour comprendre l'origine de ses fortes contraintes d'écoulement et de sa dureté. Elle a fait l'objet de nombreuses revues dans la littérature scientifique (transformation de phase, cristallographie, résistance…) [KRAUSS 1999][OLSON 1992]. Ces travaux montrent que le paramètre « microstructural » clef qui semble contrôler le comportement de la martensite est sa teneur en carbone. La dépendance de la dureté de la martensite en fonction de la teneur en carbone est certainement la relation microstructure/propriétés la mieux acceptée de la littérature, mais la moins bien expliquée! Les corrélations avec d'autres défauts microstructuraux et leurs dimensions associées (taille de lattes, de paquets, ancien joint de grain austénitique) et les limites d'élasticité de ces aciers sont largement discutées sans consensus.

Les mécanismes d'écrouissage et leur modélisation ont par contre tout bonnement été négligés, ce qui est très surprenant compte tenu de l'importance industrielle croissante de ces aciers [16]. On retrouve bien dans la littérature quelques modèles phénoménologiques polynomiaux ou simplifiés de comportement, mais pas de modèle réellement à base microstructurale.

La première version de notre plateforme de modélisation était basée sur le modèle de comportement de martensite inspiré de celui des aciers ferritiques à grains ultrafins [COBO 2008]. Il était ajusté empiriquement et souffraient de graves lacunes (description des taux d'écrouissage initiaux en particulier). Depuis 2009, nous avons décidé de revisiter les mécanismes fondamentaux du comportement de la martensite, en collaboration avec M. Takahashi de NSC, F. Danoix du GPM et dans le cadre de la thèse de G. Badinier à l'UBC (directeur de thèse : C. Sinclair). Ces travaux ont donné lieu à des publications récentes [BADINIER 2011][ALLAIN 2013], en particulier notre nouveau modèle de comportement pour la martensite [ALLAIN 2012].

---

[16] Certaines universités japonaises restent toutefois très actives dans ce domaine [NAKASHIMA 2007] [LHUISSIER 2011] [NAMBU 2009] avec des travaux de nature principalement expérimentale (MET, RX) sur la plasticité des aciers martensitiques.

Dans la suite de cet exposé, nous reviendrons sur les caractéristiques principales de ce modèle, ses performances et ses limites. Il ouvre un angle d'investigation original et des perspectives nouvelles pour l'analyse de ces structures martensitiques.

Cette nouvelle approche a été suggérée par l'analyse préalable de nombreuses courbes de traction et d'essais mécaniques d'aciers martensitiques. Ces constatations se limitent au cas des aciers faiblement alliés présentant une structure principalement en lattes [SHERMAN 1983]. Les constatations principales sont les suivantes :
- Tous les aciers martensitiques, à l'état brut de trempe, présentent une limite de microplasticité faible et constante, de l'ordre de 400 MPa (en traction ou compression). Ceci suggère que la martensite contient une fraction significative de « zones molles » ayant une contrainte d'écoulement faible et indépendante de la teneur en carbone.
- Tous les aciers présentent un fort taux d'écrouissage initial après cette limite de microplasticité, croissante avec la teneur en carbone de l'acier considéré. Ce résultat explique pourquoi les limites d'élasticité conventionnelle augmentent dans ces aciers avec la teneur en carbone, malgré une limite de microplasticité constante.
- Cette dépendance du taux d'écrouissage à la teneur en carbone est maintenue après de grandes déformations.
- L'écrouissage des aciers martensitique est principalement d'origine cinématique, comme le suggère les rares essais Bauschinger disponibles dans la littérature.

Ce résultat suggère que le comportement des aciers martensitiques ne peut s'interpréter par un simple mécanisme de stockage des dislocations. Les taux d'écrouissage sont en effet très largement supérieurs à $Y/100$ – $Y$ le module d'Young [SEEGER 1963][KEH 1963][ANSELL 1963]. Le mécanisme de déplétion (exhaustion) en dislocations mobiles suggéré par [NAMBU 2009][ [NAKASHIMA 2007] ne peut non plus expliquer les forts niveaux de contraintes internes. Par contre, toutes ces caractéristiques suggèrent que la martensite ne doit pas être considérée comme une phase homogène, mais plutôt comme un composite hétérogène continu, constitué d'un mélange de phases « molles » et « dures », dont les fractions respectives varient avec la teneur en carbone. Cette idée a été suggérée récemment aussi par Hutchinson *et al* [HUTCHINSON 2011]. Les phases « molles » contrôlent le seuil de microplasticité alors que les phases « dures » en restant élastique assurent le taux d'écrouissage macroscopique (une fraction élevée du module d'Young). Le comportement de la martensite peut donc se résumer à une transition élasto-plastique très étendue.

### 3.3.2.2. *Approche Composite Continue*

Le comportement de la martensite a donc été décrit grâce à un modèle de Masing généralisé à un continuum de phases élastiques-parfaitement plastiques en interaction. Chacune des phases du composite est décrite par son module d'Young $Y$ (le même pour toutes les phases) et sa contrainte d'écoulement respective. Le composite est donc défini de manière univoque par sa fonction continue de distribution de densité de probabilité f(σ) de trouver une phase ayant une contrainte d'écoulement σ. Cette distribution est appelée dans la suite spectre de contrainte et représente finalement la distribution des contraintes d'écoulement dans la microstructure. On définit alors sa fonction cumulée F(σ) comme :

$$F(\sigma) = \int_{-\infty}^{\sigma} f(\zeta) d\zeta \tag{63}$$

La Figure 69 montre un exemple de spectre de contraintes et de sa fonction cumulée.

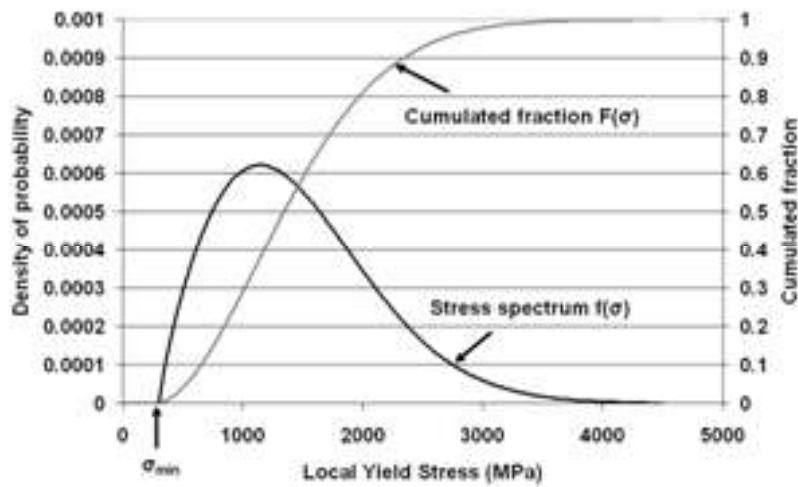

**Figure 69 : Exemple de « spectre de contraintes» f et sa fonction cumulée F.**

Pour des raisons de cohérence, ces fonctions doivent respecter les conditions suivantes :

$$\forall \sigma, f(\sigma) \geq 0$$
$$F(+\infty) = \int_{-\infty}^{+\infty} f(\zeta) d\zeta = 1 \tag{64}$$

Si on appelle $\sigma_{min}$ la contrainte d'écoulement de la phase la plus molle de ce composite, on a alors

$$\forall \sigma \leq \sigma_{min}, f(\sigma) = 0 \text{ et } F(\sigma) = 0 \tag{65}$$

Elle correspond à la phase du composite qui va contrôler le seuil de microplasticité macroscopique (environ 400-500 MPa). Une fois cette distribution connue, on peut estimer la contrainte macroscopique $\sigma^m$ de ce composite en fonction de la déformation macroscopique appliquée $\varepsilon^m$ :

$$\sigma^m = \int_{\sigma_{min}}^{\sigma_L} f(\sigma)\sigma d\sigma + \sigma_L \int_{\sigma_L}^{+\infty} f(\sigma)d\sigma \qquad (66)$$

Le premier terme de l'intégrale correspond aux phases déjà plastifiées et le second à celles restant élastiques sous une contrainte limite $\sigma_L$ qui dépend du chargement et de l'interaction entre les phases. Dans le cas général, cette équation ne peut être résolue analytiquement. Par contre, il est possible d'estimer le module tangent de cette loi, indépendamment de la fonction F :

$$\begin{aligned} d\sigma^m &= f(\sigma_L)\sigma_L d\sigma_L + \left(-f(\sigma_L)\sigma_L d\sigma_L + \left(\int_{\sigma_L}^{+\infty} f(\sigma)d\sigma\right)d\sigma_L\right) \\ &= (1-F(\sigma_L))\times d\sigma_L = (1-F(\sigma_L))\times \frac{d\sigma^m + \beta d\varepsilon^m}{1+\frac{\beta}{Y}} \qquad \text{avec} \quad \sigma_L = \frac{\sigma^m + \beta\varepsilon^m}{1+\frac{\beta}{Y}} \\ \Rightarrow \frac{d\sigma^m}{d\varepsilon^m} &= \frac{1}{\frac{1}{Y}+\frac{F(\sigma_L)}{\beta}}(1-F(\sigma_L)) \end{aligned} \qquad (67)$$

Avec $\beta$, un paramètre constant contrôlant l'interaction entre les phases.

$$\beta = -\frac{\sigma^m - \sigma}{\varepsilon^m - \varepsilon} \qquad (68)$$

Ce paramètre permet de gérer des scénarios de localisation caricaturaux allant de l'iso-déformation entre toutes les phases du composite ($\beta = +\infty$ - hypothèse de localisation type Taylor) à l'iso-contrainte ($\beta = 0$ - hypothèse de localisation type Sachs). Pour tous autres détails calculatoires on se reportera à [ALLAIN 2012].

Dans les conditions d'iso-déformation ($\beta \gg Y$), l'équation précédente se simplifie :

$$\frac{d\sigma^m}{d\varepsilon^m} = Y(1-F(Y\varepsilon^m)) \qquad (69)$$

Le taux d'écrouissage devient une simple fonction des phases restant élastiques dans le composite et du module d'Young. Dans ce cas, la formulation est trop rigide pour rendre compte du comportement expérimental (contraintes internes trop élevées donc surestimation du taux d'écrouissage). Par contre, cette formule simple est largement utilisé dans le cas de la modélisation de la fatigue [POLAK 1982][HOLSTE 1980]. Dans la suite, on retiendra une valeur de β intermédiaire de Y/4 (50GPa).

On notera que la formulation retenue pour le module tangent reste aussi valable dans le domaine élastique. Si $\sigma^m \leq \sigma_{min}$ alors $\sigma_L \leq \sigma_{min}$ et donc $F(\sigma_L) = 0$. On retrouve alors le module d'Young Y :

$$\frac{d\sigma^m}{d\varepsilon^m} = Y \qquad (70)$$

Ce modèle a ensuite été appliqué pour décrire le comportement mécanique de 6 aciers martensitiques. La fonction de distribution cumulée F est décrite par une loi type Avrami à trois paramètres :

$$\text{si } \sigma < \sigma_{min} + \sigma_{friction} \text{ alors } F(\sigma) = 0 \text{ sinon } F(\sigma) = 1 - \exp\left(-\left(\frac{\sigma - (\sigma_{min} + \sigma_{friction})}{\sigma_0}\right)^n\right) \qquad (71)$$

avec $\sigma_{friction}$ la contrainte de friction due aux éléments en solution solide.

$$\sigma_{friction} = 60 + 33\ \%Mn + 81\ \%Si + 48\ \%\ Cr + 48\ \%\ Mo \qquad (72)$$

Il ressort de la procédure d'ajustement que les paramètres n et $\sigma_{min}$ peuvent prendre des valeurs constantes (n = 1.82 et $\sigma_{min}$ = 300 MPa respectivement) pour tous les aciers alors que $\sigma_0$ caractéristique de la largeur de la distribution de contrainte d'écoulement dans le composite varie significations avec la teneur en carbone $\%C_m$ :

$$\sigma_0 = 645 + 5053 \times (\%C_m)^{1.34} \qquad (73)$$

On relèvera que l'ordre de grandeur de la somme $\sigma_{min} + \sigma_{friction}$ correspond bien au 400-500 MPa attendu pour le seuil de microplasticité et pourrait correspondre avec la contrainte d'écoulement d'une structure ferritiques hautement déformée sans carbone en solution solide. On rapporte dans la littérature des valeurs de densités de dislocations de l'ordre de 1 à 5 $10^{14}$ m$^{-2}$ [SHIBATA 2009] donc des valeurs de $\sigma_{min}$ comprises entre 240 et 540 MPa environ, valeurs très cohérentes à celles déduites de cette analyse.

La Figure 70 montre que le modèle, une fois ajusté, permet de reproduire de façon précise les courbes de comportement des 6 aciers considérés, mais aussi la forme de leur taux d'écrouissage.

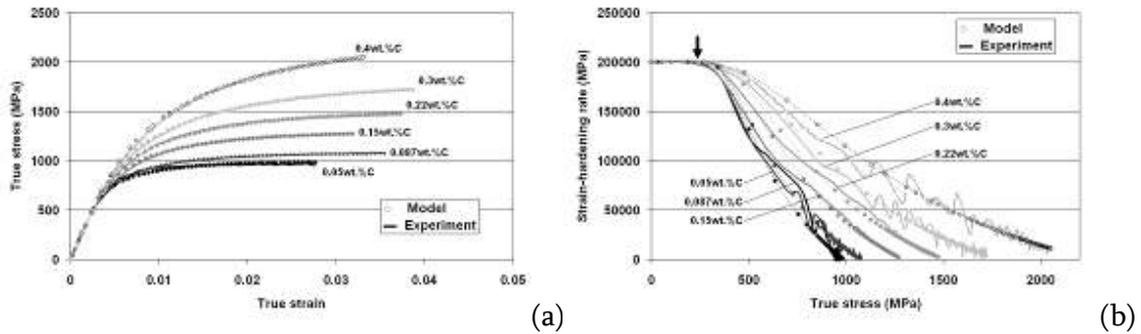

Figure 70 : (a) Courbes de traction et (b) taux d'écrouissage en fonction de la contrainte de 6 aciers martensitiques avec des teneurs en carbone nominales différentes (expérience et modèle ajusté). La flèche indique le seuil de microplasticité constant pour tous les aciers étudiés [ALLAIN 2012].

Comme suggéré par Asaro [ASARO 1975], ce type d'approche micromécanique offre d'intéressantes possibilités pour évaluer des contraintes internes (contribution cinétique de type I selon la classification d'Asaro) et de prédiction du comportement sur des changements de trajet alternés. La Figure 71 montre par exemple les prédictions du modèle lors d'un essai de Bauschinger sans aucun paramètre d'ajustement supplémentaire[17]. Peu de modèles peuvent se prévaloir d'une aussi bonne description (à la charge et la décharge) lors de chargement alternée sans paramètres d'ajustements supplémentaires.

---

[17] Comme le laisse entrevoir la Figure 71, le modèle surévalue sensiblement les contraintes internes. On attribue cette trop grande rigidité à l'absence d'écrouissage des phases du composite (comportement parfaitement plastique). Le taux d'écrouissage est donc uniquement du ressort des phases non plastifiées, dont la proportion est alors surévaluée. Cette surévaluation conduit nécessairement à des contraintes internes plus importantes. Une façon de résoudre ce problème a été proposée récemment par G. Badinier, qui consiste à introduire un écrouissage linéaire dans chacune des phases (module k'). La loi précédente devient alors :

$$\frac{d\sigma^m}{d\varepsilon^m} = \frac{k'+\beta+\beta\left(\frac{k'}{Y}-1\right)F(\sigma_L)}{\frac{1}{Y}(k'+\beta)-\left(\frac{k'}{Y}-1\right)F(\sigma_L)}$$

Quand toutes les phases du composite ont plastifié, $F(\sigma_L) \rightarrow \infty$ alors $\lim_{E\rightarrow\infty}\frac{d\sigma^m}{d\varepsilon^m} = k'$

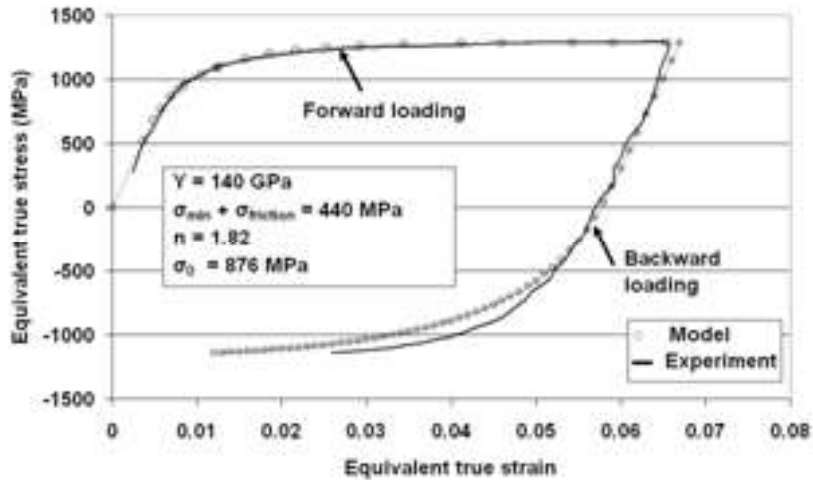

Figure 71 : Courbes de comportement expérimental et simulé d'un acier martensitique lors d'un trajet Bauschinger (chargement alterné). La phase de chargement alternée (« backward loading ») est modélisée sans paramètre supplémentaire par rapport au chargement initial (« forward loading ») [ALLAIN 2012].

Cette possibilité du modèle composite présente en soi un grand intérêt pour l'étude des aciers martensitiques. Cependant, cette description précise des chemins alternés est trop complexe pour être intégrée dans le modèle biphasé, en particulier car la description de la ferrite n'est pas aussi détaillée. En conséquence, la composante d'écrouissage cinématique de la martensite a été évaluée de façon différente. En inspirant de nos travaux préalables sur l'écrouissage des aciers ferrite-perlitiques, on se propose d'écrire :

$$\sigma_X^m = \frac{1}{2} \int_{\sigma_{min}}^{+\infty} f(\sigma) |\sigma - \Sigma| d\sigma \qquad (74)$$

On relie l'écrouissage cinétique à la moyenne des différences de comportement des phases par rapport à la moyenne. De même que l'équation précédente, cette composante cinématique ne peut être intégrée analytiquement. Par contre, sa dérivée s'exprime aussi très simplement :

$$d\sigma_X^m = F(\Sigma) d\Sigma \qquad (75)$$

La Figure 72 montre les évolutions de ce terme $\sigma^m_X$ en fonction de la déformation pour les 6 aciers considérés sur la Figure 70. Les ordres de grandeurs prévus par ce terme sont cohérents avec les mesures disponibles d'écrouissage cinématique [COBO 2008].

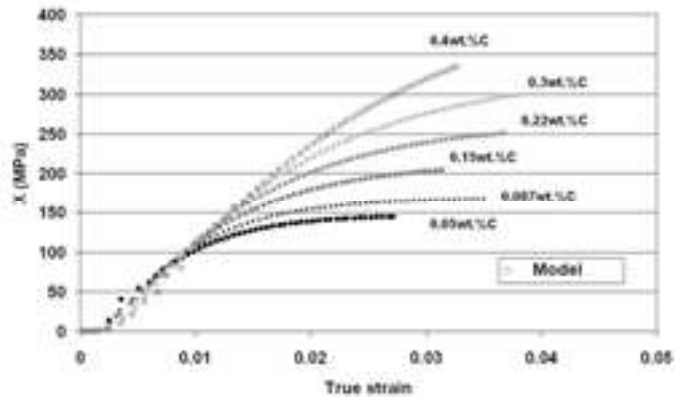

**Figure 72 : Composante cinématique de l'écrouissage d'après l'équation (75) pour les aciers de la Figure 70**

### *3.3.2.3.    Nouvelles perspectives*

Cette nouvelle approche du comportement de la martensite est donc une donnée d'entrée idéale pour le modèle biphasé. En effet, elle décrit parfaitement la transition élasto-plastique dans la martensite et son très fort taux d'écrouissage initiale mais aussi permet d'estimer une composante cinématique intrinsèque à cette phase. Ce modèle est en outre sensible à la teneur en carbone de la martensite.

Par contre, l'origine de cette dispersion de contraintes d'écoulement (spectre de contraintes) n'est pas clairement identifiée et reliée à des caractéristiques microstructurales (comme les lattes, blocs ou paquets ou bien comme les zones nanomaclées [KELLY 1963]).

Nous travaillons donc sur différentes hypothèses pour pouvoir valider cette approche. Mon axe de travail principal reste l'identification des échelles d'hétérogénéités de déformation résultantes. Elle nous permettra d'orienter nos recherches soit vers des explications structurales (i.e. en relation avec la microstructure) ou des explications d'ordre chimique (distribution de la teneur en carbone). A cette fin, je coordonne et participe actuellement à 3 études à différentes échelles (MEB, MET, SAT) sur le même acier martensitique modèle de composition Fe5Ni0.15C.

- MEB EBSD et essai de traction *in situ* en collaboration avec D. Barbier d'AM et C. Tasan et S. Zaefferer du MPI. Les premiers résultats montrent que la déformation est très hétérogène, non seulement à l'échelle des anciens grains austénitiques, mais aussi à l'échelle des lattes. La Figure 73 montre un exemple de cliché MEB avant et après 4% de déformation d'un paquet de lattes (mise en évidence par une légère attaque) très déformé. L'analyse EBSD révèle que la désorientation évolue peu à l'intérieur des lattes mais se concentre près des interfaces, comme observée aussi par [LHUISSIER 2011][NAMBU 2009]. Par contre, d'autres paquets sont très peu déformés, comme attendus par l'approche composite.

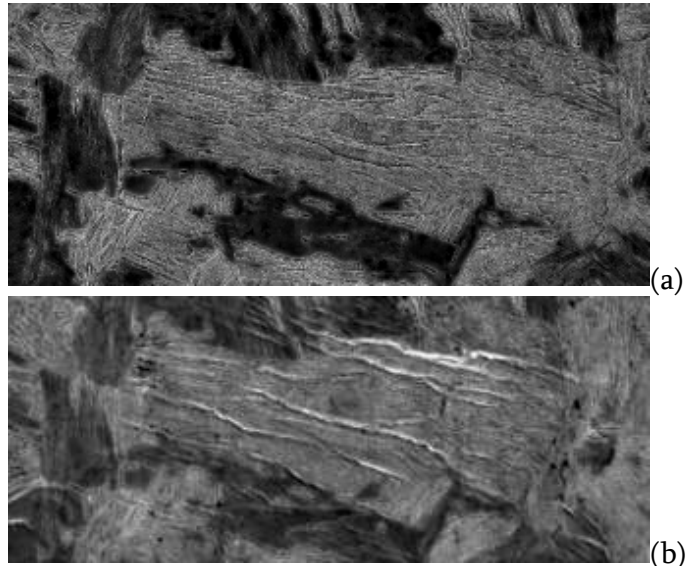

Figure 73 : Micrographie MEB après attaque d'un acier martensitique (a) dans l'état initial – mise en évidence d'un paquet de lattes et (b) après déformation *in situ* – mise en évidence de bandes de déformations (micrographies de D. Barbier et C. Tasan).

- MET en collaboration avec P. Barges d'AM. Cette étude a permis de montrer que la structure en lattes ne contient qu'une faible fraction de zones nanomaclées (environ 15%). Cette fraction est très faible en comparaison de la fraction de phases dures nécessaire pour expliquer le comportement du composite. L'hypothèse de Kelly [KELLY 1963] justifiant la dureté de la martensite sur cette seule fraction est donc insuffisante.

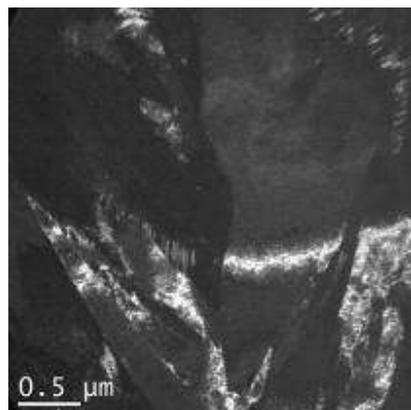

Figure 74 : Micrographie MET en champ sombre mettant en évidence des zones micromaclées dans les lattes de martensite (fraction faible) (micrographies de P. Barges).

- SAT (Sonde Atomique Tomographique 3D) en collaboration avec F. Danoix du GPM et M. Gouné de l'ICMCB. 10 pointes ont été analysées pour donner un aperçu de la distribution statistique du carbone dans cet acier martensitique aux échelles les plus fines. Contrairement aux observations sur des martensites vierges [ALLAIN 2013], les atomes de carbone ne sont jamais répartis de façon homogène, mais sont systématiquement ségrégés soit sur des défauts plans (joints de lattes ?), soit sur des rubans (dislocations) ou de façon isotropes (décomposition spinodale ?). La prochaine étape est de pouvoir extraire une pointe directement dans une zone identifiée comme dure (non déformée) par les essais de traction *in situ* sous MEB.

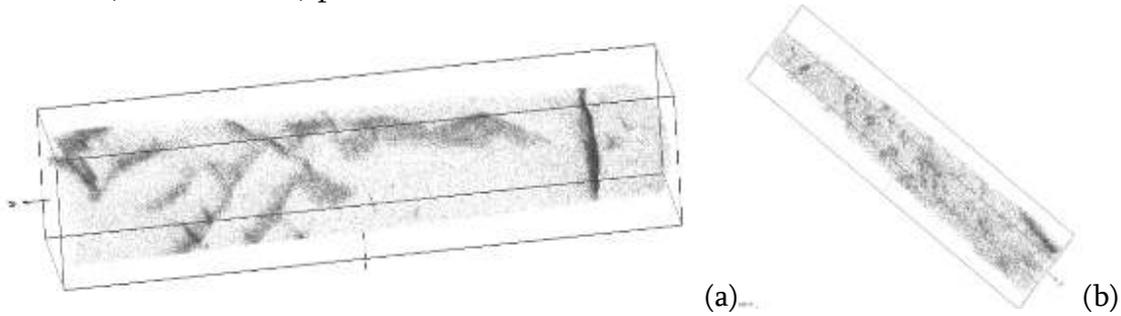

(a) (b)

Figure 75 : Distribution spatiale des atomes de carbone observée en sonde atomique tomographique 3D sur deux pointes (données de F. Danoix) (a) 65x65x290 nm3 – ségrégation des atomes de carbone sur des défauts linéaires et plans (b) 34x34x180 nm3 – ségrégation isotrope.

### 3.3.3. Comportement du composite et ajustement du modèle

#### 3.3.3.1. *Modélisation des interactions entre Ferrite et Martensite*

Dans la première et seconde partie de cette section, ont été présentés respectivement un modèle de comportement de la ferrite sensible à la taille de grain et un modèle de comportement de martensite sensible à la teneur en carbone. Afin de décrire le composé biphasé Ferrite-Martensite, il est nécessaire de suivre un schéma dit d'homogénéisation (celui de Hill dans ce cas) et d'établir des lois d'interactions entre les phases (loi de transition d'échelle type IsoW et lois de couplage spécifique).

La contrainte d'écoulement macroscopique du composite Ferrite-Martensite peut s'exprimer en fonction de la contrainte dans chacune des phases pondérée par leurs fractions relatives. De la même manière, on peut calculer la déformation macroscopique :

$$\Sigma^{DP}\left(E^{DP}\right) = F_m \sigma^m\left(\varepsilon^m\right) + \left(1 - F_m\right)\sigma^\alpha\left(\varepsilon^\alpha\right) \qquad (76)$$

$$E^{DP} = F_m \varepsilon^m + \left(1 - F_m\right)\varepsilon^\alpha \qquad (77)$$

On admet de manière implicite par cette formulation que le comportement de la martensite et de la ferrite sont les mêmes quelles que soient les échelles considérées. Nous reviendrons dans la partie discussion sur la validité de cette approximation pour la martensite.

Nous avons montré que la transition élasto-plastique était un processus clef dans le comportement de la martensite. Il est donc important de distinguer les composantes de déformations élastique et plastique et ce à toutes les échelles. La déformation macroscopique peut donc se décomposer de la façon suivante :

$$E^{DP} = E_{el}^{DP} + E_p^{DP} = \frac{\Sigma^{DP}}{Y} + E_p^{DP} \tag{78}$$

$E_p^{DP}$ la déformation plastique macroscopique, $E_{el}^{DP}$ la déformation élastique, fonction de la contrainte macroscopique et du module d'Young commun des deux phases.

Nous avons choisi comme loi de transition d'échelle l'hypothèse IsoW, qui s'exprime simplement sous la forme ;

$$\sigma^m d\varepsilon^m = \sigma^\alpha d\varepsilon^\alpha \tag{79}$$

Cette loi permet de gérer de façon continue la transition élasto-plastique.

Comme discuté au chapitre 3.3.2.1 page 111, le modèle en l'état ne traduirait l'effet DP que par un durcissement cinématique lié à la différence de comportement entre la ferrite et la martensite. Nous avons donc introduit un terme de couplage original pour une approche biphasée qui reproduit un sur-durcissement isotrope dans la ferrite. Ce terme correspond à l'introduction d'une densité de DGN d'Ashby dans l'équation (58).

$$\sigma_R^\alpha\left(\varepsilon_p^\alpha\right) = \alpha M \mu b_{111} \sqrt{\rho_\alpha + \rho^{DGN}} \tag{80}$$

Cette densité supplémentaire de dislocations s'exprime de la façon suivante en s'inspirant de l'équation (49) :

$$\rho^{DGN} = \frac{F_m}{1-F_m} \frac{1}{b_{111} L_\alpha} \frac{\left(1 - \exp\left(-\beta_{DGN}\left(\varepsilon^\alpha - \varepsilon^m + \varepsilon_0\right)\right)\right)}{\beta_{DGN}} \tag{81}$$

avec $L_\alpha$ une longueur caractéristique de la dispersion des dislocations d'Ashby dans la ferrite à partir des interfaces ferrite/martensite et β un paramètre gérant un mécanisme de saturation de l'effet. La Figure 76 illustre le mécanisme décrit par l'équation (81) et observé expérimentalement sur les Figure 65 et Figure 61. La distance $L_\alpha$ doit être inférieure à la taille de grain ferritique et de l'ordre de grandeur de la taille des ilots martensitiques.

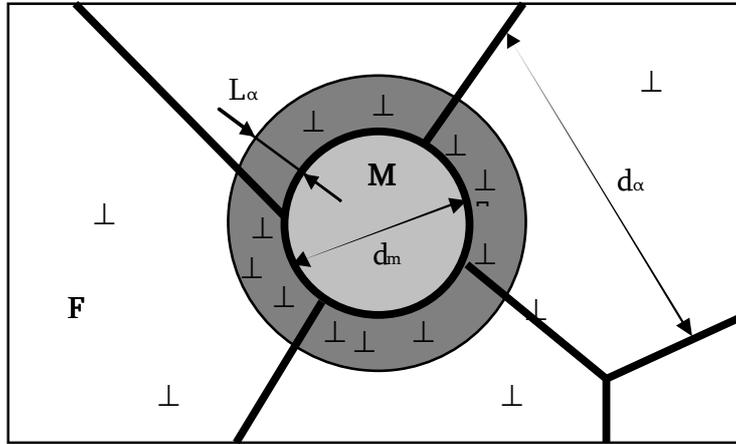

Figure 76 : Définition de la longueur $L_\alpha$ dans la ferrite (F) autour d'une particule de martensite (M), de la taille de grain ferritique $d_\alpha$ et taille d'ilot martensitique $d_m$ [ALLAIN 2005]

De manière analogue à l'équation (49), on définit une densité initiale de dislocations due la déformation induite par la transformation martensitiques grâce à la pré-déformation $\varepsilon_0$. Cette déformation initiale contribue aussi à la densité de dislocations statistiquement stockées (DSS):

$$\rho_\alpha\left(\varepsilon_p^\alpha = 0\right) = F_m \frac{M}{b_{111} d_\alpha} \varepsilon_0 \tag{82}$$

Dans la mesure où ces dislocations interagissent dans les grains ferritiques, l'équation (59) est modifiée et devient :

$$\frac{d\rho_\alpha}{d\varepsilon_p^\alpha} = M \cdot \left( \frac{\left(1 - \frac{n_\alpha}{n_0^\alpha}\right)}{b d_\alpha} + k_\alpha \sqrt{\rho_\alpha + \rho^{DGN}} - f_\alpha \rho_\alpha \right) \tag{83}$$

### 3.3.3.2. *Contraintes internes et durcissement cinématique*

Les bases de cette nouvelle approche biphasée étant posées, le comportement des aciers DP va s'expliquer globalement selon un processus classique, décrit par de nombreux auteurs [TOMOTA 1992][LI 1990_1][LIAN 1991] [HUPPER 1999] et schématisé sur la Figure 77.

- Stade 1 : Si la contrainte macroscopique appliquée est inférieure à la contrainte d'écoulement de la ferrite, les deux phases se comportent élastiquement.
- Stade 2 : La courbe de comportement s'éloigne de la linéarité à partir du moment où la ferrite plastifie. Le taux d'écrouissage macroscopique est très important de l'ordre de FxY dans la mesure où la martensite reste élastique. Les gradients de déformations entre les phases sont importants, générant beaucoup de DGN aux interfaces. La contribution cinématique du durcissement est donc importante. Cette phase caractéristique de l'effet DP est souvent qualifiée de « continous yielding ».
- Stade 3 : Toutefois, ce stade élasto-plastique est de courte durée, car la limite de microplasticité de la martensite est rapidement atteinte. Le durcissement cinématique dû à la différence de comportement entre ferrite et martensite commence à saturer. Le taux d'écrouissage diminue en conséquence.

Le comportement macroscopique d'un acier DP résulte donc d'une loi de mélange entre les états de la ferrite (durcit par les DGN d'Ashby) et de la martensite. La pente de la loi de mélange dépend du modèle de transition d'échelle retenu et peut évoluer au cours de la déformation.

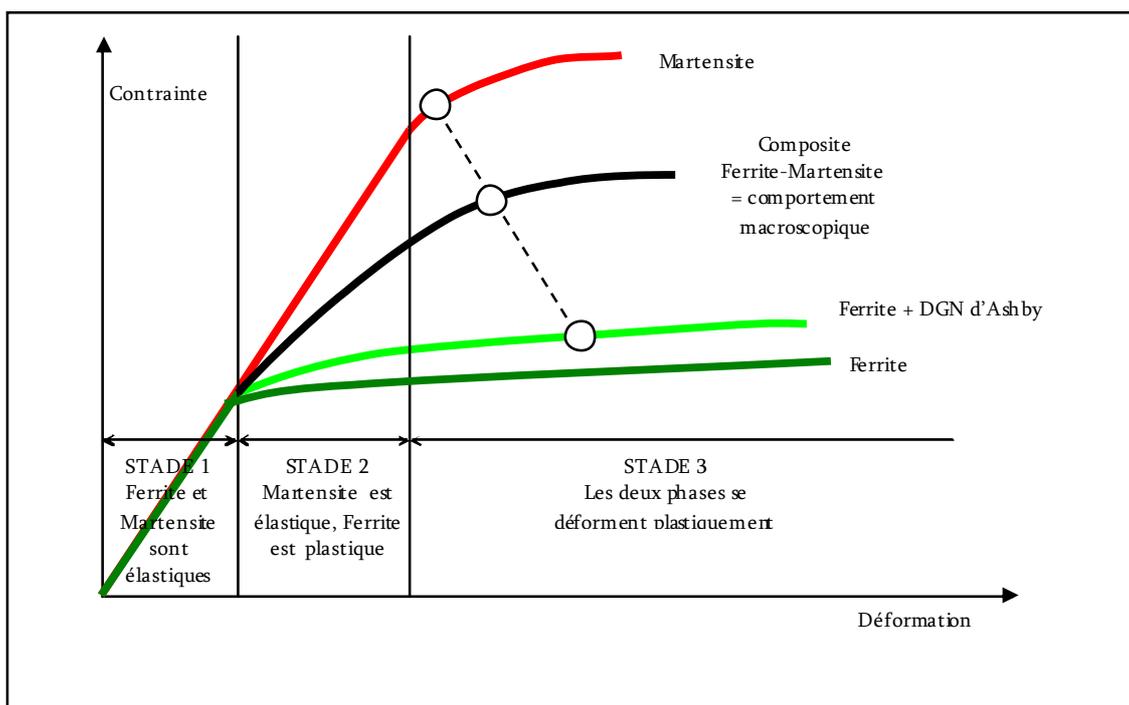

Figure 77 : Le comportement macroscopique d'un acier DP résulte d'une loi de mélange (symbolisée par la ligne pointillée) entre les états de la ferrite (durcie par les DGN d'Ashby) et de la martensite. Il se décompose en 3 stades dépendant de l'état de déformation des deux phases constitutives. La pente de la loi de mélange dépend du modèle de transition d'échelle retenu et peut évoluer au cours de la déformation.

La contribution cinématique au durcissement $X^{DP}$ est estimée a posteriori, selon la méthodologie évoquée dans le chapitre précédent (cf. 2.3.5.2 page 55), sans paramètres supplémentaires [ALLAIN 2008]. Elle tient compte de deux contributions :
- Les écrouissages cinématiques des phases respectives prises indépendamment mais pondérés par les fractions de phase,
- Un terme de mélange qui tient compte du gradient de contraintes entre les deux phases.

$$X^{DP}\left(E^{DP}\right) = F_m \sigma_X^m + \left(1 - F_m\right) \sigma_X^\alpha + F_m \left(1 - F_m\right) \left(\sigma^m - \sigma^\alpha\right) \tag{84}$$

### 3.3.3.3. *Ajustement du modèle : base de données et paramètres*

Cette démarche de modélisation a été validée grâce à de nombreux résultats obtenus dans le cadre de cette étude (aciers DP modèles) et de la littérature. Seuls ont été retenus des aciers non microalliés (précipitation), sans bainite et non revenu (non auto-revenu dans la mesure du possible). Malheureusement, la plupart des caractérisations microstructurale disponibles étaient lacunaires (fraction de martensite, taille de grain ferritique, taille des ilots de martensite, teneur en carbone de la martensite). Sauf si mesurées ou reportées par les auteurs respectifs, les fractions de martensite et tailles de grains ferritiques ont été estimées à partir des micrographies disponibles.

La teneur en carbone moyenne dans la martensite est calculée par un simple bilan carbone, en supposant que la ferrite ne contient pas de carbone :

$$\%C_m = \frac{\%C}{F_m} \tag{85}$$

Avec %C la teneur nominale de l'acier. Par soucis de cohérence, cette valeur est bornée à 0.6% dans le modèle de comportement de la martensite. Cette valeur correspond à la teneur en carbone seuil partir de laquelle on observe expérimentalement une saturation de la dureté de la martensite [KRAUSS 1999]. Cette saturation est attribuée dans la littérature à une stabilisation significative d'austénite résiduelle dans la martensite.

Le Tableau 4 montre les caractéristiques des aciers DP ayant servies à ajuster le modèle (sources, composition chimique, caractéristiques microstructurale mesurées expérimentalement et caractéristiques microstructurales utilisées pour le calcul). La Figure 78 présente la diversité des couples fraction de martensite / teneur nominale en carbone des aciers de cette base de données. Ils couvrent une large gamme sauf le coin gauche en haut de la figure car ces couples sont inaccessibles du point de vue des transformations de phases.

| Source | composition chimique | | | | | Données microstructurales (exp.) | | | | Données microstructurales (est.) | | |
|---|---|---|---|---|---|---|---|---|---|---|---|---|
| | C | Mn | Si | Cr | Mo | F (%) | Cm (%) | dm (µm) | df (µm) | Cm (%) | df (µ) | dm (µm) |
| Matlock D. & Speer J., | 0.1 | 1.52 | 1.48 | | | 28 | 0.29 | x | 6 | 0.36 | 6.0 | 2.9 |
| Matlock D. & Speer J., | 0.14 | 1.48 | 1.48 | | | 39 | 0.32 | x | 5 | 0.36 | 5.0 | 3.4 |
| Matlock D. & Speer J., | 0.21 | 1.46 | 1.47 | | | 49 | 0.35 | x | 3 | 0.43 | 3.0 | 2.8 |
| Matlock D. & Speer J., | 0.15 | 1.95 | 0.45 | | | 13 | x | x | 10 | 1.15 | 10.0 | 2.6 |
| Speich G.R. & Miller R.L. | 0.06 | 1.5 | | | | 24.8 | x | x | 8.3 | 0.24 | 8.3 | 3.6 |
| Speich G.R. & Miller R.L. | 0.2 | 1.5 | | | | 48.9 | x | x | x | 0.41 | 4.0 | 3.8 |
| Speich G.R. & Miller R.L. | 0.4 | 1.5 | | | | 89.7 | x | x | x | 0.45 | 2.0 | 13.7 |
| Al-Abbasi F.M. & Nemes J.A. | 0.09 | 1.5 | 0.9 | 0.06 | 0.04 | 34 | x | 13 | 17 | 0.26 | 17.0 | 9.9 |
| Tavares S. S. M. et al. | 0.12 | 0.82 | 0.235 | | | 37 | x | x | x | 0.32 | 50.0 | 32.2 |
| Lei T.C. & Schen H.P. | 0.23 | 0.5 | 0.28 | | | 34 | x | 6.5 | 7 | 0.68 | 7.0 | 4.1 |
| Lei T.C. & Schen H.P. | 0.23 | 0.5 | 0.28 | | | 33 | x | 7.5 | 9 | 0.70 | 9.0 | 5.1 |
| Liedl U. et Al. | 0.09 | 1.4 | 0.1 | 0.7 | | 6 | x | x | x | 1.50 | 5.0 | 0.8 |
| Liedl U. et Al. | 0.09 | 1.4 | 0.1 | 0.7 | | 11 | x | x | 3.4 | 0.82 | 3.4 | 0.8 |
| Liedl U. et Al. | 0.09 | 1.4 | 0.1 | 0.7 | | 18 | x | x | x | 0.50 | 5.0 | 1.6 |
| Prahl et al. | 0.19 | 1.5 | 0.26 | | | 85 | 0.235 | x | x | 0.22 | 5.0 | 22.7 |
| Prahl et al. | 0.04 | 1.48 | 0.85 | | | 21 | x | 40 | 40 | 0.19 | 40.0 | 14.8 |
| Jian Z. et al. | 0.2 | 0.47 | 0.3 | 0.07 | | 33 | x | 11.4 | 24.5 | 0.61 | 24.5 | 13.9 |
| Jian Z. et al. | 0.2 | 0.47 | 0.3 | 0.07 | | 33 | x | 10 | 17.8 | 0.61 | 17.8 | 10.1 |
| Jian Z. et al. | 0.2 | 0.47 | 0.3 | 0.07 | | 33 | x | 7 | 12.1 | 0.61 | 12.1 | 6.8 |
| Jian Z. et al. | 0.2 | 0.47 | 0.3 | 0.07 | | 33 | x | 5 | 8.6 | 0.61 | 8.6 | 4.9 |
| Jian Z. et al. | 0.2 | 0.47 | 0.3 | 0.07 | | 40 | x | 14.8 | 19.5 | 0.50 | 19.5 | 13.8 |
| Jian Z. et al. | 0.2 | 0.47 | 0.3 | 0.07 | | 40 | x | 12.5 | 15.2 | 0.50 | 15.2 | 10.8 |
| Jian Z. et al. | 0.2 | 0.47 | 0.3 | 0.07 | | 40 | x | 10.7 | 12.6 | 0.50 | 12.6 | 8.9 |
| Jian Z. et al. | 0.2 | 0.47 | 0.3 | 0.07 | | 40 | x | 8.5 | 9.1 | 0.50 | 9.1 | 6.4 |
| Jacques P.J. et al. | 0.29 | 1.42 | 1.41 | | | 50 | x | x | x | 0.58 | 2.0 | 1.9 |
| Lawson R.D. et al. | 0.063 | 1.29 | 0.24 | | | 21 | x | x | x | 0.30 | 15.0 | 5.6 |
| Lawson R.D. et al. | 0.063 | 1.29 | 0.24 | | | 34 | x | x | x | 0.19 | 15.0 | 8.8 |
| Hansen S.S. & Pradhan R.R. | 0.11 | 1.5 | 1.17 | | | 37.5 | x | x | 5 | 0.29 | 5.0 | 3.3 |
| Hansen S.S. & Pradhan R.R. | 0.037 | 1.55 | 1.16 | | | 7.7 | x | x | 5 | 0.48 | 5.0 | 0.9 |
| Lian J. et al. | 0.2 | 0.47 | 0.3 | | | 33 | x | 11.4 | 24.5 | 0.61 | 24.5 | 13.9 |
| Lian J. et al. | 0.2 | 0.47 | 0.3 | | | 40 | x | 14.8 | 19.5 | 0.50 | 19.5 | 13.8 |
| Lian J. et al. | 0.2 | 0.47 | 0.3 | | | 52 | x | x | x | 0.38 | 13.0 | 13.5 |
| Lian J. et al. | 0.2 | 0.47 | 0.3 | | | 66 | x | x | x | 0.30 | 8.0 | 13.6 |
| Lian J. et al. | 0.2 | 0.47 | 0.3 | | | 85 | x | x | x | 0.24 | 5.0 | 22.7 |
| Wasilkowska A. et al. | 0.1 | 1.5 | 0.1 | 0.77 | | 7 | x | x | 18 | 1.43 | 6.0 | 1.1 |
| Hasegawa K. et al. | 0.118 | 1.9 | 1.4 | | | 34 | x | x | x | 0.35 | 8.0 | 4.7 |
| Hasegawa K. et al. | 0.126 | 1.9 | 1.29 | | | 49 | x | x | x | 0.26 | 3.0 | 2.8 |
| Hasegawa K. et al. | 0.052 | 1.9 | 0.02 | | | 99 | x | x | x | 0.05 | 3.0 | 225.8 |
| Pushkareva I. | 0.15 | 1.9 | 0.215 | 0.195 | | 81.2 | x | x | x | 0.18 | 5.0 | 17.6 |
| Pushkareva I. | 0.15 | 1.9 | 0.215 | 0.195 | | 93.2 | x | x | x | 0.16 | 3.0 | 31.9 |
| Pushkareva I. | 0.15 | 1.9 | 0.215 | 0.195 | | 98.4 | x | x | x | 0.15 | 3.0 | 140.5 |
| Pushkareva I. | 0.15 | 1.9 | 0.215 | 0.195 | | 80.7 | x | x | x | 0.19 | 3.0 | 10.2 |
| Pushkareva I. | 0.15 | 1.9 | 0.215 | 0.195 | | 61.28 | x | x | x | 0.24 | 10.0 | 14.3 |
| Pushkareva I. | 0.15 | 1.9 | 0.215 | 0.195 | | 62.31 | x | x | x | 0.24 | 10.0 | 14.8 |
| Pushkareva I. | 0.15 | 1.9 | 0.215 | 0.195 | | 81.99 | x | x | x | 0.18 | 5.0 | 18.5 |
| Pushkareva I. | 0.15 | 1.9 | 0.215 | 0.195 | | 85.51 | x | x | x | 0.18 | 5.0 | 23.6 |
| Pushkareva I. | 0.15 | 1.9 | 0.215 | 0.195 | | 99.58 | x | x | x | 0.15 | 3.0 | 539.6 |
| Pushkareva I. | 0.15 | 1.9 | 0.215 | 0.195 | | 99.58 | x | x | x | 0.15 | 3.0 | 539.6 |
| Pushkareva I. | 0.15 | 1.9 | 0.215 | 0.195 | | 98.9 | x | x | x | 0.15 | 3.0 | 205.1 |
| Pushkareva I. | 0.15 | 1.9 | 0.215 | 0.195 | | 96 | x | x | x | 0.16 | 3.0 | 55.3 |
| Données AM | 0.09 | 2 | 0.2 | 0.3 | | 7.7 | x | 1 | 4.6 | 1.17 | 4.6 | 0.9 |
| Données AM | 0.087 | 1.9 | 0.15 | 0.1 | 0.05 | 98 | x | x | x | 0.09 | 2.0 | 74.7 |
| Données AM | 0.087 | 1.9 | 0.15 | 0.1 | 0.05 | 81 | x | x | x | 0.11 | 3.0 | 10.4 |
| Données AM | 0.087 | 1.9 | 0.15 | 0.1 | 0.05 | 81 | x | x | x | 0.11 | 3.0 | 10.4 |
| Données AM | 0.087 | 1.9 | 0.15 | 0.1 | 0.05 | 70 | x | x | x | 0.12 | 5.0 | 10.0 |
| Données AM | 0.087 | 1.9 | 0.15 | 0.1 | 0.05 | 56 | x | x | x | 0.16 | 5.0 | 5.9 |
| Données AM | 0.087 | 1.9 | 0.15 | 0.1 | 0.05 | 56 | x | x | x | 0.16 | 5.0 | 5.9 |
| Données AM | 0.087 | 1.9 | 0.15 | 0.1 | 0.05 | 32 | x | x | x | 0.27 | 5.0 | 2.7 |

Tableau 4 : Référence, composition, données microstructurales disponibles et estimées des aciers DP décrits dans la littérature ayant servi à l'ajustement des paramètres du modèle.

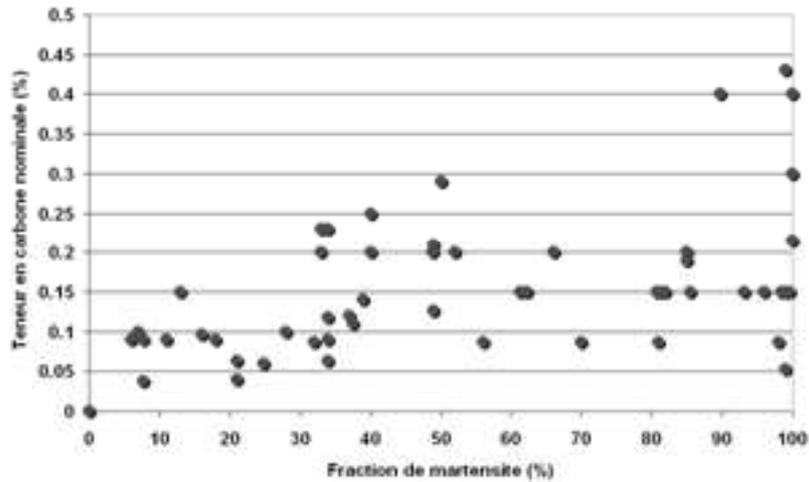

Figure 78 : Fraction de martensite et teneur en carbone nominale des aciers décrits dans le Tableau 4.

Dans la mesure où les lois de comportement de la ferrite et de la martensite ont été ajustées sur des aciers monophasés de manière indépendante, les seuls paramètres d'ajustement du modèle spécifique aux aciers DP sont les paramètres $\beta_{DGN}$, $L_\alpha$ et $\varepsilon_0$. Dans une première version du modèle, nous avions trouvé une dépendance empirique linéaire entre $L_\alpha$ et la taille des ilots $d_m$. Ce résultat avait l'avantage de donner au modèle une sensibilité naturelle à cette dimension microstructurale, à l'instar des équations du monophasé (cf. équations (51) et (52)). Toutefois, les résultats étaient dispersés et ce terme devenait négligeable aux grandes fractions de $F_m$. Comme nous le verrons, il se pose même une question sur la définition de cette dimension quand la structure martensitique est percolée, voire englobante par rapport à la ferrite.

Dans sa version la plus récente avec le modèle de martensite composite, nous avons déterminé que les trois paramètres pouvaient être choisis constants pour tous les aciers. Les valeurs retenues sont respectivement :
- $\beta_{DGN}$ = 18 (valeur cohérente avec les valeurs généralement utilisées pour l'équation (52))
- $\varepsilon_0$ = 0.05 (valeur cohérente avec les valeurs généralement utilisées pour l'équation (51))
- $L_\alpha$ = 0.8 µm – cette valeur a le même ordre de grandeur que l'épaisseur des zones dures (coques) déterminées par EF par Kadkhodapour *et al.* pour décrire les effets de taille dans ces aciers [KADKHODAPOUR 2011].

La Figure 79 montre une comparaison entre les contraintes expérimentales et modélisées pour les aciers de la base après 4 niveaux de déformation (ou moins si l'allongement réparti est faible). Le modèle permet donc de prédire de manière satisfaisante, sans biais, les tendances de cette base de données, des aciers ferritiques aux aciers martensitiques à haute teneur en carbone. On peut considérer ce résultat comme très satisfaisant compte tenu de la qualité des données d'entrée (disparates et souvent fragmentaires).

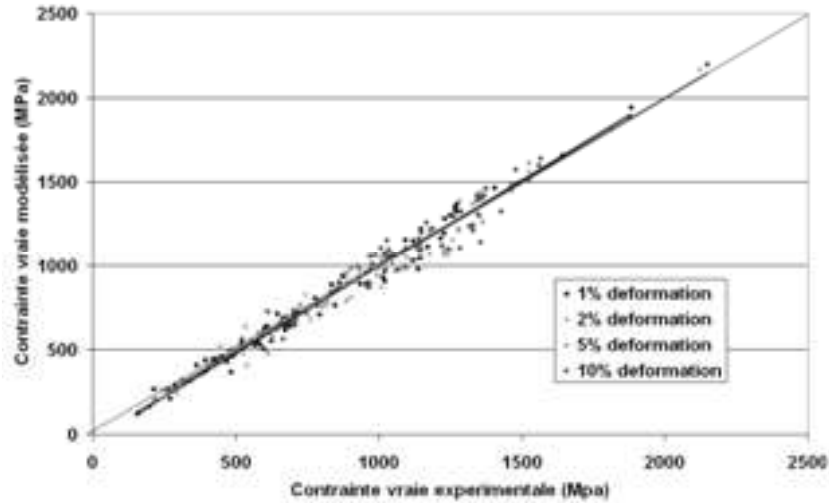

Figure 79: Comparaison des contraintes vraies expérimentales et modélisées après différents niveaux de déformation vraie pour les aciers de la base de données

### 3.3.4. Vers une démarche « d'alloy-design »

#### 3.3.4.1. *Application au DP600*

Le premier exemple d'application du modèle concerne un acier DP600 (Fe0.09C2Mn0.2Si0.3Cr), présentant une fraction de martensite faible (8%), une taille de grain ferritique de 5 µm environ et une taille d'ilots martensitiques de 1µm. La Figure 80(a) montre une micrographie après attaque de cet acier (confirmant l'absence de bainite) et la figure (b) la construction de la courbe de traction de cet acier à partir du comportement de ses constituants, ferrite et martensite. Pour quelques couples contrainte-déformation de cette courbe, sont indiquées par les lignes pointillées bleues les contraintes et déformations respectives dans les deux constituants définissant la loi de mélange. Un modèle d'iso-déformation aurait donné des lignes systématiquement verticales par exemple, mais une réponse très rigide. A contrario, avec l'IsoW, la pente de ces lignes évolue significativement au cours de la déformation à partir d'un scénario quasiment iso-déformation.

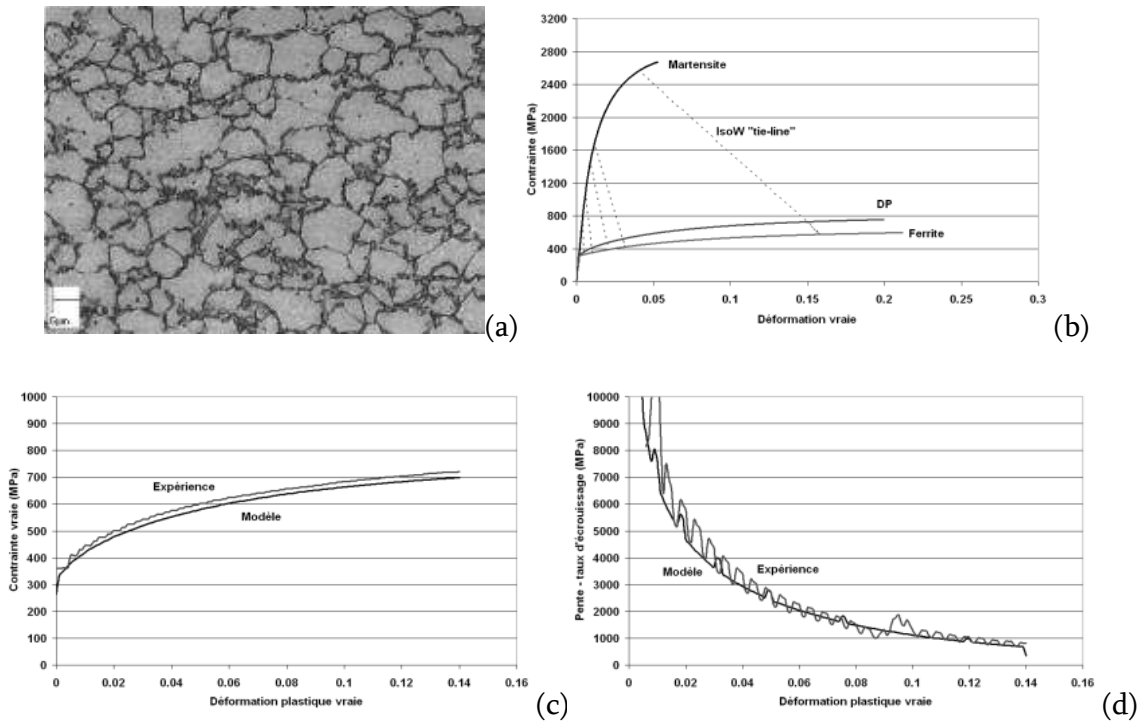

Figure 80 : (a) Micrographie optique après attaque de l'acier DP600 considéré (b) Détail de la modélisation de cet acier, en distinguant le comportement de sa ferrite et de sa martensite. La loi de mélange permettant de décrire le composite est représentée par des lignes pointillées et évolue au cours de la déformation. (c) et (d) courbes de traction et d'écrouissage expérimentales et modélisées respectivement.

L'écrouissage cinématique $X^{DP}$ modélisé (équation (84)) n'a pas pu être comparé à des données expérimentales sur cet acier en particulier. Par contre, les valeurs ont été comparées sur la Figure 81 à des données disponibles sur 2 aciers de même grade (DP 600) et présentant des courbes de comportement similaires [SADAGOPAN 2003]. L'écrouissage cinématique prédit est très cohérent, en particulier pour l'acier DP600 (2), en termes de valeurs absolues et d'évolution avec la déformation (comportement saturant après 5-10% de déformation comme attendue par les modèles analytiques monophasés – cf. équation (52)).

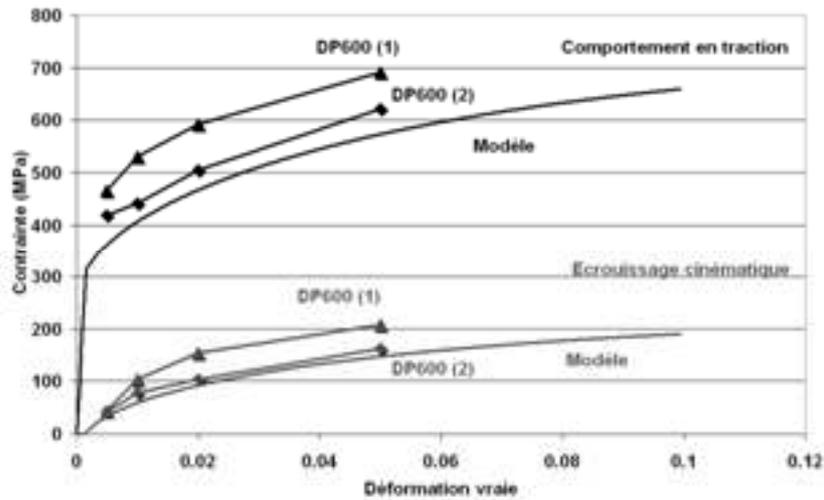

Figure 81 : Comportement en traction et valeur d'écrouissage cinématique de deux aciers DP 600 décrits par [SADAGOPAN 2003]- comparaison avec l'acier DP 600 décrit sur la Figure 80.

Cette capacité à prédire non seulement les comportements sur trajets monotones avec une grande précision mais aussi les contraintes internes générées au cours de la déformation (de nature inter- et intra-phases) est une preuve indirecte supplémentaire de la pertinence globale de l'approche et un élément discriminant fort pour le choix entre des modèles.

### 3.3.4.2. *Effet de fraction et de dilution !*

La Figure 82 montre l'évolution des propriétés de traction attendues par le modèle en fonction de la fraction de martensite, sans modifier la chimie et la taille de grain ferritique. Sont reportés les résistances mécaniques, limites d'élasticité, allongements répartis et écrouissages cinématiques à saturation calculés.

Cette investigation virtuelle révèle deux domaines de comportement distincts :
- Domaine E : Aux faibles fractions, pour une teneur nominale en carbone donné, la teneur en carbone dans la martensite est élevée. Le comportement de l'acier DP est contrôlé par l'élasticité de la martensite. La résistance mécanique et l'écrouissage cinématique augmente rapidement avec la fraction. Le taux d'écrouissage reste élevé et conduit à de forts allongements répartis.
- Domaine P : Au-delà d'une valeur critique (environ 20%), la sensibilité de résistance mécanique à la fraction diminue, signe que le comportement est contrôlé par la plasticité de la martensite. La teneur en carbone dans cette phase est alors inférieure à 0.4%. Dans la mesure où le gradient de contraintes entre les phases se réduit, l'écrouissage cinématique diminue ainsi que le taux d'écrouissage, ce qui a pour conséquence une chute rapide des allongements répartis.

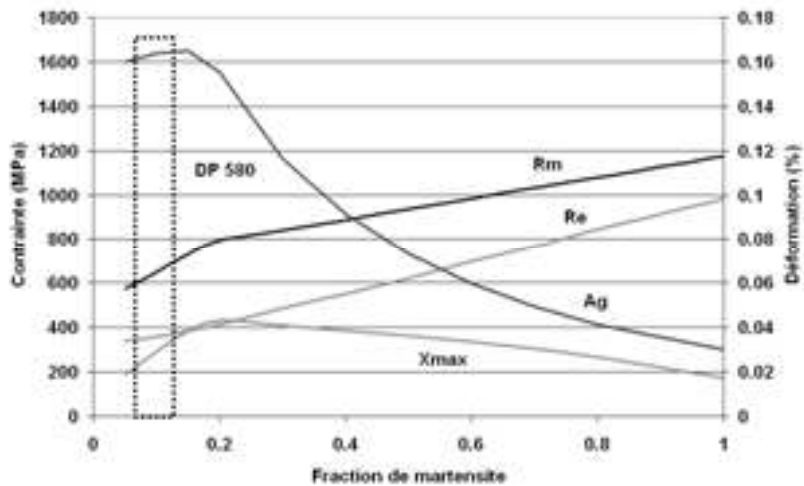

**Figure 82 :** Evolution simulée des propriétés en traction d'un acier DP 600 avec la fraction de martensite (taille de grain ferritique supposée constante). Rm = Résistance mécanique, Re = Limite d'élasticité, Ag = Allongement réparti, $X_{max}$ = Contribution cinématique à l'écrouissage à saturation.

La limite d'élasticité augmente quasiment linéairement sur les deux domaines, car elle est évaluée dans un domaine où la martensite est systématiquement élastique. On ne distingue donc pas les deux régimes. Dans la pratique, la situation est plus complexe car il est difficile de travailler à taille de grain ferritique constante. Ces domaines de comportement ont été observés bien entendu expérimentalement, comme le montre la Figure 83. Le modèle est appliqué avec un bon accord aux données internes sur un acier à Fe0.09C1.9Mn0.1Si0.3Cr0.15Mo, avec des fractions de martensite variables. Les tailles de grains ferritiques ont été estimées en microscopique optique.

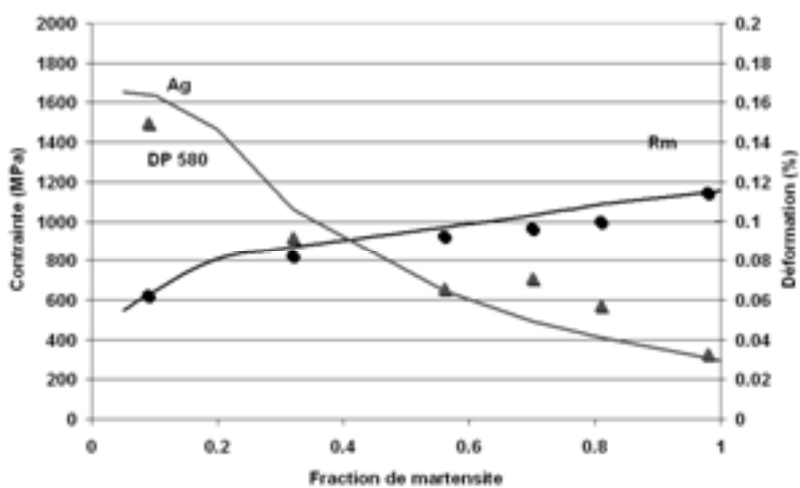

**Figure 83 :** Propriétés en traction d'un acier DP (Fe0.09C1.9Mn0.1Si0.3Cr0.15Mo) en fonction de la fraction de martensite (tailles de grain ferritique mesurées). Les points correspondent à des données expérimentales et les lignes continues aux résultats du modèle. Rm = Résistance mécanique, Ag = allongement réparti.

Le domaine E, celui contrôlé par l'élasticité de la martensite, correspond au domaine de validité des modèles analytique monophasés. La Figure 84 démontre par exemple que les équations (51) et (52) permettent de reproduire de manière très satisfaisante les courbes de traction mais aussi la composante cinématique de l'écrouissage de l'acier DP 600 décrit précédemment. Le modèle monophasé décrit correctement ce premier domaine mais s'avère trop rigide pour capter la transition vers le régime contrôlé par son comportement plastique (au-delà de 20% de martensite sur la Figure 84 (b)).

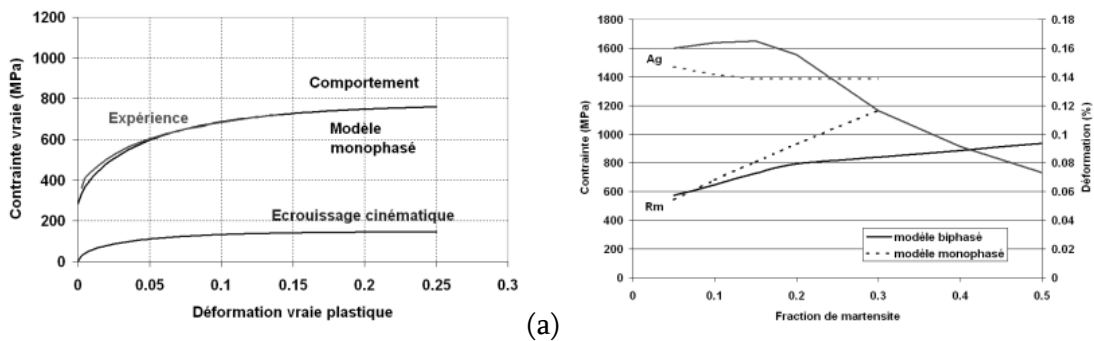

Figure 84 : (a) Prédiction du modèle monophasé (équations (51) et (52)) pour décrire la courbe de traction et l'écrouissage cinématique de l'acier DP 600 de la Figure 80. (b) Comparaison des modèles monophasés et biphasés sur une large gamme de fraction de martensite.

Cette transition entre domaines est bien entendu due à la réduction de la teneur locale en carbone dans la martensite (effet de dilution). En conséquence, la fraction critique correspondant à la transition augmente avec la teneur nominale en carbone. La Figure 85 présente un exemple de décalage de la transition entre les domaines avec la teneur en carbone. Sur la figure (b), les faibles fractions n'ont pas été considérées car inaccessibles d'un point de vue métallurgique.

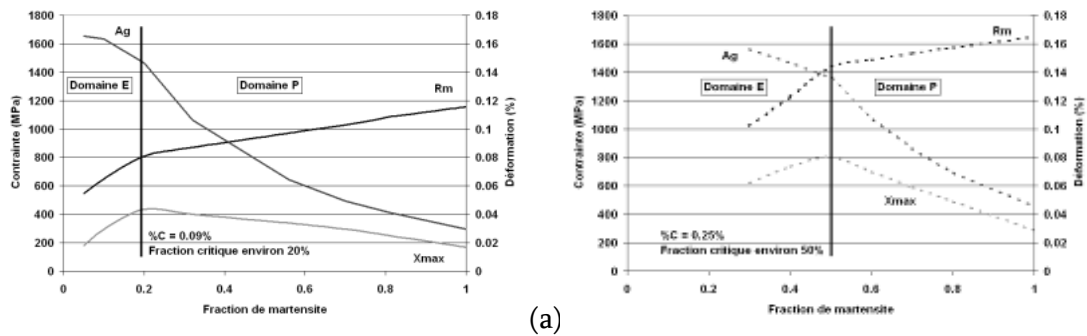

Figure 85 : Propriétés de traction simulées d'aciers DP en fonction de la fraction de martensite (taille de grain constante) et transition entre régimes contrôlés par l'élasticité (domaine E) et par la plasticité de la martensite (domaine P) pour des aciers virtuels (a) à 0.09%C et (b) à 0.25%C.

Pour finir cette discussion sur les effets de dilution, la Figure 86 présente simultanément les effets de fractions et de teneur en carbone nominale sur les résistances mécaniques, limites d'élasticité et allongements répartis prédits par le modèle. Les calculs ont été réalisés pour une taille de grain ferritique constante pour des raisons de simplicité. La transition entre domaines E et P est représentée par une ligne continue sur les différentes figures et comparée en particulier sur la figure (b) aux calculs de la teneur en carbone dans la martensite $C_m$ sur un même plan. On s'aperçoit alors que la transition correspond très exactement à l'iso-$C_m$ égale à 0.4%. Autrement dit, si un acier DP présente une teneur locale en martensite inférieure à 0.4%C, alors son comportement macroscopique sera contrôlé principalement par le comportement plastique de la martensite. A contrario, si elle est supérieure, alors on peut considérer que la martensite se comporte comme une phase élastique (approximation des modèles monophasés). Dans ce domaine, les propriétés mécaniques dépendent principalement de la fraction de martensite. A titre purement illustratif, les positionnements typiques d'aciers DP ou martensitiques industriels sont repérés sur la figure (a).

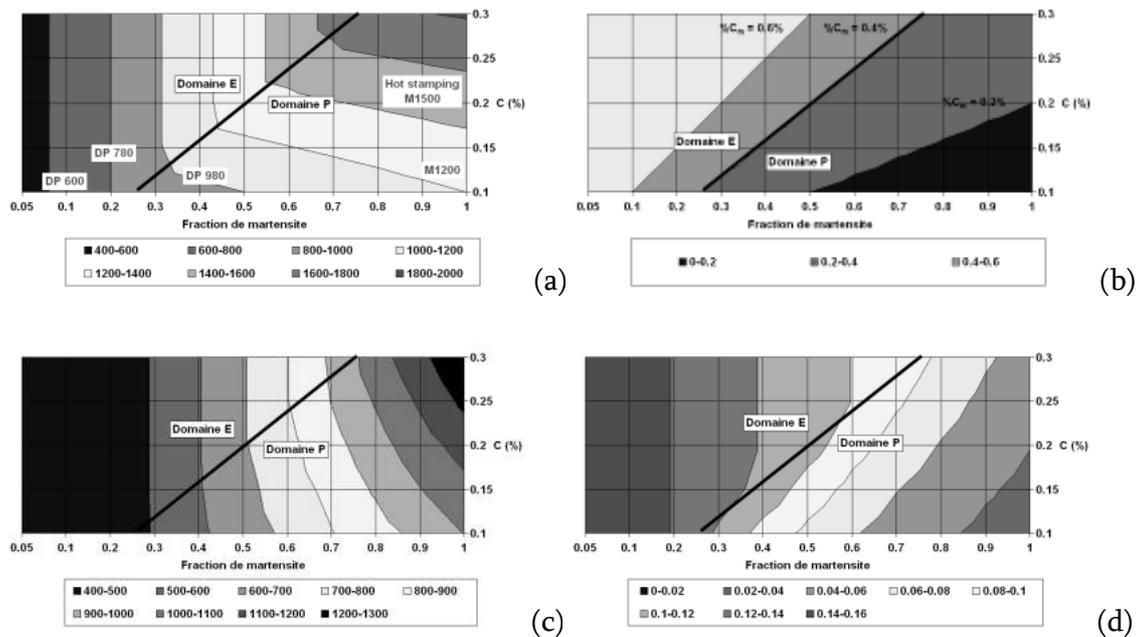

Figure 86 : Cartographie dans le plan fraction de martensite et teneur nominale en carbone des propriétés de traction modélisées (a) résistance mécanique, (b) teneur local en carbone dans la martensite, (c) limite d'élasticité et (d) allongement réparti. La transition entre les domaines E et P de comportement des aciers DP est repérée par une ligne continue noire. Sur la figure (a) sont repérés des positionnements typiques d'aciers DP ou martensitiques industriels.

C'est ici l'occasion de clarifier une idée reçue chez les « métallurgistes » qui consiste à considérer que la résistance mécanique d'un acier DP est systématiquement proportionnelle à la fraction de martensite sur toute la gamme. Cette idée est bien entendu fausse comme le montre les figures précédentes mais était soutenue par la représentation graphique des données expérimentales de Davies (reprise sur la Figure 87) [DAVIES 1978]. En fait, cette figure très populaire est biaisée par l'accumulation de données (des aciers avec des teneurs en carbone nominales différentes) et un parti pris subjectif. La figure (b) reprend en fait les résultats pour un seul acier de cette base de données, qui montrent bien une transition vers 40% de martensite et un comportement saturant.

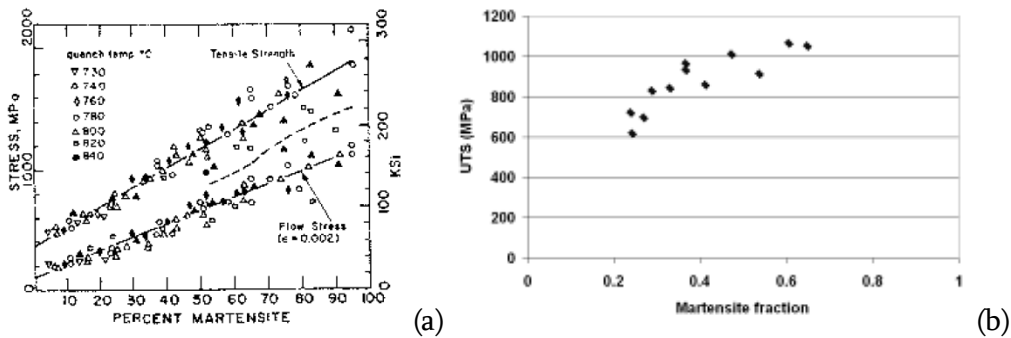

Figure 87 : (a) Evolution de la résistance mécanique et la limite d'élasticité de différentes aciers DP avec des teneurs nominales en carbone différentes, d'après [DAVIES 1978] (b) Extraction des données de résistance mécanique pour un seul acier, montrant une transition de comportement vers 40 % de martensite.

### 3.3.4.3. *Effet de la taille de grain ferritique*

Le pendant de cette analyse en termes de fractions de martensite concerne la sensibilité du modèle aux tailles de grain ferritique. La Figure 88 montre des exemples de performance du modèle pour capter ces effets, confirmant la pertinence de l'approche.

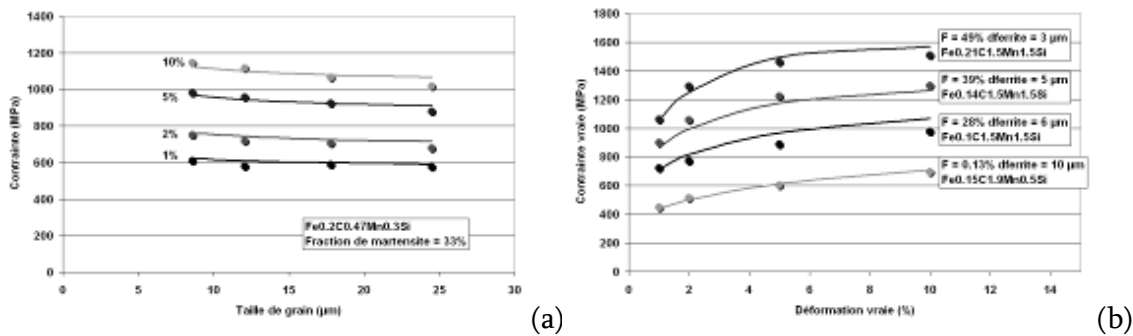

Figure 88 : Comparaison entre résultats expérimentaux (points) et modèles (lignes continues) (a) Effet de la taille de grains sur la contrainte d'écoulement pour différentes valeurs de déformation (acier Fe0.2C0.47Mn0.3Si, données de [JIANG 1992]) (b) Courbes de traction de différents aciers avec des fractions de martensite et tailles de grain ferritique différents (données de [MATLOCK 2005]).

*3.3.4.4. Vers l'utilisation en ingénierie des microstructures*

Ces différents éléments de discussion sur le rôle respectif des différentes caractéristiques de la microstructure des aciers DP nous amènent à considérer l'un des intérêts industriels majeurs de cette approche « générique », c'est-à-dire d'être un puissant outil de « d'alloy design » ou d'ingénierie des microstructures. Il permet d'identifier par exemple des solutions microstructurales non triviales pour répondre à un cahier des charges donné.

Si l'on se fixe une double contrainte technique sur deux paramètres de traction (Rm > 980 & Ag > 8 % par exemple), la Figure 89(c) montre les microstructures des aciers pouvant correspondre à ces deux critères simultanément (décrits indépendamment sur les figures (a) et (b)). Cette figure conduit à la conclusion que si l'acier doit présenter, de plus, de faibles teneurs en carbone pour des raisons de soudabilité par exemple, alors la solution métallurgique est unique (mélange de 50/50 de Ferrite et Martensite). Cet exercice d'application volontairement simpliste doit être adapté en fonction des faisabilités industrielles (taille de grains ferritiques, présence de bainite) pour être pleinement opérationnel.

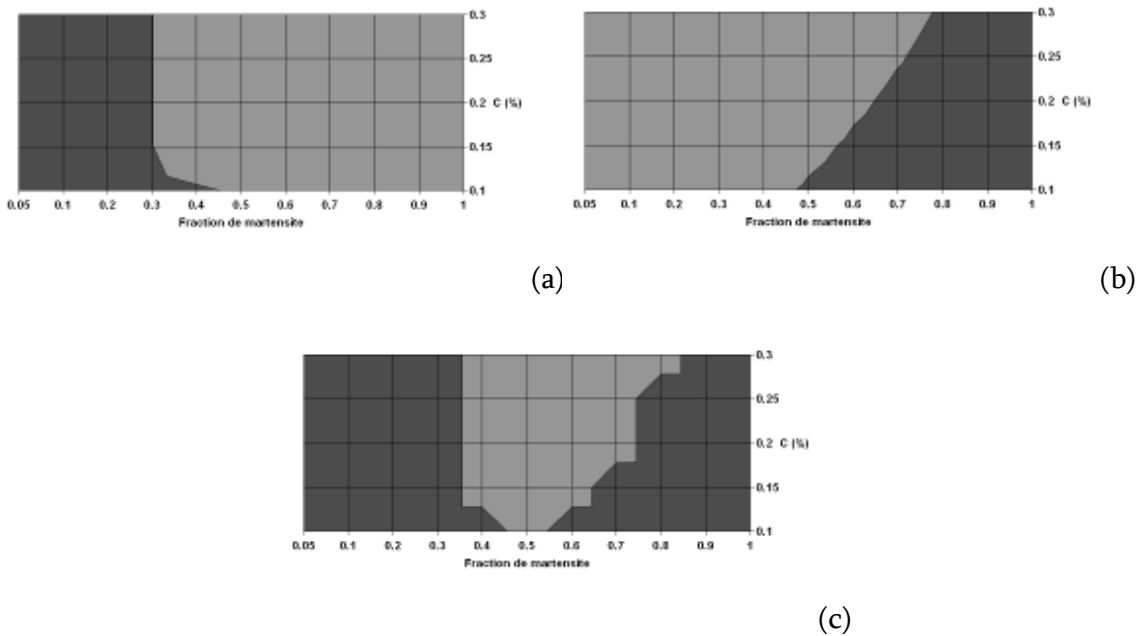

Figure 89 : Cartographie dans le plan fraction de martensite et teneur nominale en carbone des propriétés de traction (a) résistance mécanique et (b) allongement réparti. Les zones vertes correspondent aux aciers vérifiant le cahier des charges (Rm > 980 & Ag > 8%).(c) superposition des zones répondant aux deux contraintes techniques simultanément.

### 3.3.5. Limites et perspectives

Le développement présenté est capable de comprendre et de quantifier les variations non triviales des propriétés de traction des aciers Ferrito-Martensitiques, en fonction des paramètres microstructuraux pertinents d'après la littérature (fraction de martensite, taille de grains, compositions). Il repose sur une compréhension des différents mécanismes d'écrouissage de chacune des phases constituantes et de leurs interactions (gradient de contraintes et de déformations entre les phases). Cette démarche permet une excellente description des contraintes internes au cours de la déformation et conduit à un faible nombre de paramètres d'ajustement (3 pour les biphasés). Contrairement aux approches monophasées, le comportement plastique de la martensite joue un rôle clef et permet des démarches « d'alloy design » et d'études de sensibilité sur toute la gamme de fraction de martensite.

Une différence importante est apparue toutefois entre les modèles monophasé et biphasé concernant leur sensibilité à la taille d'ilot martensitique. En suivant la démarche d'ajustement proposée sur des données de la littérature disparates, il semble difficile d'identifier une tendance claire reliant par exemple le paramètre $L_\alpha$ et $d_m$. Une meilleure description de cette sensibilité dans la nouvelle approche semble un axe de recherche intéressant, avec une difficulté que nous évoquerons dans la section suivante concernant la définition de ce paramètre une fois la phase martensitique percolée.

La Figure 90 illustre une autre limite du modèle. Elle représente les écarts en contrainte entre les prédictions du modèle et l'expérience pour deux niveaux de déformation. Bien entendu, aux fractions extrêmes (0% et 100% de martensite), les prédictions présentent un accord excellent. Par contre, aux fractions intermédiaires, le modèle surestime statistiquement les taux d'écrouissage. Une explication possible est liée à la dispersion des teneurs en carbone des ilots de martensite observés expérimentalement [GARCIA 2007]. Ces travaux expérimentaux posent la question de l'utilisation comme donnée d'entrée du modèle de martensite. Celui-ci est basé sur une distribution de propriétés supposée de contraintes d'écoulement sur des aciers martensitiques massifs. D'un point de vue métallurgique, il n'est absolument pas garanti que cette distribution soit aussi représentative de celle des aciers DP. Cette question centrale est donc liée à l'imbrication de deux modèles de transition d'échelle. Les travaux engagés sur l'origine du spectre de contraintes et des échelles en jeu sur le comportement de la martensite sont un préalable important avant d'envisager une adaptation du modèle.

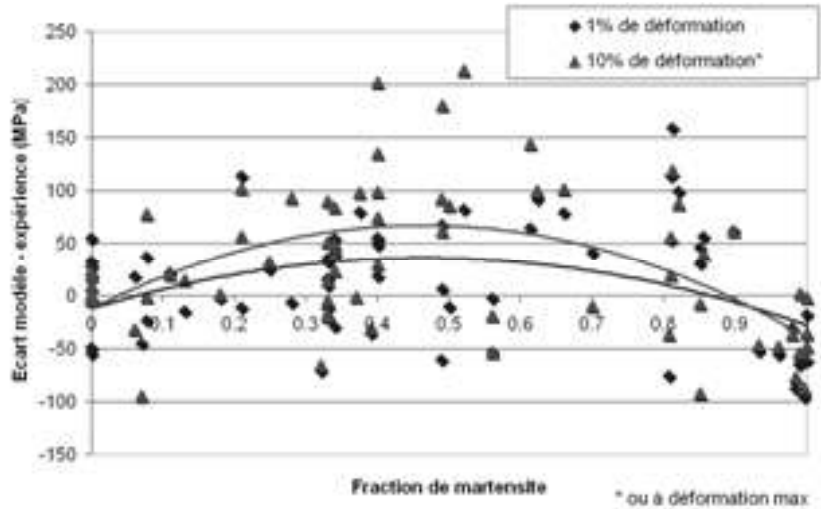

Figure 90 : Ecart entre modèle et expérience en contrainte pour deux niveaux de déformation (ou déformation maximum) en fonction de la fraction de martensite pour les différents aciers de la base de données.

A l'instar de tous les modèles à champs moyens, une limitation du modèle importante concerne ses capacités à prédire les processus d'endommagement. Ces processus sont de différentes natures dans ces aciers (germinations de cavités aux interfaces ferrite/martensite, ruptures fragile et ductile de la martensite …) et très locaux. Dans ces structures composites, ils interviennent très tôt au cours de la déformation à cause des forts gradients de contraintes et de déformations aux interfaces ferrite-martensite et conduisent à une ruine rapide. Ce comportement spécifique est caractéristique des aciers DP et limite fortement leur application industrielle.

Un modèle de décohésion des interfaces Ferrite/Martensite a été proposé récemment par Landron *et al.* [LANDRON 2010]. Ce critère inspiré par le critère d'Argon tient compte des contraintes internes (gradients de contraintes aux interfaces), force motrices pour la décohésion :

$$\chi = \sigma_{eq}\left(1 + T\left(\frac{\sigma_{eq}}{\sigma_{eq} - X}\right)\right) > \sigma_C \qquad (86)$$

Avec $\sigma_{eq}$ la contrainte d'écoulement équivalent, T la triaxialité du chargement extérieur et X la contribution cinématique au durcissement. Cette fonction a été évaluée par exemple sur la Figure 91 en fonction de la teneur en carbone et de la fraction de martensite (T = 2). La contrainte d'écoulement est prise égale à Rm et X à sa valeur à saturation (Xmax).

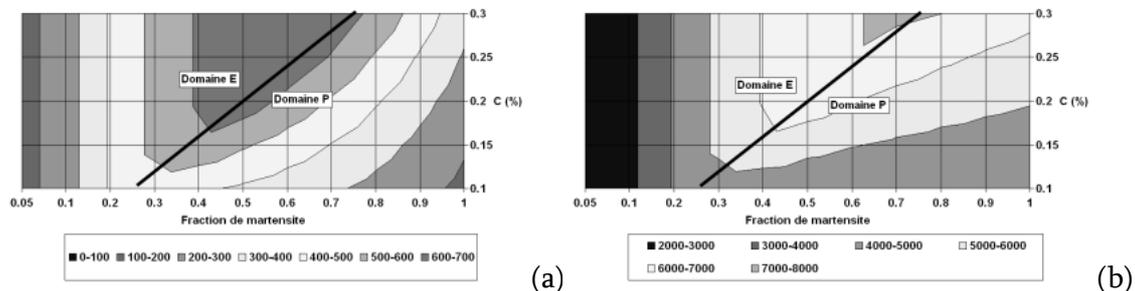

Figure 91: Cartographie dans le plan fraction de martensite et teneur nominale en carbone des propriétés modélisés d'endommagement décrites par l'équation (87) (a) Xmax, la valeur de l'écrouissage cinématique à saturation et (b) χ calculé pour une triaxialité macroscopique de T= 2. Position de la transition entre domaines E et P reprise de la Figure 86.

Ce modèle ouvre des perspectives intéressantes en décrivant l'impact d'un mécanisme d'endommagement (décohésion entre ferrite et martensite), mais ne peut être appliqué de façon générique car un seul mécanisme est décrit. De plus, ce comportement étant contrôlé par la triaxialité des contraintes, la prise en compte de la morphologie des phases semblerait indispensable.

En parallèle de ce développement d'un modèle à champs moyens, j'ai engagé une action de recherche pour mieux appréhender les effets de la dispersion et de la morphologie de la martensite sur le comportement des aciers DP grâce à des modélisations micromécaniques par Eléments Finis (EF) de Volume Elémentaire Représentatif (VER) de microstructures.

## 3.4. Modélisation par EF de VER : Effet de la morphologie de la martensite

Dans cette dernière partie, nous montrerons un exemple concret d'application sur des microstructures Dual-Phase des outils que j'ai développés pour réaliser des calculs par EF (Eléments Finis) sur VER (Volume élémentaire Représentatif) de microstructures. L'objectif ultime de ce projet est de mieux comprendre la compétition entre les mécanismes d'endommagement dans ces aciers et optimiser leur microstructures en conséquence, non seulement en terme de fraction et tailles mais aussi en terme de morphologie. L'exemple détaillé dans la suite concerne les effets de la structure en bandes sur les propriétés d'endommagement. Ces travaux ont été réalisés principalement dans le cadre de la thèse de B. Krebs au LETAM.

### 3.4.1. Problématique : Effet de la structure en bande

A l'instar de [LI 1990_2][LI 1990_1][PAUL 2012][CHOI 2009][KUMAR 2007][SODJIT 2012][PAUL 2013], nous avons modélisé le comportement d'aciers DP par la méthode des EF en sollicitant numériquement des VER de leur microstructure. La démarche de simulation n'est pas nouvelle en soi, mais nous a permis d'investiguer des questions originales :
- l'effet de la structuration dite « en bande » des DP sur les propriétés d'endommagement des aciers DP
- Effet du passage 3D / 2D sur des structures numérisées (issues de données expérimentales)

La structure dite « en bandes » des aciers DP se caractérise par une distribution spatiale anisotrope et une morphologie très allongée des ilots de martensite. Elle est due à la présence de microségrégations de manganèse issue du processus de solidification (en coulée continue ou en lingots) mais n'apparaît que dans certaines conditions de traitement thermique (vitesse de refroidissement et de taille grain austénitique) [KREBS 2009][VIARDIN 2008]. La figure suivante montre par exemple deux micrographies d'aciers DP obtenus à partir d'une même composition (Fe0.15C1.5Mn) mais ayant subi ou non un traitement thermique d'homogénéisation préalable.

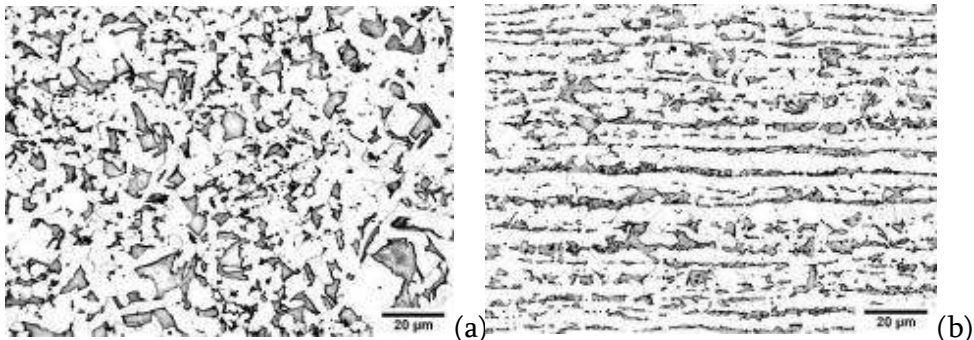

Figure 92 : Micrographie optique après attaque de deux microstructures DP obtenues à partir d'une même composition (Fe0.15C1.5Mn) (a) structures isotropes après traitement thermique d'homogénéisation des microségrégations de manganèse (b) structure dite « en bande » - Les deux aciers présentent la même fraction de martensite (environ 30%) [KREBS 2009].

Ces structures en bandes sont très communes dans les produits industriels et connues dans la littérature pour avoir un effet néfaste sur les propriétés d'endommagement, i.e. les mécanismes de rupture et de cavitation à l'échelle de la microstructure conduisant à la ruine macroscopique du matériau. Toutefois, la corrélation n'est pas systématique car la rupture des aciers DP résulte de la compétition de plusieurs mécanismes [PUSHKAREVA 2009], illustrés dans le cas d'un acier DP 600 sur la Figure 93 (reprise d'Avramovic *et al.* [AVRAMOVIC 2009]) :
- Germination de cavités sur des inclusions,
- Rupture fragile ou ductile de la martensite,
- Germination de cavités par décohésion des interfaces ferrite-martensite (ou à proximité des interfaces dans la ferrite),
- Puis, coalescence des cavités par cisaillement.

La Figure 93(g) montre un exemple de processus d'endommagement d'un acier DP présentant une forte structure en bande. On identifie sur cette figure plusieurs mécanismes opérant simultanément (rupture fragile d'ilots, décohésion et coalescence par cisaillement). Les bandes de martensite seront donc plus ou moins des sites privilégiés de rupture en fonction de leur morphologie (continuité des bandes par exemple), de leur dureté et ténacité (teneur en carbone) ou de leur facteur de forme (triaxialité du tenseur des contraintes en pointe).

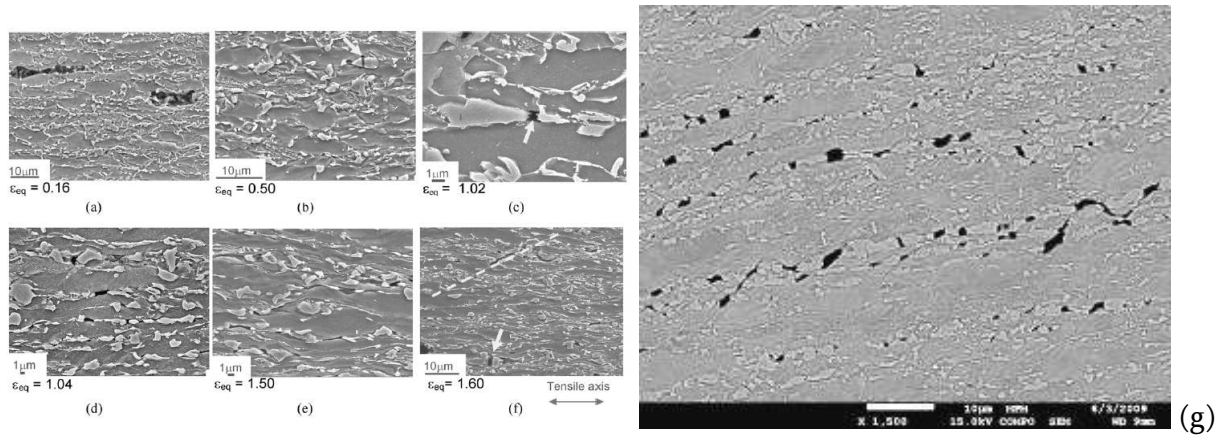

Figure 93 : Micrographies MEB après attaque montrant différents processus d'endommagement dans un acier Dual-Phase (a) germination sur des inclusions (b), (c) rupture fragile de la martensite, (d) et (e) germination des cavités par décohésion des interfaces ferrite-martensite, (f) coalescence des cavités par cisaillement d'après [AVRAMOVIC 2009], (g) Micrographies MEB après attaque d'un acier DP montrant différents mécanismes d'endommagement en relation avec une forte structure en bande (données AM).

Notre objectif était donc d'étudier numériquement la susceptibilité à l'endommagement de différentes microstructures selon leur topologie (en bandes ou visuellement isotropes). Nous nous sommes intéressés dans cette discussion à la différence entre des approches 3D et 2D sur des structures numérisées (à iso-fraction de martensite).

### 3.4.2. Caractéristiques principales des simulations

Les simulations ont été réalisées sur des volumes élémentaires reconstruits à partir d'observations microstructurales (microstructures numérisées). Ce type d'approche est maintenant assez couramment utilisé en 2D [CHOI 1999][KUMAR 2007][SODJIT 2012][PAUL 2013] (analyse d'une seule coupe métallographique) mais à notre connaissance, ce type de calcul n'a jamais été réalisé en 3D. Dans le cadre de cette étude, les microstructures ont été reconstruites grâce à la technique des coupes sériées (repositionnées et calées en épaisseur grâce à des indentations), comme le montre le schéma de principe sur la Figure 94:

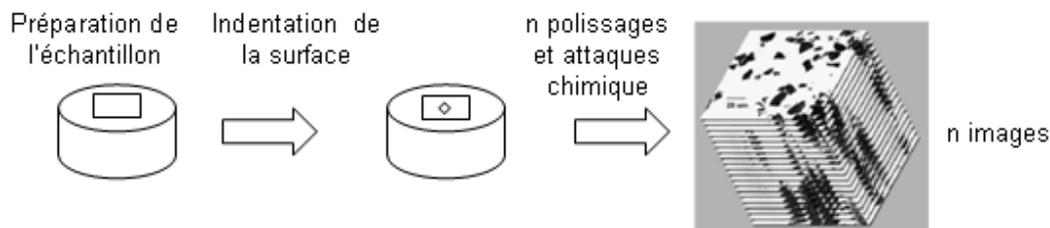

Figure 94 : Schéma de principe de la technique de coupes sériées utilisée pour reconstruire les structures DP en 3D [KREBS 2009]

Dans ce domaine, on attend donc des progrès importants des techniques d'observation EBSD couplées à un FIB (Field Ion Beam), qui permettront de reconstruire en 3D non seulement la structure martensitique (en utilisant l'indice de qualité pour distinguer ferrite et martensite) mais aussi la structure et l'orientation cristallographique des grains ferritiques.

La Figure 95 montre les coupes 2D obtenues sur les deux aciers DP étudiés, présentant des fractions volumiques de martensite similaire (30% environ) mais une distribution topologique des « ilots » différente (mêmes aciers que la Figure 92). Ces coupes métallographiques ont été segmentées pour faire apparaître la martensite en blanc et la ferrite en noir. Les reconstructions en 3D des réseaux de martensite de ces deux aciers sont représentées sur les figures (c) et (d).

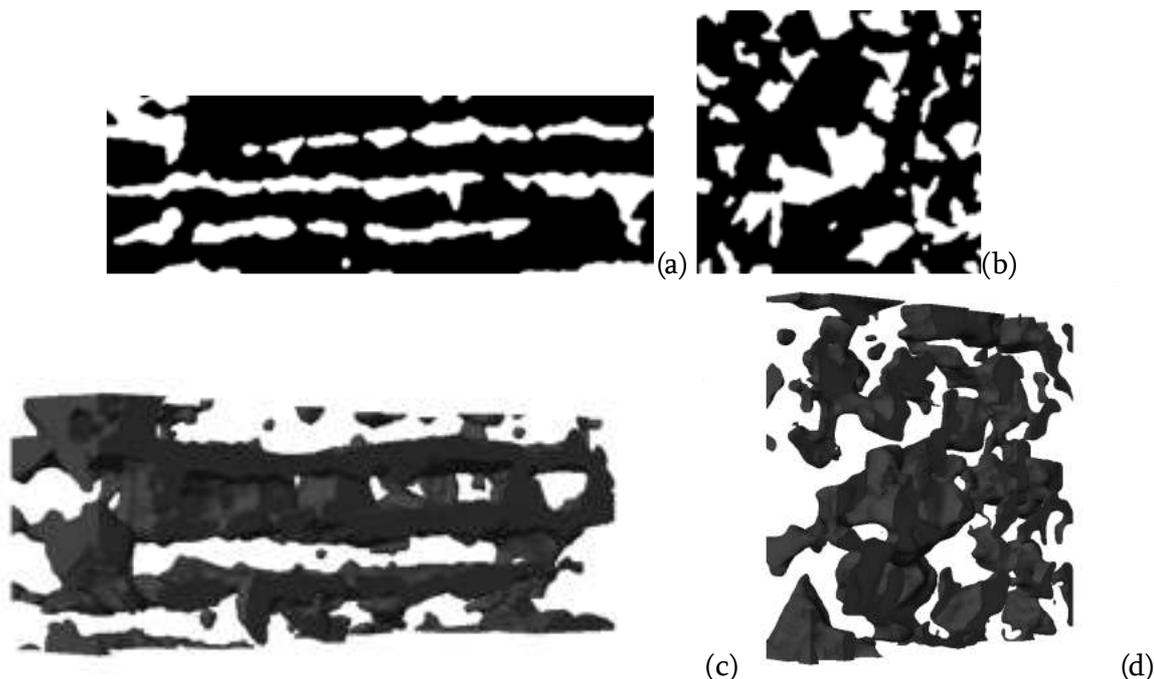

Figure 95 : (a) et (b) coupes 2D binarisées des structures DP en bandes et isotropes étudiées, la ferrite apparaît en noir et la martensite en blanc ; (c) et (d) reconstruction 3D des mêmes aciers DP – seule la martensite a été représentée en rouge pour quantifier la percolation de cette phase [KREBS 2009]. Les volumes reconstruits en (c) et (d) sont de 40 x 13 x 7.7 µm³ et 33 x 31 x 6.5 µm³ respectivement.

Dans les deux cas étudiés, la martensite forme de façon très surprenante une structure connexe. Ce résultat confirmé par des mesures de nombres de connexités négatifs malgré les observations en 2D qui laissent penser que tous les « ilots » sont disjoints. Dans le cas de la structure en bandes, il existe de claires jonctions entre les bandes. Certains grains de ferrite sont même enchâssés dans une matrice martensitique. A contrario, dans la structure homogène, les ilots semblent connectés un par un. Cette dernière configuration illustre bien la difficulté de définir une taille d'ilot représentative $d_m$, dans la mesure où le réseau de martensite percole même aux faibles fractions.

Les microstructures DP sont le fruit de processus de transformation de phase complexes, structurés et multi-échelle. Elles ne sont pas réductibles à des microstructures simples, représentables en 3D par des distributions aléatoires inspirées des coupes 2D. Cette observation justifie donc pleinement l'intérêt des simulations en 3D sur des microstructures réelles malgré les difficultés techniques de numérisation.

On pourra toutefois regretter que les volumes des structures reconstruites soient comparativement faibles et qu'ils ne constituent pas des volumes représentatifs stricto-sensu. Pour palier à ce manque de représentativité, des calculs ont aussi été conduits sur des microstructures 2D virtuelles reproduisant des volumes analysés plus importants. Ces microstructures ont été « simulées » numériquement pour reproduire des fractions de martensite de l'ordre de 30% mais des indices de structures en bandes différents (croissant sur la Figure 96).

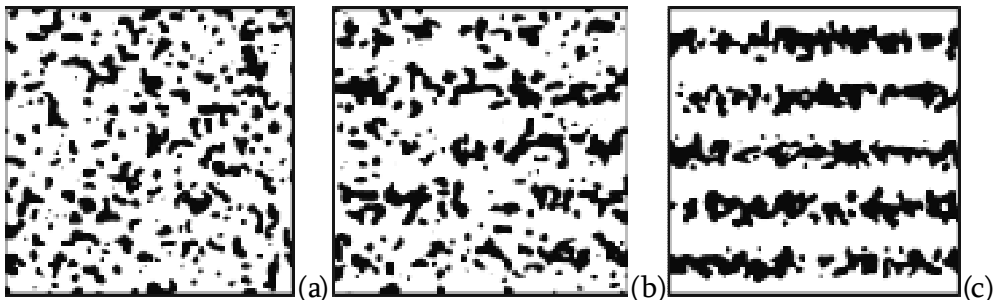

Figure 96 : Microstructures DP virtuelles en 2D reproduisant des volumes analysés plus grands que les numérisations 3D (a) structures isotropes, (b) faiblement en bandes et (c) fortement en bandes [KREBS 2009].

Le maillage des microstructures en 2D et 3D a été réalisé après segmentation sur un logiciel commercial SIMPLEWARE, détourné de son domaine d'application originel, l'analyse d'image en tomographie médicale et la micromécanique biomédicale. Le résultat de la procédure est un maillage libre 3D adaptatif (maillage plus grossier loin des interfaces) avec des éléments quadratiques et tétraédriques comme le montre les deux exemples de la Figure 97.

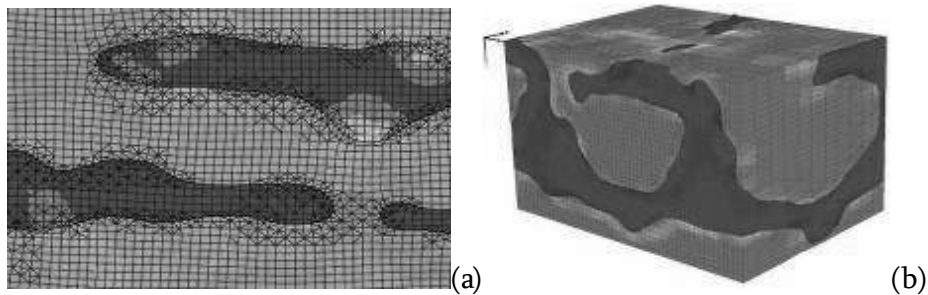

**Figure 97 : Exemple de maillage libre adaptatif réalisé sous SIMPLEWARE (a) en 2D (après calcul ABAQUS) et (b) en 3D. La taille caractéristique des éléments est plus petite que les rayons de courbures des structures martensitiques.**

La taille caractéristique des éléments est bien inférieure à celle des rayons de courbures des structures martensitiques. Cette finesse du maillage permet d'investiguer le comportement à l'intérieur de cette phase.

Dans cette première étape, les deux phases du composite sont considérées comme des milieux continus malgré les échelles considérés (corps de Von Mises élasto-plastiques). On néglige de ce fait la plasticité intra et polycristalline, ainsi que l'anisotropie élastique des phases. Les grains de ferrite ne sont donc pas définis. Les courbes de comportement des deux constituants sont représentées sur la Figure 98:

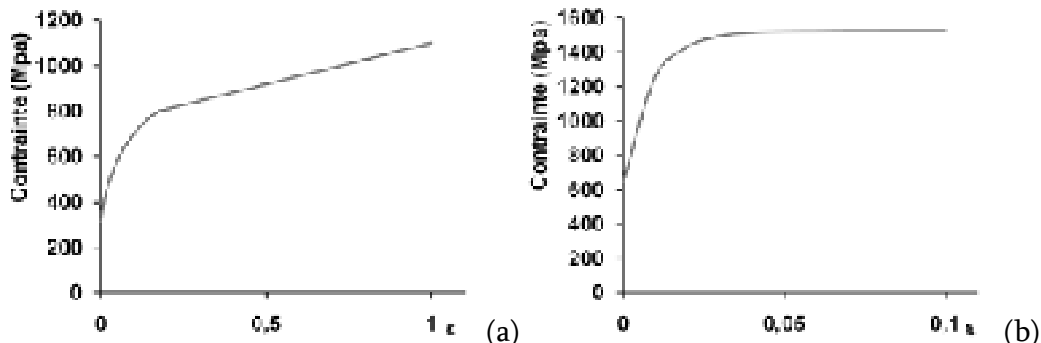

**Figure 98 : Lois de comportement (a) de la ferrite et (b) de la martensite utilisé pour les calculs EF [KREBS 2009].**

Ces courbes ont été déduites de calculs préalables avec le modèle biphasé afin d'introduire un sur-durcissement isotrope dans la ferrite dû aux DGN et extrapolées linéairement aux grandes déformations pour éviter des instabilités numériques lors de la résolution. Les volumes élémentaires ont été soumis à des chargements de type déformation plane, avec des conditions aux limites anti-périodiques. Les calculs ont été résolus implicitement en déformation imposée sous ABAQUS STANDARD.

### 3.4.3. Résultats

*3.4.3.1.  Incidence de la topologie sur les propriétés macroscopiques*

La Figure 99 montre les courbes de comportement contrainte – déformation vraie déduites de la sollicitation des différents VER décrits ci-dessus ; les 3 microstructures virtuelles 2D (appelées modèle homogène, faiblement et fortement en bandes) et les 2 microstructures numérisées 3D (appelées réelle homogène ou ségrégée). Les déformations sont représentées de 0 à 100% de la déformation imposée (10% d'allongement).

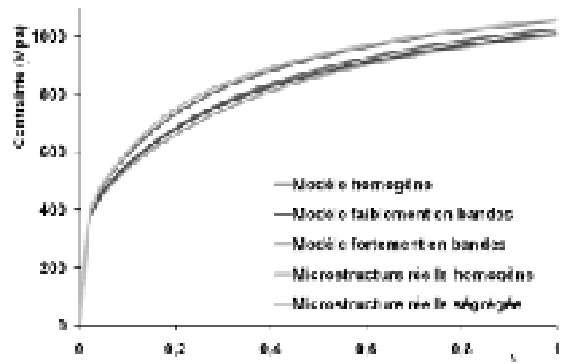

Figure 99 : Courbes de comportement déduites des calculs sur VER (3 configurations 2D virtuelles et 2 configurations 3D numérisées [KREBS 2009].

Malgré des fractions de martensite identiques, les microstructures en bandes présentent des taux d'écrouissages initiaux légèrement supérieurs aux microstructures homogènes, en 2D ou 3D. Ce résultat numérique est très cohérent par rapport aux résultats expérimentaux de la Figure 63 page 98. L'effet de la topologie de la structure martensitique reste faible en intensité par rapport aux effets de fraction même dans le cas d'une structuration très marquée, ce qui justifie pleinement l'approche développée au chapitre 3.3 page 107.

*3.4.3.2.  Hétérogénéités de la déformation plastique*

La Figure 100 montre les résultats des simulations pour 2 VER 2D virtuels (homogène et fortement en bande) après 10% d'allongement ; les contraintes équivalentes au sens de Von Mises dans la ferrite et la martensite ainsi que la déformation plastique équivalente dans les deux phases.

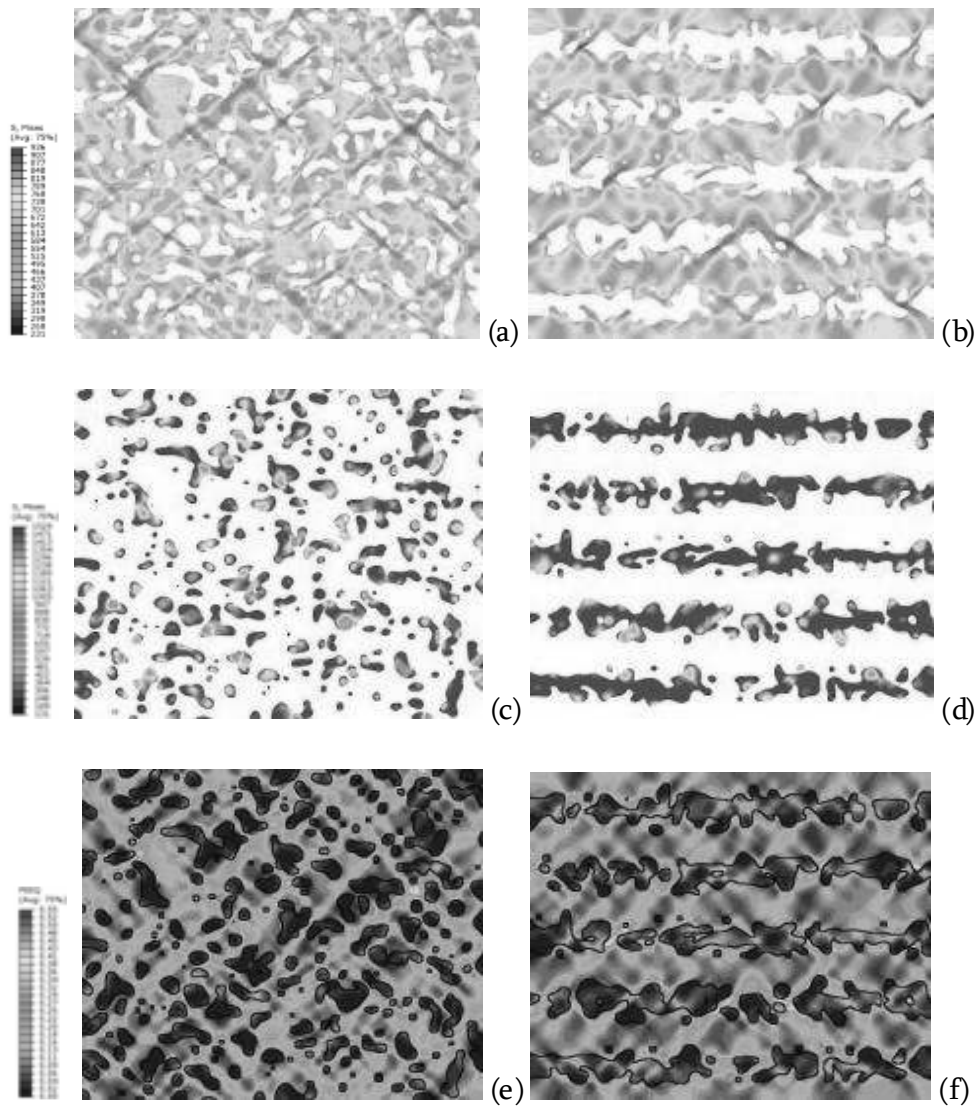

Figure 100: Résultats des calculs EF sur après 10% d'allongement sur structures 2D virtuelles isotropes et fortement en bande, (a) et (b) contraintes équivalentes de Von Mises dans la ferrite, (c) et (d) dans la martensite, (e) et (f) déformation plastique équivalente dans les deux phases [KREBS 2009].

Le comportement plastique de la ferrite est hautement hétérogène. Des bandes de déformation plastiques intenses apparaissent rapidement dans les deux types de VER (niveaux de déformation jusqu'à 35% localement). Ces bandes de déformation ont souvent été observées dans toutes les simulations de VER réalistes d'aciers DP et correspondent à des cisaillements très intenses comme le précise la Figure 101.

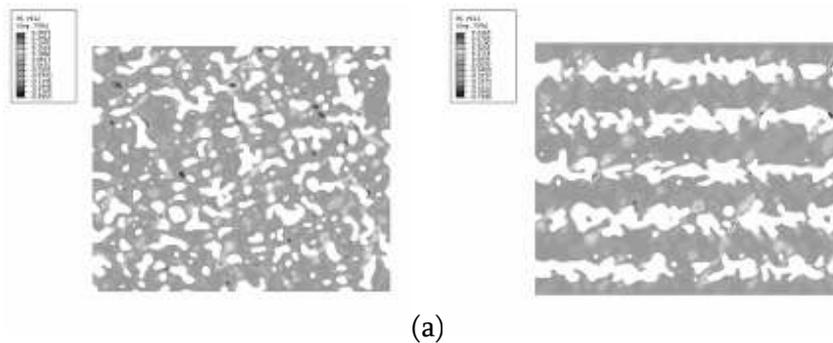

(a)                                         (b)

Figure 101 : Résultats des calculs EF après 10% d'allongement sur structures 2D virtuelles (a) isotropes et (b) fortement en bande – visualisation du champ de déformation plastique en cisaillement (PE12).

Le comportement de la martensite semble bien plus homogène comme attendu dans le cas d'une phase en inclusion (inclusion type Eshelby). Cependant statistiquement, la Figure 102 montre que la structuration en bandes conduit à une structure martensitique en moyenne plus chargée en terme de contraintes équivalentes. Ceci qui cohérent avec une contrainte d'écoulement du composite plus importante (cf. Figure 99). La distribution en termes de déformation plastique confirme aussi que plus d'ilots martensitiques seront fortement déformés.

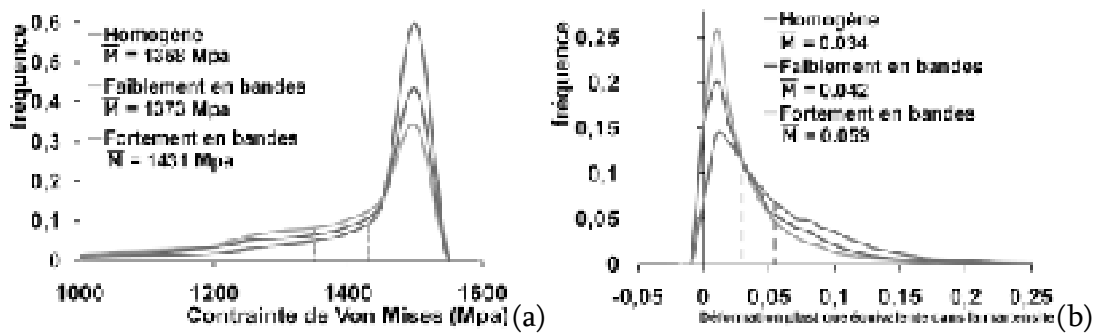

Figure 102 : Distribution en fréquence (a) des contraintes équivalentes de Von Mises et (b) déformation plastique équivalente dans la martensite dans les 3 configurations 2D virtuelles [KREBS 2009].

Un comportement strictement équivalent a été retrouvé lors des calculs sur VER 3D numérisés, à la fois en terme de déformation plastique de la ferrite ou de chargement de la structure martensitique percolée. Les bandes de déformation n'apparaissent pas stricto sensu comme des plans dans l'espace mais sont plus proches de ligaments. La Figure 103(b) et la Figure 103(c) sont deux coupes sagittales qui montrent l'extinction de certaines bandes dans l'épaisseur. Des calculs sur des volumes plus importants seraient nécessaires pour confirmer cette tendance. Dans la suite, pour faciliter les représentations, seul le cas des VER 2D sera étudié en détail.

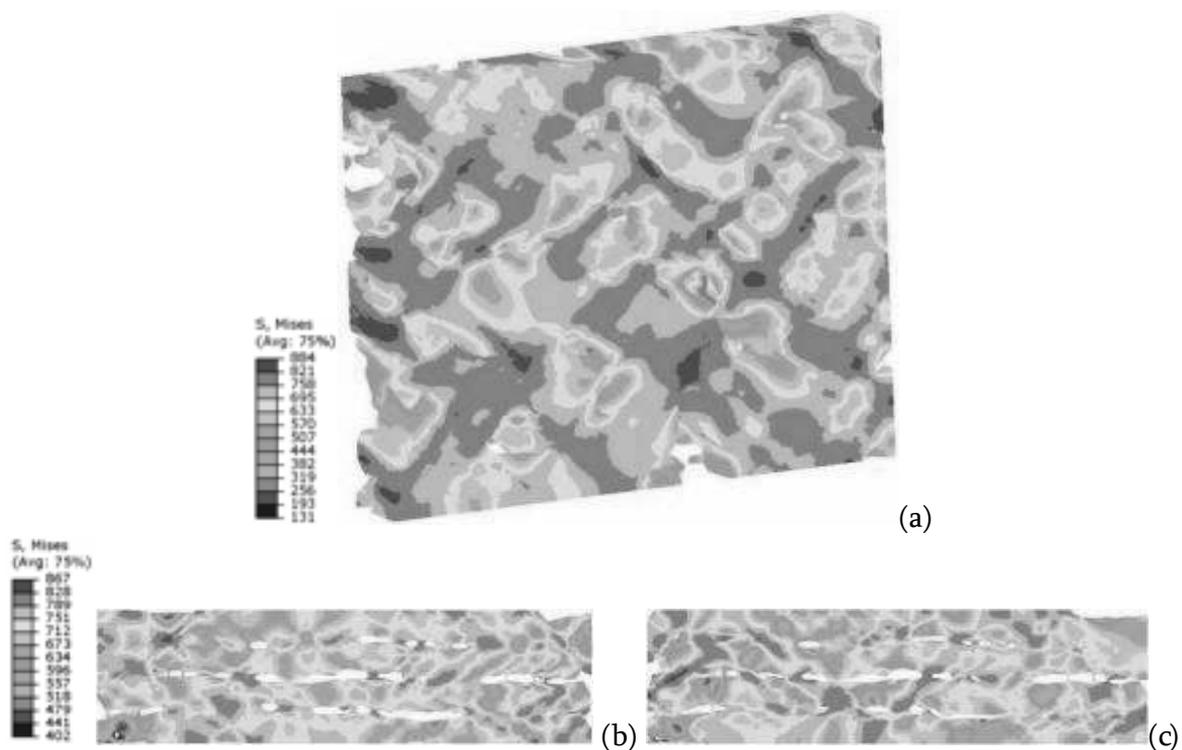

Figure 103 : Résultats des calculs EF après 10% d'allongement sur structures 3D numérisées (a) isotropes et (b) et (c) en bandes (2 plans de coupes) [KREBS 2009].

Ces bandes de déformation ne sont pas un « artefact » de simulation numérique et ont été mises en évidence récemment grâce au progrès de la corrélation d'images numériques *in situ* en MEB [TASAN 2010][GHADBEIGI 2010]. La Figure 104 montre un bel exemple de mesure de champs de déformation sur un acier DP 1000 après différentes déformations macroscopiques. Des bandes de déformation dont l'échelle est bien supérieure à la taille caractéristique des structures martensitique sont observées comme dans le cas des simulations numériques. Par contre cette structuration semble apparaitre plus tardivement (après 20% d'allongement), ce qui est peut-être un indice que les gradients de déformation aux interfaces, non pris en compte dans ce calcul par EF, peuvent jouer un rôle important dans la stabilisation des écoulements entre particules.

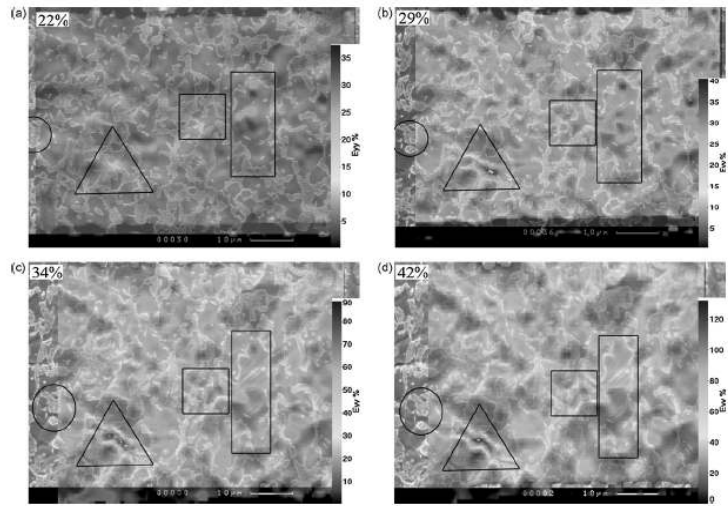

**Figure 104 : Mesures de champs de déformation plastique par une technique de corrélation d'images en MEB *in situ* sur un acier DP 1000 d'après [GHADBEIGI 2010].**

### 3.4.3.3. *Vers les mécanismes d'endommagement*

Dans la littérature, on rapporte trois mécanismes principaux de germination de l'endommagement primaire dans ces structures :
- l'endommagement ductile de la martensite
- la rupture fragile de la martensite
- la décohésion entre ferrite et martensite

Ces trois mécanismes sont associés à des critères d'apparition de différentes natures et relevant de valeurs critiques très variables selon l'état métallurgique. Le premier mécanisme peut être associé à un critère de déformation plastique maximum à rupture ; le second à des contraintes principales de traction critiques (critère type Griffith) et le troisième à des contraintes critiques normales aux d'interfaces (critère type Argon) [PUSHKAREVA 2009].

Nous avons pu vérifier sur la Figure 102(b) que la structure en bandes conduisait à des déformations plastiques plus élevées dans certains ilots de martensite (en considérant en particulier la queue de distribution). La Figure 105 montre qu'elle est aussi néfaste en termes de contraintes critiques maximum dans la martensite (cf. Figure 107(a)). Elle tend donc à favoriser la rupture fragile des ilots de martensite, mécanisme souvent observé expérimentalement. Il est assez difficile numériquement d'évaluer son impact sur le troisième critère. Par contre, nous avons aussi mis en évidence que la structuration en bande allait avoir un rôle défavorable sur les mécanismes de croissance des cavités. La Figure 106 montre qu'elle contribue à augmenter les valeurs de triaxialité du tenseur des contraintes au niveau des pointes des ilots constitutifs des bandes (cf. Figure 107(b)). Ces zones correspondent en fait aux intersections des bandes de cisaillements intenses discutées précédemment.

La topologie de la martensite contrôle donc les mécanismes de germination et de croissance des cavités dans les processus d'endommagement. Toutefois, nos résultats ne permettent pas de quantifier la compétition entre ces mécanismes et leur prévalence, dans la mesure où les critères de germination ne sont pas bien documentés dans la littérature et que ces mécanismes interagissent certainement entre eux (relaxation ou concentrations de contraintes par exemple). Un modèle micromécanique complet à cette échelle pourrait permettre de progresser significativement sur ces questions et envisager des solutions métallurgiques optimisées pour réduire la sensibilité à l'endommagent de ces structures (revenu, mécanismes de transformation de phase, ..).

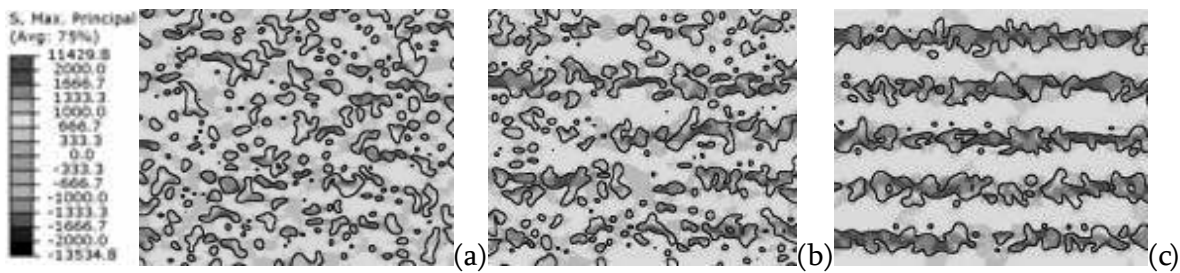

**Figure 105 : Résultats des calculs EF après 10% d'allongement sur structures 2D virtuelles (a) isotropes et (b) faiblement en bande et (c) fortement en bande – visualisation des champs de norme de la contrainte principale maximum** [KREBS 2009].

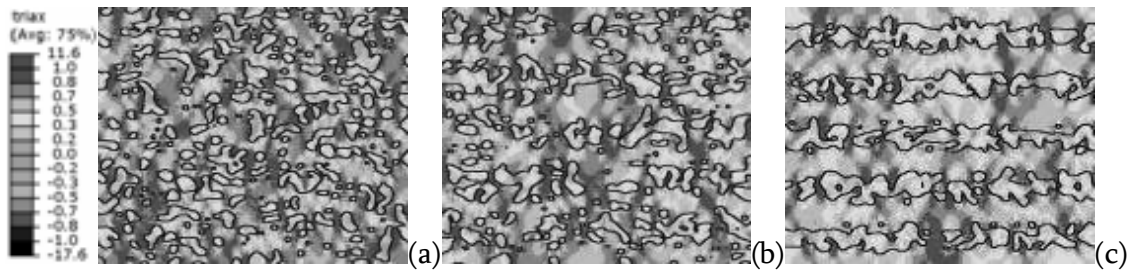

**Figure 106 : Résultats des calculs EF après 10% d'allongement sur structures 2D virtuelles (a) isotropes et (b) faiblement en bande et (c) fortement en bande – visualisation des champs scalaires de triaxialité du tenseur des contraintes** [KREBS 2009].

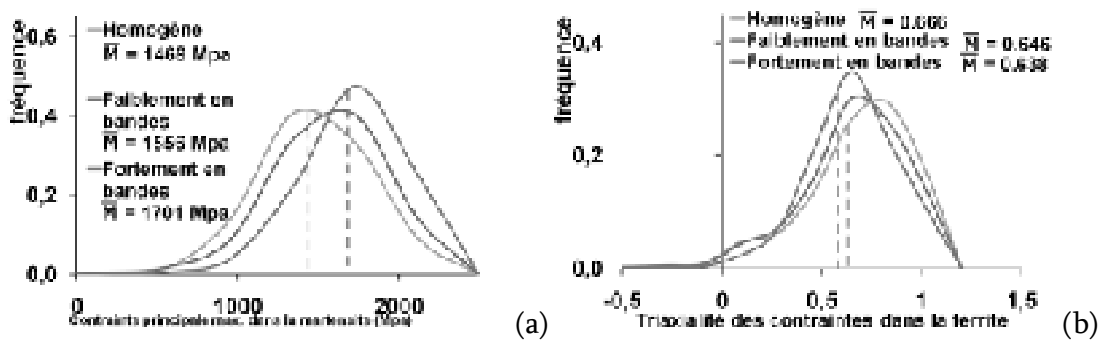

**Figure 107 : Distribution en fréquence (a) de la contrainte principale maximum dans la martensite et (b) de la triaxialité du tenseur des contraintes dans la ferrite pour les 3 configurations 2D virtuelles** [KREBS 2009].

### 3.4.4. Perspectives de cette démarche

Ces travaux de modélisation s'inscrivent en fait dans une démarche plus générale évoquée en introduction. Elle vise à développer une chaine de simulation numérique du comportement mécanique et d'endommagement des aciers DP.

Les objectifs scientifiques de cette démarche sont de quantifier et de mieux comprendre les mécanismes d'endommagement de ces structures DP et à terme leurs actionneurs métallurgiques. Ceci nécessite non seulement une connaissance fine et juste des états de contraintes et déformation des différentes phases au cours de la déformation mais aussi de pouvoir gérer la compétition entre les mécanismes d'endommagement (relaxation).

Pour répondre à ce besoin, les calculs par EF simplifiés décrits ci-dessus sont donc doublement insuffisants. Les champs locaux sont calculés sans tenir compte des mécanismes de plasticité cristalline et la susceptibilité à certains processus d'endommagement calculée a posteriori. Nous travaillons donc actuellement sur deux axes principaux :
- le développement d'une VMAT ABAQUS en collaboration avec le CEIT (D. Gonzales, JM Esnaola, J. Gil Sevillano) intégrant la plasticité cristalline. La spécificité de cette approche serait une sensibilité directe aux gradients de déformations (couplage simple). Cet outil est en train d'être validé sur des cas simples (acier ferritique mono- et polycristallin) [GONZALES 2011_1][GONZALES 2011_2] avant d'être appliqué au cas plus complexe des aciers DP. Les premiers résultats permettent de reproduire un effet « Hall et Petch » tout à fait satisfaisant et les distributions statistiques de DGN dans un polycristal de fer. Ce développement permettra de confirmer l'influence respective des tailles de grains et d'ilots sur le comportement des aciers DP et donc de faire progresser aussi les modèles à champs moyens.
- La modélisation micromécanique de l'endommagement par l'introduction d'éléments cohésifs. La Figure 108 montre un exemple d'implémentation de ces éléments autour d'un ilot de martensite dans le cas du VER 2D virtuel homogène. Ils permettent de simuler un processus de décohésion en fonction de contraintes normales (comportement de type pseudo-2D). La Figure 109 montre un exemple de simulations où les interfaces de deux des trois particules recouvertes de ces éléments s'endommagent. Il apparaît alors des bandes de cisaillement intenses entre ces particules (processus de localisation par cisaillement de type Brown-Embury).

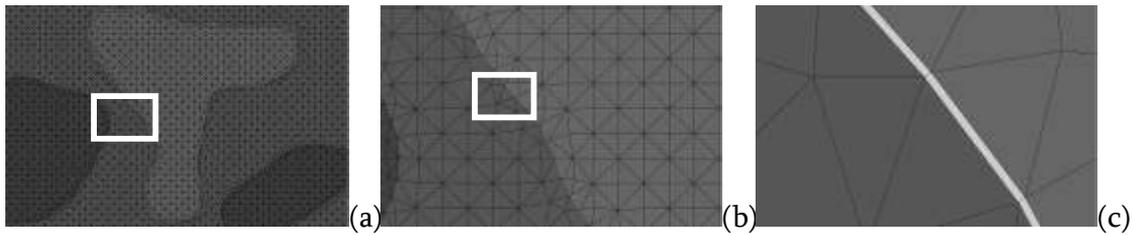

Figure 108 : Illustration de l'utilisation d'éléments cohésifs (en gris) pour reproduire le comportement de l'interface ferrite (en rouge)-martensite (en vert et bleu) dans le cas d'une configuration 2D virtuelle. (a), (b) et (c) agrandissements successifs d'une même zone.

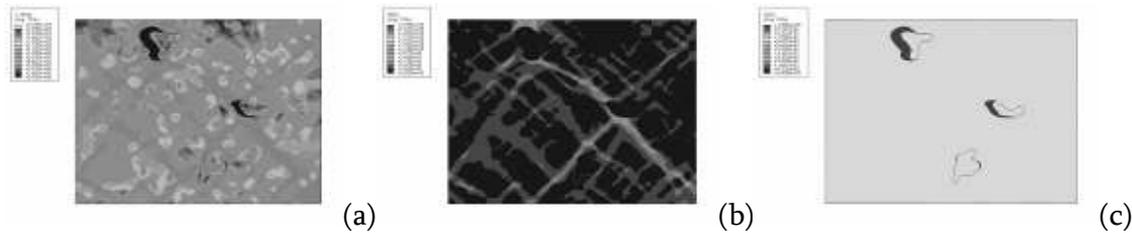

Figure 109 : Résultats des calculs EF après 10% d'allongement sur structure 2D virtuelle isotrope (a) Contraintes équivalentes de Von Mises (b) Déformation plastique équivalente et (c) Endommagement (les cavités apparaissent en rouge). 3 particules sont recouvertes d'éléments cohésifs pour simuler un processus de décohésion de l'interface ferrite-martensite.

### 3.5. Conclusions et perspectives

Mes travaux sur le comportement des aciers DP ont porté sur différents types de modélisation ayant des finalités spécifiques ;
- l'extension en plasticité polycristalline d'un modèle monophasé pour des applications en rhéologie appliquée (prévision des surfaces de charges sous sollicitations complexes).
- le développement d'un modèle biphasé quasi-analytique pour des utilisations en « alloy-design » métallurgique. Le modèle a été ajusté sur une large base de données issues de la littérature et permet de capter avec justesse les effets de fraction (dilution) et de taille sur le comportement des aciers DP.
- le développement d'une chaine de simulation en EF, de la numérisation de microstructures aux calculs sur VER sensible aux gradients de déformation et intégrant des mécanismes simulant l'endommagement (éléments cohésifs) en vue de parfaire les connaissances sur les effets de tailles, de morphologie et de topologie sur le comportement et la rupture de ces aciers composites. L'approche est encore incomplète mais permet de traiter des questions au premier ordre comme l'aspect néfaste d'une structure en bandes sur l'endommagement.

Ces travaux ouvrent de nombreuses perspectives stimulantes que nous avons pu déjà largement évoquer. Par ordre de difficulté croissante :

- Axe de recherche 1 : Effet des éléments substitutionnels sur comportement de la ferrite. Les effets de taille dans cette phase semblent largement compris et le comportement suffisamment bien reproduit à la fois par des approches analytiques monophasées ou les modèles de plasticité cristalline par EF. Toutefois, l'influence sur l'écrouissage de certains éléments d'alliages « classiques » en solution solide, comme le silicium ou même le manganèse, n'est ni comprise ni décrite.
- Axe de recherche 2 : comportement de la martensite « fraiche ». Nos travaux sur le comportement de la martensite ont ouvert un angle de recherche original sur cette phase. Sa nature composite est à préciser expérimentalement en termes d'échelle. Se pose aussi naturellement la question de la validité de cette description dans le cas des aciers DP (imbrication de deux approches de transition d'échelle). Ces résultats devraient naturellement conduire à une meilleure compréhension aussi du comportement de la martensite « revenue ».
- Axe de recherche 3 : Effet de taille et de morphologie dans les modèles biphasés. Ces questions ne pourront progresser qu'avec le développement de la chaine de simulation par EF, qui permettra de découpler numériquement les différents effets.
- Axe de recherche 4 : investiguer la compétition entre les mécanismes d'endommagement. Ce travail reposera sur la chaine de simulation par EF mais devra être validé par des résultats expérimentaux sur des structures modèles (interface ferrite-martensite modèle) et des structures « revenus ». Ce dernier traitement change les conditions de transfert de charges entre les phases, les critères d'endommagement sans modifier la morphologie et les tailles de structures.

# 4. Conclusion personnelle

Ce manuscrit détaille mes résultats de recherche sur le comportement mécanique des aciers pour la construction automobile. Ils couvrent une large gamme de mécanismes et de microstructures, des aciers austénitiques TWIP aux différents types de microstructures ferritiques. Ces systèmes présentent toutefois de surprenantes ressemblances, et je pense avoir montré, du point de vue de la démarche, la transversalité de certains concepts comme l'écrouissage par effet composite, applicable aux aciers TWIP aussi bien qu'aux aciers DP ou les effets de tailles de microstructures sur le comportement (taille de grains ou espacement entre micromacles). Ces analogies me sont particulièrement utiles pour traiter du comportement des structures bainitiques sans carbures auxquelles je m'intéresse beaucoup actuellement. Elles présentent en effet simultanément une structuration en latte à l'échelle nanométrique et un fort effet TRIP (transformation de l'austénite en martensite $\alpha'$).

Je pense avoir aussi illustré les possibles applications industrielles et technologiques actuelles de ces travaux. Toutefois, ce document est loin d'être un aboutissement et propose de nouvelles pistes de recherches dans ces domaines déduites de l'analyse de la littérature et de mes contributions personnelles. Elles permettront à terme de mieux comprendre et prévoir les comportements complexes de ces aciers, donc favoriser leurs implémentations.

J'espère enfin avoir démontré par ce document, illustrant ma démarche scientifique, mes collaborations et encadrements, mes principales réussites et questions ouvertes, ma capacité à diriger des recherches.

# 5. Références bibliographiques

[ADLER 1986]: Adler P.H., Olson G.B., Cohen M., Metallurgical Transactions A 17 (1986) 1725-1737
[AL-ABBASI 2007]: Al-Abbasi F.M, Nemes J.A., Computational Materials Science 39 2 (2007) 402
[ALLAIN 2002]: Allain S., Chateau J.P., Bouaziz O., Steel Research 73 (2002) 299-302
[ALLAIN 2004_1]: Allain S., Thèse de l'institut National Polytechnique de Lorraine, Caractérisation et modélisation thermomécaniques multi-échelles des mécanismes de déformation et d'écrouissage d'aciers austénitiques à haute teneur en manganèse – Application à l'effet TWIP, (2004)
[ALLAIN 2004_2]: Allain S., Chateau J.P., Bouaziz O., Materials Science And Engineering A 387 (2004) 143-147
[ALLAIN 2004_3]: Allain S., Chateau J.P., Bouaziz O., Migot S., Guelton N., Materials Science and Engineering A 387 (2004) 158-162
[ALLAIN 2004_4]: Allain S., Chateau J.P., Dahmoun D., Bouaziz O., Materials Science And Engineering A 387 (2004) 272-276
[ALLAIN 2005]: Allain S., Bouaziz O., Lemoine X, USINOR report IRD/AUP/2005/2476, ABC model: an extended Microstructural Advanced Behavior Laws (MABL) for Ferrite/Martensite steels (Field 5-Gear 1-phase 2)
[ALLAIN 2008_1]: Allain S, Bouaziz O., Materials Science and Engineering A 496 (2008) 329-336
[ALLAIN 2008_2]: Allain S., Cugy P., Scott C., Chateau J.P., Rusinek A., Deschamps A., International Journal of Materials Research 99 7 (2008) 734-738
[ALLAIN 2008_3]: Allain, S; Iung, T, Revue de Metallurgie-CIT 105 (2008) 10
[ALLAIN 2009]: Allain S., Bouaziz O., Lemoine X., Revue de Métallurgie-CIT 106 2 (2009) 80
[ALLAIN 2010_2] : Allain S., Chateau J.P., Bouaziz O., Proc. SHSS 2010, Verone, Italy (2010)
[ALLAIN 2010_2]: Allain S., Bouaziz O., Chateau J.P., Scripta Materialia 62 7 (2010) 500-503
[ALLAIN 2010_3]: Allain S., Bouaziz O., International Journal of Material research 101 (2010) 12
[ALLAIN 2011_2] Allain S., Bouaziz O., Lebedkina T., Lebyodkin M., Scripta Materialia 64 8 (2011) 741-744
[ALLAIN 2012]: Allain S., Bouaziz O., Takahashi M., ISIJ International 52 4 (2012) 717-722
[ALLAIN 2013]: Allain S., Danoix F., Goune M., Hoummada K., Mangelinck D., Philosophical Magazine Letters (2013), accepted
[ANDREWS 2000]: Andrews S.D., Sehitoglu H., Karaman I, Journal of Applied Physics 87 (2000) 2194
[ANSELL 1963]: Ansell G.S., Arrot A., technical report; Office of Naval Research, contract NONR 591 (15) (1963)

[ASARO 1975]: Asaro R.J., Acta Metallurgica 23(1975) 1255
[ATKINSON 1974]: Atkinson J.D., Brown M., Stobbs W.M., Philosophical Magazine (1974) 1247
[AVRAMOVIC 2009]: Avramovic-Cingara G., Ososkov Y., Jain M.K., Wilkinson D.S., Material Science and Engineering A 516 (2009) 7-16
[BADINIER 2011]: Badinier G., Sinclair C.W., Bilyk V., Sauvage X., Allain S., Wang X., Euromat 2011 Conference Montpellier (2011)
[BARBIER 2009_1]: Barbier D., Thèse de l'université Paul Verlaine de Metz, Etude du comportement mécanique et des évolutions microstructurales de l'acier austénitique Fe22Mn0.6C à effet TWIP sous sollicitations complexes. Approche expérimentale et modélisation, (2009)
[BARBIER 2009_2]: Barbier D., Gey N., Bozzolo N., Allain S., Humbert M., Journal Of Microscopy-Oxford 235 1 (2009) 67-78
[BARBIER 2009_3]: Barbier D., Gey N., Allain S., Bozzolo N., Humbert M., Materials Science and Engineering A 500 (2009) 196-206
[BARBIER 2012]:
[BAYRAKTAR 2004]: Bayraktar E., Khalid F.A., Levaillant C., Journal of Materials Processing Technology 147 (2004) 145-154
[BEAL 2011]: Béal C., Thèse de l'institut national des Sciences appliquées de Lyon, Mechanical behavior of a new automotive high manganese TWIP steel in the presence of liqui zinc (2011)
[BERBENNI 2004]: Berbenni S., Favier V., Lemoine X., Berveiller M., Material Science and Engineering A 372 (2004) 128-136
[BOUAZIZ 2001]: Bouaziz O., Guelton N., Materials Science and Engineering A 319-321 (2001) 246-249
[BOUAZIZ 2001]: Bouaziz O., Iung T., Kandel M., Lecomte C., Journal de Physique IV 11 (2001) 223-231
[BOUAZIZ 2002]: Bouaziz O., Buessler P., La Revue de Métallurgie CIT (2002) 71-77
[BOUAZIZ 2005]: Bouaziz O., HDR Université Paul Verlaine de Metz, Relations microstructure-Comportement des aciers et couplage avec les évolutions métallurgiques (2005)
[BOUAZIZ 2008_1]: Bouaziz O., Allain S., Scott C.P., Scripta Materialia 58 6 (2008) 484-487
[BOUAZIZ 2008_2]: Bouaziz O., Aouafi A., Allain S., Materials Science Forum 584-586 (2008) 605-609
[BOUAZIZ 2010]: Bouaziz O., Allain S., Estrin Y., Scripta Materialia 62 9 (2010) 713-715
[BOUAZIZ 2011_1]: Bouaziz O., Allain S., Scott C.P., Cugy P., Barbier D., Current Opinion In Solid State & Materials Science 15 4 (2011) 141-168
[BOUAZIZ 2011_2]: Bouaziz O., Zurob H., Chehab B., Embury J.D., Allain S., Huang M., Materials Science and Technology 27 3 (2011) 707-709
[BOUAZIZ 2011_3]: Bouaziz O., techniques de l'ingenieur (2011)
[BOUAZIZ 2013]: Bouaziz O., Renard K., Allain S., Jacques P.J., submitted (2013)

[BRACKE 2007_1]: Bracke L., Kestens L., Penning J., Scripta Materialia 57 5 (2007) 385-388
[BRACKE 2007_2]: Bracke L., Thèse de l'université de Ghent, Deformation behaviour of austenitic FeMn alloys by twinning and martensitic transformation (2007)
[BROWN 1971]: Brown L.M., Stobbs W.M., Philosophical Magazine (1971) 1201
[BYUN 2003]: Byun T.S., Acta Materialia 51 (2003) 3063-3071
[CAHN 53]: Cahn W., Acta Materialia 1 (1953) 49
[CANADINC 2008]: Canadinc D., Efstathiou C., Sehitoglu H., Scripta Materialia 59 (2008) 1103-1106
[CHATEAU 2009]: Chateau J.P., Lebedkina T.A., Lebyodkin M.A., Jacques A., Allain S., Proc. ICSMA 15, Dresden, Germany (2009)
[CHEN 2007]: Chen L., Birosca S., Han Kim S., Kim S. K., De Cooman B.C., Proc. Int. Conf. MST 2007
[CHOI 1999]: Choi H.C., Ha T.K., Shin H.C., Chang Y.W., Scripta Metallurgica 40 (1999) 1171
[CHOI 2009]: Choi K.S., Liu W.N., Sun X., Khaleel M.A., Metallurgical and Materials transactions A 40A (2009) 796
[CHRISTIAN 1969]: Christian J.W., in Physics of Strength and Plasticity, MIT Press, London, England, (1969) 85-95
[CHRISTIAN 1995]: Christian J.W., Mahajan S., Progress in Materials Science 39 (1995) 1-157
[CHUMLYAKOV 2002]: Chumlyakov Y., Kireeva I., Zakharova E., Luzginova N., Sehitoglu H., Karaman I., Russian Physics Journal 45 3 (2002) 274-284
[CHUNG 2011]: Chung K, Ahn K, Yoo D.H., Chung K.H., Seo M.H, Park S.H., International Journal of Plasticity 27 1 (2011) 52-81
[COBO 2008]: Cobo S., Bouaziz O., Proc. Int. Conf. on SHSS, Buenos Aires, Argentina (2008)
[COHEN 1963_1]: Cohen J.B., Weertman J., Acta Metallurgica 11 (1963) 996-998
[COHEN 1963_2]: Cohen J.B., Weertman J., Acta Metallurgica 11 (1963) 1368-1369
[COLLET 2009]: Collet J.L. Thèse de l'Institut National Polytechnique de Grenoble, Les mécanismes de déformation d'un acier TWIP FeMnC : une étude par diffraction des rayons X (2009)
[CORNETTE 2005]: Cornette D., Cugy P., Hildenbrand A., Bouzekri M., Lovato G., Revue de Métallurgie 12 (2005) 905-918
[COTES 1998]: Cotes S., Fernàndez Guillermet A., Sade M., Journal of Alloys and Compounds 278 (1998) 231-238
[COUJOU 1983]: Coujou A., Acta Metallurgica 31 (1983) 1505-1515
[COUJOU 1992]: Coujou A., Beneteau A., Clement N., Acta Metallurgica et Materialia, 40 (1992) 337-344
[COUPEAU 1999]: Coupeau C., Tranchant F., Vergnol J., Grilhé J., European Journal of Applied Physics 6 (1999) 1-6
[CURTZE 2010] : Curtze S., Kuokkala V.T., Oikari A., Talonen J., Hänninen H., Acta Materialia 59 3(2011) 1068-1076

[DAGBERT 1996]: Dagbert C., Sehili M., Gregoire P., Galland J., Hyspecka L., Acta Materialia 44 (1996) 2643
[DASTUR 1981]: Dastur Y.N., Leslie W.C., Metallurgical Transactions A 12 (1981) 749-759
[DAVIES 1978]: Davies R.G., Metallurgical transactions A 9 (1978) 674
[DECOOMAN 2008]: DeCooman B.C., Proc. Int. Conf. SHSS 2008 Buenos Aires, Argentina (2008)
[DECOOMAN 2009]: DeCooman B.C., Chen L., Kim H.S., Estrin Y., Kim S.K., Voswinckel H., in Microstructure and Texture in Steels, Springer (2009) 165
[DELINCE 2007] Delincé M., Bréchet Y., Embury J.D., Geers M.G.D., Jacques P.J., Pardoen T., Acta Materialia 55 (2007) 2337
[DILLIEN 2010_1]: Dillien S., Seefeldt M., Allain S., Bouaziz O., Van Houtte P., Material Science and Engineering 527 (2010) 947-953
[DILLIEN 2010_2]: Dillien S., Thèse de l'université catholique de Leuven, Bridging the physics-Engineering cap in Dual Phase Formability
[DUMAY 2008_1]: Dumay A., Thèse de l'institut National Polytechnique de Lorraine, Amélioration des propriétés physiques et mécaniques d'aciers TWIP FeMnxC : influence de la solution solide, durcissement par précipitation et effet composite, (2008)
[DUMAY 2008_2]: Dumay A., Chateau J.P., Allain S., Migot S., Bouaziz O., Materials Science And Engineering A 483-84 (2008) 184-187
[ESTRIN 1984]: Estrin Y., Mecking H., Acta Metallurgica 32 (1984) 57
[FARENC 1992]: Farenc S., Thèse de l'Université des Sciences Paul Sabatier de Toulouse, Etude des mécanismes de déformation dans le titane et l'alliage intermétallique TiAl (1992)
[FARENCE 1993]: Farenc S., Coujou A., Couret A., Philosophical Magazine A 67 1 (1993) 127-142
[FAVIER 2012]: Favier V., Barbier D., Scripta Materialia 66 12 (2012) 972-977
[FEAUGAS 1999]: Feaugas X., Acta Materialia 47 (1999) 3617-3632
[FERREIRA 1998]: Ferreira P.J., Müllner P., Acta Materialia 46 (1998) 4479-4484
[FISCHER 2003]: Fisher F.D., Schaden T., Appel F., Clemens H., European Journal in Mechanics 22 (2003) 709-726
[FRIEDEL 1964]: Friedel J., Dislocations, Pergamon Press Ltd., Oxford, England (1964)
[FULLMAN 1953]: R.L. Fullman, Trans. A.I.M.E. 197 (1953) 447
[GARCIA 2007]: Garcia-Junceda A., Caballero F.G., Capdevila, Garcia de Andres C., Scripta Materialia 57 (2007) 89-92
[GARDEY 2005]: Gardey B., Bouvier S., Bacroix B., Metallurgical and Materials Transactions A 36 (2005) 2937
[GEROLD 1989]: Gerold V., Karnthaler H.P., Acta Metallurgica 37 8 (1999) 2177-2183
[GHADBEIGI 2010]: Ghadbeigi H., Pinna C., Celotto S., Yates J.R., Material Science and Engineering A 527 (2010) 5026-5032
[GHOSH 2002]: Ghosh G., Olson G.B., Acta Materialia 50 (2002) 2655-2675
[GONZALES 2011_1]: González D., Martínez-Esnaola J.M., Allain S., Gil Sevillano J., Euromat 2011 Conference Montpellier (2011)

[GONZALES 2011_2]: González D., Martínez-Esnaola J.M., Allain S., Gil Sevillano J., Euromat 2011 Conference Montpellier (2011)

[GRASSEL 2000]: Grässel O., Krüger L., Frommeyer G., Meyer L.W., International Journal of Plasticity 16 (2000) 1391-1409

[GUTIERREZ 2010]: Gutierrez-Urrutia I., Zaefferer S., Raabe D., Materials Science and Engineering A 527 (2010) 3552–3560

[GUTIERREZ 2012]: Gutierrez-Urrutia I, Raabe D., Acta Materialia 60 (2012) 5791-5802

[HAMDI 2008]: Hamdi F., Asgari S., Metallurgical and Materials Transactions A 39 (2008) 294-303

[HASEGAWA 2010]: Hasegawa K., Kawamura K., Urabe T., Hosoya Y., ISIJ International 44 (2004) 605

[HELL 2011_1]: Hell J.C., Thèse de l'université Paul Verlaine de Metz, Aciers bainitiques sans carbure : Caractérisation microstructurale mitl-échelles et *in situ* de la transformation austénite-bainitique et relations entre microstructure et comportement mécanique (2011)

[HELL 2011_2]: Hell J.C., Dehmas M., Allain S., Prado J.M., Hazotte A., Chateau J.P., ISIJ International 51 10 (2011) 1724-1732

[HILLERT 1978]: Hillert M., Jarl M., Calphad 2 (1978) 227-238

[HIRTH 1970]: Hirth J.P., Metallurgical Transactions A 1 (1970) 2367-2374

[HOLSTE 1980]: Holste C., Burmeister H.-J., Physica Status solidi A 57 (1980) 269

[HONG 1986]: Hong S.I., Material Science and Engineering A 792 (1986) 1-7

[HUANG 1989]: Huang W., Calphad 3 (1989) 243-252

[HUANG 2011]: Huang M., Bouaziz O., Barbier D., S. Allain, Journal of Materials Science 46 23 (2011) 7410-7414

[HUPPER 1999]: Hüpper T., Endo S., Ishikawa N., Osawa K., ISIJ International 39 3 (1989) 288-294

[HUTCHINSON 2011]: Hutchinson B., Hagström J., Karlsson O., Lindell D., Tornberg M., Lindberg F., Thuvander M., Acta Materialia 59 14 (2011) 5845-5858

[IDRISSI 2009]: Idrissi H., Ryelandt L., Veron M., Schryvers D., Jacques P.J., Scripta Materialia 60 11 (2009) 941-944

[IDRISSI 2010_1]: Idrissi H., Renard K., Ryelandt L., Schryvers D., Jacques P.J., Acta Materialia 58 7 (2010) 2464-2476

[IDRISSI 2010_2]: Idrissi H., Renard K., Schryvers D., Jacques P.J., Scripta Materialia 63 (2010) 961

[INDEN 1981]: Inden G., Physica B 103 (1981) 82-100

[JACQUES 2007]: Jacques P.J., Furnémont Q., Lani F., Pardoen T., Delannay F., Acta Materialia 55 (2007) 3681-3693

[JIANG 1992]: Jiang Z., Liu J., Lian J., Acta Metallurgica et Materialia 40 7 (1992) 1587-1597

[JIN 1995] : Jin Z., Bieler T.R., Materials Science and Engineering A 192 (1995) 729-732

[KADKHODAPOUR 2011]: Kadkhodapour J., Schmauder S., Raabe D., Ziaei-Rad S., Weber U., Calcagnotto M., Acta Materialia 59 (2011) 4387-4394

[KARAMAN 2000_1]: Karaman I., Sehitoglu H., Gall K., Chumlyakov Y.I., Maier H.J., Acta Materialia, 48 (2000) 1345-1359

[KARAMAN 2000_2]: Karaman I. Sehitoglu H., Beaudoin A.J., Chumlyakov Y.I., Maier H.J., Tomé C.N., Acta Materialia 48 (2000) 2031-2047

[KARAMAN 2001]: Karaman I., Sehitoglu H. Chumlyakov Y.I., Maier H.J., Kireeva I.V., Scripta Materialia 44 (2001) 337-343

[KEH 1963]: Keh A., Weisman S., in Electron Microscopy and Strength of Crystals, Thomas&Washburn Ed. (1963) 231

[KELLY 1963]: Kelly P.M., in Electron Microscopy and Strength of Crystals, Thomas&Washburn Ed. (1963) 917

[KIBEY 2006]: Kibey S., Liu J.B., Johnson D.D., Sehitoglu H., Acta Materialia 55 20 (2007) 6843-6851

[KIM 1986]: Kim Y.G., Han J.M., Lee J.S., Metallurgical Transactions. A 17 (1986) 2097

[KIM 2009] Kim J.K., Chen L., Kim H.S., Kim S.K., Estrin Y., De Cooman B.C., Metallurgical and Material Transactions A 40 (2009) 3147-3158

[KIM 2012]: Kim J.H., Kim D., Barlat F., Lee M.G., Material Science and Engineering A 539 (2011) 259-270

[KOCKS 1985]: Kocks U.F., Metallurgical Transactions A 16 (1985) 2109-2129

[KOCKS 2003]: Kocks U.F., Mecking H., Progress in Materials Science 48 (2003) 171-273

[KORZEKWA 1984]: Korzekwa D.A., Matlock D.K., Krauss G., Metallurgical transactions A, 15 (1984) 1221

[KOYAMA 2012]: Koyama M., SawaguchiT., Lee T, Lee C.S., Tsuzaki K., Materials Science and Engineering A 528 24 (2011) 7310-7316

[KRAUSS 1999]: Krauss G., Materials Science and Engineering A273–275 (1999) 40

[KREBS 2009]: Krebs B., Thèse Université Paul Verlaine de Metz, Caractérisation et prévision des structures en bandes dans les aciers Dual Phase. Lien avec les propriétés d'endommagement (2009)

[KUMAR 2007]: Kumar A., Singh S.B., Ray K.K., Materials Science and Engineering A 474 (2008) 270-282

[KUNTZ 2008]: Kuntz M. Thèse de l'université de Stuttgart, Verformungsmechanismen hoch manganlegierter austenitischer TWIP-stähle (2008)

[LAI 1989_1] : Lai H.J., Wan C.M., Scripta Materialia 23 (1989) 179

[LAI 1989_2] : Lai H.J., Wan C.M., Journal of Material Science 24 (1989) 2449

[LANDRON 2010]: Landron C., Bouaziz O., Maire E., Adrien J., Scripta Materialia 63 10 (2010) 973-976

[LANI 2007]: Lani F., Furnémont Q., Van Rompaey T., Delannay F., Jacques P.J., Pardoen T., Acta Materialia 55 (2007) 3695-3705

[LEBEDKINA 2009]: Lebedkina T., Lebyodkin M., Chateau J.P., Jacques A., Allain S., Materials Science And Engineering A 519 (2009) 147-154

[LEBYODKIN 2012]: Lebyodkin M.A., Lebedkina T.A., Roth A., Allain S., Acta Physica Polonica 122 3 (2012) 478

[LECROISEY 1972]: Lecroisey F., Pineau A., Metallurgical Transactions A 3 (1972) 387-396
[LEI 1981]: Lei T.C., Schen H.P., in Fundamentals of DP Steels, (1981) 369
[LEMAITRE 2004]: Lemaitre J., Chaboche J.L., Mécanique des matériaux solides, second ed., Dunod, Paris, (2004)
[LHUISSIER 2011]: Lhuissier P., Inoue J., Koseki T., Scripta Materialia 64 (2011) 970-973
[LI 1990_1]: Li Z., Haicheng G, Metallurgical Transactions A 21 (1990) 717
[LI 1990_2]: Li Z., Haicheng G, Metallurgical Transactions A 21 (1990) 725
[LIAN 1991]: Lian J., Jiang Z., Liu J., Materials Science and Engineering A 147 (1991) 55-65
[LIEDL 2002]: Liedl U., Traint S., Werner E.A., Computational Materials Science 25 (2002) 122-128
[LIN 1997]: Lin L., Hsu T.Y., Calphad 21 (1997) 443-448
[LUBENETS 1985]: Lubenets S.V., Starsev V.I., Fomenko L.S., Physic State solidii 95 (1985) 11-55
[MA 1989]: Ma M.T., Sun B.Z., Tomota Y., ISIJ International 29 1 (1989) 74-77
[MAGEE 1971]: Magee C.L., Davies R.G., Acta Metallurgica (1971) 345
[MAGGI 2012]: Maggi S., Federici C., d'Aiuto F, Presentation Fiat-Chrystler (2012)
[MATLOCK 2005]: Matlock D., Speer J., DOE Report TRP 9904 project (2005)
[MECKING 1981]: Mecking H., Kocks U. F., Acta Metallurgica 29 (1981) 1865-1875
[MEYERS 2001]: Meyers M.A., Vöhringer O., Lubarda V.A., Acta materialia 49 (2001) 4025–4039
[MIODOWNIK 1998]: Miodownik A.P., Zeitschrift für Metallkunde 89 (1998) 840-846
[MIURA 1968]: Miura S., Takamura J., Narita N., Transactions of the Japan Institute of Metals 9 (1968) 555
[MORI 1980]: Mori T., Fujita H., Acta Metallurgica 28 (1980) 771-776
[MULLNER 2002]: Müllner P., Solid State Phenomena 87 (2002) 227-238
[MYAZAKI 1989]: Miyazaki S., Otsuka K., ISIJ International 29 5 (1989) 353-377
[NAKADA 1971]: Nakada Y., Keh A.S., Metallurgical Transactions A 2 (1971) 441
[NAKANO 2010]: Nakano J., Jacques P.J., Calphad 34 2 (2010) 167-175
[NAKASHIMA 2007]: Nakashima K., Fujimura Y., Matsubayashi H., Tsuchuiyama T., Takaki S., Tetsu-to-Hagané 93 6 (2007) 459
[NAMBU 2009]: Nambu S., Michiuchi M., Ishimoto Y., Asakura K., Inoue J., Koseki T., Scripta Materialia 60 (2009) 221-224
[OLSON 1975]: Olson G., Cohen M., Metallurgical Transactions A 6 (1975) 791-795
[OLSON 1992]: Olson G.B., Owen W.S., in Martensite A tribute to Morris Cohen, ASM International Publishers, USA (1992)
[OWEN 99]: Owen W.S., Grujicic M., Acta Materialia 47 (1999) 111
[PARK 2005]: Park K.S., Park K.T., Lee D.L., Lee C.S., ISIJ International 45 9 (2005) 1352-1357
[PAUL 2012]: Paul S.K., Kumar A., Computational Materials Science 63 (2012) 66-74
[PAUL 2013]: Paul S.K., Materials and Design 44 (2013) 397-406

[PEETERS 2001]: Peeters B., Seefeldt M., Teodosiu C., Kalidindi S. R., Van Houtte P., Acta Materialia 49 (2001) 1607-1619
[PERLADE 2003]: Perlade A., Bouaziz O., Furnémont Q., Material Science and Engineering A (2003) 145-152
[PIPARD 2009]: Pipard J.M., Nicaise N., Berbenni S., Bouaziz O., Berveiller M., Computational Materials Science 45 3 (2009) 604-610
[POLAK 1982]: Polàk J., Klesnil M., Fatigue of Engineering Materials and Structures 5 (1982) 19
[PRAHL 2007]: Prahl U., Papaefthymiou S., Uthaisanguk V., Bleck W., Sietsma J., Van der Zwaag S., Computational Materials Science 39 (2007) 17-22
[PUSCHL 2002]: Püschl W., Progress in Materials Science 47 (2002) 415-461
[PUSHKAREVA 2009]: Pushkareva I., Thèse de l'Institut National Polytechnique de Lorraine, Evolution microstructurale d'un acier Dual Phase Optimisation de la résistance à l'endommagement (2009)
[RAUCH 1997]: Rauch E., Materials Science and Engineering 234-236 (1997) 653-656
[RAUCH 2004]: Rauch E., Revue de Métallurgie 12 (2004) 1007-1019
[REMY 1975]: Rémy L., Thèse d'état de l'université Paris Sud, (1975)
[REMY 1978]: Rémy L., Acta Metallurgica 26 (1978) 443-451
[RENARD 2010]: Renard K., Ryelandt S., Jacques P.J., Materials Science and Engineering A 527 12 (2010) 2969-2977
[RENARD 2012]: Renard K., Thèse de l'université catholique de Louvain-La-Neuve, Characterization of the relationship between microstructure evolution and work hardening of FeMnC TWIP steels
[ROTH 2012]: Roth A., Lebedkina T.A., Lebyodkin M.A., Materials Science and Engineering A 539 (2012) 280-284
[SADAGOPAN 2003]: Sadagopan S., Urban D., Formability characterization of a new generation of high strength steels, rapport DOA DE-FC07-97ID13554 (2003)
[SAEED-AKBARI 2009]: Saeed-Akbari A., Imlau J., Prahl U., Bleck W., Metallurgical and Materials Transactions A 40 (2009) 3076-3090
[SATO 2011]: Sato S., Kwon E.P., Imafuku M., Wagatsuma K., Suzuki S., Materials Characterization 62 (2011) 781-788
[SCHMITT 1994]: Schmitt J.H., Shen E.L., Raphanel J.L., International Journal of Plasticity 10 5 (1994) 535-551
[SCHUMANN 1972]: Schumann V.H., Neue Hütte 17 (1972) 605-609
[SCOTT 2006]: Scott C., Allain S., Faral M., Guelton N, Revue de Metallurgie-CIT 103 6 (2006) 293-302
[SEEGER 1963] : Seeger A., Mader S., Kronmüller H., in Electron Microscopy and Strength of Crystals, Thomas&Washburn Ed. (1963) 665
[SEVILLANO 2009]: Gil Sevillano J., Scripta Materialia 60 5 (2009) 336-339
[SEVILLANO 2012]: Gil Sevillano J., de Las Cuevas F., Scripta Materialia 66 12 (2012) 978-981

[SHERMAN 1983]: Sherman A.M., Eldis G.T., Cohen M., Metallurgical Transactions 14A (1983) 995

[SHIBATA 2009]: Shibata A., Morito S., Furuhara T., Maki T., Acta Materialia 57 (2009) 483-492

[SHIEKHELSOUK 2006] : Shiekhelsouk N., Thèse de l'université Paul Verlaine de Metz, Modélisation polycristalline et étude expérimentale du comportement mécanique d'aciers Fe-Mn à effet TWIP. Prise en compte du traitement thermique d'élaboration sur le maclage et contraintes internes (2006)

[SHUN 1992]: Shun T.S., Wan C.M., Byrne J.G., Acta Metallurgica et Materialia 40 (1992) 3407

[SINCLAIR 2006]: Sinclair C.W., Poole W.J., Bréchet Y., Scripta Materialia 55 (2006) 739

[SODJIT 2012]: Sodjit S., Uthaisanguk V., Materials and Design 41 (2012) 370-379

[SPEICH 1981] Speich G.R., Miller R.L., in Fundamentals of DP Steels, (1981) 279

[STRINGFELLOW 1992]: Stringfellow R.G., Parks D.M., Olson G.B., Acta Metallurgica et Materialia 40 (1992) 1703-1716

[SUGIMOTO 1997]: Sugimoto K.I., Kobayashi M., Yasuki S.I., Metallurgical and Materials Transactions A 28 (1997) 2637

[SUZUKI 1958]: Suzuki H., Barrett C.S., Acta Metallurgica 6 (1958) 156

[TALYAN 1998] : Talyan C., Wagoner R.H., Lee J.K., Materials Transactions A 29 (1998) 2161-2171

[TASAN 2010]: Tasan C.C., Hoefnagels J.P.M., Geers M.G.D., Scipta Materialia 62 (20102) 835-838

[TAVARES 1999]: Tavares S.S.M., Pedroza P.D., Teodosio J.R., Gurova T., Scripta Materialia 40 8 (1999) 887-892

[TIMOKHINA 2007]: Timokhina I.B., Pereloma E.B., Hodgson P.D., Metallurgical and Materials Transactions A 38 10 (2007) 2442-2454

[TOMITA 1995]: Tomita Y., Iwamoto T., International Journal of Mechanical Sciences 37 (1995) 1295-1305

[TOMOTA 1986]: Tomota Y., Strum M.,. Morris Jr J.W, Metallurgical Transactions A 17 (1986) 537

[TOMOTA 1992]: Tomota Y, Umemoto M., Komatsubara N, Hiramatsu A., Nakajima N., Moriya A. Watanabe T., Nanba S., Anan G., Kunishige K., Higo Y., Miyahara M., ISIJ International 32 3 (1992) 343-349

[UTHAISANGSUK 2011]: Uthaisanguk V., Muenstermann S., Prahl U., Bleck W., Schmitz H.P., Pretorius T., Computational Materials Science 50 (2011) 1225-1232

[VENABLES 1964]: Venables J.A., in Deformation Twinning, Proceedings of a Metallurgical Society Conference, Gordon and Breach Science, New-York, E.-U.A., (1964)

[VENABLES 1974] : Venables J.A., Philosophical Magazine A 30 (1974) 1165-1169

[VERCAMMEN 2004]: Vercammen S., Blanpain B., De Cooman B.C., Wollants P., Acta Materialia 52 7 (2004) 2005-2012

[VIARDIN 2008]: Viardin A., Thèse de l'Institut National Polytechnique de Lorraine, Modélisation par Champs de Phase de la croissance de ferrite allotriomorphe dans les aciers Fe-C-Mn (2008)

[VIEWPOINT 2012]: Viewpoint Set 50: Twinning Induced Plasticity Steels, Scripta Materialia 66 12 (2012) 955-1084

[VOLOSEVICH 1976]: Volosevich P.Y., Grindnev V.N., Petrov Y.N., Physics of Metals and Metallography 42 (1976) 126–30

[YANEZ 2003]: Ros-Yanez T., Barros J., Colas R., Houbaert Y., ISIJ International 43 3 (2003) 447-453

[ZAVATTIERI 2010]: Zavattieri P.D., Savic V., Hector Jr. L.G., Fekete J.R., Tong W., Xuan Y., International Journal of Plasticity 25 12 (2009) 2298-2330

[ZHANG 2002]: Zhang Y.S., Lu X., Tian X., Qin Z., Materials Science and Engineering A 334 (2002) 19-27

[ZUIDEMA 1987]: Zuidema B.K., Subramanyam D.K., Leslie W.C., Metallurgical Transactions 18 (1987) 1629

# 6. Annexe

La question de la germination des macles nous a amenés à considérer l'influence d'une précipitation fine sur ce processus. Les interactions possibles entre précipitations et maclage ont été ainsi étudiés dans le cadre de la thèse d'A. Dumay, sujet très peu documenté dans la littérature scientifique. A titre purement illustratif, on notera que la revue pourtant très complète de Christian et Mahajan dédiée au maclage mécanique, ne discute de ces possibles interactions que sur un paragraphe de 10 lignes sur 120 pages [CHRISTIAN 1995].

Cette étude a été en partie consacrée à l'étude d'une fine précipitation de carbures de vanadium sur le comportement mécanique et la microstructure de maclage de la nuance de référence (Fe22Mn0.6C+0.2% atomique de VC). Une étude détaillée en MET des précipités a permis d'estimer leur taille moyenne (de l'ordre de 7 nm) mais surtout leur degré de cohérence (orientation cube-cube entre matrice et précipités $\{111\}_\gamma//\{111\}_{VC}$ et un faible degré de cohérence) (cf. Figure 111(a)).

La comparaison entre les courbes de traction d'un alliage avec précipités et d'un acier de composition équivalente sans précipités et une même taille de grain montre que la précipitation ne modifie pas l'écrouissage de l'alliage malgré des fractions précipitées importantes. Elle n'induit qu'un simple durcissement de type de Orowan (contribution additive à la contrainte d'écoulement) (cf. Figure 110). La précipitation des carbures de vanadium n'a donc pas d'impact sur la microstructure de maclage (cinétique et structuration spatiale). Cette vision est confortée par les observations en MET (cf. Figure 111) révélant l'absence d'interactions fortes entre micromacles et précipités (simple contournement, pas de blocage fort et systématique). Cette étude n'a pas non plus permis d'observer que les précipités pouvaient s'avérer être des sites de germination privilégiés pour le maclage.

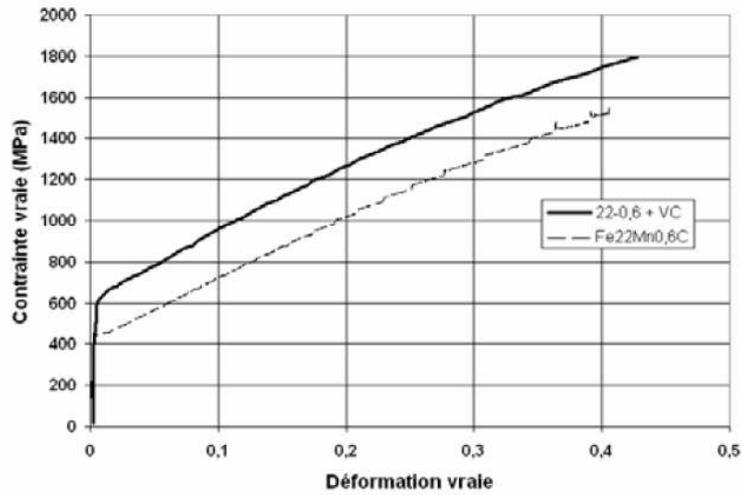

Figure 110 : Courbes de traction rationnelles de la nuance de référence et de sa version microallié au vanadium [DUMAY 2009].

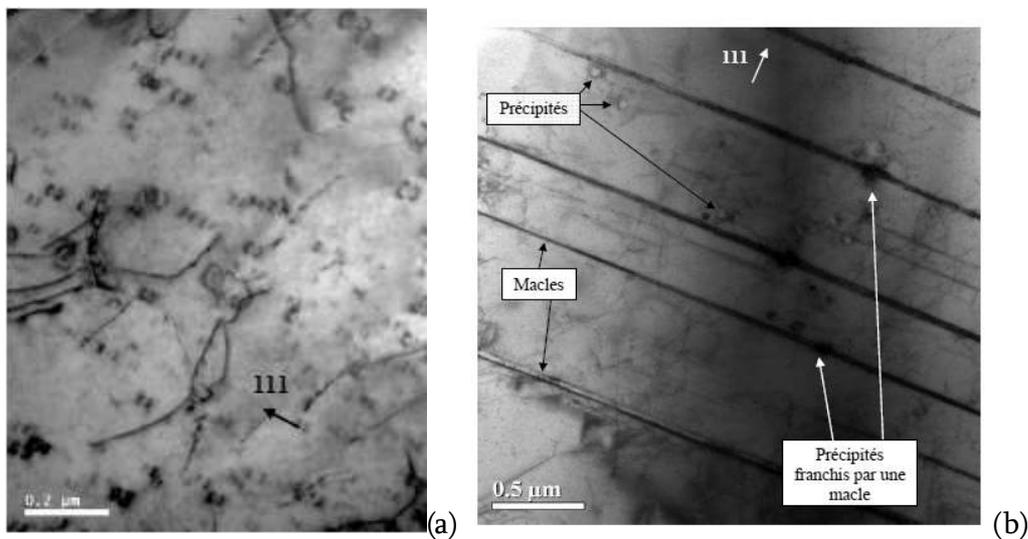

Figure 111 : Micrographies en MET – champ clair (a) observation des précipités de carbure de vanadium avant déformation – contraste en grain de café dû à la cohérence résiduelle avec la matrice (b) interactions entre nanomacles et précipités après déformation [DUMAY 2009].

# Curriculum vitae

# Rapport de soutenance

Mon activité de recherche scientifique concerne principalement la compréhension et la modélisation du comportement mécaniques des aciers ; des mécanismes fondamentaux à la déformation macroscopique. Ce mémoire est consacré en particulier à l'effet TWIP (TWinning Induced Plasticity) des aciers austénitiques FeMnC à haute teneur en manganèse et l'effet Dual-Phase des aciers Ferrite-Martensite.

L'effet TWIP est un mécanisme d'écrouissage spécifique des aciers austénitiques FeMnC lié à un processus de maclage mécanique, mécanisme de déformation compétitif au glissement des dislocations. L'accumulation de ces macles, défauts plans d'épaisseur nanométrique, crée au cours de la déformation une microstructure enchevêtrée et difficilement franchissable par les dislocations mobiles à l'intérieur des grains austénitiques. Au cours de nos travaux, ces microstructures ont été expliquées et quantifiées à différentes échelles. Nous avons ainsi pu modéliser la double contribution du maclage à l'écrouissage grâce à une augmentation de densité de dislocations statistiquement stockées et à une contribution de nature cinématique, associée à l'incompatibilité de déformation entre macles et matrice. L'influence de ce mécanisme a en conséquence été mieux comprise lors de trajets de mise en forme complexes. Afin de pouvoir optimiser le comportement mécanique de ces aciers TWIP, notre second axe de recherche a porté sur l'effet de leurs compositions chimiques sur ces mécanismes d'écrouissage, en particulier au travers de la relation entre maclage mécanique et énergie de défaut d'empilement (EDE). Ces travaux ont débouchés sur l'identification d'un « paradoxe carbone » que nous sommes en passe de résoudre.

Mes travaux ultérieurs de modélisation du comportement des aciers Dual-Phase (DP) Ferrite-Martensite se sont aussi attachés à décrire systématiquement les effets de fraction et de tailles des microstructures. Ils ont eu différentes finalités :

- l'extension en plasticité polycristalline d'un modèle monophasé analytique pour des applications en rhéologie appliquée (prévision des surfaces de charges sous sollicitations complexes).
- le développement d'un modèle biphasé générique pour des utilisations en « alloy-design » métallurgique. Le modèle intègre en outre nos travaux les plus récents sur les aciers martensitiques (Approche Composite Continu) et a été ajusté sur une large base de données issues de la littérature.
- l'approfondissement de nos connaissances sur les effets de morphologie et de topologie de la microstructure DP sur le comportement et la rupture de ces aciers composites. Il passe par le développement d'une chaîne de simulation à champs locaux par Eléments Finis (EF), allant de la numérisation aux calculs sur Volume Elémentaire Représentatif (VER) de la microstructure, sensibles aux gradients de déformation, et intégrant les mécanismes d'endommagements pertinents. L'approche est encore incomplète mais permet de traiter des questions au premier ordre comme l'aspect néfaste d'une structure en bandes sur l'endommagement.

My scientific activities aim at understand and predict the mechanical behaviour of steels, from fundamental mechanisms to macroscopic deformation. This manuscript is dedicated to the TWIP effect (TWinning Induced Plasticity) of high manganese austenitic steels and DP effect (Dual-Phase) in more conventional Ferrite-Martensite steels.

TWIP effect is a specific work-hardening mechanism of high manganese austenitic steels. It is related to a specific deformation process which enters in competition to dislocation gliding, the mechanical twinning. The accumulation of twins, which are nanometre thick planar defects, generates along with the deformation an intricate microstructure which can hardly be crossed and overcame by mobile dislocations inside austenitic grains. In our works and publications, these evolving microstructures have been quantified at different scales and explained. We thus have modelled the TWIP effect as a double contribution to work-hardening: an isotropic one related to an increase in statistically stored dislocations debris due to the decrease in the mean free paths of mobile dislocations (Mecking-Kocks-Estrin process) and a kinematical contribution associated to a strain incompatibility between the twins and the matrix. The influence of this mechanism has been better understood along with complex forming operations (change in loading directions). In order to optimize the behaviour of these typical steels, our second research axis is about the effect of alloying elements on work-hardening mechanisms, in particular throughout the relationship between mechanical twinning and Stacking Fault Energy (SFE). These works has permitted to highlight a "carbon paradox" that we are quite confident to solve in a next future.

My subsequent works about the modelling of the DP steels behaviour has permitted to capture systematically fraction and size effects in these microstructures. Three different approaches have been developed with varying objectives:

- Extension in a polycristalline plasticity framework of one dimension analytical models for applications in rheology (load surface predictions under complex loadings).
- Development of a generic multiphase mean field model for metallurgical alloy-design applications. The model integers our more recent development about martensitic steels (CCA approach) and has been adjusted on a large database from literature. The model is thus validated from IF steels to fully martensitic steels.
- Improvement of our knowledge about the morphology and topology effect on DP microstructures on behaviour and damaging processes. It has required developing a Finite Element simulation chain, from 3D reconstruction and meshing of Representative Volume Element of the microstructure, sensitive to strain gradient, and integrating local damaging mechanisms (cohesive elements). This local field approach is still incomplete but permits to handle first order questions as the detrimental effect of band structure on properties.